\documentclass{article}
\usepackage{amssymb}

\input{tcilatex}

\begin{document}

\subsection{\textbf{Manifestation and Origin of the Isotope Effect.}}

\ \ \ \ \ \ \ \ \ \ \ \ \ \ \ \ \ \ \ \ \ \ \ \ \ \ \ \ \ \ \ \ \ \ \ \ \ \
\ \ \ \textbf{\ V.G. Plekhanov}

\bigskip

\textbf{Computer Science College, Erika Street 7a, Tallinn 10416, Estonia.}

\bigskip

\textbf{Abstract.} This article reviews from one point of view the current
status of the manifestations of isotope effect in a nuclear, atomic and
molecular as well as solid state physics. Although these manifestations
vary, they have one common feature - they all depend on mass. Such view
allows to see the success and failure as well as the borders of the isotope
effect. Force constant dependence on the isotope mass as well as "mass -
independent" isotope effect explicitly gives evidence to these borders. The
dependence of the energy of zero - point vibrations on the mass of isotopes
puts forward the unique possibility of the experimental study of its, for
instance, in isotope - mixed crystals. The energy of zero - point vibrations
in isotope - mixed crystals obtained from experiments turned out, as a rule,
to be compared (excluding $^{12}$C$_{x}^{13}$C$_{1-x}$, LiH$_{x}$D$_{1-x}$
mixed crystals) to the energy of the longitudinal optical phonons. This is
the first experimental determination of the energy zero - point vibrations,
that is, on the other hand, a witness to the correct quantum electrodynamics
in solids. The analysis of the neutron $\beta $ - decay accompanying the
participation of different kind of quarks allows to trace the "disappear"
mass in this process. At present time we don't explain not only the observed
mass pattern (M$_{n}$, M$_{p}$, m$_{u}$, m$_{d}$, etc.) but also the
Standard Model can't explain the origin of the mass of elementary particles
and their hierarchy. The last one don't permit to find the origin of the
isotope effect. To explain this the New Physics as in Cosmology and Nuclear
Physics beyond the Standard Model is required.

\textbf{1. Introduction.}

\textbf{\bigskip }

\textbf{2. Sub - nucleonic structure and the modern picture of isotopes.}

\bigskip

2.1. The nucleons and its constituents.

2.1.1. Mass and nuclear binding energy.

2.1.2. The nuclear radius.

2.2.The force between nucleons.

2.3. Nucleon structure.

2.3.1. Quarks and leptons.

2.3.2. Strong and weak interactions.

2.4. Isotope effect in nuclear physics.

2.5. The origin of the mass.

2.6. New physics beyond the Standard Model.

\bigskip

\textbf{3. Isotopes in atomic and molecular physics.}

\bigskip

3.1. Some general remarks.

3.2. Motion of the nucleus - atomic isotope shift.

3.3. Separation of mass - and field shift contributions.

3.3.1. Mass isotope shift.

3.3.2. Field isotope shift.

3.4. Vibrations in a diatomic molecule.

3.4.1. Raman and IR spectra of molecules.

3.4.2. Isotope shift in molecular frequencies.

3.4.3. "Mass - independent" isotope effect.

\bigskip

\textbf{4. Isotope Effect in Solids.}

\bigskip

4.1. Elementary excitations.

4.2. Phonons.

4.3. Electronic excitations.

4.4. Effects related to isotopic disorder.

4.4.1. Thermal conductivity.

4.4.2. Disorder - induced Raman scattering.

4.4.3. Effects of isotope randomness on electronic properties and exciton
transition.

4.5. Zero - point vibration energy.

\bigskip

\textbf{5. Some modern application of the stable and radioactive isotopes.}

\bigskip

5.1. Stable isotopes.

5.1.1. Traditional applications.

5.1.1.1. Diffusion.

5.1.1.2. Neutron transmutative doping of semiconductors.

5.1.1.3. Optical fiber.

5.1.1.4. Laser materials.

5.1.2. New applications - quantum information.

5.2. Radioactive isotopes.

5.2.1. Human health.

5.2.2. Geochronology.

5.2.3. Solid state physics.

\bigskip

\textbf{6. Conclusion.}

\bigskip

\textbf{Acknowledgments.}

\textbf{\bigskip }

\textbf{7. References.}

\bigskip

\textbf{PACS:} 03.67.-a; 24.85.+p; 29.25.Rm; 31.30.Gs; 32.10.Bi; 33.20.-t;
42.65.-k; 82.20Tr; 91.80.Hj.

\bigskip

\textbf{Keywords:} physics of isotopes; zero-point energy; mass of quarks;
standard model.

\textbf{\bigskip }

\textbf{\bigskip }

\textbf{Introduction.}

\bigskip

Investigation, manufacture and application of isotopes are highly variety
and is determined by the \ different areas of science and technique. The
range of the application of isotopes is exclusively wide: starting with the
investigation of universal principle of the structure matter \ and common
normality evolution of Universe [1; 2] and finished by different biochemical
process in live organisms as well as special technical applications [3]. The
presence of isotopes and isotope effect in nature serves the bright
illustration of the mutual connection between simplicity and complexity in
science [4].

The paramount meaning has the role of isotopes in the fundamental natural
science investigations. This includes not only the study nature nuclear
interactions and, in this way, the origin of isotope effect, but also the
reconstructions of nucleogenesis process of the Universe, which could
explain the observable in nature relative spreading of chemical elements [1;
2]. As is well-known, one of the intrigue peculiarity in the dependence on
their atomic number is a sharp gap at the transition from light elements to
hard one - the region of Li, Be, B and further to carbon. The observable gap
connects with the next reason: the synthesis of the light elements is
accompanied by the couple collisions between nucleon and nucleus with
subsequent decay inside stars [5]: d + n $\longrightarrow $ T $%
\longrightarrow $ $^{3}$He + n $\longrightarrow $ $^{4}$He. Couple mechanism
cuts off \ the symmetric nucleus $^{4}$He because such nucleus $^{5}$He
doesn't exist it is therefore impossible transition to hard nuclides. This
paradox overcomes with help of three particle scheme - synthesis C nucleus
and 3$\alpha $ - particles [2]: 3$\alpha $ $\longrightarrow $ $^{12}$C. This
reaction opens the synthesis of the hard elements.

Before describing the contents of the next chapters of the present review
let's dwell on the sources of nuclear physics and its structure which would
allow to make the isotope effect model more transparent.

Investigations of the \ atomic nucleus, and the fundamental forces that
determine nuclear structure as is well-known offer fascinating insights into
the nature of the physical word [5-15]. We all well-known that the history
of the nuclear physics dates from the latter years of the nineteenth century
when Henry Becqeurel in 1896 \ discovered the radioactivity. He was working
with compounds containing the element uranium. Becqeurel \ found that
photographic plates covered to keep out light became fogged, or partially
exposed, when these uranium compounds were anywhere near the plates. Two
years after Becquerel's discovery, Pierre and Marie Curie in France and
Rutherford in England succeeded in separating a naturally occurring
radioactive element, radium (Z = 88) from the ore. It was soon revealed that
there are three, distinctly different types of radiation emitted by
radioactive substances. They were called alpha ($\alpha $), beta ($\beta $)
and gamma ($\gamma $) rays - terms which have been retained in ours days.
When a radioactive source was placed in a magnetic field, it was found that
there were three different of activity, as the trajectories of some of the
rays emitted were deflected to one direction, some to the opposite
direction, and some not affected at all. Subsequently it was found that $%
\alpha $ - rays consist of positively charged $^{4}$He nuclei, $\beta $ -
rays are made of electrons (positrons) and $\gamma $ - rays are nothing but
electromagnetic radiation that carries no net charge. The existence of the
nucleus as the small central part of an atom was first proposed by
Rutherford in 1911. Rutherford proposal that the atom does consist of a
small, heavy positively charged centre surrounded by orbiting electrons
which occupy the vast bulk of the atoms volume. The simplest atom - hydrogen
- consisted of a proton and a single orbital electron. Later, in 1920, the
radii of a few heavy nuclei were measured by Chadwick and were found to be
of the order of 10$^{-14}$ m., much smaller than the order of 10$^{-10}$ m
for atomic radii (for details, see e.g. [7-12]).

The building blocks of nuclei are neutrons and protons, two aspects, or
quantum states, of the same particle, the nucleon. Since a neutron does not
carry any net electric charge and is unstable as an isolated particle (see,
below), it was not discovered until 1932 by Chadwick, whose existence has
been anticipated by Rutherford as early as 1920. Since only positive charges
(protons) are present in nucleus, the electromagnetic force inside a nucleus
is repulsive and the nucleons cannot be held together unless there is
another source of force that is attractive and stronger than Coulomb. Here
we have our first encounter with strong interaction (see, also Table 1). In
the 1934 Hideki Yukawa proposed the first significant theory of the strong
force to explain how the nucleus holds together. As we know, with Fermi and
Yukawa's papers the modern model of the atom was complete [5-9].

Studies of the structure of the nucleus have shown that it is composed of
protons and neutrons, and more recently studies [15-17] of very high energy
collisions have shown that these protons and neutrons are themselves
composed of elusive particles called quarks. Particle physics deals with the
world of the quarks and all other \ particles still thought to be
fundamental. One may argue that, since nuclear force is only one aspect of
the strong interaction between quarks, all we need therefore to do is to
understand quantum chromodynamics (QCD)$^{1\ast )}$ (for details see [15-18]
and below). The structure of neutrons and protons is discerned only at very
high energies (see, e.g. [18]) and, for all practical purpose concerning
nuclear structure, research and nuclear physics applications in the modern
world, the neutron - proton model of the nucleus is entirely adequate.

--------------------------------------

1*)QCD is the modern theory of the strong interaction. QCD, the theory of
quarks, gluons and their interactions, is a self-contained part of the
Standard Model (see below) of elementary particles. Historically its route
are in nuclear physics and the description of ordinary \ matter -
understanding what protons and neutrons (and their structure) and how they
interact. Nowadays QCD is used to describe most of what goes at high -
energy accelerators.

------------------------------------------

Thus, our present knowledge of physical phenomena suggests that there are
four types of forces \ between physical objects:

1) gravitational;

2) electromagnetic;

3) strong and

4) weak.

Both gravitational and electromagnetic forces are infinite in range and
their interaction strength diminish with the square of the distance of
separation. Clearly, nuclear force cannot follow the same radial dependence.
Being much stronger, it would have pulled the nucleons in different nuclei
together into a single unit and destroy all the atomic structure we are
familiar with. In fact, nuclear force has a very short distance. As we know
at present time, only three particles, the proton, the electron and the
photon, are stable. Another particle, the neutron, is stable when it is
bound within a nucleus, and is unstable with life-time of 887$\pm $2 s when
it is free (for details see, also [15-17]. Since nuclei are involved in a
wide variety of applied and pure research, nuclear physics overlaps with a
number of other fields of physics: particles; astrophysics; stellar
evolution etc. Therefore, the primary aim of nuclear physics is to
understand the force between nucleons, the structure of nuclei, and how
nuclei interact with each other and with other subatomic particles. These
three questions are, to a large extent, related with each other. Much of
current research in nuclear physics (see, e.g. [4-14]) relates to the study
of nuclei under extreme conditions such as high spin and excitation energy.
Nuclei may also have extreme shapes (for instance similar to that American
footballs) or extreme neutron-to-proton ratios. Modern experimenters can
create such nuclei using artificially induced fusion or nucleon transfer
reactions, employing ion beams from different sources. Beams with even
higher energies (e.g. from accelerator) can be used to create nuclei at very
high temperatures, and there are \ signs that these experiments have
produced phase transition from normal nuclear matter to a new state, the
quarks condensate, the quark-gluon plasma, in which the quarks mingle with
one another, rather than being segregated in triplets as they are in
neutrons and protons.

If in the nuclear physics the meaning of isotope is establishing one [10,
12, 13, 18] then application isotope effect in atomic [19 - 22] and
molecular [23 - 25] physics allow to get the results, which are difficult to
overestimate so far as owing to this results it was to construct the
"building" of the science of 20 century - the quantum mechanics. The last
fifty years the isotope effect is one of the modern and power methods to
investigation of structure and properties of solids. This conclusion support
the numerous reviews (see, e.g. [26 - 28]) and first monographs [3, 19]
dedicated to isotope effect of stable isotopes. In the last years the more
and more investigations of solid state physics are conducted by using
radioactive isotopes, which give evidence already comprehensive list of
references (see, for instance [30 - 33]). It is a well-known fact large and
successful application of the radioactive elements in medicine [34 - 36],
the direction in isotope physics, which is more finance supportive in
different states (see, for example, [37] and references therein). Moreover
it is obviously a leading role of the isotope physics at the study of the
nature nuclear interactions and reconstruction of nucleogenesys process in
the Universe [1, 2, 38] which could be explained as the observable in nature
relative spreading of chemical elements.

Such wide field of isotope applications stimulate necessity for examination
and critical analysis from one point of view the microscopical nature of
isotope effect. Such approach to isotope physics allow to make known not
only the intrinsic contradiction inherent this area of physics but also
determine the borders of the effect. The expedient of such approach is not
only obvious, but also topical so far as the author of present review knows
no such similar paper. With the aim of filling up this gap undertakes the
present attempt to consider not only the isotope physics from the very
common position (from quarks to condensed matter) but also indicate the
principle of new isotope application in quantum information. A step-by-step
comparison with existing theoretical models not only revealing the degree of
agreement (or disagreement) but provide a new impulse both for the
development of new theoretical ideas and for conducting new experiments.

The present review consists of five part. First one is traditional
introduction in subject of writing. The second part devotes to the short
description of the ground of nature of isotope effect. With this aim the
detail analysis of the neutron and proton structure and their mutual
transformation in the weak interaction process was conducted. Noted, that
the main characteristics of isotope effect - the mass of free particles
(proton and neutron) doesn't conserve in the weak interaction process. This
contradiction is removed although partly if we take into account the modern
presentation [39 - 41] that the mass of proton (neutron) is created from
quark condensate (not from constituent quarks [18, 41]) which is the
coherent superposition of the states with different chirality. Thus the
elucidation of the reason of origin of the nucleon mass is taken down to
elucidation of the reason to break down the chiral symmetry in Quantum
Chromodynamics [42 - 47]. Isotope effect in atomic and molecular physics
shortly describe in third part. Here is an example of mass independent
isotope effect. In a more detailed fourth part of the review is considered
the manifestation of isotope effect in phonon and electron (exciton) states
of solids. With comparison to the change of corresponding characteristics
(for example: the lines shift in absorption, scattering, emission spectra)
at the isotope effect in atomic physics and condensed matter physics on two
orders more in solid (see, for example [48]). In this part the try to
determination of the border of isotope effect and force constant in isotope
effect on the isotope mass is made. It is underlined that taking into
account only linear part of electron - phonon interaction is not sufficient
for the description of the experimental facts on the elementary excitations
of systems consisting of light elements with isotope effect.

The subject of quantum information brings together ideas from quantum
physics, classical information theory and computer science. This topic is
devoted the fifth part of review. It is very significant that information
can be expressed in different ways without losing its essential nature,
since this leads to the possibility of the automatic manipulation of
information - a machine need only be able to manipulate quite simple things
like integers in order to do surprisingly powerful information processing.
It is easy to do from document preparation to differential calculus and even
to translating between human languages.

We should remind that quantum mechanics has developed originally as a theory
to explain behavior of large number (ensembles) of microscopic objects, such
as atoms or electrons [49]. However over the last decades, considerable
interest in the application of quantum theory to individual systems -
mesoscopic and even macroscopic systems where a small number of collective
degree of freedom show genuine quantum behavior (see, e.g. [50]). One
exciting aspect of this developing fundamental research is its technological
potential. It results that might be termed quantum information technology.
As we well know, the first deep insight into quantum information theory came
with Bell's 1964 analysis [51] of the paradoxical thought-experiment by
Einstein and coworkers in 1935 [52]. Bell's inequality draws attention to
the importance of correlations between separated quantum objects which have
interacted in the past, but which no longer influence one another. In
essence his argument shows that the degree of correlation which can be
present in such systems exceeds that which could be predicted on the basis
of any law of physics which describes particles in terms of classical
variables rather than quantum states. The next link between quantum
mechanics and information theory came about when it was realized that simple
properties of quantum systems, such as the unavoidable disturbance involved
in measurement, could be put to practical use in quantum cryptography (see,
e.g. review [53] and references therein). Quantum cryptography covers
several ideas, of which the most firmly established is quantum key
distribution. This is an ingenious method in which transmitted quantum
states are used to perform a very particular communication task. The
significant feature is that the principles of quantum mechanics guarantee a
type of conservation of quantum information, so that if the necessary
quantum information arrives at the parties wishing to establish a random
key, they can be sure it has not gone elsewhere, such as to spy. In this
part of review consider not only theory of cryptography but also its
practical application [54].

Some unsolved and difficult task of isotope physics are listed in
conclusion. From the immense volume of literature concerned with isotope
physics and its application we primarily selected those review and
monographs which contain extensive references.

\bigskip

\textbf{2. Sub - nucleonic structure and the modern picture of isotopes.}

\bigskip

2.1. The nucleons and its constituents.

\bigskip

An atom consists of an extremely small, positively charged nucleus (see Fig.
1) surrounded by a cloud of negatively charged electrons. Although typically
the nucleus is less than one ten-thounandth the size of the atom, the
nucleus contains more than 99.9\% of the mass of the atom. Atomic nucleus is
the small, central part of an atom consisting of A - nucleons, Z - protons
and N - neutrons (Fig.2). The atomic mass of the nucleus, A, is equal to
Z+N. A given element can have many different isotopes, which differ from one
another by the number of neutrons contained in the nuclei [55]. In a neutral
atom, the number of electrons orbiting the nucleus equals the number of
protons in the nucleus. As usually nuclear size is measured in fermis (1fm =
10$^{-15}$m, also called femtometers). The basic properties of the atomic
constituents can be read in Table 2.

As we can see from this Table, protons have a positive charge of magnitude e
= 1.6022$\cdot $10$^{-19}$ C (Coulombs)equal and opposite of that of the
electron. Neutron are uncharged. Thus a neutral atom (A,Z) contains Z
electrons and can be written symbolically as $_{Z}^{A}$X$_{N}$ (see also
Fig. 2). Here X is chemical symbol \ and N is neutron number and is equal N
= A - Z$^{2\ast )}$. The masses of proton and neutron are almost the same,
approximately 1836 and 1839 electron masses (m$_{e}$), respectively. Apart
from electric charge, the proton and neutron have almost the same
properties. This is why there is a common name of them: nucleon. Both the
proton and neutron are nucleons. As we well know the proton is denoted by
letter p and the neutron by n. Chemical properties of an element are
determined by the charge of its atomic nucleus, i.e. by the number protons
(electrons). It should be add, that although it is true that the neutron has
zero net charge, it is nonetheless composed of electrically charge quarks
(see below), in the same way that a neutral atom is nonetheless composed of
protons and electrons. As such, the neutron experiences the electromagnetic
interaction. the net charge is zero, so if we are far enough away from the
neutron that it appears to occupy no volume, then the total effect of the
electric force will add up to zero. The movement of the charges inside the
neutrons do not cancel however, and this is what gives the neutron its
nonzero magnetic moment.

---------------------------------

2*) Nuclei with the same N and different Z are called isotones, and nuclides
with the same mass number A are known as isobars. In a symbolic
representation of a nuclear specie or nuclide, it is usual to omit the N and
Z subscripts and include only the mass number as a superscript, since A = N
+ Z and the symbol X representing the chemical elements.

-------------------------------------------------------

Each of the atomic constituencies a spin 1/2 in units of $\hbar $ (= h/2$\pi 
$) and is an example of the class of particles of half-integer spin known as
fermions. Fermions obey the exclusion principle of Pauli (see, e.g. [12],
which determines the way electrons can occupy atomic energy states. The same
rule applies, as will be shown below, to nucleons in nuclei. Associated with
the spin is a magnetic dipole moment. Compared with the magnetic moment of
electron, nuclear moment are very small. However, they play an important
role in the theory of nuclear structure. It may be surprising that the
uncharged neutron has a magnetic moment. This reflects the fact that it has
an underlying quark substructure (see, e.g. [56]), consisting of charged
components. Electron scattering off these basic nuclear constituents (proton
and neutron) makes up for the ideal probe to obtain a detailed view of the
internal structure. A very detailed analysis using the best available data
has been carried out recently by Kelly [57], these data originate from
recoil or target polarizations experiments. In Fig. 3 the proton charge and
magnetization distribution are given. What should be noted is the softer
charge distribution compared to the magnetic one for proton. These resulting
densities are quite similar to Gaussian density distributions that can be
expected starting from quark picture (for details, see below) and, at the
same time more realistic than the exponential density distributions [57].
The neutron charge and magnetization are also given in Fig. 3. What is
striking is that magnetization distribution resembles very closely the
corresponding \ proton distribution. Since scattering on neutrons normally
carries the larger error (see, e.g. [9,10], the neutron charge distribution
is not precisely fixed. Nonetheless, one notices that the interior charge
density is balanced by a negative charge density, situated at the neutron
surface region, thereby making up for the integral vanishing of the total
charge of the neutron.

We should remind from atomic physics that the quantity e$\hbar $/2m is
called magneton. For atomic motion we use the electron mass and obtain the
Bohr magneton $\mu _{B}$ = 5.7884$\cdot $10$^{-5}$eV/T. Putting in the
proton mass we have the nuclear magneton $\mu _{N}$ = 3.1525$\cdot $10$^{-8}$%
eV/T. Note that $\mu _{N}$ $\ll $ $\mu _{B}$ owing to the difference in the
masses, thus under most circumstances atomic magnetism has much larger
effects than nuclear magnetism. Ordinary magnetic interactions of matter
(ferromagnetism, for example) are determined by atomic magnetism.

We can write

$\mu $ = g$_{\text{l}}$l$\mu _{N}$ \ \ \ \ \ \ \ \ \ \ \ \ \ \ \ \ \ \ \ \ \
\ \ \ \ \ \ \ \ \ \ \ \ \ \ \ \ \ \ \ \ \ \ \ \ \ \ \ \ \ \ \ \ \ \ \ \ \ \
\ \ (1),

where g$_{\text{l}}$ is the g - factor associated with the orbital angular
momentum l. For protons g$_{l}$ = 1, because neutrons have no electric
charge; we can use Eq. (1) to describe the orbital motion of neutrons if we
put g$_{l}$ = 0. We have thus been considering only the orbital motion of
nucleons. Protons and neutrons, like electrons, as above mentioned also have
intrinsic or spin magnetic moments, which have no classical analog but which
we write in the same form as Eq (1):

$\mu $ = g$_{s}$s$\mu _{N}$ \ \ \ \ \ \ \ \ \ \ \ \ \ \ \ \ \ \ \ \ \ \ \ \
\ \ \ \ \ \ \ \ \ \ \ \ \ \ \ \ \ \ \ \ \ \ \ \ \ \ \ \ \ \ \ \ \ \ \ \ (2),

where s = 1/2 for protons, neutrons, and electrons (see Table 2). The
quantity g$_{s}$ is known as the spin g - factor and is calculated by
solving a relativistic quantum mechanics equation (see, also [12]). For free
nucleons, the experimental values are far from the expected value for point
particles: \ proton - g$_{s}$ = 5.5856912$\pm $0.0000022 and neutron - g$%
_{s} $ = 3.8260837$\pm $0.0000018. Table 3 gives some representative values
of nuclear magnetic dipole moments according [58]. The next nonvanishing
moment is the electric quadrupole moment. The quadrupole moment eQ of a
classical point charge e is of the form e(3z$^{2}$ - r$^{2}$). If the
particle moves with spherical symmetry, then (on the average ) z$^{2}$ = x$%
^{2}$ = y$^{2}$ = r$^{2}$/3 and the quadrupole moment vanishes (for details,
see [11]). Some examples of the values of nuclear electric quadrupole
moments are presented in Table 4.

\bigskip

2.1.1. Mass and nuclear binding energy.

\bigskip

Inside a nucleus, neutrons and protons interact with each other and are
bound within (as mentioned above) the nuclear volume under the competing
influences of attractive nuclear and repulsive electromagnetic forces. This
binding energy has a direct effect on the mass of an atom. It is therefore
not possible to separate a discussion of nuclear binding energy; if it were,
then nucleon would have masses given by Z$_{m_{p}}$ + Z$_{m_{n}}$ and the
subject would hardly be of interest.

As is well-known, in 1905, Einstein presented the equivalence relationship
between mass and energy: E = mc$^{2}$. From this formula, we see that the
speed of light c is very large and so even a small mass is equivalent to a
large amount of energy. This is why in nuclear physics it is more convenient
to use a much smaller unit called megaelectronvolt (1MeV = 1.602$\cdot $10$%
^{-13}$J). On the atomic scale, 1u is equivalent to 931.5 MeV/c$^{2}$, which
is why energy changes in atoms of a few electron-volt cause insignificant
changes in the mass of atom. Nuclear energies, on the other hand, are
millions of electron-volts and their effects on atomic mass are easily
detectable. For example, the theoretical mass of $_{17}^{35}$Cl is 17 x
1.00782503 + 18 x 1.00866491 = 35.28899389 amu. Its measured (see below)
mass is only 34.96995 amu. therefore, the mass defect and binding energy of $%
_{17}^{35}$Cl are

$\Delta $ = 0.32014389 amu.

E$_{B}$ = $\frac{0.32014389\text{ x 931.5}}{35}$ = 8.520 MeV/nucleon \ \ \ \
\ \ \ \ \ \ \ \ \ \ \ \ \ \ \ (3)

and in common sense the binding energy is determined by next relation

E$_{B}$ = Zm$_{p}$ + Nm$_{n}$ - B/c$^{2}$ \ \ \ \ \ \ \ \ \ \ \ \ \ \ \ \ \
\ \ \ \ \ \ \ \ \ \ \ \ \ \ \ \ \ \ \ \ \ \ \ \ \ \ \ \ \ \ \ \ \ \ \ \ (4),

where B/c$^{2}$ is the actual nuclear mass.

As we can see below, the binding energy of the atoms of most elements have
values ranging from about 7.5 to 8.8 MeV [5-10]. The binding energy per
nucleon rises slightly with increasing mass number and reaches a maximum
value for $^{62}$Ni. Thereafter the binding energies decline slowly with
increasing mass number. The binding energies of the atoms of H, He, Li and
Be are lower than the binding energies of the other elements (see, also Fig.
5 below).

The measurement of nuclear masses occupies an extremely important place in
the development of nuclear physics. Mass spectrometry (see, e.g. [59])was
the first technique of high precision available to the experimenter, and
since the mass of a nucleus increases in a regular way with the addition of
one proton or neutron. In mass spectrometers, a flux of identical nuclei
(ions), accelerated (see, e.g. Fig. 3.13 in [17]) to a certain energy, is
directed to a screen (photoplate) where it makes a visible mark. Before
striking the screen, this flux passes through magnetic field, which is
perpendicular to velocity of the nuclei. As a result, the flux is deflected
to certain angle. The greater mass, the smaller is the angle. Thus,
measuring the displacement of the mark from the center of the screen, we can
find the deflection angle and then calculate the mass. The example of a
mass-spectrum of a different isotopes of krypton is shown in Fig. 4. From
the relative areas of the peaks it can be determine the abundance of the
stable of krypton (for details see [58]).

Relative masses of nuclei can also be determined from the results of nuclear
reactions or nuclear decay. For example, if a nucleus is radioactive and
emits an $\alpha $ - particle, we know from energy conservation that it mass
must be greater than that of decay products by the amount of energy released
in the decay. Therefore, if we measure the letter, we can determine either
of the initial or final nuclear masses if one of them is unknown. An example
of this is presented briefly below. At present we shall illustrate some
typical reactions, bridging the gap between "classical" methods and the more
advanced "high-energy" types of experiments (see, also [10, 57]).

The possible, natural decay processes can also brought into the class of
reaction processes with the conditions: no incoming light particle $\alpha $
and Q $>$ 0. We list them in the following sequence:

$\alpha $ - decay:

$_{Z}^{A}$X$_{N}$ $\rightarrow $ $_{Z-2}^{A-4}$Y$_{N-2}$ \ + $_{2}^{4}$He$%
_{2}$ \ \ \ \ \ \ \ \ \ \ \ \ \ \ \ \ \ \ \ \ \ \ \ \ \ \ \ \ \ \ \ \ (5).

$\beta $ - decay:

$_{Z}^{A}$X$_{N}$ $\rightarrow $ $_{Z-1}^{A}$Y$_{N+1}$ + e$^{+}$ + $\nu _{e}$
\ (p $\rightarrow $ n - type) \ \ \ \ \ \ (6)

$_{Z}^{A}$X$_{N}$ $\rightarrow $ $_{Z+1}^{A}$Y$_{N-1}$ + e$^{-}$ + $%
\overline{\nu }_{e}$ \ (n $\rightarrow $ p - type) \ \ \ \ \ (6')

$_{Z}^{A}$X$_{N\text{ +e}^{-}}$ + e$^{-}$ $\rightarrow $ $_{Z-1}^{A}$Y$%
_{N+1} $ + $\nu _{e}$ (e$^{-}$ - capture) \ (6").

Here e$^{-}$, e$^{+}$, $\nu _{e}$ and $\overline{\nu }_{e}$ are electron,
positron, neutrino and antineutrino.

$\gamma $ - decay:

$_{Z}^{A}$X$_{N}^{\ast }$ $\rightarrow $ $_{Z}^{A}$X$_{N}$ + h$\nu $ \ \ \ \
\ \ \ \ \ \ \ \ \ \ \ \ \ \ \ \ \ \ \ \ \ \ \ \ \ \ \ \ \ \ \ \ \ \ \ \ \ \
\ \ \ \ (7).

Here X$^{\ast }$ is excited nuclei.

Nuclear fission:

$_{Z}^{A}$X$_{N}$ $\rightarrow $ $_{Z_{1}}^{A_{1}}$Y$_{N_{1}}$ + $%
_{Z_{2}}^{A_{2}}$U$_{N_{2}}$ + x$\cdot $n \ \ \ \ \ \ \ \ \ \ \ \ \ \ \ \ \
\ \ \ \ \ \ \ \ \ \ \ \ \ (8).

Since mass and energy are equivalent (see Einstein formula above), in
nuclear physics it is customary to measure masses of all particles in the
units of energy (MeV). Examples of masses of subatomic particles are given
in Table 5.

As was noted above, nuclear binding energy increases with the total number
of nucleons A and, therefore, it is common to quote the average binding
energy per nucleon (B/A) The variation of B/A with A is shown in Fig. 5.
Several remarkable features are immediately apparent. First of all, the
curve is relatively constant except for the very light nuclei. The average
binding energy of most nuclei is, to within 10\%, about 8 MeV per nucleon.
Second, we note that the curve reaches peak near A = 60, where the nuclei
are most tightly bound, light and very heavy nuclei are containing less
bound nucleons. Thus, the source of energy production in fusion of light
nuclei or fusion of very heavy nuclei can be source of energy [16, 17].

In concluding this paragraph we should remind that it is often stated that $%
^{56}$Fe is the most tightly bound nucleus, this is not correct since $^{62}$%
Ni is more bound by a difference of 0.005 MeV/nucleon (for details see [61]
and references therein). In conclusion, it is very interesting note, that
one cubic millimeter of nuclear material, if compressed together, would have
a mass around 200 000 tonnes. Neutron stars are composed of such material.

\bigskip

2.1.2. The nuclear radius.

\bigskip

As will be shown in this paragraph nuclei vary from about one to a few
fermis \ in radius. Recall that the Bohr radius of hydrogen is the order\ 10$%
^{-10}$ meters , so \ the nucleus at present time, despite its small size
the nucleus has about, as was noted above, 99.9\% of the mass of the atom
(see, also [5, 6]). Electron scattering off nuclei is, for example, one of
the most appropriate \ methods to deduce radii. The results of this
procedure for several different nuclei are shown in Fig. 6. One remarkable
conclusion is obvious - the central nuclear charge density is nearly the
same for all nuclei. Nucleons do not congregate near the center of the
nucleus, but instead have a fairly constant distribution out to the surface.
The conclusion from measurements of the nuclear matter distribution is the
same [62]. Under this assumptions of saturation and charge independence each
nucleon occupies an almost equal size within the nucleus. Calling r$_{0}$ an
elementary radius for a nucleon in the nucleus, a most naive estimate gives
for the nuclear volume

V = 4/3$\pi $r$_{0}^{3}$A \ \ \ \ \ \ \ \ \ \ \ \ \ \ \ \ \ \ \ \ \ \ \ \ \
\ \ \ \ \ \ \ \ \ \ \ \ \ \ \ \ \ \ \ (9)

or

R = r$_{0}$A$^{1/3}$ \ \ \ \ \ \ \ \ \ \ \ \ \ \ \ \ \ \ \ \ \ \ \ \ \ \ \ \
\ \ \ \ \ \ \ \ \ \ \ \ \ \ \ \ \ \ \ \ (10).

This relation describes the variation of the nuclear radius, with value of r$%
_{0}$ $\simeq $ 1.2 fm when deducing a "charge" radius and a "value of r$%
_{0} $ $\simeq $ 1.4 fm for the full matter" radius (see also Figs. 3.5 and
3.9 in [8]). In simple way the nuclear radius is defined as the distance at
which the effect of the nuclear potential is comparable to that of the
Coulomb potential (see Fig. 7).

We should indicate another way to determine the nuclear \ charge radius is
from direct measurement of the Coulomb energy differences of nuclei.
Consider, for example, $_{1}^{3}$H$_{2}$ and $_{2}^{3}$He$_{1}$. To get from 
$^{3}$He$_{1}$ to $^{3}$H$_{1}$ we must change a proton into a neutron. As
we know, there is strong evidence which suggest that the nuclear force does
not distinguish between protons and neutrons. Changing proton into a neutron
should therefore not affect the nuclear energy of the three nucleon system:
only \ the Coulomb energy should change, because the two protons in $^{3}$He$%
_{1}\exp $experience a repulsion that is not present in $^{3}$H. The energy
difference between $^{3}$He and $^{3}$H is thus a measure of the Coulomb
energy of the second proton, and the usual formula for \ for the Coulomb
repulsion energy can be used to calculate the distance between the protons
and thus the size of the nucleus.

\bigskip

2.2.The force between nucleons.

\bigskip

The interactions between two nucleons (NN) is one of the central questions
in physics and its importance goes beyond the properties of nuclei. Nucleons
can combine to make four different few-nucleon systems, the deuteron (%
\textit{p + n}), the triton (\textit{p} + 2\textit{n}), the helion (2\textit{%
p + n}) and the $\alpha $ - particle (2\textit{p} + 2\textit{n}) (see, e.g.
[63 - 65]). These particles are grouped together because they are stable
(excluding from the radioactive triton which have a half-life of about
twelve years and so may be treated as a stable entity for most practical
purpose), have no bound excited states (except the $\alpha $ - particles
which has two excited states at about 20 and 22 MeV), and are frequently
used as projectiles in nuclear investigations. The absence of stable
particles of mass of five provides a natural boundary between few - nucleon
systems and heavier nuclei [1, 2, 64]. Few nucleon systems provide the
simplest systems to study nuclear structure. The deuteron provides important
information about the nucleon-nucleon interaction.

Even before describing any further experimental and theoretical results to
study the force between two nucleons, we can already guess at a few of the
properties of the N-N force:

1. At short distances it is stronger than the Coulomb force; the nuclear
force can overcome the Coulomb repulsion (see also Fig. 7) of protons in the
nucleus.

2. At long distances, of the order of atomic sizes, the nuclear force is
negligibly feeble. The interaction among nuclei in a molecule can be
understood based only on the Coulomb force.

3. Some fundamental particles are immune from the nuclear force. At present
time we have not any evidence from atomic structure, for example, that
electrons feel the nuclear force at all.

4. The N -N force seems to be nearly independent of whether the nucleons are
neutrons or protons. As is well-known this property is called charge
independence.

5. The N-N force depends on whether the spins of the nucleons are parallel
or antiparallel.

6. The N - N force includes a repulsive term, which keeps the nucleons at a
certain average separation.

7. The N -N force has a noncentral or tensor component. This part of the
force does not conserve orbital angular momentum, which is a constant of the
motion under central forces.

We should add that with knowledge of the N - N interaction provided by%
\textit{\ p - p }\ and \textit{p - n }scattering and by the deuteron (see,
also [66 - 68]), one can try to calculate the properties of the triton and
the helion. The principal properties of few-nucleon systems are summarized
in Table 6.

\textbf{Deuteron.} The deuteron is a very unique nucleus in many respects.
It is only loosely bound, having a binding energy much less than the average
value ($\curlyeqprec $ 8 Mev [1,2]) between a pair of nucleons in all other
stable nuclei. We have seen in Eq. 94) that the binding energy E$_{B}$ 0f a
nucleus is given by the mass difference between the neutral atom and the sum
of the masses of free neutrons and protons in the form of hydrogen atoms.
For a deuteron, as we can see from Table 6, the mass M$_{d}$ is 1876.1244
Mev/c$^{2}$. The binding energy is then difference between M$_{d}$ and the
sum of those for a neutron m$_{n}$ and a hydrogen atom m$_{H}$ ( = m$_{p}$):
m$_{n}$c$^{2}$ = 939.565; m$_{H}$c$^{2}$ = 938.7833 MeV and m$_{n}$ + m$_{H}$
\ = 1878.3489 MeV. We can write according Eq. (4): E$_{B}$ = m$_{n}$ + m$%
_{H} $ - M$_{d}$ = 2.224 MeV$.$ A more precise value, E$_{B}$ = 2.22457312
MeV is obtained from radioactive capture of a neutron by hydrogen. In this
reaction p(n,$\gamma $)d, a slow neutron is captured by a hydrogen atom
followed by the emission of a $\gamma $ - ray (for details see [69]).

To simplify the analysis of the deuteron binding energy, we will assume that
we can represent the N - N potential os three - dimensional square well, as
shown in Fig. 8.

V(r) = - V$_{0}$, for r \TEXTsymbol{<} R (= 2.1 fm.)

\ \ \ \ \ \ \ = 0, for r \ \TEXTsymbol{>} R \ \ \ \ \ \ \ \ \ \ \ \ \ \ \ \
\ \ \ \ \ \ \ \ \ \ \ \ \ \ \ \ \ \ \ \ \ \ \ \ \ \ \ \ \ \ \ \ \ \ \ \ \
(11).

This is of course an oversimplification, but is sufficient for at last some
qualitative conclusions. \ In Eq. (11) r represents the separation between
the proton and the neutron, so R is in effect a measure of the diameter of
the deuteron (Fig. 9). If we express the energy, corresponding to the ground
state value E \ = -E$_{B}$,the Schrodinger equation becomes for the
one-dimensional, radial problem with zero angular moment, just like the
lowest energy state of hydrogen atom.

$\frac{\text{d}^{2}\text{u}}{\text{dr}^{2}}$ + k$^{2}$u = 0, \ \ r 
\TEXTsymbol{<} R (see, Fig. 10)

$\frac{\text{d}^{2}\text{u}}{\text{dr}^{2}}$ - $\alpha ^{2}$u = 0, \ \ r 
\TEXTsymbol{>} b \ \ \ \ \ \ \ \ \ \ \ \ \ \ \ \ \ \ \ \ \ \ \ \ \ \ \ \ \ \
\ \ \ \ \ \ \ \ \ \ \ \ \ \ \ \ \ \ \ \ (12),

defining

k$^{2}$ = $\frac{\text{m}_{n}}{\hbar ^{2}}$(u - E$_{B}$), \ $\alpha ^{2}$ = $%
\frac{\text{m}_{n}}{\hbar ^{2}}$ E$_{B}$ \ \ \ \ \ \ \ \ \ \ \ \ \ \ \ \ \ \
\ \ \ \ \ \ \ \ \ \ \ \ \ \ \ \ \ \ \ (13)

and using the radial solution

u(r) = rR(r). \ \ \ \ \ \ \ \ \ \ \ \ \ \ \ \ \ \ \ \ \ \ \ \ \ \ \ \ \ \ \
\ \ \ \ \ \ \ \ \ \ \ \ \ \ \ \ \ \ \ \ \ \ \ \ \ \ \ \ \ \ \ \ \ \ \ (14).

Approximate solutions in the two regions became

u(r) = Asinkr, \ \ \ \ \ \ \ \ \ \ \ \ \ r \TEXTsymbol{<} R and

u(r) = Be$^{-\alpha \text{(r - R)}}$ \ \ \ \ \ \ \ \ \ r \TEXTsymbol{>} b \
\ \ \ \ \ \ \ \ \ \ \ \ \ \ \ \ \ \ \ \ \ \ \ \ \ \ \ \ \ \ \ \ \ \ \ \ \ \
\ \ \ (15).

Matching the logarithmic derivatives at r = R gives

kcotankr = -$\alpha $ \ \ \ \ \ \ \ \ \ \ \ \ \ \ \ \ \ \ \ \ \ \ \ \ \ \ \
\ \ \ \ \ \ \ \ \ \ \ \ \ \ \ \ \ \ \ \ \ \ \ \ \ \ \ \ \ \ \ \ \ \ \ \ \
(16)

and matching the wave functions at r = R gives

AsinkR = B \ \ \ \ \ \ \ \ \ \ \ \ \ \ \ \ \ \ \ \ \ \ \ \ \ \ \ \ \ \ \ \ \
\ \ \ \ \ \ \ \ \ \ \ \ \ \ \ \ \ \ \ \ \ \ \ \ \ \ \ \ \ \ \ \ \ \ (17).

These two relations lead to the condition

k$^{2}$A$^{2}$ = (k$^{2}$ + $\alpha ^{2}$)B$^{2}$ \ \ \ \ \ \ \ \ \ \ \ \ \
\ \ \ \ \ \ \ \ \ \ \ \ \ \ \ \ \ \ \ \ \ \ \ \ \ \ \ \ \ \ \ \ \ \ \ \ \ \
\ \ \ (18).

The normalization of the wave function 4$\pi \int $u$^{2}$(r)dr = 1 becomes

$\frac{\text{A}^{2}}{\text{2k}}$(2kR - sin2kR) + $\frac{\text{B}^{2}}{\alpha 
}$ = $\frac{\text{1}}{\text{2}\pi }$ \ \ \ \ \ \ \ \ \ \ \ \ \ \ \ \ \ \ \ \
\ \ \ \ \ \ \ \ \ \ \ \ \ \ \ \ \ \ \ \ (19).

Eliminating A$^{2}$ from the last two equations, gives the value for B as

B $\simeq $ $\sqrt{\frac{\alpha }{\text{2}\pi }}$e$^{-\alpha \text{R/2}}$ \
\ \ \ \ \ \ \ \ \ \ \ \ \ \ \ \ \ \ \ \ \ \ \ \ \ \ \ \ \ \ \ \ \ \ \ \ \ \
\ \ \ \ \ \ \ \ \ \ \ \ \ \ \ \ \ \ \ \ \ \ \ (20).

Knowing the binding energy E$_{B}$ (see, Eq. (13)), we can determine the
value $\alpha $ = 0.232 fm$^{-1}$. A best value for R can be determined from
proton - neutron scattering (see, e.g. [63, 64]) as R = 1.93 fm. This then
gives u = 38.5 MeV. One can show that this value of u and the value for R
just give rise to a single, bound 1s state, all other higher-lying 1p, 1d,
2s being unbound. Since we also have

A $\simeq $ B \ \ \ \ \ \ \ \ \ \ \ \ \ \ \ \ \ \ \ \ \ \ \ \ \ \ \ \ \ \ \
\ \ \ \ \ \ \ \ \ \ \ \ \ \ \ \ \ \ \ \ \ \ \ \ \ \ \ \ \ \ \ \ \ \ \ \ \ \
\ \ \ \ \ \ \ \ \ (21)

we obtain the final wave functions

u(r) = $\sqrt{\frac{\alpha }{\text{2}\pi }}$e$^{-\alpha \text{R/2}}$sinkr, \
\ \ \ \ \ \ \ \ \ \ \ \ \ \ r \TEXTsymbol{<} R and

u(r) = $\sqrt{\frac{\alpha }{\text{2}\pi }}$e$^{-\alpha \text{R/2}}$e$%
^{-\alpha \text{r}}$ \ \ \ \ \ \ \ \ \ \ \ \ \ \ \ \ \ \ r \TEXTsymbol{>} b
\ \ \ \ \ \ \ \ \ \ \ \ \ \ \ \ \ \ \ \ \ \ \ \ \ \ \ \ (22).

A potential which gives a satisfactory account of the properties of the
deuteron given in Table 6 is shown in Fig. 1o. We should add that in all
deuteron potentials the tensor term is very sizeable part of the two-nucleon
potential, and is characterized by a somewhat larger range than the centra
potential (see Fig. 10) being appreciably different from zero even when the
central potential is already negligible.

\textbf{Proton - proton and proton - neutron interactions. }Most of the
present theories (see, also [64] and references therein) of nuclear
structure and nuclear reactions are based on the assumption that nuclear
properties depend mainly on two-body interactions between its constituents.
Three-body forces or many-body forces are expected to play only a minor role$%
^{3\ast )}$. It is thus of paramount importance to describe as accurately as
possible the two-nucleon interaction. At the fundamental level this
interaction is a consequence of the quark structure of the nucleons and
should be described by QCD [65] in terms of the quark-gluon field (see, also
[39, 47, 83]).

-----------------------------------------------

3*) If the two-body potential has an average strength of 20 MeV, then the
three-body one would have a strength of about 1 MeV. We should add that all
models have a one-pion exchange character at long range, which gives rise to
a spin-spin central potential and a tensor term ( for details see [56, 92]).

----------------------------------------------

However this approach is still in its infancy and therefore we are still far
from solution. There are also many indications [63] that at interaction
energies below a few hundred MeV it is possible to describe the N - N
interaction in terms of the exchange of various types of mesons [71 - 73].

In principle there are four types of scattering measurements involving two
nucleons that can be carried out. The scattering of an incident proton off a
proton (\textit{pp} - scattering) is the simplest one of the four from an
experimental point of view, as it is relatively easy to accelerate protons
and to construct targets containing hydrogen. For neutron scattering, there
are two major sources for incident beam. At low energies , neutrons from
nuclear reactors may be used. At higher energies, one can make use of
neutrons produced by a beam of protons, for instance, through a (\textit{p,n}%
) reaction on a $^{7}$Li target. However, both the intensity and the energy
resolution of neutron beams obtained in these ways are much more limited
than those for proton beams. \ As a result, neutron scattering is, in
general, a more difficult experiment than those with protons. In addition to 
\textit{pp} - and \textit{np} - measurements, one can, in principle, carry
out \textit{pn }- and \textit{nn}- scattering experiments as well. Here,
instead of using protons as the target,, a neutron target is used. As we
know, free neutrons are unstable (see above), with a half-life on the order
of 10 min. It is therefore impossible to construct a fixed neutron target,
in contrast to protons where material consisting of hydrogen may used..
There are, in principle [72], two methods of getting around this
limitations. One way is to carry out a colliding beam experiment. In place
of a target fixed in the laboratory, a second neutron beam is used and,
instead of having an incident beam scattering from a fixed target, two beams
of particles are directed toward each other. Scattering takes place when the
particles in the two beam collide. To be practical, such an experiment
requires high intensities in both beams, and currently highly intense beams
of neutrons are not easily available. The other way is to simulate a fixed
neutron target using deuterium. Since the deuteron is a loosely bound system
of a neutron and a proton, the desired \textit{pn} - or \textit{nn}-
scattering results can be obtained by carrying out the corresponding \textit{%
pd} - or \textit{nd }- scattering experiments. The contribution due to
protons \ in the deuterium target may be removed by substracting from the
measured values the corresponding results obtained in \textit{pp} - or 
\textit{np} - scattering. The information obtained from pn - and nn -
scattering may not be any different from that in np and pp - scattering. For
example, the only difference between \textit{pn} - and \textit{np} -
scattering is whether the neutron or the proton is the target. Under
time-reversal invariance, these two arrangements are expected to give
identical results. As early to simplify the notation, we shall use the
symbol NN from now on to represent a system of two nucleons, as early, when
there is no need to differentiate between neutrons and protons and the
symbol \textit{np} to represent both \textit{np} - and\textit{\ pn-} unless
further distinction is required by the occasion. Futher more, we shall
assume that Coulomb contribution, where present, has already been taken out
and we can therefore ignore it in the discussion.

The quantity measured in a scattering experiment is the number of counts
registered by a detector ($\theta $, $\varphi $) (see, e.g. [60]). The
counting rate depends on the solid angle subtended by the detector at the
scattering center, the intensity of the incident beam, the number of target
nuclei involved, and the differential cross-section d$\sigma $/d$\Omega $.
Naturally our primary interest is in d$\sigma $/d$\Omega $, a function of
the bombarding energy as well as the scattering angle. For simplicity we
shall consider first only elastic scattering, and as a result, the wave
number k in the center of mass of the two particles has the same magnitude
before and after the scattering. The differential scattering cross-section
at angles ($\theta $, $\varphi $) is given by next equation

$\frac{\text{d}\sigma }{\text{d}\Omega }$($\theta $, $\varphi $) = $%
\left\vert f\text{(}\theta ,\varphi \text{)}\right\vert ^{2}$ \ \ \ \ \ \ \
\ \ \ \ \ \ \ \ \ \ \ \ \ \ \ \ \ \ \ \ \ \ \ \ \ \ \ \ \ \ \ \ \ \ \ \ \
(23).

Here $f$($\theta ,\varphi $) is the scattering amplitude. As shown in Fig.
11 the geometry of 0 - scattering arrangement is such that it is coordinate
system at the center of the scattering region and take the direction of the
incident beam as the positive direction along the z- axis. The incident wave
vector $\overrightarrow{k}$ and the scattered vector $\overrightarrow{k}%
^{\prime }$ define a plane, the scattering plane.

For a central potential, the relative angular momentum $\overrightarrow{l}$
between the two scattering nucleons is a conserved quantity. Under such
conditions, it is useful to expand the wave function as a sum over the
contributions from different partial waves, each with a definite l - value

$\Psi $(r, $\theta $) = $\dsum\limits_{l\text{ = 0}}^{\infty }$a$_{l}$Y$%
_{l0} $($\theta $)R$_{l}$(k, r) \ \ \ \ \ \ \ \ \ \ \ \ \ \ \ \ \ \ \ \ \ \
\ \ \ \ \ \ \ \ \ \ \ (24).

Here a$_{l}$ are the expansion coefficients. Only spherical harmonics Y$%
_{lm} $($\theta $, $\varphi $) with m = 0 appears in the expansion since,in
the absence of polarization, the wave functions is independent of the
azimutal angle $\Phi $. We have explicitly included the wave number k have
in the arrangement of the radial wave function R$_{l}$(k, r) so as to
emphasize the dependence of energy.

For a free particle, V = 0, and the radial wave function reduces to

R$_{l}$(k, r) $\rightarrow $ $\frac{\text{1}}{\text{kr}}$sin(kr - $\frac{%
\text{1}}{\text{2}}$l$\pi $), \ \ \ \ \ \ \ \ \ \ \ \ \ \ \ \ \ \ \ \ \ \ \
\ \ \ \ \ \ \ \ \ \ \ \ \ \ \ \ (25)

where k = $\sqrt{\text{2}\mu \text{E}}$/$\hbar $ and j$_{l}$($\rho $) is the
spherical Bessel function of order l. If only elastic scattering is allowed
by the potential, the probability current density in each partial-wave
channel is conserved. The only effect the potential can have on the wave
function is a change in the phase angle. In other words

R$_{l}$(k, r) (scatt/r$\rightarrow \infty $) $\rightarrow $ \ $\frac{\text{1}%
}{\text{kr}}$sin(kr - $\frac{\text{1}}{2}$l$\pi $ + $\delta _{l}$), \ \ \ \
\ \ \ \ \ \ (26)

where $\delta _{l}$ is the phase shift in the l -th partial - wave channel.

After that, the scattering amplitude may be expressed in terms of $\delta
_{l}$ as

f(0) = $\frac{\sqrt{\text{4}\pi }}{k}\dsum\limits_{\text{l = 0}}^{\infty }%
\sqrt{\text{2l + 1}}$e$^{i\delta _{i}}$sin$\delta _{l}$Y$_{l0}$($\theta $) \
\ \ \ \ \ \ \ \ \ \ \ \ \ \ \ \ \ \ \ (27).

In such case the differential scattering cross-section may be written in
terms of the phase shift

$\frac{\text{d}\sigma }{\text{d}\Omega }$ = $\frac{\text{4}\pi }{\text{k}^{2}%
}\left\vert \dsum\limits_{\text{l = 0}}^{\infty }\sqrt{\text{2l + 1}}\text{e}%
^{i\delta _{i}}\text{sin}\delta _{l}\text{Y}_{l0}\text{(}\theta )\right\vert
^{2}$ \ \ \ \ \ \ \ \ \ \ \ \ \ \ \ \ \ \ \ (28).

The scattering cross-section, the integral of $\frac{\text{d}\sigma }{\text{d%
}\Omega }$ over all solid angles, becomes

$\sigma $ = $\int \frac{\text{d}\sigma }{\text{d}\Omega }$d$\Omega $ = $%
\frac{\text{4}\pi }{\text{k}^{2}}\dsum\limits_{\text{l = 0}}^{\infty }$(2l +
1) sin$^{2}\delta _{l}$(k) \ \ \ \ \ \ \ \ \ \ \ \ \ \ \ \ \ \ \ (29).

Decomposition into partial waves is a useful way to analyze the scattering
results for a given bombarding energy. In particular, only a few of the
low-order partial waves can contribute to the scattering at low energies, as
shown in Fig. 12. For realistic nuclear potential, the orbital angular
momentum is not conserved.

Since we are dealing with identical fermions, the scattering of two nucleons
can take place only in a state that is totally antisymmetric with respect to
a permutation of the two particles, in the same way as for deuteron. For 
\textit{pp} - scattering, we have T = 1$^{4}$*$^{)}$ and the two nucleons
are symmetric, as for as their total isospin wave function [56] is
concerned. If the intrinsic spins of the two protons are coupled together to
S = 0 (antisymmetric state) and, as a result, only even l-values are
allowed. For S = 0, we have J = l (we remind that $\overrightarrow{J}$ = $%
\overrightarrow{l}$ + $\overrightarrow{S}$), \ and the partial waves for the
lowest two orders \ of \textit{pp} - scattering are $^{1}$S$_{0}$ (l = 0)
and $^{1}$D$_{2}$ (l = 2). The phase shifts extracted from measured \textit{%
pp} - scattering data \ for these two partial waves of bombardering energy
less than 300 MeV, in the laboratory are shown in Fig. 13 as illustrative
examples (for details see [74]). Only the real part of the phase shift are
given. At laboratory energy less than 300 MeV, contributions from inelastic
scattering are still relatively unimportant and the imaginary parts of the
phase shifts extracted from measured scattering cross-section are small.

----------------------------

4*) In 1932 Heisenberg suggested [77] on the basis of the approximate of the
proton and neutron mass (see \ also Table 2) that these particles might be
considered as two different charge states of a single entity, the nucleon,
formally equivalent to the up and down states of a spin 1/2 particle. To
exploit this hypothesis the nucleon wave function in addition to a space and
a spin component also has an isotopic spin (isospin) component (see, also
e.g. [10]).

----------------------------

By the same token, partial waves for triplet (S = 1) \textit{pp} -
scattering have odd l - values. The lowest order in this case is a p - wave
(l = 1). When l = 1 is coupled with S = 1, three states, with J = 0, 1, 2
are produced. The phase shifts for two of the triplet of states, $^{3}$P$%
_{0} $ and $^{3}$P$_{1}$, are also shown in Fig. 13$^{a}$. \ There is no
admixture between the two J = 0 states, $^{3}$P$_{0}$ and $^{1}$S$_{0}$as
they are of different parity. As a result, we find that both l and S are
good quantum numbers here by default (for details see [76] and references
therein).

The \textit{np} - system may be coupled together to either isospin T = 0 or
T = 1. For T = 0 the two nucleons are antisymmetric in isospin. In this case
the S = 0 states must have odd l - values in order to be antisymmetric in
the total wave function. The lowest order partial wave here is l = 1 and the
phase shifts for $^{1}$P$_{1}$ - scattering extracted from experimental data
are shown in Fig. 13$^{c}$. In order for p - wave \textit{np} - scattering
to be in the S = 1 state, it is necessary for the total isospin to be T = 1.
The phase shift in this case are expected to be identical to those found in 
\textit{pp} - scattering, if nuclear force is charge independent and Coulomb
effects are removed. An examination of the two sets of empirical p - wave
phase shifts, $^{3}$P$_{0}$ and $^{3}$P$_{1}$ given in Fig.14$^{b}$, shows \
that they are only slightly different from corresponding values given in
Fig. 14$^{a}$ for \textit{pp} - scattering. It is not clear whether the
small differences come from the way the phase shifts are extracted from
experimental scattering cross - section or they are indications of a weak
charge dependence in the nuclear force.

The other T = 0 phase shift in the \textit{np }- system, shown in Fig. 14$%
^{c}$, are for triplet (S = 1), even l - scattering. This is the first time
we encounter a mixing of different l - partial waves. Up to now, each phase
shift has been characterized by a definite l - value (as well as J - and S -
values) even though the orbital angular momentum is not fundamentally a good
quantum number. Mixing of different l - partial waves has not taken place
because of parity and other invariance conditions, however the tensor force
can mix two triplet of the same J but different in l by two units (l = J $%
\pm $1) (see also [76]).

\bigskip

2.3. Nucleon structure.

\bigskip

Our present knowledge of nuclear physics suggests that there are two main
families of particles leptons and hadrons (baryons and mesons). The hadrons
were first thought to be elementary like the leptons, but soon a very large
number of hadrons were discovered, which suggests that they are not
elementary (see, also 8, 62, 78]). As we can see from Table 6 the leptons
are fundamental particles, but hadrons are not. They are made up of quarks
[79, 80] (for details see below). The only hadrons found in normal matter
are the proton and the neutron. Quarks are one of the two basic constituents
of matter which is described QCD. QCD [47, 81, 82] is the theory of the
strong interaction, a fundamental force describing the interactions of the
quarks and gluons found in nucleons such as the proton and the neutron. QCD
is an important part of the Standard Model (SM)$^{5\ast )}$ of particle
physics (see, also [81, 82]). In the present SM [47] there are six "flowers"
of quarks (see, below Table 9) most familiar baryons are the proton and
neutron, which are each constructed up and down quarks [83 - 85]. Quarks are
observed to occur only in combinations of two quarks (mesons), three quarks
(baryons), and the recently discovered with five quarks (pentiquarks [83]).

----------------------------------

5*) As is well-known, the Standard Model [45, 47, 82, 92, 93] is a unified
gauge theory of the strong, weak and electromagnetic interactions, the
content of which is summarized by the group structure SU(3)xSU(2)xU(1),
where SU(3) refers to the theory of strong interactions,QCD, and latter two
factors [SU(2)xU(1)] describe the theory of electroweak interactions.
Although the theory remains incomplete, its development \ represents a
triumph for modern physics (for details see [93] and below).

----------------------------------------

\bigskip

2.3.1. Quarks and leptons.

a) \textbf{Quarks. }We now know that all the known properties of the hadrons
(their quantum numbers, mass, charge, magnetic moment), their excited states
and their decay properties (see, also below) may be explained by assuming
that the mesons are made of quark - antiquark pairs, the baryons of three
quarks and the antibaryons of three antiquarks [81 - 83]. To obtain this
picture we need sic quarks: up (u), down (d), charm \ (c), strange (s), top
(t) and bottom (beauty) (b) (see Table 9). These six particles may be
arranged according to their masses into three pairs, with one number of each
pair having a charge +2/3e and the other -1/3e as shown in Table 9.Since
quarks have not been observed in isolation - they appear either as bound
quark - antiquark$^{6}$*$^{)}$ pairs in the form of mesons or bound groups
in the form of baryons (see, also Fig. 15 below) - the name assigned to
them, up, down, strange, etc., are only mnemonic symbols \ to identify of
different species. The word "flavor" is used, for convenience, to
distinguish between different types of quark. Besides flavor, quarks also
come in three \ different color, for example,

---------------------------

6*) The first question that occur is whether the quarks actually exist
inside the hadrons or whether they are merely a convenient mathematical
ingredient leading to the geometrical symmetry [10]. A substantial clue in
this direction is obtained in deep inelastic scattering from nucleons [14 -
16]. The nucleon appears to be made up of to regions in the asymptotic free
regime [86 - 88], and the outer region of the meson cloud where pions and
other heavy mesons can exist (see, also [89]). A number of early results on
the internal proton structure became accessible through highly inelastic
electron scattering carried out at the Stanford Linear Accelerator Center
(SLAC). Later work of Kendell et al. [14 -16] helped to identify these
structures with quarks inside the proton (for details see also [90]).

----------------------------------

red, green, and blue. Color and flavor are quantum - mechanical labels, or
in other words, quantum numbers, very similar to spin and parity. Since
there are no classical analogous to flavor and color degrees of freedom,
there are no observables that can be directly associated with them. In this
respect, they are similar to the parity label of a state which must be 
\TEXTsymbol{<}observed\TEXTsymbol{>} through indirect evidence.

At now we know that color charge is the charge associated with strong
interaction. Color is whimsically named attribute of quarks and gluons [90]
that cannot be seen. Gluons have one color and one anticolour [91]. There
are, however, only eight types of gluons [8], not the nine as we might
expect. Quarks and gluons are only found inside hadrons. The quarks inside a
hadron are bathed in a sea of gluons (and additional quark-antiquark pairs)
that are responsible for the binding forces in the hadron. Quarks
continually emit and absorb gluons. Color charge is conserved in every such
process. The color - mathematics always work out so that at any instant the
entire hadron system is color neutral.

For quarks, the interaction is very strong at low energies where nuclear
physics operates and where most of the experimental observations are made.
Because of what is generally known as asymptotic freedom [86 - 88], the
quark - quark interaction is weak only at extremely high energies. As a
result, perturbational techniques apply to QCD only at such extremes, far
beyond the realm of nuclear physics and low-lying hadron spectroscopy. Since
\ quarks are not observed in isolation, their mutual interaction must have a
component that grows stronger as the distance of separation between them
increases. This is opposite to our experience in the macroscopic world,
where interactions, such as gravitational and electromagnetic, grow weaker
as the distance of separation between the interacting objects is increased
(and the relation is given by the inverse square low).

From the above text it has become clear that protons and neutrons are no
longer considered as elementary (see, also Fig 15) but are composed of
quarks in a bound state. The binding forces are quit distinct from
electromagnetic, gravitational forces: at very short distance, the quarks
appear to move freely but, with increasing separation, the binding forces
increase in strength too. So it is not possible to separate the nucleon into
its constituent quarks$^{7\ast )}$. From this picture is followed that
quarks are to be able to exist only in combination with other quarks
(baryons) or with antiquarks (mesons) [56, 90]. This picture has also
modified our ultimate view of a system of densely packed nucleons. For
composite

----------------------------

7*) As we know, nonrelativistic quark model use constituent quark masses,
which are \ of order 350 MeV for u- and d-quarks. Constituent quark \ masses
model the effect of dynamical chiral symmetry breaking are not related to
the quark mass parameters m$_{q}$ of the QCD Lagrangian.

\bigskip

nucleons, interpenetration will occur if the density is increased high
enough and each quark will find many other quarks in its immediate vicinity
(see Fig. 16). The concepts of a nucleon and of nuclear matter become ill -
defined at this high - energy limit and a new state of matter might
eventually be formed: a quark plasma whose basic constituents are unbound
quarks [91; 92]. Starting with matter of vanishing baryon density, the
energy density of a non - interacting gas of massless quarks and gluons is
(see, also [92])

E $\simeq $ 12T$^{4}$ \ \ \ \ \ \ \ \ \ \ \ \ \ \ \ \ \ \ \ \ \ \ \ \ \ \ \
\ \ \ \ \ \ \ \ \ \ \ \ \ \ \ \ \ \ \ \ \ \ \ \ \ \ \ \ \ \ \ \ \ \ \ \ \
(30),

here T is temperature. Just like in the Stefan - Boltzman for a proton gas,
the numerical factor in (30) is determined by the number of degrees of
freedom of the constituent particles: their spins, colors, and flavors. The
energy density for quarks plasma via computer simulations obtained in [92].
The transition temperature from the mesonic regime to the plasma regime is
around 200 MeV which means an energy density of at least 2.5 GeVfm$^{-3}$ in
order to create a quark - gluon plasma.

b)\textbf{\ Leptons.} Leptons are fundamental particles that have no strong
interactions The six known types of leptons are shown in the Table 10.The
family of leptons is further subdivided into three generation; the electron
and the electron neutrino (e, $\nu _{e}$), the muon and the muon neutrino ($%
\mu $, $\nu _{\mu }$), the $\tau $ and the $\tau $ - neutrino ($\tau $, $\nu
_{\tau }$). There are also six antilepton types, one for each lepton [60,
78, 80, 83]. Leptons, or light particle (from greek), are not made of
quarks. They, as is indicated above, in electromagnetic and weak
interactions \ but not in strong interaction. As we can see from Table 10
the neitrinos are known to be much lighter and their rest masses may even be
zero. A large amount of effort has been devoted in recent years to measuring
the mass $\nu _{e}$. The best estimate at the moment is that it is a few eV
or less, although much larger values for the upper limit have also been
reported. As we can see from Table 10 for the other two types of neutrinos,
only the upper limits of their masses are known: m$_{\nu _{\mu }}$ 
\TEXTsymbol{<} 0.25 Mev and m$_{\nu _{\tau }}$ \TEXTsymbol{<} 30 MeV. In
nuclear physics, leptons make their presence felt through nuclear $\beta $ -
decay and other weak transition. In general, only electrons and electron
neutrinos are involved; occasionally muons may enter, such as in the case of
a muonic atom where a muon replace one of the electrons in atom.

In addition to the properties listed in Table 10, each lepton has a lepton
number L = 1 (-1 for the corresponding antilepton) and the numbers of each
generation are further characterized by a new number, the electron - lepton L%
$_{e}$ = 1, the muon - lepton number L$_{\mu }$ = 1, and the tau - lepton
number L$_{\tau }$ = 1, respectively. Within each generation the anrilepton
have L$_{e}$ = -1, L$_{\mu }$ = -1, L$_{\tau }$ = -1. These numbers, the
lepton number and the lepton number generation, are \ conserved in all
reactions involving leptons (see, e.g. 83, 90]).

The leptons are often produced in hadron decay, that is when a heavier
particle subject to the strong interaction spontaneously transforms into
lower mass particles. When this occurs, the decay is due to the weak
interaction and the corresponding particle life - time is enormously longer
than that corresponding to a strong interaction decay. The first of these
decays to be discovered was the transformation of a neutron into a proton
(Fig. 17). Charge conservation requires that together with proton a
negatively charged particle be produced, and conservation of the energy
requires that this particle be an electron. However, this is not sufficient
to conserve angular momentum because the three particles are all fermions
with spin 1/2 and a system made by a proton \ and an electron alone has
integer spin. The decay into a proton and an electron alone also disagrees
with the observation that the proton and the electron do not have definite
energies \ as would happen in a two - body decay. It was then postulated by
Pauli [see, e.g.[5]) principle that \ another particle very difficult to
observe is emitted at the same time. It is fermion with spin 1/2, which was
called the neutrino. The neutron decay was thus written

n $\rightarrow $ p + e$^{-}$ + $\widetilde{\nu _{e}}$ \ \ \ \ \ \ \ \ \ \ \
\ \ \ \ \ \ \ \ \ \ \ \ \ \ \ \ \ \ \ \ \ \ \ \ \ \ \ \ \ \ \ \ \ \ \ \ \ \
\ \ (31),

where $\widetilde{\nu _{e}}$ is an electron antineutrino. This decay
illustrates some of this conservation lows which govern particle decays.

The proton in the product satisfies the conservation of baryon number, nut
the emergence of the electron unaccompanied would violate conservation of
lepton number. The third particle must be an electron antineutrino to allow
the decay to satisfy lepton number conservation The electron has lepton
number 1 and the antineutrino has lepton number -1. However a proton bound
in a nucleus may also transform into neutron by emitting a positron and a
neutrino. This process is a bound as $\beta ^{+}$ - decay and discussed in
any textbooks (see, e.g. [9, 10]). Also for this transformation the above
consideration hold and the proton transformation into a neutron was written

p$_{bound}$ \ $\rightarrow $ n + e$^{+}$ + $\nu _{e}$ \ \ \ \ \ \ \ \ \ \ \
\ \ \ \ \ \ \ \ \ \ \ \ \ \ \ \ \ \ \ \ \ \ \ \ \ \ \ \ \ \ \ \ \ \ \ \ \
(32),

where $\nu _{e}$ \ is an electron neutrino. In conclusion of this part, we
should note, that the lepton number conservation rule is applied to all
cases it is found to work.

As we well-known at present time all hadrons are subdivided into two
classes, baryons and mesons (see Fig. 15). Baryons are distinguished by the
fact that they are fermions, particles that obey Fermi - Dirac statistics.
Because of this property, two baryons cannot occupy the same quantum -
mechanical state. The fact that baryons are fermions implies that quarks
must also be fermions, as it is impossible to construct fermions except from
odd numbers of fermions. Furthermore, if we accept that a quark cannot exist
as a free particle, the lightest fermion in the hadron family must be made
of three quarks (see also Fig. 15). Among the baryons, we are mostly
concerned with the lightest pair, the neutron and the proton. From charge
conservation alone, it can be deduced that a proton carrying a charge +e,
must be made of two u - quarks, each having a charge of 2/3e (Table 9), and
one d - quark, -1/3e. The quark wave function of a proton may be represented
as

$\mid $p\TEXTsymbol{>} = $\mid $uud\TEXTsymbol{>} \ \ \ \ \ \ \ \ \ \ \ \ \
\ \ \ \ \ \ \ \ \ \ \ \ \ \ \ \ \ \ \ \ \ \ \ \ \ \ \ \ \ \ \ \ \ \ \ \ \ \
\ \ \ \ \ \ \ \ \ \ \ \ \ \ \ (33).

Similarly, the quark wave function of a neutron is

$\mid $n\TEXTsymbol{>} = $\mid $udd\TEXTsymbol{>} \ \ \ \ \ \ \ \ \ \ \ \ \
\ \ \ \ \ \ \ \ \ \ \ \ \ \ \ \ \ \ \ \ \ \ \ \ \ \ \ \ \ \ \ \ \ \ \ \ \ \
\ \ \ \ \ \ \ \ \ \ \ \ \ \ \ (34)

so that the total charge of a neutron in units of e is 2/3 -1/3 - 1/3 = 0.

Boson particles obeyng Bose - Einstein statistics may be made from even
number of fermions. This means \ that mesons are constructed of an even of
quarks. Since, on the one hand, bosons can be created or annihilated under
suitable conditions and, on the other hand, the number of quarks is
conserved in strong interaction processes, a meson must be made of an equal
number of quarks (see, also Fig. 15). The simplest meson is, therefore made
of quark - antiquark. For example, pions ($\pi $), the lightest members
among \ the mesons, are made of a quark, either u or d and an antiquark,
either $\overline{u}$ or $\overline{d}$ (see, e.g. [90]).

\bigskip

2.3.2. Strong and weak interactions.

\bigskip

As was indicated above, our present knowledge of physical phenomena suggests
that there are four types of forces between physical bodies:

1) gravitational;

2) electromagnetic;

3) strong;

4) weak.

The gravitational and the electromagnetic forces vary in strength as the
inverse square of the distance and so are able to influence the state of an
object even at very large distances whereas the strong and weak forces fall
off exponentially and so act only at extremely short distances.

One common feature of the nuclear interactions is their exchange and tensor
character: they occur through exchange of quanta, virtual particles that
exist only for a very short time and cannot be detected experimentally. As
we know the short range character of the strong and weak interactions is due
to the finite mass (in opposite of electrodynamics with massless photon) of
the quanta of the respective fields. These are various kind of meson in the
case of the strong force and W$^{\pm }$ \ and Z \ particles in the case of
the weak interaction. In fact if the exchanged virtual particle has a mass
m, its minimum energy is rest mass energy mc$^{2}$. Thus

$\Delta $t $\leq $ $\frac{\hbar }{\text{mc}^{2}}$ \ \ \ \ \ \ \ \ \ \ \ \ \
\ \ \ \ \ \ \ \ \ \ \ \ \ \ \ \ \ \ \ \ \ \ \ \ \ \ \ \ \ \ \ \ \ (35)

and the range of the interaction is always smaller than

R $\leq $ $\frac{\hbar \text{c}}{\text{mc}^{2}}$ = $\frac{\hbar }{\text{mc}}$
\ \ \ \ \ \ \ \ \ \ \ \ \ \ \ \ \ \ \ \ \ \ \ \ \ \ \ \ \ \ \ \ \ \ \ \ \ \
(36).

In the case of the strong interaction R $\preceq $ 1.4 fm, so that the
quanta of the field have a mass m $\geq $ 140 MeV/c$^{2}$ [47; 72; 75; 92].
In the case of the weak interaction, the high mass of the W$^{\pm }$ and Z
particles (m$_{W^{\pm }}$ = 80 MeV/c$^{2}$ and \ m$_{Z}$ = 91 MeV /c$^{2}$)
makes R $\simeq $ 10$^{-3}$ fm. \ As already mentioned the basic
interactions (four forces) have different strength. Loosely speaking this
means that when two of these interactions act together the stronger
dominates. A simple criterion to judge the relative strengths is the value
of the coupling constants which characterize the corresponding fields, which
are of the order of unity for the strong, 10$^{-2}$ for electromagnetic, 10$%
^{-5}$ for the weak and 10$^{-39}$ for the gravitational forces (see Table
1). However, this does not mean that, for instance, the strong interaction
is more effective than the electromagnetic interaction in producing bound
states between two interacting particles. For example, the hydrogen atom
comprising a proton and an electron bound by the electromagnetic force has
an infinity of bond states, whereas the deuteron, the bound system formed by
a proton and a neutron interacting through the strong force, has only one
bound state.

Nuclear $\beta $ - decay is one and the oldest of the example of weak
interactions. The basic reaction involving weak interaction in nuclei may be
characterized by the delay of a neutron and a (bound) proton (see (31, 32)
above). on a more fundamental level, $\beta $ - decay of hadrons may be
viewed as the transformation of one type of quark to another through the
exchange of charged weak currents [72, 78, 95]. In a proton, the quark takes
about 300 MeV to flip the spin of a quark (for example u $\rightarrow $ d
quarks transformations). The flavor of quarks is conserved in strong
interactions [92, 94]. However, through weak interactions, it is possible
for quarks to change flavor, for example, by transforming from d - quark to
a u - quark:

d $\rightarrow $ u + e$^{-}$ $\widetilde{\nu _{e}}$ \ \ \ \ \ \ \ \ \ \ \ \
\ \ \ \ \ \ \ \ \ \ \ \ \ \ \ \ \ \ \ \ \ (37).

This is what takes place in the $\beta $ - decay of a neutrons. In terms of
quarks, Eq. (31) may be written

(udd) $\rightarrow $ (uud) + e$^{-}$ + $\widetilde{\nu _{e}}$ \ \ \ \ \ \ \
\ \ \ \ \ \ \ \ \ \ (38).

Similarly, the $\beta ^{+}$ - decay of a bound proton to a neutron involves
the transformation of u -quark to a d - quark

u $\rightarrow $ d + e$^{+}$ + $\nu _{e}$ \ \ \ \ \ \ \ \ \ \ \ \ \ \ \ \ \
\ \ \ \ \ \ \ \ \ \ \ \ \ \ (39).

Diagrammatically\TEXTsymbol{<} the processes given by Eqs. (31, 32) are
represented by diagrams such as those shown in Fig. 17.

Now QCD is known to be the theory of the strong interaction, classification
of the strongly - interacting particles by their transformations properties
under the flavor group SU (3) (and the isospin subgroup SU (2)) follows
naturally [45]. Of course, the flavor symmetries are only approximate (see
e.g. [96]), as they are broken by the differences in the quark masses (see
Table 9). The impact that these mass differences have on hadronic spectra
and interactions is determined by how they compare to the scale of strong
interactions. The SU (2) symmetry f isospin is a very good symmetry of
nature, being violated only at the percent level by both the electromagnetic
interactions and the mass difference between the up and down quarks. The
approximate SU (3) symmetry is violated typically at the $\sim $ 30\% level
due to the large mass difference between the strange quark and the up and
down quarks as compared to the scale of the strong interaction ($\sim $
1GeV). Owing to QCD asymptotic-free \ nature (see [86 - 88] and references
therein), perturbation theory can be applied at short distances; the
resulting predictions have achieved a remarkable success, explaining a wide
range of phenomena where large momentum transfers are involved. In the low -
energy domain (see, for example Fig. 1 in Weise paper [96]), however the
growing of the running QCD coupling and the associated confinement of quarks
and gluons make very difficult to perform a through analysis of the QCD
dynamics in terms of these fundamental degrees of freedom. A description in
terms of the hadronic states seems more adequate; unfortunately, given the
richness of the hadronic spectrum, this is also a formidable task.

At very low energies, a great simplification of the strong - interaction
dynamics occurs. In this energy domain, this simplification allowed the
development of a powerful theoretical framework, Chiral Perturbation Theory
(ChPT) [96], to systematically analyze the low - energy implications of the
QCD symmetries. This formalism is based on two key ingredients: the chiral
symmetry properties of QCD and the concept of effective field theory (for
details see [96]).

So, Yukawa's original paper [97] which then because the pion is still, more
than seventy years after its first release, a generic starting point for our
understanding of nuclear systems and interactions. Indications of ChPT so
far are promising that this framework, constrained by the symmetry breaking
pattern of low - energy QCD, can serve as foundation for a \ modern theory
of the nucleus.

Thus, the gauge theory of the electroweak interaction has made it clear that
the quark and lepton masses do not have the status of basic constant of
nature: the masses of, say, the electron not occur in the Lagrangian of the
theory (for details see [95, 96]).

\bigskip

2.4. Isotope effect in nuclear physics.

\bigskip

In this paragraph we'll describe the influence of neutrons on the charge
distribution. This influence has been studied using isotopic shift, the
difference in the charge distributions of nuclei with the same number of
protons but a different number of neutrons. If charge distribution in a
nucleus is independent of neutrons, we expect the isotopic difference to be
negligible. The measured results (see, e.g. [98-100]) indicate that, in
general, the shifts are small but nonzero. Isotope shift (of spectral lines)
can be divided into two classes, that caused by the mass effect and that
resulting from the field effect. The mass effect consists of two parts,
normal and specific, and results from the nucleus having a finite mass. The
normal mass effect can be calculated exactly, while the specific (for
details see, also Chapter 3) mass effect \ present in spectra of atoms with
more than one electron, is very difficult to calculate precisely. Both of
this effects decrease with increasing Z. The field effect, which increase
with increasing Z, arises because of the deviation of the nuclear electric
field from a Coulomb field and can be used to study details of nuclear
structure. this is probably the most important consequence of isotope shift
studies.

In the very light elements the mass effect dominates and can account
qualitatively for the observed shifts. In the heaviest elements the mass
effect is negligible and the field effect can roughly account for the
observed shift. In the element of intermediate mass the two effects are
comparable. As a result, the shifts observed are small because the mass and
field effects within the levels are often in such a direction as to oppose
one another. In order to use the field effect in the demonstration of
nuclear properties, it is necessary that the contribution of the mass and
field effects to be observed shifts be known.

\textbf{Isotope effect in calcium isotopes. }The isotopic shift data [98],
obtained from electron scattering are summarized in Table 11. As we can see,
the difference in the root-mean-square radius $\left\langle
r^{2}\right\rangle ^{1/2}$ between the isotopes given in Table 11 are quite
small. However, the good accuracies achieved in the measured values indicate
a genuine difference among them. Since the radius decrease by 0.01fm in
going from $^{40}$Ca to $^{48}$Ca, it means that the addition of neutrons to
calcium isotopes reduces the size of the charge distribution of the same 20
protons when neutron number is increased from 20 to 28. If we take the
simple view that charges were distributed eventualy throughout the nuclear
volume, the charge radius should have increased by 6\% based on simple R = r$%
_{0}$A$^{1/3}$ relation. This is found to be true in the case of $^{48}$Ti
[98], a nucleus with two more protons and six more neutrons than $^{40}$Ca.
Here, the size of the charge distribution in increased by 0.1 fm for $^{48}$%
Ti not far from the expectation of an A$^{1/3}$ dependence, instead of
decreasing for $^{48}$Ca. There are two possible explanations for the
decrease in the charge radius with increasing neutron number among even
calcium isotopes. The first is that addition of neutrons makes the protons
more tightly bound and, hence, the charge radius is smaller. this is,
however, not true for nuclei in general (see, e.g. [ 8, 12]). A second
explanation is based on the charge distribution within a neutron (see, Fig.
3). One possible model for the charge distribution in a neutron is that the
central part is positive and the region \ near the surface is negative, as
shown in Fig. 3. The detailed charge distribution is not well known, because
of the difficulty in measuring the small charge for-factor (see, also [99]).
However, a small excess of negative charge in the surface region can produce
about a third of decrease in the charge radius in going from $^{40}$Ca to $%
^{48}$Ca, as suggested by the authors of paper [104]. The other two-thirds
may be attributed to the spin dependence in interactions of protons with
other nucleons in the nucleus (see, e.g. [63, 72]). Regardless of the exact
cause of the isotopic shift among calcium isotopes, it is clear that
neutrons have a definite influence on the measured charge distribution of a
nucleus (for details see [98, 99]). The same effect can also observed in
other measurement, for example, such as the energy scattering of x-rays from
muonic atoms [101-103].

\textbf{Isotopic shift in muonic atoms.} As was shown above, a muon is a
lepton with properties very similar to an electron. For this reason, it is
possible to replace one of the electrons on an atom by a (negative) muon to
form muonic atom. However, since the mass of a muon is 207 times larger than
that of an electron, the radii of the muonic orbits are much smaller than
those of electrons.

According atomic physics (see, e.g. [101]) hydrogen-like atom with Z protons
in the nucleus and only a single electron outside, the radius of the n-th
orbit is given by

r$_{n}$(e$^{-}$) = $\frac{\text{n}^{2}\hbar ^{2}}{\alpha \hbar \text{cZm}_{e}%
}$ \ \ \ \ \ \ \ \ \ \ \ \ \ \ \ \ \ \ \ \ \ \ \ \ \ \ \ \ \ \ \ \ \ \ \ \ \
\ \ (40)

here, m$_{e}$ is the mass of an electron, $\alpha $ \ is the fine structure
constant. For hydrogen atom (Z = 1), the ground state (n = 1) radius is
well-known Bohr radius (see, also [101]);

a$_{0}$ = $\frac{\hbar }{\alpha \text{cm}_{e}}$ = 5.29$\cdot $10$^{-11}$m \
\ \ \ \ \ \ \ \ \ \ \ \ \ \ \ \ \ \ \ \ \ (41).

Using (40), we can obtain the analogous results for a muonic atom by
replacing m$_{e}$ by m$_{\mu }$

r$_{n}$($\mu ^{-}$) = a$_{0}\frac{\text{n}^{2}\text{m}_{e}}{\text{Zm}_{\mu }}
$ \ \ \ \ \ \ \ \ \ \ \ \ \ \ \ \ \ \ \ \ \ \ \ \ \ \ \ \ \ \ \ \ \ \ \ \ \
\ \ (42).

Using a muon mass m$_{\mu }$ = 106 MeV/c$^{2}$, we obtain for a heavy
nucleus, such as $^{208}$Pb (Z = 82) the radius of the lowest muonic orbit

r$_{1}$($\mu ^{-}$) $\simeq $ 3.1$\cdot $10$^{-15}$m \ \ \ \ \ \ \ \ \ \ \ \
\ \ \ \ \ \ \ \ \ \ \ \ \ \ \ \ \ \ \ \ \ (43)

or 3.1 fm. This is actually smaller than the value of 7.1 fm for the radius
of $^{208}$Pb, estimated using R = r$_{0}$A$^{1/3}$ with r$_{0}$ = 1.2 fm
[11]. A more elaborate calculation [102] shows that the muon spends inside a
heavy nucleus. Being very close to the nuclear surface, the low-lying muonic
orbits are sensitive to the detailed charge distribution of the different
isotopes. The resulting changes in the energy levels may be observed as
shifts in position of lines. Detailed investigation of isotopic shift was
done on the different isotopes of Fe in paper by Shera et al. [102]. Muonic
x-ray spectra from three isotopes of Fe obtained in this paper are shown in
Fig. 18. The isotope shift is large compared with isotope shift of
electronic x-rays, which is typically 10$^{-2}$eV per unit change in A.

\bigskip

2.5. The origin of the mass.

\bigskip

As we well know, that in a nucleus the protons and neutrons, collectively
known as nucleons, are bound together by the strong nuclear force. At a
fundamental level these interactions are described by Quantum Chromodynamics
(QCD), a theory of quarks and gluons carrying color charges that are
asymptotically-free at short distances. However, the quarks and gluons in a
nucleus are very far from being asymptotically-free. Instead they comprise
individual, colorless, nucleons, which largely retain their identity in the
many-body system. The color-singlet nucleons are then bound to each other by
what can be thought of as 'residual' QCD strong interactions. This sketch of
nuclear dynamics from the QCD point of view - brief as it is - makes it
clear that from this standpoint the nucleus is an incredibly complicated,
nonperturbative, quantum-field-theoretic, infinite-body problem.

Understanding the nucleon mass and its dependence of light quark masses is
clearly one of the most fundamental issues [40; 44; 94] in nuclear and
particle physics (see, also [105-110]). A key question concerns the origin
of the nucleon mass: how do almost massless u and d quarks and massless
gluons cooperate dynamically to form a localized baryonic compound with a
mass of almost 1 GeV? An equally fundamental issue is the origin of the
nucleon spin: how is the total angular momentum of the nucleon in its rest
frame distributed between its quarks and gluons and in turn between their
spin and orbital angular momentum? We will not discuss the last question
here further (for details, see e.g. [110-111]).

As we all know, almost all of the mass of the visible Universe is determined
by the mass of the sum of the masses of nucleons in the cosmos. The gluonic
energy density in the presence of three localized valence quarks obviously
plays a decisive role in generating the nucleon mass [44, 108]. Basic QCD
symmetries and the corresponding conserved currents as a guiding principle
to construct effective Lagrangians which represent QCD at low energies and
momenta. A rapidly advancing approach to deal with non-perturbative QCD is
Lattice Gauge Field Theory (see, e.g. [96, 112]).Considerable progress is
being made solving QCD on a discreditsed Euclidean space-time lattice using
powerful computers (for details, see [109] and references therein).Lattice
QCD has progressed to the joint that it can give reliable results concerning
this issue, but with input quark masses still typically an order of
magnitude larger than the actual current quark masses entering the QCD
Lagrangian. Combining CHPT with lattice QCD has thus become a widely used
routine in recent years (see, e.g. Fig. 1 in [109]).

For better understand the origin of the mass we should analyze the QCD
condensates. In QCD by condensates there are called the vacuum mean values 
\TEXTsymbol{<}0$\mid 
\mathbb{Q}
\mid $0\TEXTsymbol{>} of the local (i.e. taken at a single point of
space-time) of the operator $%
\mathbb{Q}
_{i}$(x) which are due to non-perturbative effects. When determining vacuum
condensates one implies the averaging only over non-perturbative
fluctuations. If for some operator $%
\mathbb{Q}
_{i}$ the non-zero vacuum mean value appears also in the perturbation
theory, it should not be taken into account in determination of the
condensate. In other words when determining condensates the perturbative
vacuum mean values should be substracted in calculation of the vacuum
averages. As we know, the perturbation theory series in QCD are asymptotic
series. So, vacuum mean operator values appear due to one or another summing
of asymptotic series. The vacuum mean values of such kind are commonly to be
referred to vacuum condensates [113]. The non-zero value of quark condensate
means the transition of left-hand quark fields into right-hand ones and is
not small value would mean to chiral symmetry violation in QCD. Quark
condensate may be considered as an order parameter in QCD corresponding to
spontaneous violation of the chiral symmetry \ [96, 112].

For quark condensate \TEXTsymbol{<}0$\mid \overline{q}$q$\mid 0>$ (q = u, d
are the fields of u and d quarks) there holds the Gell-Mann-Oakes-Renner
(GMOR) relation [114]

\TEXTsymbol{<}0$\mid \overline{q}$q$\mid 0>$ = -$\frac{\text{1}}{\text{2}}%
\frac{\text{m}_{\pi }^{2}\text{f}_{\pi }^{2}}{\text{m}_{u}\text{+ m}_{d}}$ \
\ \ \ \ \ \ \ \ \ \ \ \ \ \ \ \ \ \ \ \ \ \ \ \ \ (44).

Here m$_{\pi }$, f$_{\pi }$ are the mass and constant of $\pi ^{+}$ - meson
decay (m$_{\pi }$ = 140 MeV, f$_{\pi }$ \ = 92 - 131 MeV for different
authors), m$_{u}$ and m$_{d}$ are the masses of u - and d- quarks. Relation
(44) is obtained in the first order of m$_{u}$, m$_{d}$, m$_{s}$ (for its
derivation see, e.g. [44]). To estimate the value of quark condensate one
may use the values of quark masses m$_{u}$ + m$_{d}$ = 13 MeV [115].
Substituting these values into (44) we get for quark condensate

\TEXTsymbol{<}0$\mid \overline{q}$q$\mid 0>$ = - (0.23 GeV)$^{3}$ $\simeq $
- 1.6 fm$^{-3}$ \ \ \ \ \ \ \ \ \ \ (45).

This condensate is a measure, as note above, \ of spontaneous chiral
symmetry breaking. The non-zero pion mass , on the other hand, reflects the
explicit symmetry breaking by the small quark masses, with m$_{\pi }^{2}$ $%
\sim $ m$_{q}$. It is important to note that m$_{q}$ and \TEXTsymbol{<}0$%
\mid \overline{q}$q$\mid 0>$ are both scale dependent quantities. Only their
product m$_{q}$\TEXTsymbol{<}0$\mid \overline{q}$q$\mid 0>$ is scale
independent, i.e. invariant under the renormalization group.

The appearance of the mass gap $\Gamma $ $\sim $ 1 GeV in the hadron
spectrum is thought to be closely linked to the presence of chiral \
condensate \TEXTsymbol{<}0$\mid \overline{q}$q$\mid 0>$ in the QCD ground
state. Ioffe formula [116], based on QCD sum rules, connects the nucleon
mass M$_{N}$ directly with quark condensate in leading order

M$_{N}$ = - $\left[ \frac{\text{8}\pi ^{2}}{\Lambda _{B}^{2}}<0\mid 
\overline{q}q\mid 0>\right] ^{1/3}$ \ + \ .................. \ (46),

where $\Delta _{B}$ $\sim $ 1 GeV is an auxiliary scale (the Borel mass
[109]) which separates "short" and "long" distance in the QCD sum rule
analysis. While Ioffe's formula needs to be improved by including
condensates of higher dimensions, it nevertheless demonstrates the close
connection between dynamical mass generation and spontaneous chiral symmetry
breaking n QCD. Taking into account the value of quark condensate from Eq.
(45) we get for M$_{N}$

M$_{N}$ = 986.4 MeV \ \ \ \ \ \ \ \ \ \ \ \ \ \ \ \ \ \ \ \ \ (47).

The obtained value of M$_{N}$ differs from experimental meaning of M$_{N}$ =
940 MeV on the 5\%. For nuclear physics, Eqs. (46-47) give important hint:
the change of the quark condensate with increasing baryon density implies a
significant reduction of the nucleon mass in the nuclear medium.

In the chiral effective theory, the quark mass dependence of M$_{N}$
translates into dependence on the pion mass at leading order. The systematic
chiral expansion [96] of the nucleon mass gives an expression of the form

M$_{N}$ = M$_{0}$ + cm$_{\pi }^{2}$ + dm$_{\pi }^{4}$ - $\frac{\text{3}\pi }{%
\text{2}}$g$_{A}^{2}$m$_{\pi }\left( \frac{\text{m}_{\pi }}{\text{4}\pi 
\text{f}_{\pi }}\right) ^{2}\left( \text{1 - }\frac{\text{m}_{\pi }^{2}}{%
\text{8M}_{0}^{2}}\right) $ + ................ \ \ \ \ \ \ \ \ \ \ \ \ (48),

where the coefficients c and d multiplying even powers of the pion mass
include low-energy constants constrained by pion - nucleon scattering. Note,
that the coefficient d also involves a logm$_{\pi }$ term.

In conclusion of this paragraph we should note, that the fact m$_{d}$ is
larger than m$_{u}$ by a few MeV implies, that the neutron is heavier than
the proton by a few MeV. As is well-known, the experimental neutron-proton
mass difference of M$_{n}$ - M$_{p}$ = 1.2933317$\pm $0.0000005 MeV [60, 79]
receives an estimated electromagnetic contribution of [105] M$_{n}$ - M$%
_{p}\mid ^{em}$ = - 0.76 $\pm $ 0.30 MeV and the remaining mass difference
is due to a strong isospin breaking contribution M$_{n}$ - M$_{\text{p}}\mid
^{d\text{ - u}}$ = 2.05 $\pm $ 0.30 MeV. Recently Bean et al. [110] have
performed the first lattice calculation of the neutron-proton mass
difference arising from the difference between the mass of the up and down
quarks (see, also [95] and find M$_{n}$ - M$_{p}\mid ^{d\text{ - u}}$= 2.26 $%
\pm $ 0.57 MeV. This value is good agreement with the experimental result
quoted above. Concluding we should note, that we do not know why the
observed mass pattern (M$_{n}$, M$_{p}$, m$_{u}$, m$_{d}$ etc.) looks like
this, but nuclear physics can analyze the consequence of this empirical fact.

\bigskip

2.6. New physics beyond the Standard Model.

\bigskip

A major challenge for physics today is to find the fundamental theory beyond
the Standard Model [45] (the "Theory of everything"). In a nutshell, the
Standard Model (SM) is a unified gauge theory of the strong, weak and
electromagnetic interactions, the content which is summarized by the group
theory

SU(3)$_{C}$ x SU(2)$_{L}$ x U(1)$_{Y}$ \ \ \ \ \ \ \ \ \ \ \ \ \ \ \ \ \ \ \
\ \ \ \ \ \ \ \ (49),

where the first factor refers to the theory of strong interactions, or
Quantum Chromodynamics (QCD) [96, 108], and the latter two factors describe
the theory of electroweak interactions (see, also [39, 45, 85, 117, 118,
119, 120, 121]). However, we have the difficulty that the vast majority of
the available experimental information, at least in principle [118 - 121],
explained by the SM. Also, until now, there has been no convincing evidence
for existence of any particles other \ than those of the SM and states
composed \ of SM particles. All accelerator physics seems to fit well with
the SM, except for neutrino oscillations [39]. Apart from neutrino masses
and mixing angles the only phenological evidence for going beyond the SM
comes from cosmology and astrophysics [122]. It is well-known that the pure
SM predicts a too low value for the baryon number resulting from the Big
Bang [1]. Apart from these astrophysical problems, there is only very weak
experimental evidence for effects which do not match the SM extended to
include neutrino mass as well as hierarchy of elementary particles mass etc.
From these standpoints, the SM has been an enormously successful theory.
Nevertheless, there exist many reasons for believing that the SM is not the
end of the story. Perhaps the most obvious is the number of independent
parameters that must be put in by hand. For example, the minimal version of
the SM has 21 free parameters, assuming massless neutrino s and not
accounting electric charge assignments [45]. Most physicists believe that
this is just too much for the fundamental theory. The complications of the
SM can also be described in terms of a number of problems, which we list
briefly below.

1. Coupling Unification.

There exists a strongly held belief among particle physicists and
cosmologists that in the first moments of the life of the Universe, all
forces of nature were "unified", that is they all fit into a single gauge
group structure whose interaction strengths were described by a single
coupling parameter, g$_{u}$ (see, Fig. 2. in [40]). As we can see from this
Fig., that the three SM coupling almost meet at a common point around 3$%
\cdot $10$^{16}$ GeV.

2. The Hierarchy Problem.

As we know, all matter under ordinary terrestial conditions can be
constructed on of the fermions ($\nu _{e}$, e$^{-}$, u, d) of the first
family (see Table 9). Yet we know also from laboratory studies that there
are two families ($\nu _{\mu \text{, }}\mu ^{-}$, c, s) and ($\nu _{\tau }$, 
$\tau ^{-}$, t, b) are heavier copies of the first family with no obvious
role in nature. The SM gives no explanation for the existence of these
heavies families. Futher more, there is no explanation or prediction of the
fermion masses, which over at least five orders of magnitude:

M$_{W,Z}$ $\sim $ m$_{top}$ \TEXTsymbol{>}\TEXTsymbol{>}m$_{b}>>$ \ m$_{\tau
}$ \TEXTsymbol{>}\TEXTsymbol{>} m$_{e}$ \TEXTsymbol{>}\TEXTsymbol{>} m$_{\nu
}$\ \ \ \ \ \ \ \ \ \ \ \ \ \ \ \ \ \ \ \ \ \ (50).

How does one explain this hierarchy of masses? The SM gives us no clue as to
how to explain the hierarchy problem.. Really, the problem is just too
complicated. Simple grand unified theory (GUT) don't help very much with
this (for details see e.g. [45] and references therein). We should repeat
that the non-vanishing neutrino masses and mixings are direct evidence for
new physics beyond the SM.

3. Discrete Symmetry Violation.

By construction, the SM is maximally parity-violating, it was built to
account for observations that weak c.c. processes involve left-handed
particles (or right-handed antiparticles). But why this mismatch between
right-handness and left-handness? Again no deeper reason for the violation
of parity is apparent from the SM. It would be desirable to have answer to
this question, but it will take some new framework to provide them.

4. Baryon Asymmetry of the Universe.

Why do we observe more matter than anti-matter? This is problem for both
cosmology and the SM.

5.Graviton Problem.

Gravity is not fundamentally unified with other interactions in the SM,
although it is possible to graft on classical general relativity by hand.
However, this is not a quantum theory, and there is no obvious \ way to
generate one within the SM context. In addition to the fact that gravity is
not unified and not quantized there is another difficulty, namely the
cosmological constant (for details, see [122] and references therein). The
cosmological constant can be thought of as energy of the vacuum. The energy
density induced by spontaneous symmetry breaking is some $\sim $ 120 orders
of magnitude larger than the observational upper limit. This implies the
necessity of severe fine-tuning between the generated and bare pieces, which
do not have any a priory reason to be related (see, also [123]).

6. Quantization of Electric Charge.

The SM does not motivate electromagnetic charge quantization (for example,
for quarks), but simply takes it as an input. The deeper origin of charge
quantization is not apparent from the SM.

To summarize, despite the triumphant success of the SM, there exist
conceptual motivations for believing that there is something more, that the
high energy desert is not so barren after all.

\bigskip 

\bigskip 

\ \ \ \ \textbf{3. Isotopes in atomic and molecular physics.}

\ \ \ \ \ \ \ \ \ \ \ \ \ \ \ \ \ \ \ \ \ \ \ \ \ \ \ \ \ \ \ \ \ \ \ \ \
3.1. Some general remarks.

\bigskip

The interpretation of atomic isotope shifts relies partly on the knowledge
of nuclear structure. Conversely it can provide some information on the
structure nuclei. This relation between the two fields has been for many
years the main reason for the interest in isotope shifts of optical
(electronic) transitions (see, e.g. reviews and monographs [19, 21, 22, 62,
99, 124, 125, 126]).

The word "atom" introduced by Democritus more than 2400 years ago in Greak
means "inseparable". Atomism as understood by modern science was first
discovered for matter, then for electricity and finally for energies. The
atomism of matter, the recognition of the fact that all the chemical
elements are composed of atoms, followed from chemical investigations. Only
whole atoms react with one another. The first atomic model (at the beginning
of 19 century) assumed that the atoms of all elements are put together out
of hydrogen atoms. As a heuristic principle this hypothesis finally led to a
scheme for ordering the elements based on their chemical properties, the
periodic system of D.I. Mendeleev. More about this subject may be found in
introductory textbooks on chemistry [127]. Continuous investigations of
gases in the course of the 19 century led to the atomism of heat, that is,
to the explanation of teat in general and of the thermodynamic laws in
particular as consequences of atomic motion and collisions. The atomism of
electricity was discovered in 19 century by the English scientist M.
Faraday. Based on the quantitative evaluation of exceedingly careful
measurements of the electrolysis of liquid, M. Faraday concluded "There are
"atoms" of electricity". \ These "atoms" of electrilicity - the electrons
are bound to atoms in matter.

The discovery of the atomism of energy can be dated exactly: on December 14,
1900 (see, e.g. [128]) M. Planck announced the derivation of his laws for
black body radiation in a lecture before the Physical Society in Berlin. In
order to derive these laws, M. Planck assumed that the energy of harmonic
oscillators can only take on discrete values - quite contradictory to the
classical view, in which the energy values form a continuum. This date can
be called the birth date of quantum theory [49].

Our knowledge of the structure of atoms was influenced strongly by the
investigation of optical spectra [129, 130]. The most important sources of
information about the electronic structure and composition of atoms are
spectra in the visible, infrared (IR), ultraviolet (UV) frequency ranges
[101]. Optical spectra are further categorized as line, band and continuous
spectra [129]. Continuous spectra are emitted by radiand solids or high -
density gases. Band spectra consist of groups of large numbers of spectral
lines which are very close to one another. They are generally associated
with molecules [23 - 25]. Line spectra, on the other hand, are typical of
atoms (see, also [20 - 22]). They consist of single lines, which can be
ordered in characteristic series.

The founders of spectroscopic analysis, Kirchhoff and Bunsen, were the first
to discover in the mid - 19$^{th}$ century that each elements possesses its
own characteristic spectrum. Hydrogen is the lightest element, and the
hydrogen atom is the simplest atom, consisting of a proton and an electron.
The spectra of hydrogen atom have played an important role again and again
over the last (20$^{th}$) century in the development of our understanding of
the laws of atomic structure and the structure of matter.

The emission spectrum of atomic hydrogen (Fig. 19) shows three
characteristic lines in the visible region at 6563, 4861 and 4340 \AA\ (H$%
_{\alpha }$, H$_{\beta }$, H$_{\gamma }$, respectively). As we can see from
Fig. 19, the most intense of these lines was discovered in 1853 by Angstr%
\"{o}m; it is now called the H$_{\alpha }$ line. In the near UV region,
these three lines are followed by a whole series of further lines, which
fall closer together in a regular way as they approach a short - wavelength
limit (H$_{\infty }$). During this period empirical regularities in line
spectra were being found. The best known of these was Balmer's simple
formula (1885) for the wavelengths of the visible lines of the hydrogen
spectrum. For the wavenumbers ($\nu $ = 1/$\lambda $) of the lines we write
the Balmer formula

$\nu $ = R$_{H}$($\frac{\text{1}}{\text{2}^{2}}$ - $\frac{\text{1}}{\text{n}%
^{2}}$) \ \ \ \ \ \ \ \ \ \ \ \ \ \ \ \ \ \ \ \ \ \ \ \ \ \ \ \ \ \ \ \ \ \
\ \ \ \ \ \ \ \ (51),

where n - integer and equals n = 3, 4, 5,....... The quantity R$_{H}$ is
called the Rydberg constant and has the numerical value R$_{H}$ = 109677.581
cm$^{-1}$ [101]. The series limit is found for n $\longrightarrow $ $\infty $
to be $\nu $ = R$_{H}$/4. This empirical discoveries of spectral
regularities reach their culmination in the clear establishment of the Ritz
combination principle. This came in 1898 after a decade of important work on
the study of spectral series [130]. According to this result each atom may
be characterized by a set of numbers called terms (R$_{H}$/4), dimensionally
\ like wavenumbers, such that the actual wavenumbers of the spectral lines
are given by the differences between these terms. A comparison of the
calculated obtained from the Balmer formula (Eq. (51)) with the observed
lines [130] shows that the formula is not just a good approximation: the
series is described with great precision. The combination principle suggests
the existence of lines given more generally by

$\nu $ = R$_{H}$($\frac{\text{1}}{\text{m}^{2}}$ - $\frac{\text{1}}{\text{n}%
^{2}}$) \ \ \ \ \ \ \ \ \ \ \ \ \ \ \ \ \ \ \ \ \ \ \ \ \ \ \ \ \ \ \ \ \ \
\ (52),

where m \TEXTsymbol{<} n being integer. The numbers m and n are called
principal quantum numbers. For m = 3 and n = 4, 5, 6, .........the lines
fall in the IR. Paschen found them at the predicted places. Lyman also found
in the UV three lines corresponding to m = 1 and n = 2, 3, 4, ....... Table
12 contains some of the lines from the first four series and thus
illustrates the Ritz Combination Principle. As we can see, the difference of
the frequencies of two lines in a spectral series is equal to the frequency
of a spectral line which actually occurs in another series from the same
atomic spectrum. For example, the frequency difference of the first two
terms in the Lyman series is equal to the frequency of the first line of the
Balmer series. To concluding this paragraph we should repeat that the
frequencies (or wavenumbers) of all spectral lines can be represented as
differences of two terms of form R/n$^{2}$. These are just the energy levels
of the electron in a hydrogen atom. The model of the hydrogen atom consisted
of an electron and proton describing orbits about their centre of mass
according to classical mechanics under their mutual attraction as given by
the Coulomb inverse - square law. The allowed circular orbits were
determined simply by the requirement \ (an additional postulate of quantum
theory [49]) that the angular momentum of the system be an integral multiple
of $\hbar $ = h/2$\pi $. This condition yields an equation from which R$%
_{\infty }$ can be calculated (for details see [101])

R$_{\infty }$ = $\frac{\text{m}_{0}\text{e}^{4}}{8\varepsilon _{0}^{2}\hbar
^{3}\text{c}}$. \ \ \ \ \ \ \ \ \ \ \ \ \ \ \ \ \ \ \ \ \ \ \ \ \ \ \ \ \ \
\ \ \ \ \ \ \ \ \ (53)

From (53) we can find for the Rydberg constant R$_{\infty }$ the numerical
value

R$_{\infty }$ = (109737.318 $\pm $ 0.012)cm$^{-1}$. \ \ \ \ \ \ \ \ \ \ \ \
\ \ \ (54)

This may compared with empirical value in (51). In hydrogen model R is just
the ionization energy of the ground state of the atom with n = 1. Next part
of our review gives an account of what can be understood in experimental
isotope shifts before (or without) separating the two types of contributions
(mass - and field - shift), in other words, the way in which the isotopes
shift changes from one level to the other in a given spectrum. As will be
shown below the latter problem is purely problem of atomic structure.

\bigskip

\ \ \ \ \ \ \ \ \ \ \ \ \ \ \ \ \ \ \ \ \ \ 3.2. Motion of the nucleus -
atomic isotope shift.

\bigskip

The spectroscopically measured quantity R$_{H}$ (see paragraph 3.1) does not
agree exactly with the theoretical quantity R$_{\infty }$ (see [54]). The
difference is about 60 cm$^{-1}$ [130]. The reason for this is the motion of
the nucleus during revolution of the electron, which was neglected in the
above model calculation. As we remember, this calculation was made on the
basis of an infinitely massive nucleus. Now we must take the finite mass of
nucleus into account [129].

As we know from classical mechanics, the motion of two particles, of masses
m and M and at distance r from one another, takes place around the common
centre of gravity. If the centre of gravity is at rest, the total energy of
both particles is that of a fictitious particle which orbits about the
centre of gravity at a distance r and has the mass

$\frac{\text{1}}{\mu }$ = $\frac{\text{1}}{\text{m}}$ + $\frac{\text{1}}{%
\text{M}}$; \ \ \ $\mu $ = $\frac{\text{mM}}{\text{m + M}}$, \ \ \ \ \ \ \ \
\ \ \ \ \ \ \ \ \ \ \ \ \ \ \ \ \ (55)

referred to as the reduced mass. Replace the mass of the orbiting electron, m%
$_{0}$ by $\mu $ and obtain, in agreement with experiment

R = R$_{\infty }$ ($\frac{\text{1}}{\text{1 + m/M}}$). \ \ \ \ \ \ \ \ \ \ \
\ \ \ \ \ \ \ \ \ \ \ \ \ \ \ \ \ \ \ \ \ \ \ \ \ \ \ \ (56)

Here m = m$_{0}$\ \ \ \ - the mass of the orbiting electron, and M, the mass
of the nucleus. The energy corrections due to motion of the nucleus decrease
rapidly with increasing nuclear mass (see Table 13). This observation makes
possible a spectroscopic determination of the mass ratio M/m$_{0}$\ \ \ 

M$_{proton}$/m$_{electron}$ = 1836.15 \ \ \ \ \ \ \ \ \ \ \ \ \ \ (57)

Due to the motion of the nucleus, different isotopes of the same element
have slightly different the frequency of spectral line. This so - called
isotope displacement to the discovery of heavy hydrogen with the mass number
A = 2 (deuterium). It was found that each line in the spectrum of hydrogen
was actually double. The intensity of the second line of each pair was
proportional to the content of deuterium [19, 131]. Fig. 20 shows the H$%
_{\beta }$\ line with the accompanying D$_{\beta }$ at a distance of about 1 
\AA\ in a 1;1 mixture of the two gases. The nucleus of deuterium contains a
neutron in addition to the proton. There are easily measurable differences
in the corresponding lines of the H and D Lyman series:

R$_{H}$\ \ = 109677.584 cm$^{-1}$; \ \ \ R$_{D}$ \ = 109707.419 cm$^{-1}$ \
\ \ \ \ \ \ \ \ \ \ \ \ \ \ \ \ \ (58).

The more precisely the differences in hydrogen - deuterium 1s - 2s isotope
shift was done recently by Huber et al. [132]. These authors exceeds the
accuracy of earlier experiment by more than 2 orders of magnitude. Fig. 21
shows the frequency chain that gas been described in part in [133], where an
absolute measurement of the 1s - 2s frequency has been reported. Authors
[132, 133] determine the isotope shift by fitting a pair of parallel lines
to the hydrogen and deuterium data, thus accounting for a linear frequency
drift of the standard. After averaging of 10 measurements line the one shown
in Fig. 22, these authors obtained the experimental result for the 1s - 2s H
- D isotope shift

$\Delta $f$_{\exp }$ = 670994334.64 (15) kHz \ \ \ \ \ \ \ \ \ \ \ (59).

The uncertainty of 150 kHz \ is dominated \ by the frequency fluctuations of
the CH$_{4}$ - stabilized He - Ne standard. At the precision level the
theoretical contributions to the isotope shift must be reanalyzed. Most of
the H - D isotope shift of the 1s - 2s interval is caused by the different
masses of the nuclei (for details see, also [132]).

\bigskip

\ \ \ \ \ \ \ \ \ \ \ \ \ \ \ 3.3. Separation of mass- and field-shift
contributions.

\bigskip

Before the early sixties, for scientists interested in atomic isotope shift,
the Periodic Table was implicitly divided into three regions (see, e.g.
[19,22]).:

1) the light elements, with approximately Z $\leq $ 30, where mass isotope
shift only was considered to be present;

2) the heavy elements (Z $\geq $ 58), with field isotope shift only, and;

3) between these two regions, namely, beyond the 3d series and before the 4f
series, medium - weight elements with small isotope shifts, difficult to
measure accurately.

Thus mass - and field - shift effects were considered almost independently.
For mass - shift effects, the paper by Vinti [134] was a reference: it
essentially indicates the formal way in which the mass isotope shift changes
from one pure Russel - Saunders (RS) term to another [129]. For field
isotope shift , the basic papers were by Rosenthal and Breit [135] and Racah
[136], both of which considered the case of multielectronic atom: in the
case of multielectronic spectra, the field - shift was considered \ inside a
given configuration, and the way it changes from one configuration to
another was described through the use of "screening factor" introduced by
authors of paper [137].

So, the isotope shift of an optical transition is the sum of two terms: the
mass effect and the field effect. If only one isotope pair is available, the
experiment only yields this sum but not the respective contributions of the
two effects. The situation is then much less favorable than for hyperfine
structure (see, e.g. [20, 21]), because, in this latter case, the Casimir
formula allows a separation of the magnetic - dipole and electric -
quadrupole contributions [138]. Unfortunately, if the mass effect and field
effect contributions cannot be separated, the theoretical interpretations of
experimental result necessary remain rather crude (compare [139]).

\bigskip

\ \ \ \ \ \ \ \ \ \ \ \ \ \ \ \ \ \ \ \ \ \ \ \ \ \ \ \ \ \ \ \ \ \ \ \ \ \
\ \ \ 3.3.1. Mass isotope shift.

\bigskip

The theory for a nucleus of finite mass in an N - electrons has been
considered for the first time by Hughes and Eckart [140]. The correction to
the atomic energy levels due to the nuclear mass motion is given in the non
- relativistic limit by the kinetic energy of the nucleus due to its motion
about \ the centre of mass of the atom [129]

$\Delta $E = $\frac{\text{1}}{\text{2M}}$($\dsum\limits_{i}\overrightarrow{%
p_{i}}^{2}$ + $\dsum\limits_{i\text{ }\neq \text{ j}}\overrightarrow{p_{i}}%
\cdot \overrightarrow{p_{j}}$) \ \ \ \ \ \ \ \ \ \ \ \ \ \ \ \ \ \ \ \ \ \ \
\ (60)

Here M is the nuclear mass and $\overrightarrow{p_{i}}$ - the momentum of
the i-th electron. The sum is over all the electrons in the atom. The effect
of the square terms on the total energy can be evaluated exactly by taking
them together with the corresponding kinetic energy terms of the electrons,
giving the well - known reduced mass correction (see Eq. (55). The
consequent term and line shifts can be simply calculated, and are referred
to as "normal mass shifts" (NMS) [19]. It is usually it will be assumed that
the normal effect has been allowed for in all shifts discussed (see, e.g.
[62]). Shifts arising from the second term are called "specific mass shifts"
(SMS) [140]. The term contains cross products of the momenta of different
electrons and is not susceptible of exact calculation, though in light
elements results in moderate agreement with experiment have been obtained
[19, 22]. For the heavier elements, many electrons are involved and the
calculations become rapidly more complex [125]; no simple rules appear to
exist which would allow even the crudest estimates of SMS for such elements.
Generally, therefore plausibility arguments or semi - empirical methods are
used (see, also [22]).

In the region of the medium - heavy elements, semi - empirical methods for
SMS have been applied, though it appears that such allowance should often be
made also in heavier elements [125]. The total observed shift can be written

$\Delta $E$_{total}$ \ = $\Delta $E$_{\text{NMS}}$ = $\Delta $E$_{\text{SMS}%
} $ + $\Delta $E$_{FS}$, \ \ \ \ \ \ \ \ \ \ \ \ \ \ \ \ \ \ \ \ \ \ \ \ \ \
(61)

where the values of $\Delta $E apply to the differences in the upper and
lower levels involved in a transitions, i.e. they refer to line shifts. From
(61), a relation between the mass shifts in different lines of an element
may be obtained. We use the symbol m for the SMS in a line on the addition
of one neutron and put $\Delta $E$_{total}$ - $\Delta $E$_{NMS}$ = T. Then
the values of T between two isotopes with neutron numbers N and N' \ for two
spectral lines (suffixes 1 and 2) are given by

T$_{1}^{\text{N,N'}}$ = (N' - N)m$_{1}$ + F$_{1}$C$^{N,N^{\prime }};$

T$_{2}^{N,N^{\prime \prime }}$ = (N' -N)m$_{2}$ + F$_{2}$C$^{\text{N,N'}}$.
\ \ \ \ \ \ \ \ \ \ \ \ \ \ \ \ \ \ \ \ \ \ \ \ \ \ \ \ \ \ \ \ \ \ \ \ \
(62)

Here F; C are positive constant and functions of size and shape of the
nucleus [62]. In the last equations it was assumed that m for a heavy
elements independent of N. We should not, that the slight dependence can, of
course, be taken into account, but (62) is sufficiently accurate for most
purposes [125]. The superscript N,N; denotes dependence on these quantities,
the Z dependence has been omitted to make the formulae less cumbersome,
since the argument concerns only one element. From (62) we have

T$_{1}^{\text{N,N'}}$ = T$_{2}^{N,N^{\prime }}\frac{\text{F}_{1}}{\text{F}%
_{2}}$ + (N' - N)(m$_{1}$ - m$_{2}\frac{\text{F}_{1}}{\text{F}_{2}}$). \ \ \
\ \ \ \ \ \ \ \ \ \ \ \ \ \ \ \ \ \ \ (63)

Thus, if two or more corresponding shifts are measured in each of two lines
of an element, F$_{1}$/F$_{2}$ can be found and also (N' - N)(m$_{1}$ - m$%
_{2}\frac{\text{F}_{1}}{\text{F}_{2}}$). King [22] suggested simply plotting
the shifts in one line against the corresponding shifts in another, first
dividing each shift by \ (N' - N). Then the points should lie on a straight
line (see, for example Figs. 2 - 3 in [125]), and the slope and intercept
are F$_{1}$/F$_{2}$ \ and (m$_{1}$ - m$_{2}\frac{\text{F}_{1}}{\text{F}_{2}}$%
) respectively. Whatever the number of lines for which results are
available, it is, of course, impossible by this method to determine any of
of the mass shift uniquely, however, plausibility arguments based on the
nature of the specific mass effect can be used to make estimates, and the
greater the number of lines foe which the total shifts are known, the more
restricted becomes the range of reasonable values of the contributions of
the specific shifts. Further, as King [22] pointed out, if the mass shift
could be computed for one line, the procedure described enables it to be
found for the others from the experimental results [125].

\bigskip

\ \ \ \ \ \ \ \ \ \ \ \ \ \ \ \ \ \ \ \ \ \ \ \ \ \ \ \ \ \ \ \ \ \ \ \ \
3.3.2. Field isotope shift.

\bigskip

In optical and electronic x - ray transitions the field shift is very nearly
proportional$^{8\ast )}$ \TEXTsymbol{<}r$^{2}$\TEXTsymbol{>}$^{1/2}$ and it
is convenient to express this in terms of a "standard shift" based on an
equivalent uniform charge density of radius R$_{eq}$ = r$_{0}$A$^{1/3}$ fm.
This standard unit of isotope shift does not have a fundamental significance
but does represent approximately the overall variation of R$_{eq}$ for
stable nuclei. Because a change \ in neutron number \ is either a more
towards or away from the region of stability there is no reason to expect
isotope shifts to be the same as the standard shift and in fact they are
usually smaller [62].

-----------------------------------

8*) The standard shift calculated on the basis of \TEXTsymbol{<}r$^{2}$%
\TEXTsymbol{>}$^{1/2}$ dependence can be corrected by dividing by the
quantity [1 + 1.2 $\cdot $ 10$^{-5}$Z$^{2}$]. This correction takes into
account the effect of \TEXTsymbol{<}r$^{4}$\TEXTsymbol{>} and \TEXTsymbol{<}r%
$^{6}$\TEXTsymbol{>} terms and is based on the results of Seltzer [141].

\textbf{a) isotope shift in optical spectra.} The nuclear electrostatic
potential acting on the electron depends on the nuclear charge distribution;
if this changes from one isotope to another, the energy of an electron which
penetrates the nucleus will also \ change in the two cases. In the early
works [135, 136], spherically symmetrical nuclear charge distributions of
simple form were assumed, and the values E$_{field}$ and $\Delta $E$_{field}$
were calculated by perturbation treatment, recognized and allowed for to
some extent the results of [135], led to evaluation of $\Delta $E$_{field}$
\ by more rigorous methods (see, also [19]). The foundation for this latter
work were laid by author of paper [142]. He considered the solutions to the
radial Dirac equation \ for the electron: 1) with the nuclear charge assumed
concentrated at a point, and 2) with extended nuclear charge distribution
(for details see [130]). The modern description of the field isotope shift
have the next form:

$\Delta $E$_{field}^{\text{N,N' }}$ = F$_{i}\lambda ^{\text{N,N'}}$ \ \ \ \
\ \ \ \ \ \ \ \ \ \ \ \ \ \ \ \ \ \ \ \ \ \ \ \ \ \ \ \ \ \ \ \ \ \ \ \ \ \
\ \ \ (64)

and

$\lambda ^{\text{N,N'}}$ = $\delta $\TEXTsymbol{<}r$^{2}$\TEXTsymbol{>}$^{%
\text{N,N'}}$ + $\frac{\text{C}_{2}}{\text{C}_{1}}\delta $\TEXTsymbol{<}r$%
^{4}$\TEXTsymbol{>}$^{\text{N,N'}}$ + $\frac{\text{C}_{3}}{\text{C}_{1}}%
\delta $\TEXTsymbol{<}r$^{6}$\TEXTsymbol{>}$^{\text{N,N'}}$ +..............
\ \ \ \ \ \ \ \ (65)

$\lambda $ is the nuclear parameter [99] and F$_{i}$ = E$_{i}$f(z) -
electronic factor. Values of f(z) calculated from the isotope shift
constants are given in Table II of [99]. The ratios C$_{n}$/C$_{1}$ have
been calculated by Seltzer [141]. The procedure of the evaluation of $%
\lambda $ was as follows[99]: first the SMS were estimated, either according
above formula of from King plots [22] of the optical isotope versus results
from muonic isotope shifts [143] or electronic x - ray shifts [19, 124, 125].

\textbf{b) x - ray isotope shift.} As was shown above, atomic isotope shifts
are measured in both optical and x - ray spectra (see Fig. 18). Since the
Coulomb potential at the nucleus is so much larger than the binding energy,
the s - wave function is largely independent of the principal quantum
number. For instance, as was shown by Seltzer [141], the change in $C_{2}$/C$%
_{1}$ (Eq.(65)) is going from 1s to 2s amounts to only a few tenths of
percent. Beyond the 2s level, there is very little further change. $C_{2}$/C$%
_{1}$ for a p$_{1/2}$ level is within a few percent of $C_{2}$/C$_{1}$ for a
s$_{1/2}$ level (4\% for Z = 80 [141]). This means that both atomic optical
and x - ray isotope shifts measure the same parameter $\lambda $. Moreover,
the largest contributions in heavy atoms to the energy shift observed on
going from one isotope to another comes from the modification of the nuclear
charge distribution. The total \ mass shift for K x - ray transitions is
usually less than a few percent of the total isotope shift [144]. The SMS
for optical transitions in heavy atoms is usually neglected. Unlike the x -
ray case, the SMS in optical transitions can be a large part of the total
isotope shift. (for details see [144]).

\bigskip

\ \ \ \ \ \ \ \ \ \ \ \ \ \ \ \ \ \ \ \ \ \ \ \ \ \ \ 3.4. Vibrations in a
diatomic molecule.

\bigskip

Since a diatomic molecule has two atoms and must be linear, it has \ one
degree of vibrational freedom. By convention the z - axis is placed
coincident with the molecular axis. Movement of the atoms in the x - and y -
directions can then be dismissed as molecule rotations. Let us choose a
diatomic molecule AB whose atoms have mass m$_{A}$ (coordinate z$_{A}$) and m%
$_{B}$(z$_{B}$). Moreover, let us now suppose that the atoms are like metal
balls and the bond between them is like a simple spiral spring. If we then
allow the distance between A and B to change by a quantity q = (z$%
_{A}^{^{\prime }}$ - z$_{B}^{^{\prime }}$) - (z$_{A}$ - z$_{B}$), while
keeping the centre of mass at the origin of axes, we could say that the
force (F) exerted by the spring \ on the particles is related to q by (see,
e.g. [26])

F = fq \ \ \ \ \ \ \ \ \ \ \ \ \ \ \ \ \ \ \ \ \ \ \ \ \ \ \ \ \ \ \ \ \ \ \
\ (66).

This is to say the spring obeys Hook's law (linear approximation), and we
shall see that this analogy is a good one. The force exerted by the spring
will cause the atoms to return to their original positions. If we denote the
velocity with which do so by dq/dt \ and the acceleration d$^{2}$q/dt$^{2}$
we may apply Newton's second law of motion and equate the force with mass
times an acceleration. At this point we shall write the reduced mass M (M = $%
\frac{\text{m}_{A}\text{m}_{B}}{\text{m}_{A}\text{ + m}_{B}}$)

-fq = M$\frac{\text{d}^{2}\text{q}}{\text{dt}^{2}}.$ \ \ \ \ \ \ \ \ \ \ \ \
\ \ \ \ \ \ \ \ \ \ \ \ \ \ \ \ \ (67)

This is a second order differential equation which may be solved by making
the substitution

q = Acos (2$\pi \nu $t + $\rho $), \ \ \ \ \ \ \ \ \ \ \ \ \ \ \ (68)

where $\nu $ is a frequency, $\rho $ is a phase factor and A - a maximum
vibration amplitude. The n we have

$\frac{\text{d}^{2}\text{q}}{\text{dt}^{2}}$ \ = - 4A$\pi ^{2}\nu ^{2}$\
cos(2$\pi \nu $t + $\rho $) = -4$\pi ^{2}\nu ^{2}$q. \ \ \ \ \ \ \ \ \ \ \ \
\ \ \ \ (69).

Substituting this back into Equation (67) we obtain

(- 4$\pi ^{2}\nu ^{2}$M + f)q \ \ = 0 \ \ \ \ \ \ \ \ \ \ \ \ \ \ \ \ \ \ \
\ \ \ \ \ \ \ \ \ \ \ \ \ \ \ \ \ \ \ \ \ \ \ \ \ \ \ \ \ \ \ (70)

and assuming q $\neq $ 0, we have

$\nu $ = $\frac{\text{1}}{2\pi }\sqrt{\frac{\text{f}}{\text{M}}}$ \ \ \ \ \
\ \ \ \ \ \ \ \ \ \ \ \ \ \ \ \ \ \ \ \ \ \ \ \ \ \ \ \ \ \ \ \ \ \ \ \ \ \
\ \ \ \ \ \ \ \ \ \ \ \ \ \ \ \ \ \ \ \ \ \ \ \ (71)

This equation is the equations of simple harmonic oscillators.\ \ \ \ \ \ \
\ 

In the last equation \ (71) f is \ proportionality constant. This
proportionality constant is known as a force constant . It could be defined
in terms of Hook's law (Eq. (66)), but this is not convenient. Alternatively
it may be defined in terms of the vibrational potential energy [23]. Before
considering the quantum mechanics of the vibrations in a diatomic molecule,
we must first discuss two assumptions that are implicit in the treatment so
far: that classical mechanics provides an adequate description of a
vibrating molecule, and that Hook's law is valid. The term molecular
coordinate \ is one which we shall meet below in various form. Unlike the
Cartesian coordinates a molecular coordinate does not define the position of
the atoms absolutely, but defines the change in coordinates from some
initial position. Thus, q defined the change in bond length in the diatomic
molecule, but did not define the actual bond length. The term coordinate
will be understood to define some change in the molecular configuration. the
origin of this coordinate system is given by the average or equilibrium
position of the atoms in the molecule, ignoring translation and rotation
(for details see, e.g. [25]). \ \ 

Perhaps the most surprising thing about molecular vibrations is that the
frequencies of vibration may be correctly calculated by means of classical
mechanics. Intuitively\ \ one feels that this is due to the wave nature of
the vibrations. However, quantum mechanics does provide greater inside into
aspect other than the frequencies. To simplify our discussion we will only
consider the simplest case, the harmonic oscillator (see, also [145]).

The wave equation for a one - dimensional oscillator such as a diatomic
molecule is [146]

$\frac{\text{d}^{2\Psi }}{\text{dq}^{2}}$\ \ + $\frac{\text{8}\pi ^{2}\text{M%
}}{\text{h}^{2}}$ (E - V)$\Psi $ \ = 0, \ \ \ \ \ \ \ \ \ \ \ \ \ \ \ \ \ \
\ \ \ \ \ \ \ \ \ (72)

where E is the total vibration energy. In the harmonic oscillator the
potential energy V is given as $\frac{\text{1}}{\text{2}}$fq$^{2}$\ \ and
the wave equation becomes

$\frac{\text{d}^{2\Psi }}{\text{dq}^{2}}$\ \ + $\frac{\text{8}\pi ^{2}\text{M%
}}{\text{h}^{2}}$ (E - $\frac{\text{1}}{\text{2}}$fq$^{2}$)$\Psi $ = 0. \ \
\ \ \ \ \ \ \ \ \ \ \ \ \ \ \ \ \ \ \ \ \ \ \ (73)

The solution of this equation are given in most texts on quantum mechanics
(see, e.g [49]). \ Thus the energy E is given as

E = (n + $\frac{\text{1}}{\text{2}}$) $\frac{\text{h}}{\text{2}\pi }\sqrt{%
\frac{\text{f}}{\text{M}}}$, \ \ \ \ \ \ \ \ \ \ \ \ \ \ \ \ \ \ \ \ \ \ \ \
\ \ \ \ \ \ \ \ \ \ \ \ \ \ \ \ \ \ (74)

where n is a quantum number that may take the values 0, 1, 2, 3, ......
Therefore when the molecule changes from one vibrational state to the next
nearest by one the change in energy is given by

$\Delta $E = $\frac{\text{h}}{\text{2}\pi }\sqrt{\frac{\text{f}}{\text{M}}}$%
, \ \ \ \ \ \ \ \ \ \ \ \ \ \ \ \ \ \ \ \ \ \ \ \ \ \ \ \ \ \ \ \ \ \ \ \ \
\ \ \ \ \ \ \ \ \ \ \ \ \ \ \ \ (75)

and since $\Delta $E = h$\nu $, the frequency of a photon associated with
this change is

$\nu $ = $\frac{\text{1}}{2\pi }\sqrt{\frac{\text{f}}{\text{M}}}$ , \ \ \ \
\ \ \ \ \ \ \ \ \ \ \ \ \ \ \ \ \ \ \ \ \ \ \ \ \ \ \ \ \ \ \ \ \ \ \ \ \ \
\ \ \ \ \ \ \ \ \ \ \ \ \ \ (76)

which is the same result as given by classical mechanics. It can then be
shown that in the general case classical and quantum mechanics give the same
answer for the vibration frequencies; hence we are fully justified in using
classical mechanics to calculate them (see, also [24]). If n = 0 in equation
(74) the molecule has an energy of $\frac{\text{h}}{\text{4}\pi }\sqrt{\frac{%
\text{f}}{\text{M}}}.$ This energy is known as the zero point energy and has
no counterpart in classical mechanics [49].

For a diatomic molecule we may expand the vibrational potential energy as a
Maclaurin series about the position of minimum energy, V$_{0}$ [147,148]

V = V$_{0}$ + ($\frac{\text{dV}}{\text{dq}}$)$_{0}$q + $\frac{\text{1}}{%
\text{2}}$($\frac{\text{d}^{2}\text{V}}{\text{dq}^{2}}$)$_{0}$q$^{2}$ + $%
\frac{\text{1}}{\text{6}}$($\frac{\text{d}^{3}\text{V}}{\text{dq}^{3}}$)$%
_{0} $q$^{3}$ + $\frac{\text{1}}{\text{24}}$($\frac{\text{d}^{4}\text{V}}{%
\text{dq}^{4}}$)$_{0}$q$^{4}$ + ............ \ \ \ \ \ \ \ \ \ \ (77)

The subsript zero indicates the position of minimum potential energy, so
that ($\frac{\text{dV}}{\text{dq}}$)$_{0}$ = 0. The quantity V$_{0}$ is a
constant independent of q and may be ignored since it does not affect the
vibrational frequencies. We have already met the force constant definition f
= $\frac{\text{d}^{2}\text{V}}{\text{dq}^{2}}$, so we may rewrite the
potential energy as

V = $\frac{\text{1}}{\text{2}}$fq$^{2}$ + $\frac{\text{1}}{\text{6}}$($\frac{%
\text{d}^{3}\text{V}}{\text{dq}^{3}}$)$_{0}$q$^{3}$ + $\frac{\text{1}}{\text{%
24}}$($\frac{\text{d}^{4}\text{V}}{\text{dq}^{4}}$)$_{0}$q$^{4}$ +
............ \ \ \ \ \ \ \ \ \ \ \ \ \ \ \ \ \ \ \ \ \ \ \ \ \ \ \ \ \ \ \ \
\ \ \ \ \ \ \ \ \ \ \ \ \ \ \ \ (78)

The calculations in the previous text were based on the approximation that
the terms in power of q higher than two may be ignored. Of course the true
potential energy of a molecule must be more complex, if only because for
large values of q the molecules must dissociate. Fig 23 shows scheme of the
potential energy of typical diatomic molecule as a function of q, while the
dotted line is the parabola calculated for V = $\frac{\text{1}}{\text{2}}$fq$%
^{2}$. The true curve is steeper than the parabola at small internuclear
distances, because of the interatomic repulsion energy\TEXTsymbol{<} but at
large internuclear distance the true potential energy tends asymptotically
to a constant value, \ the dissociation energy Q. Near the equilibrium
internuclear separation (q = 0) the parabola is quite a good approximation
to the potential energy. For this reason, the approximation V = $\frac{\text{%
1}}{\text{2}}$fq$^{2}$ is quite good. Taking V = $\frac{\text{1}}{\text{2}}$%
fq$^{2}$ is known as the harmonic approximation [24], and the potential
energy is said to be quadratic. The force constant f defined as $\frac{\text{%
d}^{2}\text{V}}{\text{dq}^{2}}$ is said to be quadratic force constant. The
addition of the cubic term in Eq. (77) gives a better fit with the true
potential energy curve at the minimum and may be used in accurate work.
However, the equations of motion obtained by using a cubic term in potential
energy are not easy to handle.

One approximate solution to the Schr\"{o}dinger equation that may be found
expresses the energy in terms of the harmonic frequency $\nu _{e}$ and an
anharmonic constant x$_{e}$, thus (see, e.g [148])

E = h$\nu _{e}$(n + $\frac{\text{1}}{\text{2}}$) - h$\nu _{e}$x$_{e}$(n + $%
\frac{\text{1}}{2}$)$^{2}$ \ \ \ \ \ \ \ \ \ \ \ \ \ \ \ \ \ \ \ \ \ \ \ \ \
\ \ \ (79)

Therefore if two transitions corresponding, to $\Delta $n = 1 and $\Delta $n
= 2, can be observed both $\nu _{e}$ and x$_{e}$ may be calculated. The
quantity $\nu _{e}$ may be regarded as the frequency the molecule would have
if it was a harmonic oscillator, so that x$_{e}$ supplies a means of
adjusting the observed frequency. In practice, then, the observed
frequencies are adjusted to give the harmonic frequencies and the theory of
the harmonic oscillator is then applied [23, 147]. It is conventional to
express harmonic frequencies cm$^{-1}$, and to give them the symbol $\omega
_{e}$, where $\omega _{e}$ = $\nu _{e}$/c and c is the velocity of light.
Anharmonic affects molecular vibrations in two important ways [24]. Firstly,
the selection rule (see, also below) derived for the harmonic oscillator, $%
\Delta $n = $\pm $1, ceases to be a rigorous selection rule, and transitions
with, for example, $\Delta $n = $\pm $2 become allowed. Secondly, the
vibrational energy levels are not spaced apart equally by the quantity h$\nu 
$. Thus, not only may it be possible to observe a transitions with $\Delta $%
n = $\pm $2, but this transitions will not have exactly double the frequency
of the transition for which $\Delta $n = $\pm $1 (see, e.g. Fig. 48 in [24]).

\ \ \ \ \ \ \ \ \ \ \ \ \ \ \ \ \ \ \ \ \ \ \ \ \ \ \ 3.4.1. Raman and IR
spectra of molecules.

\bigskip

Vibrational spectroscopy involves the use of light to probe the vibrational
behavior of molecular systems, usually via an absorption and light
scattering experiment. When light interacts with matter, the photons which
make up the light may be absorbed or scattered, or may not interact with the
material and may pass straight through it. If the energy of an incident
photon corresponds to the energy gap between the ground state of a molecule
and an excited state, the photon may be absorbed and the molecule promoted
to the higher energy excited state. \ It is this change which is measured in
absorption spectroscopy by the detection of the loss of that energy of
radiation from the light. however, it is also possible for the photon to
interact with the molecule and scatter from it. In this case there is no
need for the photon to have an energy which matches the difference between
two energy levels of the molecule. The scattered photons can be observed by
collecting light at an angle to the incident light beam, and provided there
is no absorption from any electronic transitions which we have similar
energies to that of the incident light. This process is called the process
of Raman scattering.

Fig. 24 illustrates one key difference between IR and Raman scattering. As
described above\TEXTsymbol{<} IR absorption would involve direct excitation
of the molecule from state m to state n by a photon of exactly the energy
difference between them. In contrast\TEXTsymbol{<} Raman scattering uses
much higher energy radiation and measures the difference in energy between n
and m by substracting the energy of the scattered photon from that of the
beam.

Before it will demonstrate IR and Raman spectra of some molecules, we should
say \ about selection rule for these processes. As we all know, a triatomic
molecule will have three modes of vibrations. They are symmetrical stretch,
a bending \ or deformation mode and an asymmetrical of water (H$_{2}$O) \
shown in Fig. 25. These diagrams use spring and ball models. The spring
represents the bond or bonds between the atoms. The stronger the bod has the
higher frequency. The balls represent the atoms and the heavier they are the
lower the frequency (see Eq. (71)). Thus, it is clear that strong bonds and
light atoms will give higher frequencies of vibration and heavy atoms and
weak bonds will give lower ones. This simple model is widely used to
interpret vibrational spectra. If either molecule vibrates, the electron
cloud will alter as the positive nuclei change position and depending on the
nature of the change, this can cause a change of dipole moment or
polarization. In these triatomic molecules, the symmetrical stretch causes
large polarization changes and hence strong Raman scattering with weak or no
dipole change and hence \ weak or no IR absorption. \ The deformation mode
cause a dipole change but little polarization change and hence strong IR
absorption and weak or non-existent Raman scattering.

Figure 26 shows a comparison of the IR and Raman spectra for bensoic acid
(see, e.g [149] and references therein). The x - axis is given in
wavenumbers for which the unit is cm$^{-1}$. For IR absorption each peak \
represents an energy of radiation absorbed by the molecule. The y - axis
gives the amount of the light absorbed (\% - unit) and is usually shown with
the maximum absorbance as the lowest point on the trace. Raman scattering is
presented only as the Stokes spectrum and is given as a shift in energy from
the energy of the laser beam. This is obtained by substracting the scattered
energy from the laser energy. In this way the difference in energy
corresponding to the ground and excited vibrational states (n and m in Fig.
24). This energy difference is what is measured directly by infrared. The
scattering is measured as light detected by the spectrometer and the maximum
amount of light detected is the highest point on the trace. Strictly
speaking, Raman scattering should be expressed as a shift in energy from
that of the exciting radiation and should be referred to as $\Delta $ cm$%
^{-1}$ but it is often expressed simply as cm$^{-1}$.

\bigskip

\ \ \ \ \ \ \ \ \ \ \ \ \ \ \ \ \ \ \ \ \ \ \ \ \ 3.4.2. Isotope shift in
molecular frequencies.

\bigskip

The study \ of the spectra of molecules in which one or more of their atoms
are substituted by the corresponding isotope can often furnish information
about the structure of the molecule \ which cannot at all, or only with
difficulty, be obtained in any other way. This is especially true for those
molecules in which a hydrogen atom is replaced \ by its heavy isotope
deuterium, because for this substitution the relative change in the masses
is so much greater than for all other isotopic substitutions. In order to
make full use of the material which can be obtained in this way it is
necessary to know exactly the changes which must be expected in the
structure of the corresponding energy levels and wave functions.

When an atom of a molecule is replaced by an isotopic atom of the same
element, it is assumed that the potential energy function and configuration
of the molecule are changed by negligible amount (see, e.g. [24]). The
frequencies of vibration may, however, be appreciably altered because of the
change in mass involved (Eq. (71)). This especially, as delined above, true
if hydrogen is the atom in question because of the large percentage change
in mass. This shift or isotopic effect is very useful for several purposes.
In the first place it may be used to help assign spectral lines to modes of
vibration. Thus a normal mode of vibration in which the hydrogen atom in
question is oscillating with a large relative amplitude will suffer a
greater isotopic change in frequency then a normal mode in which a small
relative amplitude. In the limiting case in which only hydrogen atoms are
moving, replacement of all of them by deuterium atoms should decrease the
corresponding fundamental frequency by the factor 1/$\sqrt{\text{2}}$, this
being the square root of the ratio of masses. The totally symmetric (A$_{1}$%
) vibration of methane [149] is an example of this situation (see, also
[150]). For CH$_{4}$, the frequency is 2914.2 cm$^{-1}$ which decrease to
2084.2 cm$^{-1}$ for the case CD$_{4}$. The ratio $\omega _{CD_{4}}$/$\omega
_{CH_{4}}$ thus has experimental value of $\rho $ = $\nu ^{i}$/$\nu $ = $%
\sqrt{\mu /\mu ^{i}}$ = 0.715, compared to a theoretically expected value of
0.707 [151]. The discrepancy is attributed to the fact that the observed
frequencies are influenced by cubic and quartic terms in the potential
energy (see above Eq. (77)), so that the vibration is nor strictly harmonic
as has been assumed in the theoretical development (for details, see [151]).

As was pointed out above, the potential energy functions of two isotopic
molecules are identical to a high degree of approximation since they depend
only on the motions of the electrons and the Coulomb repulsion of the nuclei
[25]. The latter, of course, is entirely independent of these masses (see
below). Not only the form of the potential curves, but also the relative
positions of the potential curves \ of different electronic energies E$_{e}$
are the same for two isotopic molecules [152]. The mass difference affects
only, as noted above, the vibrational and rotational energy of the molecule
in each electronic state. Restricting our considerations to the non-rotating
molecule, we have for the band systems of two isotopic molecules, neglecting
cubic and higher terms (in Eq. (77))

$\nu $ = $\nu _{e}$ + $\omega ^{\prime }$(v' +$\frac{\text{1}}{\text{2}}$) - 
$\omega _{e}^{\prime }$x'$_{e}$(v' + $\frac{\text{1}}{\text{2}}$)$^{2}$ - [$%
\omega _{e}^{\prime \prime }$(v\textquotedblright\ + $\frac{\text{1}}{2}$) - 
$\omega _{e}^{\prime \prime }$x\textquotedblright $_{e}$(v\textquotedblright%
\ + $\frac{\text{1}}{\text{2}}$)$^{2}$] \ \ \ \ \ \ \ \ \ \ \ \ \ \ \ \ \ \
\ \ \ \ \ (80)

and for isotope molecule

$\nu ^{\prime }$ = $\nu _{e}$ + $\rho \omega ^{\prime }$(v' +$\frac{\text{1}%
}{\text{2}}$) - $\rho ^{2}\omega _{e}^{\prime }$x'$_{e}$(v' + $\frac{\text{1}%
}{\text{2}}$)$^{2}$ - [$\rho \omega _{e}^{\prime \prime }$%
(v\textquotedblright\ + $\frac{\text{1}}{2}$) - $\rho ^{2}\omega
_{e}^{\prime \prime }$x\textquotedblright $_{e}$(v\textquotedblright\ + $%
\frac{\text{1}}{\text{2}}$)$^{2}$] . \ \ \ \ \ \ \ (81)

Here \ $\nu _{e}$\ is the difference in energy of the minima of the
potential curves \ of the two electronic states involved which is the same
to a very good approximation for the two isotopic molecules (see, however
[28]). In the last two formula (80) and (81) one and two stroke \ label the
upper and lower states. The equations (80) and (81) can be written in the
approximation form

$\nu $ = $\nu _{e}$\ + $\nu _{v}$\ ; \ \ \ \ \ \ \ \ \ $\nu ^{i}$\ \ = $\nu
_{e}$\ + $\rho \nu _{v}$\ . \ \ \ \ \ \ \ \ \ \ \ \ \ \ (82).

Following to the quantum mechanival formulae (80 - 81) do give a shift for
the 0 - 0 band, owing to the fact that the zero - point vibration energies
in the upper and lower states, in general, have different magnitudes and are
different for the two isotopic molecules. A shift of the 0 - 0 band is
directly observed in many isotope molecules (see, e.g. [25, 152]). Thus the
existence of zero - point vibration is proved.

a)\textbf{\ isotope effect in water molecule. }Before to describing this
effect we should remind the symmetry of the vibrations of water (see Fig.
25). We reproduce the C$_{2v}$ point group (see, e.g. [153]) in Table 14
which would be correct point group for a single molecule of water. In this
table, the symmetry elements are shown across the top. The first column
contains a series of letters and numbers. The first one we see is A$_{1}$.
This is a way of describing a vibration, or for that matter an electronic
function. It describes what happens to the vibration with each symmetry
element of the molecule. These symbols are called irreducible
representations and the top line always contains the one which refers to the
most symmetrical vibration in terms of its behavior when it is rotated or
reflected \ by the symmetry operations. In higher symmetry point groups
[153] where there is a centre of symmetry, there would also be a g or a u
subscript. There are four possible letters, A,B, E and T. A and B mean that
the vibration is singly degenerate. E means it is doubly degenerate and T
means it is triply degenerate. In the C$_{2v}$ point group all vibrations
are singly degenerate. A is more symmetric than B. Across the line from the
symbols representing the irreducible representations, there are a series of
numbers for each. The numbers are either 1 or - 1 and 1 is more symmetric
than -1. For example, in the Table 14, an A$_{1}$irreducible representation
gives the value of 1 for every symmetry element. Fig. 25 shows three
vibrations of water. For the stretching vibration, when the molecule is
rotated about C$_{2}$ axis, the direction of the arrow representing a
vibration does not change. This is the highest symmetry and is denoted as 1.
In the asymmetric stretching vibration the sign of the arrow is reversed for
C$_{2}$ and one plane. When this happens this is given the number -1. Thus,
this last vibration belongs to a lower symmetry representation. It is
conventionally given the irreducible representation B$_{1}$. the main
advantage of this assignment is that these tables also contain information
that enables us to work on whether the vibration will be allowed by symmetry
or not. For IR, this is done by multiplying the irreducible representation
of the vibration by the irreducible \ representation of x,y or z which is
given in the end column of the point group Table 14 in most, but not all,
layouts. These correspond to three Cartesian coordinates of the molecule and
are irreducible representation of the dipole operator. A similar approach is
adopted for Raman scattering but in this case we look for the more complex
quadratic functions x$^{2}$,y$^{2}$z$^{2}$xy, x$^{2}$ - y$^{2}$ etc., in the
Table14 and these are multiplied by the symmetry representation of the
vibration.

Water is the main absorber of the sunlight. As we can see above, water
molecule (with the molecular formula H$_{2}$O) are symmetric with two mirror
planes of symmetry and a 2 - fold rotation axis. Its molecular diameter is
about 2.75 \AA\ [154]. The water molecule consists of two light atoms (H)
and a relatively heavy atom (O). The approximately 16 - fold difference in
mass gives rise to its ease of rotation and the significant relative
movements of the hydrogen nuclei, which are in constant and significant
relative movement.

The water molecule may vibrate in a number ways (see, Fig. 25). In the gas
state the vibrations involve combination of symmetric stretch ($\nu _{1}$),
asymmetric stretch ($\nu _{3}$) and bending ($\nu _{2}$ of the covalent
bonds with absorption intensity (H$_{2}^{16}$O) $\nu _{1}$;$\nu _{2}$;$\nu
_{3}$ = 0.07; 1.47; 1.0 (see [154] and references therein). The frequencies
of the main vibration of water isotopologues are shown in Table 15. The
dipole moments of the molecule of water change in the direction of the
movement of the oxygen atoms as shown by the arrows on the Fig. 25. As the H
atoms are light, the vibrations have large amplitudes. The water molecule
has a very small moment of inertia on rotation which gives to rich combined
vibrational - rotational spectra in vapor containing tens of thounds to
millions of absorption lines [23, 153]. In the liquid rotations tend to be
restricted by hydrogen bonds, giving the librations. Also, spectral lines
are broader causing overlap of many of the absorption peaks. The main
stretching band in liquid water is shifted to a lower frequency ($\nu _{3}$
= 3490 cm$^{-1}$ \ and $\nu _{1}$ = 3280 cm$^{-1}$) and the bending
frequency increased $\nu _{2}$ 1644 cm$^{-1}$ by hydrogen bonding (see, e.g.
[154]). Isotope shift of water molecular frequencies is shown in Table 16.
IR spectra of ordinary and heavy water are depicted in Fig. 27 (see, also
[156] and references therein). In liquid water the IR spectrum is far more
complex than the vapor due to vibrational overtones and combinations with
librations (see Table 16). The librations depend on the moment of inertia
such that the almost doubling of the moments of inertia of D$_{2}$O,
relative to H$_{2}$O, reduces the frequencies by about a factor of $\sqrt{%
\text{2}}$ (see, however below).

\textbf{b) isotope effect in "fullerene" molecule.} The discovery [157] of
the new fullerene allotropes of carbon, exemplified by C$_{60}$ and soon
followed by an efficient method for their synthesis [158], led to a burst of
theoretical and experimental activity on their physical properties. Much of
this activity concentrated on the vibrational properties of C$_{60}$ and
their elucidation by Raman scattering [159 - 161]. Comparison \ between
theory and experiment (see, e.g. [161]) was greatly simplified by the high
symmetry (I$_{h}$), resulting in only ten Raman active modes for the
isolated molecule and the relative weakness of solid state effect [159],
causing the crystalline C$_{60}$ (c - C$_{60}$) Raman spectrum at low
resolution to deviate only slightly from that expected for the isolated
molecule [160]Since the natural abundance of $^{13}$C is 1.11\% (see, e.g.
[3]), almost half of all \ C$_{60}$ molecules made from natural graphite
contain one or more $^{13}$C isotopes. If the squared frequency of a
vibrational mod in a C$_{60}$ molecule with n$^{13}$C isotopes is written as
a series $\ \omega ^{2}$ = $\omega _{(0)}^{2}$ + $\omega _{(1)}^{2}$ + $%
\omega _{(2)}^{2}$ + $\omega _{(3)}^{2}$ + ...... in the mass perturbation
(where $\omega _{\left( 0\right) }$ is an eigenmode frequency in a C$_{60}$
molecule with 60 $^{12}$C atoms), nondegenerate perturbation theory predicts
for the two totally symmetric A$_{g}$ modes a first - order correction given
(see, e.g. [161])

$\frac{\omega _{(1)}^{2}}{\omega _{(0)}^{2}}$ = - $\frac{\text{n}}{\text{720}%
}$. \ \ \ \ \ \ \ \ \ \ \ \ \ \ \ \ \ \ \ \ \ \ \ \ \ \ \ \ \ (83)

This remarkable result, independent of the relative position of the isotopes
within the molecule and equally independent of the unperturbed eigenvector,
is a direct consequence of the equivalence of all carbon atoms in
icosahedral C$_{60}$. To the same order of accuracy within nondegenerate
perturbation theory, the Raman polarizability derivatives corresponding to
the perturbed modes are equal to their unperturbed counterparts, since the
mode eigenvectors remain unchanged. These results lead to the following
conclusion [161]: The A$_{g}$ Raman spectrum from a set of noninteracting C$%
_{60}$ molecules will mimic their mass spectrum if the isotope effect on
these vibrations can be described in terms of first - order nondegenerate
perturbation theory. It is no means obvious that C$_{60}$ will meet the
requirements for the validity of this simple theorem. A nondegenerate
perturbation expansion is only valid if the A$_{g}$mode is sufficiently
isolated in frequency from its neighboring modes. Such isolation is not, of
course, required by symmetry. Even if a perturbation expansion \ converges,
there is no a priori reason \ why second - and higher - order correction to
Eq. (83) should be negligible. As was shown in paper [161] the experimental
Raman spectrum (see below) of C$_{60}$ does agree with the prediction of Eq.
(83). Moreover, as was shown in quoted paper, experiments with isotopically
enriched \ samples display the striking correlation between mass and Raman
spectra predicted by the above simple theorem. Fig. 28 shows a high -
resolution Raman spectrum at 30 K in an energy range close to the high -
energy pentagonal - pinch A$_{g}$(2) vibration according to [161]. Three
peaks are resolved, with integrated intensity of 1.00; 0.95; and 0.35
relative to the strongest peak. The insert of this figure shows the
evolution of this spectrum as the sample is heated. The peaks cannot be
resolved beyond the melting temperature of CS$_{2}$ at 150 K. The
theoretical fit [161] yields a separation of 0.98 $\pm $ 0.01 cm$^{-1}$
between two main peaks and 1.02 $\pm $ 0.02 cm$^{-1}$ between the second and
third peaks. The fit also yields full widths at half maximum (FDWHM) of
0.64; 0.70 and 0.90 cm$^{-1}$, respectively. The mass spectrum of this
solution shows three strong peaks (Fig. 28$^{b}$) corresponding to mass
numbers 720; 721 and 722, with intensities of 1.00; 0.67 and 0.22
respectively as predicted from the known isotopic abundance of $^{13}$C. The
authors [161] assign the highest - energy peak at 1471 cm$^{-1}$ to the A$%
_{g}$(2) mode of isotopically pure C$_{60}$ (60 $^{12}$C atoms). The second
peak at 1470 cm$^{-1}$ is assigned to C$_{60}$ molecules with one $^{13}$C
isotope, and the third peak at 1469 cm$^{-1}$ to C$_{60}$ molecules with \
two $^{13}$C isotopes. The separation \ between the peaks is in excellent
agreement with the prediction from Eq. (83), which gives 1.02 cm$^{-1}$. In
addition, the width of the Raman peak at 1469 cm$^{-1}$, assigned to a C$%
_{60}$ molecule with two $^{13}$C atoms, is only 30 \% larger than the width
of the other peaks. This is consistent with the prediction of Eq. (83) too,
that the frequency of the mode will be independent of the relative position
of the $^{13}$C isotopes within the molecule. The relative intensity between
two isotope and one isotope Raman lines agrees well with the mass spectrum
ratios. Concluding this part we stress that the Raman spectra of C$_{60}$
molecules show remarkable correlation with their mass spectra. Thus the
study of isotope - related shift offers a sensitive means to probe the
vibrational dynamics of C$_{60}$.

\bigskip

\ \ \ \ \ \ \ \ \ \ \ \ \ \ \ \ \ \ \ \ \ \ \ \ \ 3.4.3. "Mass -
independent" isotope effect.

\bigskip

More than the quarter of a century [162\} Thiemens and Heidenreich [163]
demonstrated essentially equal effects of isotopic substitution of $^{18}$O
and $^{17}$O for $^{16}$O on the rates [164, 165] of formation of ozone by
an electric discharge \ in oxygen (see, also reviews [166 - 168]). Thiemens
and Heidenreich called this observation a "non - mass dependent" or "mass -
independent" isotopic effect. The observed non - mass dependent oxygen \
isotope formation is a kinetic isotope effect [164]. The observed effect was
a remarkable deviation from the accepted theory (see, e.g. [169]) of isotope
effects, which would predict that the effect of $^{18}$O would be
approximately double that of $^{17}$O. The mechanism for the effect remains
uncertain. Since that first publication, the ozone reaction has been
extensively investigated both experimentally and theoretically. A number of
other reactions have been labeled "mass - independent" isotope effect [166].
Mass - independent isotopic composition have been observed in O$_{3}$, CO$%
_{2}$, N$_{2}$O and CO in Earth's atmosphere and in carbonate from a martian
meteorite, which suggests a role for mass - independent processes in the
atmosphere of Mars [170]. According to [167], the observed mass -
independent meteoritic oxygen and sulfur isotopic (see, also [171])
compositions may derive from chemical processes in the presolar nebula and
their distributions could provide insight into early solar system evolution
(see, also [1 - 2]). Although the magnitude and direction of variation of
the isotope ratios for these processes vary, they have one common feature -
they all depend on mass.

The most complete set of experimental data for oxygen isotope effects in the
photochemical formation of ozone has been obtained by the Mauersberger group
[172 - 174]. These authors managed to sort out rate constants (see below)
for many reactions of labeled O atoms and/or labeled O$_{2}$ molecules
providing a severe test for theory (see, also review [168]). At present
time, the most successful theory in terms of agreement with experiment is
that of Marcus group [175 - 179], which is based on the RRKM (Rice -
Ramsperger - Kessel - Marcus) theory of recombination reactions, and this
will be described.

Quantum mechanical resonance calculations as performed by Babikov et al.
[180 - 182] are the ultimate tool for investigating the formation of ozone,
but these attempts are the first in the direction of "first - principle"
solutions of the ozone isotope effect [171]. Resonance lifetimes for large
total angular momenta are required and their determination is a formidable
task. What is the impact of the shallow van der Waals - like wells and the
long - lived states they support? \ Do they contribute to the stabilization
process or are these states so fragile that the first collision with M
(third body) destroys them? What is the role of the excited electronic
states which correlate with ground state products? Moreover, how the
vibrational energy of the excited complex is removed by the bath atom or
molecule is largely not understood.

On the basis of the assumption that any deviation from mass dependence must
reflect of a nuclear process, Hulston and Thode [183] suggested that
deviations from mass - dependent isotopic composition may be used to
distinguish nuclear from chemical and physical processes in meteoritic
measurements [170].The observation by Clayton et al. [184] of a non - mass
dependent isotopic composition in meteoritic substances was thus interpreted
as indicating that the meteorite contained residual primary grains from a
nucleosynthesis process [1 - 2]. Subsequently, the authors of the paper
[163] reported a chemically produced mass - independent isotopic composition
in the formation of ozone from molecular oxygen. The observed fractionation
pattern was the same as observed in meteoritic substances and thus it was
suggested that the observed isotopic composition in a meteoritic material
could reflect a chemical instead a nuclear process.

Gellene [185] modeled the isotope fractionation based on nuclear symmetry.
In his approach, symmetry restrictions arise for homonuclear diatomic (for
example, $^{16}$O$^{16}$ and $^{18}$O$^{18}$O) involved in the O + O$_{2}$
collision because a fraction \ of their rotational states (f - parity)
correlate with those of the corresponding ozone molecule. In contrast, in
the case of heteronuclear oxygen molecules (for example, $^{16}$O$^{18}$O),
all of their rotational states (e and f parity) correlate with those of the
resulting ozone molecule. Gellen's model can reproduce the general features
of the enrichment pattern quite well. A number of others attempts have been
made [186] to find an explanation of the ozone isotope anomaly; none were
able to account for the experimental results [168].

The role of molecular symmetry was questioned by Anderson et al. [172], who
presented rate coefficients of four selected ozone formation channels.
Whereas three channels

$^{16}$O + $^{16}$O + $^{16}$O + M \ \ $\longrightarrow $ \ \ $^{16}$O + $%
^{16}$O + $^{16}$O + M \ \ (84),

$^{18}$O + $^{16}$O + $^{16}$O + M \ \ \ $\longrightarrow $ \ \ $^{18}$O + $%
^{16}$O + $^{16}$O + M \ \ \ (85),

$^{18}$O + $^{18}$O + $^{18}$O + M \ \ \ $\longrightarrow $ \ \ $^{18}$O + $%
^{18}$O + $^{18}$O + M \ \ \ \ (86)

had similar rates of formation, which were consistent with a value of $\sim $
6$\cdot $ 10$^{-34}$ cm$^{6}$s$^{-1}$ [168], the fourth channel

$^{16}$O + $^{18}$O + $^{16}$O + M \ $\longrightarrow $ \ \ \ \ \ $^{16}$O + 
$^{18}$O + $^{16}$O + M \ \ \ \ (87)

resulted in a rate coefficient that was 50\% higher than the other three.
Here M represents a third - body molecule as was indicate early. The
difference in the rate coefficients of reactions 2 and 4 was unexpected.
Thus, the difference in the rate coefficients of reactions 2 and 4 does not
support an important role of symmetry in the isotope enrichment process.
Later, a similar conclusion was reached by Sahested et al. [187] \ who
performed rate coefficient studies for the $^{16}$O - $^{18}$O system on
dual channel processes, using CO$_{2}$ and A as third - body molecules, but
without isolating the process contributing to the enrichments.

More details studies [174] are shown:

1) molecular symmetry cannot explain the ozone enrichment process;

2) a collision between a light oxygen atom and a heavier molecule will
result in a rate coefficient that is higher than the coefficient from
reactions involving only one isotope.

Thus, experimental results presented provide new insights into the puzzling
ozone isotope effect.

To concluding this part we should mentioned some theoretical results of
Marcus et al. Marcus et al. recognized the fact that the ozone isotope
effect in this reaction is very pressure dependence (see, Fig. 29), as
indeed is the rate constant for recombination of an oxygen molecule with an
oxygen atom to form ozone. The reaction may be written

O + O$_{2}$ + M $\ \ \leftrightarrows $ \ \ O$_{3}$* + M \ \ \ \ \ \ \ \ \ \
\ \ \ \ \ \ \ (88),

O$_{3}^{\ast }$ + M \ \ $\longrightarrow $ \ \ O$_{3}$ + M. \ \ \ \ \ \ \ \
\ \ \ \ \ \ \ \ \ \ \ \ \ \ \ \ \ \ (89)

In this reaction \ mechanism, the ozone molecule initially formed with
excess vibrational energy (O$_{3}^{\ast }$) will revert to reactants unless
it is stabilized by energy transfer by collision with another molecule
(third body), M. This is the source of the pressure dependence of the rate
constant. Marcus and co - workers [175 - 179] found two factors important in
the RKKM analysis of the isotope effect in this pressure - dependent region:

a) zero - point energy difference between isotopic species in competing
reactions of asymmetrical molecules, such as

$^{16}$O + $^{16}$O$^{18}$O \ \ $\rightleftarrows $ \ \ $^{16}$O$^{16}$O$%
^{18}$ \ $\rightleftarrows $ \ $^{16}$O$^{16}$O \ + $^{18}$O. \ \ \ \ \ \ \
\ \ \ \ \ \ \ \ \ \ (90)

The lower zero - point energy [164] of the heavier molecule increases the
region of phase space accessible to the transition state, favoring this
channel.

b) a "non RKKM" effect that precludes certain regions of phase space to a
symmetrical transition state, thus preventing the complete randomization of
vibrational energy distribution (for details see [167, 168]).

In addition to the formation of ozone, a number of other three - isotope
systems involving oxygen or sulfur isotopes have been found to exhibit non -
mass dependent isotope effects [166, 171]. The bright example of this
situation a three - isotope system which have $^{32}$S - $^{33}$S - $^{34}$S
combination.

Thus, in the formation of ozone from molecular oxygen by photolysis or in an
electric discharge, the kinetic isotope effects of $^{17}$O and $^{18}$O are
essentially equal, in spite of the predictions of the accepted theory. This
is clearly \ a non - mass dependent isotope effect.

\bigskip

\bigskip \textbf{4. Isotope Effect in Solids.}

\bigskip

4.1. Elementary excitations.

\bigskip

The modern view of solid state physics is based on the presentation of
elementary excitations, having mass, quasiimpuls, electrical charge and so
on (see, e.g [188]). According to this presentation the elementary
excitations of the non - metallic materials are electrons (holes), excitons
(polaritons [189]) and phonons [190]. The last one are the elementary
excitations of the crystal lattice, the dynamics of which is described in
harmonic approximation. as is well - known, the base of such view on solid
are the multiparticle approach. In such view, the quasiparticles of solid
are ideal gas, which described the behavior of the system, e.g.
noninteracting electrons. We should add such approach to consider the theory
of elementary excitations as suitable model for the application of the
common methods of the quantum mechanics for the solution solid state physics
task. In this part of our review will be consider not only the
manifestations of the isotope effect in different solids, but also will
bring the new accurate results, showing the quantitative changes of
different characteristics of phonons and electrons (excitons) in solid with
isotopical substitution (see, also [191]). By the way, isotopic effect
become more pronounced when we dealing with solids. For example, on
substitution of H with D the change in energy of the electron transition in
solid state (e.g. LiH ) is two orders of magnitude larger than in atomic
hydrogen (see, e.g. [48]). In the use of an elementary excitations to
describe the complicated motion of many particles has turned out to be an
extraordinary useful device in contemporary physics, and it is view of a
solid which we describe in this part of review,

The basic Hamiltonian which our model of the solid is of the form [191]

H = H$_{ion}$ + H$_{electron}$ + H$_{electron-ion}$ \ \ \ \ \ \ \ \ \ \ \ \
\ \ \ \ \ \ \ \ \ \ (91)

where

H$_{ion}$ = $\dsum\limits_{i}\frac{\text{p}_{i}^{\text{2}}}{\text{2m}}$ + $%
\frac{\text{1}}{\text{2}}\dsum\limits_{i\neq j}$V(R$_{i}$ - R$_{j}$) \ \ \ \
\ \ \ \ \ \ \ \ \ \ \ \ \ \ \ \ \ \ \ (92),

H$_{electron}$ = $\dsum\limits_{i}\frac{\text{p}_{i}^{\text{2}}}{\text{2m}}$
+ $\frac{\text{1}}{\text{2}}\dsum\limits_{i\neq j}\frac{\text{e}^{\text{2}}}{%
\left\vert \text{r}_{i}\text{ - r}_{j}\right\vert }$ \ \ \ \ \ \ \ \ \ \ \ \
\ \ \ \ \ \ \ \ \ \ \ (93),

H$_{electron-ion}$ = $\dsum\limits_{i,j}$v(r$_{\text{i}}$ - R$_{\text{j}}$)
\ \ \ \ \ \ \ \ \ \ \ \ \ \ \ \ \ \ \ \ \ \ \ \ \ \ \ \ \ \ \ \ \ \ \ (94)

H$_{ion}$ describes a collection of ions (of a single species) which
interact through a potential V(R$_{i}$ - R$_{j}$) which depends only on the
distance between ions. \ By ion we mean a nucleus plus the closed - shell,
or core, electrons, that is, those electrons which are essentially unchanged
when the atoms are brought together to make a solid. H$_{electron}$ presents
the valence electrons (the electrons outside the last closed shell), which
are assumed to interact via a Coulomb interaction. Finally, H$%
_{electron-ion} $ describes the interaction between the electrons (excitons)
and the ions, which is again assumed to be represented by a suitable chosen
potential.

In adopting (91) as our basic Hamiltonian, we have already made a number of
approximation in a treatment of a solid. Thus, in general the interaction
between ions is not well - represented by a potential V(R), when the
coupling between the closed - shell electrons \ on different ions begins to
play an important role (see, e.g. [192, 193]). Again, in using a potential
to represent electron - ion interaction, we have neglected the fact that the
ions possess a structure (the core electrons); again, when the Pauli
principle plays an important role in the interaction between the valence
electrons, that interaction may no longer be represented by a simple
potential. It is desirable to consider the validity of these approximations
in detail (for detail see, e.g. \ [193]). In general one studies only
selected parts of the Hamiltonian (91). Thus, for example, the band theory
of solids is based upon the model Hamiltonian [192, 194].

H$_{B}$ = $\dsum\limits_{i}\frac{\text{p}_{i}^{\text{2}}}{\text{2m}}$ + $%
\dsum\limits_{i,j}$v(r$_{\text{i}}$ - R$_{\text{j0}}$) + V$_{\text{H}}$ (r$%
_{i}$) ,\ \ \ \ \ \ \ \ \ \ \ \ \ \ \ \ \ \ \ \ \ \ \ \ \ \ (95)

where the R$_{\text{j0}}$ represents the fixed equilibrium positions of the
ions and the potential V$_{\text{H}}$ describes the (periodic) Hartree
potential of the electrons. One studies the motion of a single electron in
the periodic field of the ions and the Hartree potential, and takes the
Pauli principle into account in the assignment of one - electron states. In
so doing \ one neglects aspects other than the Hartree potential of the
interaction between electrons. On the other hand, where one is primarily
interested in understanding the interaction between electrons in metals, it
is useful to consider only (93), replacing the effect of the ion cores by a
uniform distribution of positive charge [195]. In this way one can
approximate the role that electron interaction plays without having present
the additional complications introduced by the periodic ion potential. Of
course one wants finally to keep both the periodic ion potential and the
electron interactions, and to include as well the effects associated with
departure of the ions from the equilibrium positions, since only in this way
does not arrive at a generally adequate description of the solid. Usually
for the elementary excitations in solids by first considering various
different parts of the Hamiltonian (91) and then taking into account the
remaining terms which act to couple different excitations.

\bigskip

4.2. Phonons.

\bigskip

The simplest kind of motion in solids is the vibrations of atoms around the
equilibrium point. The interaction of the crystalforming particles with the
one another at the move of the one atom entanglements neighbor atoms [26].
The analysis of this kind motion \ shows that the elementary form of motion
is the wave of the atom displacement. As is well - known that the
quantization of the vibrations of the crystal lattice and after introduction
of the normal coordinates, the Hamiltonian of our task will be have the
following relation (see, e.g. [196])

H(Q,P) = $\dsum\limits_{i,q}\left[ -\frac{\hslash ^{2}}{2}\frac{\partial ^{2}%
}{\partial Q^{2}\text{(}\overrightarrow{q}\text{)}}\text{ + \ \ \ }\frac{%
\text{1}}{2}\omega _{j}^{2}\text{Q}_{j}^{2}\text{(}\overrightarrow{q}\text{)}%
\right] $ \ \ \ \ \ \ \ \ \ \ \ \ \ \ \ \ \ \ \ (96).

In this relation, the sum, where every addend means The Hamiltonian of
linear harmonic oscillator with coordinate Q$_{j}$($\overrightarrow{q}$),
the frequency $\omega _{j}$($\overrightarrow{q}$) and the mass, which equals
a unit. If the Hamiltonian system consists of the sum, where every \ addend
depends on the coordinate and conjugate its quasiimpuls, then according to
quantum mechanics [49] the wave function of the system equals the product of
wave functions of every appropriate addend and the energy is equal to the
sum of assigned energies. Any separate term of the Hamiltonian (96)
corresponds, as indicate above, the linear oscillator

- $\frac{\hslash ^{2}}{\text{2}}\frac{\partial ^{2}\Psi }{\partial \text{Q}%
^{2}}$ + $\frac{\text{1}}{\text{2}}\omega ^{2}$Q$^{2}\Psi $ = $\varepsilon
\Psi $. \ \ \ \ \ \ \ \ \ \ \ \ \ \ \ \ \ \ \ \ \ \ \ \ \ \ \ \ \ \ \ \ \ \
\ \ \ \ \ \ \ \ \ \ \ \ (97)

Solving last equation and finding the eigenvalues and eigenfunctions and
then expressing explicitly the frequency, we will obtain for model with two
atoms in primitive cell (with masses M$_{1}$ and M$_{2}$) the following
equation

$\omega ^{2}$ $\simeq $ 2C($\frac{\text{1}}{\text{M}_{1}}$ + $\frac{\text{1}%
}{\text{M}_{2}}$) \ \ \ \ \ \ \ \ \ \ \ \ \ \ \ \ \ \ \ \ \ \ \ \ \ \ \ \ \
\ \ \ \ \ \ \ \ \ \ \ \ \ \ \ \ \ \ \ \ \ \ \ \ \ \ \ \ \ \ (98)

and

$\omega ^{2}$ $\simeq $ $\frac{\text{C}}{\text{2(M}_{1}\text{ + M}_{2}\text{)%
}}$K$^{2}$a$^{2}$. \ \ \ \ \ \ \ \ \ \ \ \ \ \ \ \ \ \ \ \ \ \ \ \ \ \ \ \ \
\ \ \ \ \ \ \ \ \ \ \ \ \ \ \ \ \ \ \ \ \ \ \ \ \ \ \ \ (99)

Taking into account that K$_{\max }$ = $\pm $ $\pi /a$, where a is a period
of the crystal lattice, i.e. K$_{\max }$ respond the border of the first
Brillouin zone

$\omega ^{2}$ = $\frac{\text{2C}}{\text{M}_{1}}$ and $\omega ^{2}$ = $\frac{%
\text{2C}}{\text{M}_{2}}$ \ \ \ \ \ \ \ \ \ \ \ \ \ \ \ \ \ \ \ \ \ \ \ \ \
\ \ \ \ \ \ \ \ \ \ \ \ \ \ \ \ \ \ \ \ \ \ \ \ \ \ (100)

Formula (98) describes the optical branch of vibrations whereas (99) -
acoustical branch of vibrations \ \ \ (see, e.g. Fig. 5.17$^{a}$ in [197]).
Usually the last formula is written in this way

$\omega $ = $\sqrt{\frac{\alpha }{\text{M}}}$, \ \ \ \ \ \ \ \ \ \ \ \ \ \ \
\ \ \ \ \ \ \ \ \ \ \ \ \ \ \ \ \ \ \ \ \ \ \ \ \ \ \ \ \ \ \ \ \ \ \ \ \ \
\ \ \ \ \ \ \ \ \ \ \ \ \ \ \ \ \ \ \ (101)

where $\alpha $ is so - called the force constant (see, also formula (76)).
Here, as early M is the mass of vibrated atom (ion). From the preceding
relation it is clear that, as in molecular physics, in solid the isotope
effect directly manifests in vibration spectrum, which depends on the
symmetry [198] measures either in IR - absorption or in Raman scattering of
light. Before analyzing Raman scattering spectra of different solids we
briefly consider the classical approximation of the mechanism of Raman
effect [199].

Historically, Raman scattering denotes inelastic scattering of light by
molecular vibrations or by optical phonons in solids. In a macroscopic
picture, the Raman effect in crystals is explained in terms of the
modulation of polarizability by the quasi - particle under consideration. \
The assumption that the polarization depends linearly upon the electric
field strength [ 145] is a good approximation and is invariably used when
discussing \ the scattering of light by crystal excited by lasers. However,
the approximation is not valid for large strength such as can be obtained
from pulsed lasers [147]. The polarization may then be expressed as

P = $\alpha $E + $\frac{\text{1}}{\text{2}}\beta $E$^{2}$ + $\frac{\text{1}}{%
\text{6}}\gamma $E$^{3}$ + $\frac{\text{1}}{\text{24}}\delta $E$^{4}$ +
..........., \ (102)

where $\beta $, the first hyperpolarizability coefficient, plays an
important part for large values of E, since it responsible for the
phenomenon of optical harmonic generation using Q - switched lasers.
Isolated atoms have $\beta $ = 0, since, like $\mu $ the dipole moment, it
arises from interactions between atoms. A simplified theory of Rayleigh
scattering, the Raman effect, harmonic generation and hyper Raman scattering
is obtained by setting (see, e.g. [147])

E = E$_{0}$cos$\omega _{0}$t, \ \ \ \ \ \ \ \ \ \ \ \ \ \ \ \ \ \ \ \ \ \ \
\ \ \ \ \ \ \ \ \ \ \ \ \ \ \ \ \ \ \ \ \ \ \ \ \ \ \ \ (103)

$\alpha $ = $\alpha _{0}$ + ($\frac{\partial \alpha }{\partial \text{Q}}$%
)Q,\ \ \ \ \ \ \ \ \ \ \ \ \ \ \ \ \ \ \ \ \ \ \ \ \ \ \ \ \ \ \ \ \ \ \ \ \
\ \ \ \ \ \ \ \ \ \ \ \ \ (104)

$\beta $ \ = $\beta _{0}$ + ($\frac{\partial \beta }{\partial \text{Q}}$)Q,
\ \ \ \ \ \ \ \ \ \ \ \ \ \ \ \ \ \ \ \ \ \ \ \ \ \ \ \ \ \ \ \ \ \ \ \ \ \
\ \ \ \ \ \ \ \ \ \ \ (105)

Q = Q$_{0}$ + cos$\omega _{v}$t. \ \ \ \ \ \ \ \ \ \ \ \ \ \ \ \ \ \ \ \ \ \
\ \ \ \ \ \ \ \ \ \ \ \ \ \ \ \ \ \ \ \ \ \ \ \ (106).

Here Q is a normal coordinate, $\omega _{v}$ is the corresponding
vibrational frequency and $\omega _{0}$ is the laser frequency. After that
we have

P = $\alpha _{0}$E$_{0}$cos$\omega _{v}$t + $\frac{\text{1}}{\text{2}}$($%
\frac{\partial \alpha }{\partial \text{Q}}$)Q$_{0}$E$_{0}$cos$\omega _{0}$%
tcos$\omega _{v}$t + $\frac{\text{1}}{\text{2}}\beta _{0}$E$_{0}^{2}$cos$%
^{2}\omega _{\text{0}}$t + $\frac{\text{1}}{\text{2}}$($\frac{\partial \beta 
}{\partial \text{Q}}$)Q$_{0}$E$_{0}^{2}$cos$^{2}\omega _{0}$tcos$\omega _{v}$%
t. \ \ \ \ \ \ (107)

Then, after small algebra, we obtaine

P = $\alpha _{0}$E$_{0}^{2}$cos$\omega _{v}$t + $\frac{\text{1}}{\text{2}}$($%
\frac{\partial \alpha }{\partial \text{Q}}$)Q$_{0}$E$_{0}$cos($\omega _{0}$
- $\omega _{v}$)t + cos($\omega _{0}$ + $\omega _{v}$)t + $\frac{\text{1}}{%
\text{2}}\beta _{0}$E$_{0}^{2}$ + $\frac{\beta _{0}}{\text{4}}$E$_{0}^{2}$%
cos2$\omega _{0}$t +

+ $\frac{\text{1}}{\text{2}}$Q$_{0}$E$_{0}^{2}$($\frac{\partial \beta }{%
\partial \text{Q}}$)cos(2$\omega _{0}$ - $\omega _{v}$)t + cos(2$\omega _{0}$
- $\omega _{v}$)t. \ \ \ \ \ \ \ \ \ \ \ \ \ \ \ \ (108).

In last relation the first term describes the Rayleigh scattering, second -
Raman scattering, third - d.c. polarization, fourth - frequency doubling and
the last - hyper Raman effect. Thus the hyper Raman effect is observed with
large electric field strength in the vicinity of twice the frequency of the
exciting line with separations corresponding to the vibrational frequencies. 
$\alpha $ and $\beta $ are actually tensors and $\beta $ components $\beta
_{\alpha \beta \gamma }$ which are symmetrical suffixes [200].

Semiconducting crystals (C, Si, Ge, $\alpha $ - Sn) with diamond - type
structure present ideal objects for studying the isotope effect by the Raman
light - scattering method. At present time this is facilitated by the
availability of high - quality crystals grown from isotopically enriched
materials (see, e.g [27] and references therein). In this part our
understanding of first - order Raman light scattering spectra in
isotopically mixed elementary and compound (CuCl, GaN, GaAs) semiconductors
having a zinc blende structure is described. Isotope effect in light
scattering spectra in GE crystals \ was first investigated in the paper by
Agekyan et al. [ 201]. A more detailed study of Raman light scattering
spectra in isotopically \ mixed Ge crystals has been performed by Cardona
and coworkers [27].

It is known that materials having a diamond structure are characterized by
the triply degenerate phonon states in the $\Gamma $ point of the Brillouin
zone ($\overrightarrow{\text{k}}$ = 0). These phonons are active in the
Raman scattering spectra, but not \ in the IR absorption one [198]. Figure 30%
$^{a}$ demonstrates the dependence of the shape and position of the first -
order line of optical phonons in germanium crystal on the isotope
composition at liquid nitrogen temperature (LNT) [202]. The coordinate of
the center of the scattering line is proportional to the square root of the
reduced mass of the unit cell, i.e $\sqrt{\text{M}}$. It is precisely this
dependence that is expected in the harmonic approximation. An additional
frequency shift of the line is observed for the natural and enriched
germanium specimens and is equal, as shown in Ref. [27] to 0.34 $\pm $ 0.04
and 1.06 $\pm $ 0.04 cm$^{-1}$, respectively (see, e.g. Fig. 7 in Ch. 4 of
Ref. [203]).

First - order Raman light - scattering spectrum in diamond crystals also
includes one line with maximum at $\omega _{LTO}$($\Gamma $) = 1332.5 cm$%
^{-1}$. In Fig. 30$^{b}$ the first - order scattering spectrum in diamond
crystals with different isotope concentration is shown [204]. As shown
below, the maximum and the width of the first - order scattering line in
isotopically - mixed diamond crystals are nonlinearly dependent on the
concentration of isotopes x. The maximum shift of this line is 52.3 cm$^{-1}$%
, corresponding to the two limiting values of x = 0 and x = 1. Analogous
structures of first - order light scattering spectra and their dependence on
isotope composition has by now been observed many times, not only in
elementary Si, and $\alpha $ - Sn, but also in compound CuCl and GaN
semiconductors (for more details see reviews [26, 27]). Already short list
of data shows a large dependence of the structure of first - order light -
scattering spectra in diamond as compared to other crystals (Si, Ge). This
is the subject detailed discussion in [205].

Second - order Raman spectra in natural and isotopically mixed diamond have
been studied by Chrenko [206] and Hass et al. [207]. Second - order Raman
spectra in a number of synthetic diamond crystals with different isotope
composition shown \ in Fig. 31 are measured wit resolution ($\sim $ 4 cm$%
^{-1}$) worse than for first - order scattering spectra. The authors of
cited work explain this fact by the weak signal in the measurement of second
- order Raman scattering spectra. It is appropriate to note that the results
obtained in [207] for natural diamond \ (C$_{13_{C}}$ \ = 1.1\%), agree well
with the preceding comprehensive studies of Raman light - scattering spectra
in natural diamond [208]. As is clearly seen from Fig. 31 \ the structure of
second - order light scattering "follows" the concentration of the \ $^{13}$%
C isotope. It is necessary to add \ that in the paper by Chrenko [206] one
observes a distinct small narrow peak above \ the high - frequency edge of
LO phonons and the concentration of $^{13}$C x = 68\%. Note is \ passing \
that second - order \ spectra in isotopically mixed \ diamond crystals were
measured in the work \ by Chrenko [206] with a better resolution than the
spectra shown in Fig. 31. Second - order Raman light scattering spectra and
IR absorption spectra in crystals of natural and isotopically enriched $%
^{70} $Ge can be found in [26].

A comprehensive interpretation of the whole structure of second - order
Raman light - scattering spectra in pure LIH (LiD) crystals is given in [29,
205, 209]. Leaving this question, let us now analyze the behavior of the
highest frequency peak after the substitution of hydrogen for deuterium
(see, also [209]).

Absorption behavior of an IR - active phonon in mixed crystals with a change
in the concentrations of the components can be classified into two main
types: one and two - mode (see, e.g. the review [210]). Single - mode
behavior means that one always has a band in the spectrum with a maximum
gradually drifting from one endpoint to another. Two - mode behavior is
defined by the presence, in the spectrum, of two bands characteristic of
each components lead not only to changes in the frequencies of their maxima,
but mainly \ to a redistribution of their intensities. In principle, one and
the same system can show different types of behavior at opposite ends.[211].
The described classification is qualitative and is rarely realized in its
pure form (see, also [212]). The most important necessary condition for the
two - mode behavior of phonons (as well as of electrons [213]) is considered
to be the appearance of the localized vibration in the localized defect
limit. In the review [210] \ a simple qualitative criterion for determining
the type of the IR absorption behavior \ in crystals with an NaCl structure
type has been proposed (see also [213]). Since the square of the TO ($\Gamma 
$) phonon frequency is proportional to the reduced mass of the unit cell M,
the shift caused by the defect is equal to

$\Delta $ = $\omega _{\text{TO}}^{2}$(1 - $\frac{\overline{\text{M}}}{%
\overline{\text{M}}^{\prime }}$). \ \ \ \ \ \ \ \ \ \ \ \ \ \ \ \ \ \ \ \ \
\ \ (109)

This quantity \ is compared in [210] with the width of the optical band of
phonons which, neglecting acoustical branches and using the parabolic
dispersion approximation, is written as

W = $\omega _{\text{TO}}^{2}$($\frac{\varepsilon _{0}\text{ - }\varepsilon
_{\infty }}{\varepsilon _{0}\text{ + }\varepsilon _{\infty }}$). \ \ \ \ \ \
\ \ \ \ \ \ \ \ \ \ \ \ \ \ \ \ \ (110)

A local or gap vibrations appears, provided the condition $\left\vert \Delta
\right\vert $ \TEXTsymbol{>} (1/2)W is fulfilled. As mentioned, however, in
[210] in order for the two peaks to exist up to concentrations \ on the
order of $\sim $ 0.5, a stronger condition $\left\vert \Delta \right\vert $ 
\TEXTsymbol{>} W has to met. Substituting the numerical values from Tables 1
and 2 of [205] into formulas (109) and (110) shows that for LiH (LiD) there
holds (since $\Delta $ = 0.44$\omega _{\text{TO}}^{2}$ and W = 0.58$\omega _{%
\text{TO}}^{2}$) the following relation:

$\left\vert \Delta \right\vert $ \TEXTsymbol{>} (1/2)W. \ \ \ \ \ \ \ \ \ \
\ \ \ \ \ \ \ \ \ \ \ \ \ \ \ \ \ \ \ (111)

Thereby it follows that at small concentrations the local vibration should
be observed. This conclusion is in perfect agreement with earlier \
described experimental data [209]. As to the second theoretical relation $%
\Delta $ \TEXTsymbol{>} W, one can see from the above discussion that for
LiH (LiD) crystals the opposite relation, i.e. W \TEXTsymbol{>} $\Delta $,
is observed [214].

Following the results of [215], in Fig. 32 we show the second - order Raman
scattering spectra in mixed LiH$_{x}$D$_{1-x}$ crystals at room temperature.
In addition to what has been said \ on Raman scattering spectra at high
concentration [215], we note that as the concentration grows further (x 
\TEXTsymbol{>} 0.15) one observes in the spectra a decreasing intensity in
the maximum of 2LO ($\Gamma $) phonons in LiD crystal with a simultaneous
growth in intensity of the highest frequency peak in mixed LiH$_{x}$D$_{1-x}$
crystals. The nature of the latter is in the renormalization of LO($\Gamma $%
) vibrations in mixed crystal [216]. Comparison of the structure \ of Raman
scattering spectra (curves 1 and 2 in Fig. 32) allows us, therefore, to
conclude that in the concentration range of 0.1 \TEXTsymbol{<} x \TEXTsymbol{%
<} 0.45 the Raman scattering spectra simultaneously contain peaks of the LO($%
\Gamma $) phonon of pure LiD and the LO($\Gamma $) phonon of the mixed LiH$%
_{x}$D$_{1-x}$ crystal. For further concentration growth (x \TEXTsymbol{>}
0.45) one could mention \ two effects in the Raman scattering spectra of
mixed crystals. The first is related to an essential reconstruction of the
acoustooptical part of the spectrum. This straightforwardly follows from a
comparison of the structure of curves 1 -3 in Fig. 32. The second effect
originates from a further shift of the highest frequency peak \ toward still
higher frequencies, related to the excitation of LO($\Gamma $) phonons. The
limit of this shift is the spectral location of the highest frequency peak
in LiH. Finishing our description of the Raman scattering spectra, it is
necessary to note that a resonance intensity growth of the highest frequency
peak is observed at x \TEXTsymbol{>} 0.15 in all mixed crystals (for more
details see [217]).

Once more reason of the discrepancy between theory and results of the
experiment may be connected with not taking into account in theory the
change of the force-constant at the isotope substitution of the smaller in
size D by H ion [218]. We should stress once more that among the various
possible isotope substitution, by far the most important in vibrational
spectroscopy is the substitution of hydrogen by deuterium. As is well-known,
in the limit of the Born-Oppenheimer approximation the force-constant
calculated at the minimum of the total energy depends upon the electronic
structure and not upon the mass of the atoms. It is usually assumed that the
theoretical values of the phonon frequencies depend upon the force-constants
determined at the minimum of the adiabatic potential energy surface. This
leads to a theoretical ratio $\omega \left( \text{H}\right) $/$\omega \left( 
\text{D }\right) $of the phonon frequencies that always exceed the
experimental data. Very often anharmonicity has been proposed to be
responsible for lower value of this ratio. In isotope effect two different
species of the same atom will have different vibrational frequencies only
because of the difference in isotopic masses. The ratio p of the optical
phonon frequencies for LiH and LiD crystals is given in harmonic
approximation by:

p = $\frac{\omega \left( \text{H}\right) }{\omega \left( \text{D }\right) }$
= $\sqrt{\frac{\text{M}\left( \text{LiD}\right) }{\text{M}\left( \text{LiH}%
\right) }}\simeq $ $\sqrt{\text{2 }}$ \ \ \ \ \ \ \ \ \ \ \ \ \ \ \ \ \ \ \
\ \ \ \ \ \ \ \ \ \ \ \ \ \ \ \ \ \ \ \ (112)

while the experimental value (which includes anharmonic effects) is 1.396 $%
\div $ 1.288 (see Table1 in Ref. [218]). In this Table there are the
experimental and theoretical values of p according to formula (112), as well
as the deviation $\delta $ = $\frac{\text{P}_{Theory}\text{ - p}_{\exp }}{%
\text{p}_{theory}}$ of these values from theoretical ones. Using the least
squares method it was found the empirical formula of ln($\delta $\%) $\sim $
f(ln[$\frac{\partial \text{E}}{\partial \text{M}}]$) which is depicted on
Fig. 33. As can be seen the indicated dependence has in the first
approximation a linear character:

ln($\delta $\%) = -7.5 + 2ln($\frac{\partial \text{E}}{\partial \text{M}}$).
\ \ \ \ \ \ \ \ \ \ \ \ \ \ \ \ \ \ \ \ \ \ \ \ \ \ \ \ \ \ \ \ \ \ \ \ \ \
\ \ \ \ \ \ (113)

From the results of Fig. 33, it can be concluded that only hydrogen
compounds (and its isotope analog - deuterium) need to take into account the
force-constant changes in isotope effect. It is also seen that for
semiconductor compounds (on Fig. 33 - points, which is below of Ox line) the
isotope effect has only the changes of the isotope mass (details see [218]).

Thus, the experimental results presented in this section provide, therefore,
evidence of, first, strong scattering potential (most importantly, for
optical phonons) and, second, of the insufficiency of CPA model for a
consistent description \ of these results [207].

\bigskip

4.3. Electronic excitations.

\bigskip

Isotopic substitution only affects the wavefunction of phonons; therefore,
the energy values of electron levels in the Schr\"{o}dinger equation ought
to have remained the same. This, however, is not so, since isotopic
substitution modifies not only the phonon spectrum, but also the constant of
electron-phonon interaction (see above). It is for this reason that the
energy values of purely electron transition in molecules of hydride and
deuteride are found to be different [24]. This effect is even more prominent
when we are dealing with a solid [219]. Intercomparison of absorption
spectra for thin films of LiH and LiD at room temperature revealed that the
longwave maximum (as we know now, the exciton peak [220]) moves 64.5 meV
towards the shorter wavelengths when H is replaced with D. For obvious
reasons this fundamental result could not then receive consistent and
comprehensive interpretation, which does not be little its importance even
today. As will be shown below, this effect becomes even more pronounced at
low temperatures (see, also [26]).

The mirror reflection spectra of mixed and pure LiD crystals cleaved in
liquid helium are presented in Fig. 34. For comparison, on the same diagram
we have also plotted the reflection spectrum of LiH crystals with clean
surface. All spectra have been measured with the same apparatus under the
same conditions. As the deuterium concentration increases, the long-wave
maximum broadens and shifts towards the shorter wavelengths. As can clearly
be seen in Fig. 34, all spectra exhibit a similar long-wave structure. This
circumstance allows us to attribute this structure to the excitation of the
ground (Is) and the first excited (2s) exciton states. The energy values of
exciton maxima for pure and mixed crystals at 2 K are presented in Table 17.
The binding energies of excitons E$_{\text{b}}$, calculated by the
hydrogen-like formula, and the energies of interband transitions E$_{\text{g}%
}$ are also given in Table 17.

Going back to Fig. 34, it is hard to miss the growth of $\Delta _{\text{12}}$%
, [221], which in the hydrogen-like model causes an increase of the exciton
Rydberg with the replacement of isotopes (see Fig. 90 in [26]). When
hydrogen is completely replaced with deuterium, the exciton Rydberg (in the
Wannier-Mott model) increases by 20\% from 40 to 50 meV, whereas E$_{\text{g}%
}$ exhibits a 2\% increase, and at 2 $\div $ 4.2 K is $\Delta $E$_{\text{g}}$
= 103 meV. This quantity depends on the temperature, and at room temperature
is 73 meV, which agrees well enough with $\Delta $E$_{\text{g}}$ = 64.5 meV
as found in the paper of Kapustinsky et al. Isotopic substitution of the
light isotope ($^{\text{32}}$S) by the heavy one ($^{\text{34}}$S) in CdS
crystals [222] reduces the exciton Rydberg, which was attributed to the
tentative contribution from the adjacent electron bands (see also [228]),
which, however, are not present in LiH .The single-mode nature of exciton
reflection spectra of mixed crystals LiH$_{\text{x}}$D$_{\text{1-x}}$ agrees
qualitatively with the results obtained with the virtual crystal model (see
e.g. Elliott et al. [210]; Onodera and Toyozawa [223]), being at the same
time its extreme realization, since the difference between ionization
potentials ($\Delta \zeta $) for this compound is zero. According to the
virtual crystal model, $\Delta \zeta $ = 0 implies that $\Delta $E$_{\text{g}%
}$ = 0, which is in contradiction with the experimental results for LiH$_{%
\text{x}}$D$_{\text{1}}$-$_{\text{x}}$ crystals. The change in E$_{\text{g}}$
caused by isotopic substitution has been observed for many broad-gap and
narrow-gap semiconductor compounds (see also below).

All of these results are documented in Table 21 of Ref.[26], where the
variation of E$_{\text{g}}$, E$_{\text{b}}$, are shown at the isotope
effect. We should highlighted here that the most prominent isotope effect is
observed in LiH crystals, where the dependence of E$_{\text{b}}$ = f (C$_{%
\text{H}}$) is also observed and investigated. To end this section, let us
note that E$_{\text{g}}$ decreases by 97 cm$^{\text{-1}}$ when $^{\text{7}}$%
Li is replaced with $^{\text{6}}$Li.

Further we will briefly discuss of the variation of the electronic gap (E$_{%
\text{g}}$) of semiconducting crystals with its isotopic composition. In the
last time the whole raw of semiconducting crystals were grown. These
crystals are diamond , copper halides , germanium , silicon [321], CdS and
GaAs . All numerated crystals show the dependence of the electronic gap on
the isotope masses (see, reviews [26, 27]).

Before we complete the analysis of these results we should note that before
these investigations, studies were carried out on the isotopic effect on
exciton states for a whole range of crystals by Kreingol'd and coworkers
(see, also [203]). First, \ the following are the classic crystals Cu$_{%
\text{2}}$O [224, 225] with the substitution $^{\text{16}}$O $\rightarrow $ $%
^{\text{18}}$O and $^{\text{63}}$Cu $\rightarrow $ $^{\text{65}}$Cu.
Moreover, there have been some detailed investigations of the isotopic
effect on ZnO crystals , where E$_{\text{g}}$ was seen to increase by 55 cm$%
^{\text{-1}}$ ($^{\text{16}}$O $\rightarrow $ $^{\text{18}}$O) and 12 cm$%
^{-1}$ ( at \ $^{\text{64}}$Zn $\rightarrow $ $^{\text{68}}$Zn) [226, 227].
\ In [222] it was shown that the substitution of a heavy $^{\text{34}}$S
isotope for a light $^{\text{32}}$S isotope in CdS crystals resulted in a
decrease in the exciton Rydberg constant (E$_{\text{b}}$ ), which was
explained tentatively [228] by the contribution from the nearest electron
energy bands, which however are absent in LiH crystals.

More detailed investigations of the exciton reflectance spectrum in CdS
crystals were done by Zhang et al. [229]. Zhang et al. studied only the
effects of Cd substitutions, and were able to explain the observed shifts in
the band gap energies, together with the overall temperature dependence of
the band gap energies in terms of a two-oscillator model provided that they
interpreted the energy shifts of the bound excitons and n = 1 polaritons as
a function of average S mass reported as was noted above, earlier by
Kreingol'd et al. [222] as shifts in the band gap energies. However,
Kreingol'd et al. [222] had interpreted these shifts as resulting from
isotopic shifts of the free exciton binding energies (see, also [221]), and
not the band gap energies, based on their observation of different energy
shifts of features which they identified as the n = 2 free exciton states
(for details see [222]). The observations and interpretations, according
Meyer at al. [230], presented by Kreingol'd et al. [222] are difficult to
understand, since on the one hand a significant band gap shift as a function
of the S mass is expected [229], whereas it is difficult to understand the
origin of the relatively huge change in the free exciton binding energies
which they claimed. Very recently Meyer et al. [230] reexamine the optical
spectra of CdS as function of average S mass, using samples grown with
natural Cd and either natural S ($\sim $ 95\% $^{32}$S), or highly enriched
(99\% $^{34}$S). These author observed shifts of the bound excitons and the
n = 1 free exciton edges consistent with those reported by Kreingol'd et al.
[222], but, contrary to their results, Meyer et al. observed essentially
identical shifts of the free exciton excited states, as seen in both
reflection and luminescence spectroscopy. The reflectivity and
photoluminescence spectra i polarized light ($\overrightarrow{E}$ $\bot $ $%
\overrightarrow{C}$) over the A and B exciton energy regions for the two
samples depicted on the Fig. 35. For the $\overrightarrow{E}$ $\bot $ $%
\overrightarrow{C}$ polarization used in Fig. 35 both A and B excitons have
allowed transitions, and therefore reflectivity signatures. Fig. 35 also
reveals both reflectivity signatures of the n = 2 and 3 states of the A
exciton as well that of the n = 2 state of the B exciton.

In Table 18 Meyer et al. summarized the energy differences $\Delta $E = E (Cd%
$^{34}$S) - E (Cd$^{nat}$S), of a large number of bound exciton and free
exciton transitions, measured using photoluminescence, absorption, and
reflectivity spectroscopy, in CdS made from natural S (Cd$^{nat}$S, 95\% $%
^{32}$S) and from highly isotopically enriched $^{34}$S (Cd$^{34}$S, 99\% $%
^{34}$S). As we can see, all of the observed shifts are consistent with a
single value, 10.8$\pm $0.2 cm$^{-1}$. Several of the donor bound exciton
photoluminescence transitions, which in paper [230] can be measured with
high accuracy, reveal shifts which differ from each other by more than the
relevant uncertainties, although all agree with the 10.8$\pm $0.2 cm$^{-1}$
average shift. These small differences in the shift energies for donor bound
exciton transitions may reflect a small isotopic dependence of the donor
binding energy in CdS. This value of 10.8$\pm $0.2 cm$^{-1}$ shift agrees
well with the value of 11.8 cm$^{-1}$ reported early by Kreingol'd et al.
[222] for the B$_{n=1}$ transition, particularly when one takes into account
\ the fact that enriched $^{32}$S was used in that earlier study\TEXTsymbol{<%
} whereas Meyer et al. have used natural S in place of an isotopically
enriched Cd$^{32}$S (for details see [230]).

Authors [230] conclude that all of the observed shifts (see Table 18) \
arise predominantly from an isotopic dependence of the band gap energies,
and that the contribution from any isotopic dependence of the free exciton
binding energies is much smaller. On the basis of the observed temperature
dependencies of the excitonic transitions energies, together with a simple
two-oscillator model, Zhang et al. [229] earlier calculated such a
difference, predicting a shift with the S isotopic mass of 950 $\mu $eV/amu
for the A exciton and 724 $\mu $eV/amu for the B exciton. Reflectivity and
photoluminescence study of $^{nat}$Cd$^{32}$S and $^{nat}$Cd$^{34}$S
performed by Kreingol'd et al. [222] shows that for anion isotope
substitution the ground state (n = 1) energies of both A and B excitons have
a positive energy shifts with rate of $\partial $E/$\partial $M$_{S}$ = 740 $%
\mu $eV/amu. Results of Meyer et al. [230] are consistent with a shift of $%
\sim $710 $\mu $eV/amu for both A and B excitons. Finally, it is interesting
to note that the shift of the exciton energies with Cd mass is 56 $\mu $%
eV/amu [229], an order of magnitude less than found for the S mass.

\ \ \ The present knowledge of the electronic band structure of Si stems
from experimental observation of electronic transitions in transmission,
reflectivity, or cyclotron resonance, on the one hand, and theoretical
calculations, e.g. those based on pseudopotential or $\overrightarrow{k}%
\cdot \overrightarrow{p}$methods (for details see [231 - 235] and references
therein). In this manner it has been established that the fundamental,
indirect band gap of Si occurs between \ the $\Gamma _{8}^{+}$ valence band
maximum and the $\Delta _{0}$ conduction band minima along (100).

Recently, Lastras-Martinez et al. [233] performed ellipsometric measurements
on isotopically enriched $^{28}$Si and $^{30}$Si and deduced the isotopic
dependence of E$_{1}$ from the analysis of the data in reciprocal (Fourier
inverse) space. However, these measurements did not resolve (see, also
[226]) the nearly degenerate \ E'$_{0}$ and E$_{1}$ transitions and the
isotopic shift was assigned solely to the stronger E$_{1}$ transitions (see,
however, Fig. 36). We should add that in papers [235] very recently was
studied the dependence of indirect band gap in Si on the isotopic mass.
Photoluminescence and wavelength-modulated transmission spectra displaying
phonon assisted indirect excitonic transitions in isotopically enriched $%
^{28}$Si, $^{29}$Si, $^{30}$Si as well as in natural Si have yielded the
isotopic gap E$_{gx}$ which equals 1213.8$\pm $1.2 meV. This is purely
electronic value in the absence of electron-phonon interaction and volume
changes associated with anharmonicity (for details see [235] and below).

Returning to Fig. 36, we can see that the spectrum contains two characteric
signatures, attributed to the excitonic transitions across the E'$_{0}$ and E%
$_{1}$ gaps. Isotopic dependence of the E'$_{0}$ and E$_{1}$ is displayed in
Fig. 36, where the photomodulated reflectivity spectra of $^{28}$Si, $^{29}$%
Si, and $^{30}$Si are shown for the spectral range 3.3 $\leq $ E $\leq $
3.58 eV. The E'$_{0}$ and E$_{1}$ excitonic band gaps determined in paper
[235] from the line-shape analysis. Linear least-squares fit yielded the
corresponding isotopic dependences E'$_{0}$ = (3.4468 - 03378 M$^{-1/2}$) eV
and E$_{1}$ = (3.6120 - 0.6821 M$^{-1/2}$) eV. In concluding, we should note
that the spin-orbit interaction depends in Ge in contrast to that in Si
[235].

\QTP{Body Math}
As is well known ago, the fundamental energy gap in silicon, germanium, and
diamond is indirect (see, e.g. [ 232]). While the conduction band minima in
Si and diamond are located at the $\Delta $ point along \TEXTsymbol{<}100%
\TEXTsymbol{>}, with $\Delta _{6}$ symmetry, those of germanium with L$%
_{6}^{+}$ symmetry occur at the \TEXTsymbol{<}111\TEXTsymbol{>} zone
boundaries [235]. The onset of the absorption edge corresponds to optical
transition from the $\Gamma _{8}^{+}$ valence band maximum to the L$_{6}^{+}$
conduction band minima in Ge, and the $\Delta _{6}$ in Si and diamond; for
wavector conservation, these indirect transitions require the emission or
absorption of the relevant phonons. In Si and C, transverse acoustic (TA),
longitudinal acoustic (LA), transverse optic (TO), or longitudinal optic
(LO) phonons of $\Delta $ symmetry must be simultaneously emitted or
absorbed. In Ge (see, also above), the wavector conserving phonons are TA,
LA, TO or LO phonons with L symmetry. At low temperatures, these indirect
transitions are assisted by phonon emission. In this case we should expect
at low temperatures four excitonic derivative signatures at photon energies E%
$_{gx}$ + $\hbar \omega _{\overrightarrow{q}}$,$_{j}$ in modulated
transmission experiments and in photoluminescence at the photon energies E$%
_{gx}$ - $\hbar \omega _{\overrightarrow{q}}$,$_{j}$. Here E$_{gx}$ is the
excitonic band gap and j corresponds to a wave vector preserving phonon. In
Fig. 37 - A the photoluminescence and wavelength - modulated spectra of $%
^{30}$Si M = 2.81 amu) are displayed; the labels n = 1 and 2 designate the
ground and the first excited states of the indirect TA and and TO excitons.
From the energies of the photoluminescence and wavelength-modulated
excitonic signatures in all isotopic specimens (see [235]) cited authors
deduce E$_{gx}$ \ as well as the energies of the participating TO, LO and TA
\ phonons , shown in Fig. 37 - B as function of M$^{-1/2}$. The excitonic
band gap data are fitted well with expression E$_{gx}$(M) = E$_{gx}$($\infty 
$) - CM$^{-1/2}$, yielding E$_{gx}$ ($\infty $) = (1213.8$\pm $1.2) meV and
C = (313.7 $\pm $5.3) meV/amu. A linear fit in M can be made over small
range of available masses (see, Fig. 37 - B) with a slope ($\partial $E$%
_{gx} $/$\partial $M)$_{P,T}$ 1.01$\pm $0.04 meV/amu, which agrees with the
results of bound exciton photoluminescence of Karaiskaj et al. [234]. The
experiments in papers [235] also indicate that separation of the n = 2 and n
= 1 excitons is isotope mass independent, implying, according these authors,
the excitonic binding energy is independent on isotope mass within
experimental error. In concluding this part we should note that recent high
- resolution spectroscopic studies of excitonic and impurity transition in
high - quality samples of isotopically enriched Si have discovered the
broadening of bound exciton emission (absorption) lines connected with
isotope - induced disorder as well as the depend of their binding energy on
the isotope mass [234, 235]. The last effect was early observed on the bound
excitons in diamond [236, 237], and earlier on the free excitons [238] in LiH%
$_{x}$D$_{1-x}$ mixed crystals (see, e.g. [151] and references therein).

\bigskip

4.4. Effects related to isotopic disorder.

\bigskip

4.4.1. Thermal conductivity.

\bigskip

In insulators and semiconductors (at T \TEXTsymbol{<} $\theta _{D}$) the
thermal conduction is effected by phonons, predominantly acoustic ones
[239]. Thermal conductivity of crystals has been subject of many
experimental theoretical studies \ (see, e.g. reviews and monographs [195,
239 -246]). The first experimental results (see, e.g. [243]) have already
pointed out the existence of maximum of the thermal conductivity coefficient
k$_{m}$ at about T $\approx $ 0.05 $\theta _{D}$, where $\theta _{D}$ is the
Debye temperature. The growth of k at low temperatures has been related to
phonon scattering due to Umklap (U -) - type processes [245; 26]. In the
vicinity of k$_{m}$ thermal conductivity is quite sensitive to impurities
and defects in the specimen. The scattering of phonons dynamic isotope
disorder is independent of temperature and lattice anharmonicity. The role
of isotopes as an additional channel of phonon scattering and their
influence on thermal conductivity were first theoretically studied by
Pomeranchuk [247] \ in 1942, and were experimentally studied using Ge in
1958 [248]. According to the results of the latter reference, for a Ge
specimen (having 95.8\% $^{74}$Ge), a threefold growth of the thermal
conductivity coefficient as compared to the specimen of germanium with
natural isotope composition was observed. Later, the influence of isotopes
on diamond thermal conductivity was studied many times [ 249 - 251].

It is generally assumed (see, e. g. [26]) that at not too high temperatures,
the dominant interacting among phonons involve three phonons. In a "normal"
(N -) process the wave vectors $\overrightarrow{q}$ of the phonons are
conserved and such process tend to restore a disturbed phonon distribution
to one which can be described as a displaced Planck distribution (see, e.g.
Fig. 5.2 in [242]) which is unaffected by N - \ processes and corresponds to
a heat flow. By themselves, therefore, N - processes would not lead to a
thermal resistance.

In Umklap (U -) - process [245] the wave vectors are not conserved and, as
in other resistive processes, they tend to restore a disturbed phonon
distribution to the equilibrium Planck distribution which corresponds to
zero heat flow, and thus lead to a finite conductivity (for more details,
see, review [26]). The Debye expression [252] for the conductivity k(T) is
derived from an adoptional of the simple kinetic theory

k(T) = $\frac{\text{1}}{\text{3}}\left\langle \text{v}_{ph}\right\rangle
\ell _{ph}$(T)C$_{p}$(T), \ \ \ \ \ \ \ \ \ \ \ \ \ \ \ \ \ \ \ \ \ \ \ \ \
\ \ \ \ \ \ \ \ \ \ \ \ \ \ \ (114)

where $\left\langle \text{v}_{ph}\right\rangle $ is an average phonon
velocity, $\ell _{ph}$(T) their mean free path and C$_{p}$(T) the
corresponding specific heat (for diamond see [253]). A theory of k(T)
requires basically the calculation of

$\ell _{ph}$(T) = $\tau _{ph}$(T)$\left\langle \text{v}_{ph}\right\rangle $
\ \ \ \ \ \ \ \ \ \ \ \ \ \ \ \ \ \ \ \ \ \ \ \ \ \ \ \ \ \ \ \ \ \ \ \ \ \
\ \ \ \ \ \ \ \ \ \ \ \ \ (115)

a rather formidable task since several scattering mechanisms (normal -, \ u
- processes, boundary of sample, isotope scattering) [26] contribute to
determining the mean free path. In formula (115) $\tau _{ph}$(T) is the
phonon relaxation time. The simplest of these mechanisms, and the one that
can be varied for a given material of the acoustic phonons by isotopic mass
fluctuations. This scattering is equivalent to Rayleigh scattering (of
photons) at point defect. Within Debye approximation, we will have

k(T) = $\frac{\text{k}_{B}}{\text{2}\pi \nu }\left( \frac{\text{k}_{B}}{%
\hbar }\right) ^{3}$T$^{3}\dint\limits_{0}^{\frac{\theta _{D}}{\text{T}}%
}\tau $(x)$\frac{\text{x}^{4}\text{e}^{x}}{\text{(e}^{x}\text{ - 1)}^{2}}$%
dx. \ \ \ \ \ \ \ \ \ \ \ \ \ \ \ \ \ \ \ \ \ \ (116)

In last expression k$_{B}$ is the Boltzmann constant. Klemens [239] was the
first to try to take the role of N - processes into account. Using
perturbation theory Klemens [239] developed the following expression for the
scattering rate $\tau _{isotope}^{-1}$:

$\tau _{isotope}^{-1}$ = $\frac{\text{x(x -1)V}_{0}}{\text{4}\pi
\left\langle \text{v}_{ph}\right\rangle \ ^{3}}\left( \frac{\Delta \text{M}}{%
\text{M}}\right) ^{2}\omega ^{2}$, \ \ \ \ \ \ \ \ \ \ \ \ \ \ \ \ \ \ \ \ \
\ \ \ \ \ \ \ \ \ \ \ \ \ \ \ \ (117)

where V$_{0}$ is a volume per atom (for diamond 5.7$\cdot $10$^{-24}$ cm$%
^{3} $) and $\omega $ is phonon frequency$.$ Callaway approach [244]
successfully introduces normal phonon scattering ($\tau _{N}^{-1}$) and
resistive scattering ($\tau _{R}^{-1}$) (see formula (4.51) in [26]).

In Fig. 38 present the results of Wei et al. [251]. The solid curves are the
results of fitting the Callaway theory [244], using a single set of fitting
parameter. In this paper, Wei et al. have measured a record thermal
conductivity of 410 Wcm$^{-1}$K$^{-1}$ at 104 K for a 99.9\% $^{12}$C
enriched diamond. These authors predict that a 99.999\% $^{12}$C diamond
should have a peak value of thermal conductivity exceeding 2000 Wcm$^{-1}$K$%
^{-1}$, at about 80K, assuming, of course, that is not limited by point
defect scattering mechanisms other than minority isotopes. Similar results
have very recently been reported by Olson et al. [250]. We should stress
that none of the currently existing theories accurately takes into account
all the possible scattering processes.

Thermal conductivity studies have also performed on very highly enriched,
ultra - pure $^{70}$Ge (see, reviews [26, 27]). The maximum value of k$_{m}$
= 10.5 kWm$^{-1}$K$^{-1}$ was observed, in the vicinity of T = 16.5 K, for
the $^{70}$Ge specimen of 99.99\% purity, which is significantly higher than
the value for sapphire (6 kWm$^{-1}$K$^{-1}$ around T$_{m}$ 35K) and
comparable to the value for silver (11kWm$^{-1}$K$^{-1}$ near T$_{m}$ =
15.4K). Comparison of experimental results shows [256] that, at its maximum
(see, e.g. Fig. 6$^{a}$ [205]), the thermal conductivity of the $^{70/76}$Ge
(91.91\%) specimen is 14 times less than that of $^{70}$Ge (91.91\%). An
increase in k reaches however, only 30\% at T = 300K (see, also [26, 27]).

The thermal conductivity of monoisotopic and isotopically mixed specimens of
silicon crystals has been studied in following papers [257 - 260]. Since the
most detailed results have been obtained by the authors of [259], we
restrict ourselves to their consideration. It is well - known that natural
silicon consists of three isotopes: $^{28}$Si ($\sim $ 92\%), $^{29}$Si ($%
\sim $ 5\%), and $^{30}$Si ($\sim $ 3\%). The use of monoisotopic silicon
(for example $^{28}$Si) can substantially reduce the value of dissipated
energy scattered in electronic elements made of silicon (e.g. in the memory
of electronic computers [261]). The results studies of the thermal
conductivity of monoisotopic and isotopically mixed crystals are shown in
Fig. 39. According to the results presented in this Fig., for SI284 specimen
k = 237(8) Wm$^{-1}$K$^{-1}$ at 300K, whereas for the SINI (natural Si)
specimen it is equal to 150 Wm$^{-1}$K$^{-1}$. This means that at 300K the
thermal conductivity of a monoisotopic $^{28}$Si specimen grows, as compared
to the natural silicon, by 60 \% (later - 10\%, see, erratum). At the same
time, at about 20 K (in the vicinity of the maximum of the silicon thermal
conductivity curve) k reaches the value of 30000 $\pm $ 5000 Wm$^{-1}$K$%
^{-1} $, which is 6 times higher the value k = 5140 Wm$^{-1}$K$^{-1}$ for
natural specimen (ee, also [241]).

The thin solid and dashed lines in Fig. 39 correspond to the results of
theoretical computations of thermal conductivity for monoisotopic specimen
SI284 and for a specimen with natural silicon isotope composition. In these
calculations, the model of the Ge thermal conductivity developed in [256]
with modified Debye temperature and phonon mean free path has been used. For
fitting, the authors have used the low - temperature results, where the
thermal conductivity is described by the T$^{3}$ law. Calculations presented
in Fig. 39 were performed, for the natural specimen, for free mean path
equals 5.0 mm (dashed line). For comparison, let us point out that in Ref.
[257] the analogous quantity was equal to 5.7 mm, and for isotopically pure
SI284 specimen the corresponding value was 14.0 mm (thin solid line). As
seen from Fig. 39, there is agreement between theory and experiment, which
has also been mentioned by the authors of [259] themselves. They have also
pointed out good agreement between their experimental results and
calculations made in [262], except for the domain of U - processes. Beside
that, Ruf and co - authors have mentioned an unsatisfactory agreement
between theory and experiment in the domain of high - temperatures (300 -
400K), especially for the specimen with natural isotope composition. They
think that this disagreement can occur due to fundamental reasons that
require further study. In particular, taking into account the fine structure
of the nonequilibrium phonon distribution function could bring theoretical
and experimental results much closer. A qualitative comparison of the
influence of the isotope effect on the thermal of germanium, silicon and
diamond is given in Table 19.

In concluding we should remark, that until recently all theories on thermal
conductivity had a strongly phenomenological flavor, making use of the
relaxation time approximation. In recent years, considerable progress
towards an ab initio theory has been made [262; 263]. These authors used two
- and three - body potentials obtained by fitting phonon dispersion
relations and related the anharmonic properties with a single average Gr\"{u}%
neisen parameter. In this manner they determined the third - order coupling
coefficients for all possible three - phonon combinations. They then solved
iteratively the Boltzmann equation for phonon transport without using the
relaxation - time approximation. A scattering time must, however, still be
used to describe boundary scattering in the lowest temperature region. In
this manner they reproduced rather well the thermal conductivities of Ge, Si
and diamond and the observed isotope effects (for details see [262; 263]).

\bigskip

4.4.2. Disorder - induced Raman scattering.

\bigskip

The frequencies of vibrational modes in a solid depend on the interatomic
forces and the atomic masses. By changing the mass of atoms by isotopic
substitution the frequencies of modes are changed in a small but
characteristic way that can be monitored by Raman spectroscopy. In
isotopically pure crystals the width $\Gamma _{0}$ of the Raman line is
determined - aside from experimental resolution - by the phonon lifetime
which is governed by the spontaneous anharmonic decay into phonon of lower
energy [239]. In an isotopically disordered material an additional
contribution $\Gamma _{isotope}$ to the linewidth comes from the elastic
scattering of phonons via mass fluctuation and has been observed for many
semiconductors (see, review [27] and reference therein). Line shift and line
broadening are theoretically obtained as real and imaginary parts of a
complex self - energy which can be calculated in the framework of a coherent
potential approximation (CPA) in the case of weak \ phonons scattering [26].
This theory describes, for example, frequency shift and line broadening \
very well in isotopically disordered diamond [264], Ge [265] and $\alpha $ -
Sn [266]. A mass perturbation theory of the harmonic lattice dynamics for
calculating $\Gamma _{isotope}$ has been developed by Tamura and applied to
Ge [267], GaAs and InSb [268]. For the complex self - energy of the Raman
phonon of a semiconductor with diamond structure, the second - order term (n
= 2) contains the real part [269, 26]

$\Delta _{2}$($\omega $,x) = $\frac{\text{g}_{2}\text{(x)}}{4}\omega $($%
\frac{\text{1}}{\text{6N}_{c}}$)$\dsum\limits_{i}\omega _{i}\frac{\omega 
\text{ - }\omega _{i}}{\text{(}\omega \text{ - }\omega _{i}\text{)}^{2}+%
\text{ }\gamma ^{2}}$\ \ \ \ \ \ \ \ \ (118)

and the imaginary part (see, also [267, 268])

$\Gamma _{2}\left( \omega \text{, x}\right) $ = $\frac{\text{g}_{2}\text{(x)}%
}{4}\omega $($\frac{\text{1}}{\text{6N}_{c}}$)$\dsum\limits_{i}\omega _{i}%
\frac{\gamma }{\text{(}\omega \text{ - }\omega _{i}\text{)}^{2}+\text{ }%
\gamma ^{2}}$. \ \ \ \ \ (119)

In \ the last two formulae N$_{c}$ is the number of unit cells. For $\gamma $
$\longrightarrow $ 0 Eq. (119) simplifies to a sum over $\delta $ -
functions, which represents the one - phonon density of state $\rho $($%
\omega $) (compare to [268]),

$\Gamma _{2}\left( \omega \text{, x}\right) $\ \ \ = $\Gamma _{iso}$ = $\tau
_{iso}^{-1}$ = $\frac{\pi }{\text{6}}\omega ^{2}$g$\left\vert 
\overrightarrow{e}\right\vert \rho \left( \omega \right) $. \ \ \ (120)

From this relation we can see, that \ $\Gamma _{iso}$ depends on three
factors:

i) the relative mass variance g$_{2}$;

ii) the phonon density of states $\rho $($\omega $) at the frequency $\omega 
$ of the Raman mode, and

iii) a relevant phonon eigenvector $\overrightarrow{e}$.

4.4.2.1. Disorder shift and broadening of the lines in the Raman spectra.

In modern language, as say above, phonons are referred to as quasiparticles,
with a complex self-energy $\Sigma $ = $\Sigma _{r}$ + i$\Sigma _{i}$
induced in insulators (semiconductors) by anharmonic phonon-phonon
interactions and \ in crystals with several isotopes of a given element also
by isotopic mass disorder. In metals and heavily doped semiconductors one
must also take into account the self-energy which corresponds to the
interaction of the phonon with the conduction electrons. The purpose of this
section is to discuss the isotopic disorder contributions to the self-energy
of phonons, in the first step in semiconductors of the tetrahedral variety
with special emphasis on the quantum effects observed at low temperatures
(especially in diamond - where isotopic effects dominates over the
anharmonic ones - as well as in germanium where anharmonic effects are
larger [26]).

The definition of the average mass $\overline{\text{m}}$ = $\dsum\limits_{%
\text{i}}$c$_{\text{i}}$m$_{\text{i}}$ implies that g$_{\text{1}}$ = 0. $%
\bigskip $In further discussion we display a compilation of the
disorder-induced self-energies for the Raman phonons of elemental and
compound crystals (diamond, Si, Ge, $\alpha $-Sn and LiH$_{\text{x}}$D$_{%
\text{1-x}}$, 6H-Sic polytype) which have been obtained either by Raman
spectroscopy or from theoretical calculations by several research groups
during the last two decades. Raman studies that address the variation of the
self-energy with the isotopic composition have been conducted for LiH$_{%
\text{x}}$D$_{\text{1-x}}$ (see, for example review [214]), diamond [204,
264, 270, 271], \ and Si [272]. The coherent potential approximation (CPA)
has been employed for diamond [264, 270] and Si [272], while ab initio
electronic structure based calculations have been performed for diamond and
Ge [273].

Fig. 40 displays the Raman frequencies of diamond versus the $^{13}$C
concentration. The points (open symbols) represent experimental values. The
dashed curve represents the approximately linear dependence expected in the
VCA. The upward curvature of the experimental data (with respect \ to the
VCA line) clearly demonstrates the existence of an isotopic-disorder-induced
self-energy as emphasized by the solid line, which is a fit with Eq. (2.39)
for n = 2, 3 of Ref. [26]. It is difficult to see with the nakes eye in Fig.
40 the asymmetric behavior versus x, which may arise from third-order
perturbation terms. The asymmetry appears, however, rather clearly when the
difference between the measured (or the calculated) behavior and the VCA
line is plotted, as shown in Fig.41. In this figure, the solid line also
represents the fit to all experimental data, the dot-dashed line represents
CPA calculations while the dotted line is a fit to the asterisks which
indicate points obtained in the ab initio calculations [192; 201]. All data
in Fig. 44 show a similar asymmetric behavior, with a maximum of $\Delta _{%
\text{dis}}\left( \text{x}\right) $ at x $\approx $ 0.6. These results
allows us to conclude that the real part of the self-energy due to isotopic
disorder is well understood for these systems, including the superposition
of second-order and third-order perturbations terms in the case of the
diamond. Similar degree of understanding has been reached for $\Gamma _{%
\text{dis}}\left( \text{x}\right) $ as shown in Fig. 42. The x position of
the maxima of $\Delta _{\text{dis}}$ and $\Gamma _{\text{dis}}$ determined
from the experimental data agree with those obtained by perturbation theory
\ $\left[ \text{Eq. }\left( \text{2.42}\right) \text{ of Ref. [26]}\right] $
and also with the CPA and ab initio calculations (details see [269]). \
Concerning the other elemental semiconductors, detailed experimental results
with sufficient values of x to reach quantitative conclusions of the type
found for diamond, are \ only available for Si. \ These data for Si are
shown in Figs. 43 and 44. (filled circles) together with the results of CPA
calculations (filled squares). The latter show for $\Delta _{\text{dis}}$ \
a clear asymmetry with a maximum at x$_{\text{max,}\Delta }$ $\approx $
0.56. The quality and the number \ of the experimental points are not
sufficient to conclude that an asymmetry exists but they cannot exclude it
either. The measured absolute values of $\Delta _{\text{dis}}$ \ almost (not
quite) agree within error bars with the calculated ones. The corresponding
experimental; values of $\Gamma _{\text{dis}}\left( \text{x}\right) $ (see
Fig. 44) \ are about a factor of 2 lower than the calculated ones, although
both show the asymmetric behavior (x$_{\text{max,}\Gamma }$ $\approx $ 0.62)
predicted by theory. The reason for the discrepancy between the calculated
and measured $\Gamma _{\text{dis}}$ is to be sought in the mechanism
responsible for it in Si [69]. Within harmonic approximation $\Gamma _{\text{%
dis}}$ = 0 for Si, Ge and $\alpha $-Sn, because the Raman frequency is at
the maximum of the spectrum and thus corresponds to zero density of
one-phonon states. The rather small, but not negligible, observed value of $%
\Gamma _{\text{dis}}$ results from DOS induced at the $\Gamma $ point by the
anharmonic interactions responsible for the linewidths of the isotopically
pure crystals. Thus, the widths observed for Si, as well as for Ge and $%
\alpha $-Sn, correspond to fourth-order (twice disorder and twice
anharmonicity) and higher-order terms.

As was shown in the papers [268, 269], the isotope effects of a disordered
sublattice in a compound is different from that for the corresponding \
monatomic crystal. Widulle et al. apply Eqs. (2.50a) and (2.50b) of Ref.
[26] \ to the Raman spectroscopic results on a variety of $^{\text{nat}}$Si$%
^{\text{12}}$C$_{\text{1-x}}$ $^{\text{13}}$C$_{\text{x}}$ polytypes,
recently reported by Rohmfeld et al. [275]. They have performed a fit Eq.
(2.40) for n = 2, 3 \ of Ref. [26]to the linewidths of the transverse optic
(TO) modes of the 6H-Sic polytype measured in [275] for $^{\text{13}}$C \
concentration ranging from x = 0.15 to x = 0.40 Fig. 45). The fit performed
in Ref. [275] was based on the asymmetric curve obtained by CPA calculations
for diamond [264]. This curve was first fitted to the data points of the TO
(2/6) mode and further adjusted to the TO (0) and TO (6/6) modes \ by
multiplication \ with the constant eigenvectors. Instead, authors [269] have
considered each TO mode separately and performed fits with Eq. (2.40) \ for
n = 2, 3 of Ref. [26] in the same manner as for elemental semiconductors
(see above). Widulle et al. used, however, parameters appropriate to SiC,
not to diamond. In this way, they conclude that the behavior of $\Gamma _{%
\text{dis}}$ \ versus x is asymmetric (Fig. 45). This fact cannot, according
Widulle et al, derived from the data of the results paper [275]. The latter
can be fitted equally with either a symmetric or an asymmetric curve.

As has been mentioned many times, the isotopic disorder in the crystal
lattice lifts the forbidenness imposed by the quasiimpuls conservation law,
thus allowing a contribution to the half - width of the scattering line from
other phonons from the domain with the maximum density of states, especially
from the TO branches of Ge. The two structures observed in the spectrum of
first - order Raman scattering near 275 and 290 cm$^{-1}$ correspond to the
maximum of the density of states of TO phonons (see [27] and references
therein), which become active because of the violation of the quasi -
momentum conservation law by the isotopic disorder in the crystal (see, for
instance Fig. 20 in [ 205]). The effect of the development of an additional
structure in Raman scattering spectra was observed relatively long ago [276]
in isotopically mixed LiH$_{x}$D$_{1-x}$ (Fig. 46). The effects caused by
isotopic disorder in the crystal lattice in isotopically mixed \ are
analogous to those described above (see, also [205]) There exist, however,
principal differences. In contrast to Ge and C, in which the first - order
spectra exhibit a one - mode character, the second - order spectra of LiH$%
_{x}$D$_{1-x}$ crystals have one - and two - mode characters for LO($\Gamma $%
) phonons, and also contain a contribution from the local excitation at
small values of x. Fig. 46 demonstrates the dependence of the half - width
of the line of LO($\Gamma $) phonons in light - scattering spectra on the
concentration of isotopes. One clearly sees a substantial growth (by factor
2 - 4) of the half - width of the line with increasing concentration of
isotopes, as well as the existence of a short - wavelength structure that
has already been related in Ref. [276] to the excitation of TO phonons in
isotopically disordered crystal lattice (for more details see [277]).

To conclude this part of our review, we should note that in contrast
elemental semiconductors, where isotope scattering potential is weak, in the
case isotope-mixed crystals LiH$_{\text{x}}$D$_{\text{1-x}}$ isotope
scattering potential is very strong and CPA approximation in such simple
version does not describe the Raman spectra of these crystals.

$\bigskip $

4.4.3. Effects of isotope randomness on electronic properties and exciton
transition.

\bigskip

As follows from Fig. 34, excitons in LiH$_{\text{x}}$D$_{\text{1-x}}$
crystals display a unimodal character, which facilitates the interpretation
of their concentration dependence. Figure 47 shows the concentration
dependence of the energy of interband transitions E$_{\text{g}}$. Each value
of E$_{\text{g}}$ was found by adding together the energy of the long-wave
band in the reflection spectrum and the binding energy of the exciton. The
latter was found from the hydrogen-like formula using the experimental
values of the energies levels of Is and 2s exciton states. We see that the
100\% replacement of hydrogen with deuterium changes E$_{\text{g}}$ by $%
\Delta $E$_{g}$ = 103 meV at T = 2 K(see, e.g. [221]). This constitutes 2\%
of the energy of the electron transition, which is two orders of magnitude
greater than the value corresponding to the isotopic replacement of atomic
hydrogen with deuterium reported earlier [278].

The nonlinear concentration dependence of E$_g$ can be sufficiently well
approximated with a second order polynomial

E$_{\text{g}}\left( \text{x}\right) $ = E$_{\text{b}}$ + (E$_{\text{a}}$ - E$%
_{\text{b}}$ - b)x - bx$^{\text{2}}$, \ \ \ \ \ \ \ \ \ \ \ \ \ \ (121)

where E$_{\text{a}}$, E$_{\text{b}}$ are the values of E$_{\text{g}}$ for
LiD and LiH respectively, and b is the curvature parameter equal to 0.046
eV. This result generally agrees with the published data (see also Elliott
and Ipatova [279] and references therein). For comparison let us indicate
that in the case of isotopic substitution in germanium the energy E$_{\text{g%
}}$ depends linearly on the isotopic concentration for both direct (E$_{%
\text{0}}$, E$_{\text{0}}$+$\Delta _{\text{0}}$, E$_{\text{1}}$+$\Delta _{%
\text{1}}$) and indirect electron transitions [280]. Unfortunately, today
there is no information on the form of the function E$_{\text{g}}$ $\propto $
f(x) for isotopic substitution in C, ZnO, CdS, CuCl, Cu$_{2}$O, GaAs,GaN,
Si, etc. crystals, although, as noted above, the values of E$_{\text{g}}$
have been measured for isotopically pure crystals. However, we should add,
that isotopic substitution in Ge leads not only to the shift of the
luminescence spectrum, but also to the nonlinear concentration dependence of
the emission line half - width, as in the case of lithium hydride (see,
below) was attributed to isotopic disordering of the crystal lattice [281].

According to the results depicted on Fig. 88, the addition of deuterium
leads not only to the short-wave shift of the entire exciton structure (with
different rates for Is and 2s states [221], but also to a significant
broadening of the long-wave exciton reflection line. This line is broadened
1.5 - 3 - fold upon transition from pure LiH to pure LiD. The measure of
broadening was the halfwidth of the line measured in the standard way (see
e.g. [282]) as the distance between the maximum and the minimum in the
dispersion gap of the reflection spectrum, taken at half-height. The
concentration dependence of the halfwidth ($\Delta $E$^{\text{R}}$) of the
long-wave band in the exciton reflection spectrum at 2 K is shown in Fig.
48. Despite the large spread and the very limited number of concentrations
used, one immediately recognizes the nonlinear growth of $\Delta $E$^{\text{R%
}}$ with decreasing x. A similar concentration dependence of $\Delta E^{%
\text{R}}$ in the low-temperature reflection spectra of solid solutions of
semiconductor compounds A$_{\text{2}}$B$_{\text{6}}$ and A$_{\text{3}}$B$_{%
\text{5}}$ has been reported more than once (see e.g. the review [279] and
references therein). The observed broadening of exciton lines is caused by
the interaction of excitons with the potential of large-scale fluctuations
in the composition of the solid solution. Efros and colleagues (see e.g.
[283]) used the method of optimal fluctuation [284] to express the formula
for the concentration dependence of the broadening of exciton reflection
lines:

$\Delta $E$^{\text{R}}$ = 0.5$\alpha $ $\left[ \frac{\text{x}\left( \text{1-x%
}\right) }{\text{Nr}_{\text{ex}}}\right] ^{\text{1/2}}$. \ \ \ \ \ \ \ \ \ \
\ \ \ \ \ \ \ \ \ (122)

where $\alpha $ = dE$_{\text{g}}$/ dx; r$_{\text{ex}}$ is the exciton radius
which varies from 47 A to 42 \AA\ upon transition from LiH to LiD [49]. The
value of coefficient $\alpha $ was found by differentiating Eq. (121) with
respect to x - that is, dE$_{\text{g}}$/ dx = $\alpha $ = E$_{\text{a}}$ - E$%
_{\text{b}}$ - b + 2bx. The results of calculation according to Eq. (122)
are shown in Fig. 48 by a full curve.

The experimental results lie much closer to this curve than to the straight
line plotted from the virtual crystal model. At the same time it is clear
that there is only qualitative agreement between theory and experiment at x $%
>$ 0.5. Nevertheless, even this qualitative analysis clearly points to the
nonlinear dependence of broadening on the concentration of isotopes, and
hence to the isotopic disordering. Since isotopic substitution only affects
the energy of optical phonon, and, as a consequence, the constant of
exciton-phonon interaction (in the first place, the Fr\"{o}hlich interaction
g$_{\text{F}}^{\text{2}})$, the nonlinearity of functions $\Delta $E$_{\text{%
g}}$ $\propto $ f(x), $\Delta $E$^{\text{R}}$ $\propto $ f(x) is mainly
related to the nonlinear behavior of g$_{\text{F}}^{\text{2}}\propto $ f(x)
. In this way, the experimental study of the concentration dependence of the
exciton-phonon interaction constant may throw light on the nature and
mechanism of the large-scale fluctuations of electron potential in
isotopically disordered crystals (see also [277]).

A principal matter for further theoretical development is the question
concerning the effect of crystal lattice disordering on the binding energy E$%
_{\text{B}}$ of Wannier-Mott exciton [151] . This problem has been treated
theoretically in the papers of Elliott and coworkers [285, 286], where they
study the effect of weak disordering on E$_{\text{B}}$ (the disordering
energy is comparable with E$_{\text{B}})$. The binding energy indicated in
the papers was calculated under the coherent potential approximation by
solving the Bethe-Salpeter equation [290] as applied to the problem of
Wannier-Mott exciton in disordered medium. One of the principal results of
this paper [285] is the nonlinear dependence of E$_{\text{B}}$ on the
concentration. As a consequence, the binding energy E$_{\text{B}}$ at
half-and-half concentrations is less than the value derived from the virtual
crystal model. The exciton binding energy is reduced because the energy E$_{%
\text{g}}$ is less owing to the fluctuation smearing of the edges of the
conduction and valence band. This conclusion is in qualitative agreement
(although not in quantitative agreement, the discrepancy being about an
order of magnitude (see also [285]) with the experimental results for the
mixed crystal GaAs$_{\text{1-x}}$P$_{\text{x}}$ with x = 0.37, where the
reflection spectra exhibited two exciton maxima (see also Fig. 34) used for
finding the value of E$_{\text{b}}$(see Nelson et al. [287] and references
therein). Let us add that the pivotal feature of the model of Elliott and
coworkers is the short-range nature of the Coulomb potential (for more
details see [49]).

Before the comparison of our experimental results with the theory developed
by Elliott and Kanehisa, it would be prudent to briefly review main
properties of their theoretical model. According to Ref. 285 this model
considers an exciton with a direct gap of a semiconductor alloy. Such a
system consists of an electron (particle 1) in the conduction band (c) with
mass m$_{\text{c}}$ and a hole (particle 2) in the valence band (v) with
mass m$_{\text{v}}$. The problem of the exciton in disordered systems is to
solve the Hamiltonian

H = $\overrightarrow{\text{p}}^{\text{2}}$/2m$_{\text{c}}$ +$\overrightarrow{%
\text{p}}^{\text{2}}$/2m$_{\text{v}}$+ u($\overrightarrow{\text{r}}_{\text{1}%
}$ - $\overrightarrow{\text{r}}_{\text{2}}$) + V$_{\text{c}}$($%
\overrightarrow{\text{r}}_{\text{1}}$) + V$_{v}$($\overrightarrow{\text{r}}_{%
\text{2}}$), \ \ \ \ (123)

with both the Coulomb interaction u and the potential V$_{v}$ due to
disorder ($\nu $ = c,v). Reference 362 neglected disorder - induced
interband mixing. As it is well known, in place of the electron-hole
coordinates, ($\overrightarrow{\text{r}}_{\text{1}}$, $\overrightarrow{\text{%
p}}_{\text{1}}$) and ($\overrightarrow{\text{r}}_{\text{2}}$, $%
\overrightarrow{\text{p}}_{\text{2}}$), one may introduce the center-of-mass
and relative coordinates, ($\overrightarrow{\text{R}}$, $\overrightarrow{%
\text{P}}$) and ($\overrightarrow{\text{r}}$, $\overrightarrow{\text{p}})$
to rewrite (123) as

H = $\overrightarrow{\text{p}}^{\text{2}}$/2m$_{\text{r}}$ + u($%
\overrightarrow{\text{r}}$) + $\overrightarrow{\text{P}}^{\text{2}}$/2M + V$%
_{\text{c}}$($\overrightarrow{\text{R}}$ + m$_{\text{v}}\overrightarrow{%
\text{r}}$/M) + V$_{\text{v}}$($\overrightarrow{\text{R}}$ - m$_{\text{c}}%
\overrightarrow{\text{r}}$/M), \ \ \ \ (124)

where m$_{r}$ and M are the reduced and total masses, respectively. Because
of the random potential, the translational and relative degrees of freedom
cannot be decoupled. This is essentially difficult when considering the
two-body problem in a disordered system [288]. However, when the exciton
state in question is well separated from other states so that the energy
spacing is much larger than the translational width and disorder, one can
forget about the relative motion and just apply any single-particle alloy
theory (see, e.g. Ref. [289 and references therein) solely to their
translational motion. For each exciton state the translational part of
Hamiltonian in this case is

H$_{t}$ = $\overrightarrow{\text{P}}^{\text{2}}$/2M + \={V}$_{\text{c}}$($%
\overrightarrow{\text{R}}$) + \={V}$_{\text{v}}$(\={R}) \ \ \ \ (125).

Here \={V}$_{\text{c}}$ and \={V}$_{\text{v}}$ are averages of V$_{\text{c}}$
and V$_{\text{v}}$. This approach is very similar to the Born-Oppenheimer
adiabatic approximation. Such situations hold in some mixed alkali halide
crystals and probably A$_{\text{2}}$B$_{\text{6}}$ crystals. On the
contrary, when the exciton binding energy is comparable to the disorder
energy, the adiabatic approximation breaks down, and it is essential to take
into account the effect of disorder on both the translational and relative
motions. This is the case with the Wannier-Mott exciton in A$_{\text{3}}$B$_{%
\text{5}}$ alloys, for which the Elliott and Kanehisa model was developed.
In this case the solution task is to start from the independent electron and
hole by neglecting u in (124) and then to take into consideration the
Coulomb interaction between the average electron and average hole. A further
simplified approach adopted in the literature [289] in solving the
Bethe-Salpeter [290] equation is to suppose a free-electron-like one
particle Green's function with a built-in width to allow for the random
potential due to disorder. In the cited theoretical model, the average (or
\textquotedblright virtual crystal\textquotedblright ) gap is given by

E$_{\text{g}}^{\text{vc}}$(x) = E$_{\text{0}}$ + ($\delta _{\text{c}}$ - $%
\delta _{\text{v}}$)(x - 1/2), \ \ \ \ (126)

where E$_{\text{0}}$ is average gap, $\delta _{\text{c}}$, $\delta _{\text{v}%
}$ are the values of the fluctuation broadening of the conduction and
valence bands, respectively. Reference [285] also assumed the Hubbard
density of states for both the conduction and valence bands with width W$_{%
\text{c}}$ and W$_{\text{v}}$, respectively, as well as similar dispersion
in both bands. With this assumption the exciton binding energy has been
calculated according to the CPA model (see, also [210]). It should be added
here the ley feature of the model developed in Ref. [285] is the short-range
nature of the Coulomb potential.

The data from Table 17 and other published sources (see [49, 221, 277]) were
used for plotting the energy E$_{\text{B}}$ as a function of isotopic
concentration x in Fig. 49. The values of binding energy E$_{\text{B}}$ were
calculated using the hydrogen-like formula (see below) with the energies of
exciton levels of Is and 2s states being found from the reflection spectra
(see Fig. 34). Theoretical description of the binding energy of Wannier-
Mott excitons as a function of x was based on the polynomial derived by
Elliott and coworkers [285]:

E$_{\text{b}}$ = E$_{\text{b}}^{\text{crys}}$- E$_{\text{bow}}\left[ \frac{%
\text{1-W}}{\text{2U}_{\text{0}}}\right] $-E$_{\text{eff}}$, \ \ \ \ (127)

E$_{\text{eff}}$ = x$\left( \text{1-x}\right) \frac{\delta _{\text{c}}\delta
_{\text{v}}}{\text{W}}$, \ \ \ \ \ \ \ \ \ \ \ \ \ \ \ \ \ \ \ \ \ \ \ \
(128)

E$_{\text{b}}^{\text{crys}}$ = U$_{\text{0}}$ + $\frac{\text{W}}{\text{2U}_{%
\text{0}}}$ - W. \ \ \ \ \ \ \ \ \ \ \ \ \ \ \ \ \ \ (129)

where W = W$_{\text{c}}$ + W$_{\text{v}}$, and W$_{\text{c}}$ and W$_{\text{v%
}}$ are the widths of the conduction band and the valence band which are
equal to 21 eV [291] and 6 eV [292, 293] respectively. Here E$_{\text{bow}}$
is the curvature parameter found from the function E$_{\text{g}}$ $\propto $
f(x); $\delta _{\text{c}}$ and $\delta _{\text{v}}$ are the magnitudes of
the fluctuation smearing of the valence band and the conduction band edges , 
$\delta _{\text{c}}$ = 0.103 eV and $\delta _{\text{v}}$ = - 0.331 eV. As
follows from Fig. 49, these values of the parameters give a good enough
description of the nonlinear dependence of the binding energy of
Wannier-Mott exciton in disordered medium. This agreement between theory and
experiment once again proves the inherent consistency of the model proposed
by Kanehisa and Elliott, since the isotopic substitution affects the
short-range part of the interaction potential.

In this way, the nonlinear dependence of the binding energy of Wannier-Mott
exciton is caused by isotopic disordering of the crystal lattice. As is see
from Fig. 49 the exciton binding energy decreasing (relative linear law
(VCA)-see dashed line in Fig. 49) in the vicinity of the middle meaning
concentration really calls out the fluctuated broadening of the edge of the
conduction and valence bands. In accordance with the theoretical model the
last reason gives rise to the reduced E$_{\text{g}}$ and there by the
shallowing of the exciton levels and, respectively, the reduction of E$_{%
\text{b}}$ (for more details see [151, 277]). \ 

\bigskip

4.5. Zero - point vibration energy.

\bigskip

The short history of zero - point vibration energy can be found in
Rechenberg's paper [294]. According to Rechenberg M. Panck used this
conception at a Berlin meeting of German Physical Society at 1911 [295]
after that it used numerously (see, e.g. [252, 296]). For the first time in
solid state physics the conception of zero - point vibration energy probably
have used \ by Fan [297] and for the interpretation of experimental results
by Kreingol'd et al. [224, 225, 227], Cardona et al. (see, e.g. 298, 27]) as
well as Agekyan and coworkers [201]. The classical definition of this
conception have proposed by Baym [299]...." The minimum amount of kinetic
energy coming from the uncertainty principle is called zero - point energy"
(see, also [123]).

The effect of changing the atomic mass M is to change the phonon frequencies 
$\omega $ according to Equation (101)

$\omega $ = $\sqrt{\frac{\alpha }{\text{M}}}$, \ \ \ \ \ \ \ \ \ \ \ \ \ \ \
\ \ \ \ \ \ \ \ \ \ \ \ \ \ \ \ \ \ \ \ \ \ \ \ \ \ \ \ \ \ \ \ \ \ \ \ \ \
\ \ \ \ \ \ \ \ \ \ \ \ \ \ \ \ \ \ \ \ \ \ \ \ \ \ \ \ \ \ \ \ \ \ \ \ \
(101')

where $\alpha $ is a force constant characteristic of the phonon under
consideration (see above). The change in atomic mass implies, at low
temperatures (see below), a change in the average atomic displacement for
each phonon mode. In the case of one atom per primitive cell the mean
squared phonon amplitude $\langle $u$^{2}\rangle $ is given by [284]:

$\langle $u$^{2}\rangle $ = $\langle \frac{\hbar ^{2}}{\text{4M}\omega }%
\left[ 1\text{ + 2n}_{B}\text{(}\omega \text{)}\right] \rangle $ = $\langle 
\frac{\hbar }{\text{4M}^{1/2}\alpha ^{1/2}}\left[ 1\text{ + 2n}_{B}\text{(}%
\omega \text{)}\right] \rangle $, \ \ \ \ \ \ \ \ \ \ \ \ \ \ \ \ \ \ \ \ \
\ \ (130)

where n$_{B}$($\omega $) is the Bose - Einstein statistical factor, $\omega $
is the frequency of a given phonon and $\langle $...$\rangle $ represents an
average over all phonon modes. \ The average in r.h.s. of (130) is often
simplified by taking the value inside $\langle $...$\rangle $ at an average
frequency $\omega _{D}$ which usually turns out to be close to the Debye
frequency. We should distinguish between the low temperature ($\hbar \omega $
\TEXTsymbol{>}\TEXTsymbol{>} k$_{B}$T) and the high temperature ($\hbar
\omega $ \TEXTsymbol{<}\TEXTsymbol{<} k$_{B}$T) limits and see:

($\hbar \omega $ \TEXTsymbol{>}\TEXTsymbol{>} k$_{B}$T), \ \ \ \ \ \ \ \ \ \ 
$\langle $u$^{2}\rangle $ = $\frac{\hbar }{\text{4M}\omega _{D}}$ $\sim $ M$%
^{-1/2}$ \ \ independent of T and

($\hbar \omega $ \TEXTsymbol{<}\TEXTsymbol{<} k$_{B}$T), \ \ \ \ \ \ \ \ \ \ 
$\langle $u$^{2}\rangle $ = $\frac{\text{k}_{B}\text{T}}{\text{2M}\omega ^{2}%
}\sim $ T \ \ \ \ \ \ \ \ \ independent of M \ \ \ \ \ \ \ \ \ \ (131).

Using Eq. (130) we can find from last equations that $\langle $u$^{2}\rangle 
$, the zero-point vibrational amplitude, is proportional to M$^{-1/2}$ \ at
low temperatures: it thus decrease with increasing M and vanishes for M $%
\longrightarrow \infty $. For high T, however, we find that \ $\langle $u$%
^{2}\rangle $ is independent of M and linear in T (details see [26] and
references therein).

The temperature and isotopic mas dependence of a given energy gap E$_{g}$(T,M%
$_{i}$) can be described by average Bose-Einstein statistical factor n$_{B}$
corresponding to an average phonon frequency $\theta _{i}$ as (see also
[300, 298, 233]

E$_{g}$(T,M$_{i}$) = E$_{bar}$ - a$_{r}$($\frac{M_{nat}}{M_{i}}$)$^{1/2}%
\left[ \text{1 + 2n}_{\text{B}}\right] $, \ \ \ \ \ \ \ \ \ \ \ \ \ \ \ \ \
\ \ \ \ \ \ \ \ \ \ \ \ \ \ \ \ \ \ \ \ \ \ \ \ \ \ (132)

where n$_{B}$ \ = 1/[exp$\left( \frac{\theta _{i}}{\text{T}}\right) $ -1]
and E$_{bar}$ and a$_{r}$ the unrenormalized (bare) gap and the
renormalization parameter, respectively. In the low-temperature limit, T%
\TEXTsymbol{<}\TEXTsymbol{<}$\theta _{i}$, equation (132) reduces

E$_{g}$(T,M$_{i}$) = E$_{bar}$ - a$_{r}$($\frac{M_{nat}}{M_{i}}$)$^{1/2}$ \
\ \ \ \ \ \ \ \ \ \ \ \ \ \ \ \ \ \ \ \ \ \ \ \ \ \ \ \ \ \ \ \ \ \ \ \ \ \
\ \ \ \ \ \ \ \ \ \ \ \ \ \ \ \ \ \ \ \ \ \ (132)

Here E$_{g}$(T,M$_{i}$) is independent of temperature and proportional to
(1/M$_{i}$)$^{1/2}$, whereas a$_{r}$ is the energy difference between the
unrenormalized gap (M$_{i}$ $\rightarrow \infty $) and the renormalized
value [297, 301].

In the high-temperature limit, T\TEXTsymbol{>}\TEXTsymbol{>}$\theta _{i}$
and Eq. (132) can be written as

E$_{g}$(T,M$_{i}$) = E$_{bar}$ - 2T$\frac{a_{r}}{\theta }$, \ \ \ \ \ \ \ \
\ \ \ \ \ \ \ \ \ \ \ \ \ \ \ \ \ \ \ \ \ \ \ \ \ \ \ \ \ \ \ \ \ \ \ \ \ \
\ \ \ \ \ \ \ \ \ \ \ \ \ \ \ \ \ \ \ \ \ \ (133)

and E$_{g}$(T,M$_{i}$) is independent of M$_{i}$ [5]. The extrapolation of
Eq. (133) to T = 0K can be used to determine the unrenormalized gap energy E$%
_{bar}$ , i.e., the value that corresponds to atoms in fixed lattice
position without vibrations (frozen lattice [278]), from the measured
temperature dependence of E$_{g}$(T) in the high-temperature (i.e. linear in
T) region. Fig. 50 shows the indirect gap of silicons T [303].It is
important note that the experimental T range of Fig. \% (confined to T 
\TEXTsymbol{<} $\theta $) does not suffice to determine the asymptotic
behavior for T $\longrightarrow $ $\infty $. The experimental range is
usually limited (to 300K in Fig 50) by the broadening of optical spectra
with increasing T. As already mentioned, the zero - point renormalization of
a gap can also be estimated from the corresponding isotope effect. In the
case of E$_{ind}$ for silicon, a gap shift of - 1.04 meV/amu is found for T 
\TEXTsymbol{<}\TEXTsymbol{<} $\theta $ [302] On the basis of the Eq. (133),
the isotope effect leads to a renormalization of 60$\pm $1 meV for the E$%
_{ind}$ in silicon, in reasonable agreement with the value $\sim $ 57 meV
found from the E$_{ind}$ (T) asymptote. The isotope shift, and the
corresponding zero - point renormalization, have also been determined for
the indirect \ gap of Ge (see Fig. 51 [26]). They amount to (at T $\simeq $
6K): isotope shift equals 0.36 meV/amu, renormalization equals -53 meV. The
value of - 370 meV found for the zero - point renormalization of E$_{ind}$
in diamond (see, Fig. 52) can be used to obtain the dependence on isotopic
mass by means of the M$^{-1/2}$ rule (Eq.(132)) [27]

$\frac{\text{dE}_{ind}\text{(0)}}{\text{dM}}$ = $\frac{\text{1}}{2}$ $\cdot $
370 $\cdot $ $\frac{\text{1}}{\text{13}}$ = 14.2 meV/amu \ \ \ \ \ \ \ \ \ \
\ \ \ \ \ \ \ \ \ \ (134).

The last value is in excellent agreement with the value estimated above. A
recent semiempirical LCAO calculation results in a renormalization of 600
meV [298], even larger than the experimental values. The value of the
exciton energy shown in Fig. 52, 5.79 eV, can be compared with ab initio
calculations of the \ indirect gap \ [27], which yield an indirect gap \ of
5.76 eV. This value must be compared with that from Fig. 52 adding the
exciton binding energy (5.79 + 0.07 = 5.86 eV). This experimental gap value
according to [27] is in rather good agreement with ab initio calculations.
The agreement becomes considerably worse if the zero - point renormalization
of 370 meV is not taken into account (5.48 vs. 5.76) All these arguments
support \ the large zero - point renormalization found for the indirect gap
of diamond.

Using Eq. (132) it can be written the difference in energy $\Delta $E$_{g}$
between a given energy gap in isotopically pure material (LiH) and its
isotope analogue (LiD)

$\Delta $E$_{g}$ = E$_{g}$(M$_{i}$) - E$_{g}$(M$_{nat}$) = a$_{r}$ $\left[ 
\text{1 - }\left( \frac{M_{nat}}{M_{i}}\right) ^{1/2}\right] $. \ \ \ \ \ \
\ \ \ \ \ \ \ \ \ \ \ \ \ \ \ \ \ \ \ \ \ (135)

As can be seen from Table 1 and results of [301] $\Delta $E$_{g}$ at 2K
equals $\Delta $E$_{g}$ = 0.103 eV and E$_{g}$(LiH, T = 0K) =5.004 eV
(linear approximation and E$_{g}$(LiH, T = 300K) = 4.905 eV then using Eq.
(133) we obtain a$_{r}$ = 0.196 eV. This magnitude is very close
(approximately 84\%) to the value of 0.235 eV of zero (see Fig. 53)
vibration renormalization of the energy band gap in LiH crystals (details
see [301, 203]). Using Eq. (135) we obtain $\Delta $E$_{g}$(theor) = 0.134
eV that is very close, on the other hand, to observed experimental value
equals 0.103 eV. The discrepancy between these values may be caused by the
negligible contribution of the isotopic lattice expansion to the band gap
renormalization as well as small temperature range. We should add, so far
LiH and LiD have different temperature coefficients $\frac{\text{dE}_{n\text{
= 1s}}}{\text{dT}}$(LiH) = 0.19 meVK$^{-1}$ and $\frac{\text{dE}_{n\text{ =
1s}}}{\text{dT}}$(LiD) = 0.25 meVK$^{-1}$ [277], we may conclude that the
isotope effect at high temperature is disappeared ( Fig. 54 [303]) (see,
also [304]).

Several groups have conducted low-temperature studies of the direct and
indirect band gaps of natural and isotopically controlled Ge single
crystals. Parks et al. [280] have used piezo - and photomodulated
reflectivity spectra of four monoisotopic and one natural Ge crystals. These
techniques do not require the extreme sample thinning which is necessary for
optical-absorption measurements and the derivative nature of the spectra
emphasizes the small changes. The excellent signal - to - noise ratio and
the superb spectral resolution allowed a very accurate determination of the
dependence of E$_{DG}$ on isotopic mass. At very low temperatures an inverse
square-root dependence accurately describes band-gap dependence \ E$_{DG}$ =
E$_{DG}^{\infty }$ + $\frac{C}{\sqrt{M}}$ .

A fit through five data points yields E$_{DG}^{\infty }$ = 959 meV and C =
-606 meV/amu$^{1/2}$. Written as a linear dependence for the small range of
isotopic masses, Parks et al. find dE$_{DG}$/dA = 0.49 meV/amu, in perfect
agreement with the results of other researchers [27]. Parks et al. also
determined the isotope mass dependence of the sum of the direct gap and the
split-off valence band $\left( \Delta _{0}\right) $ and find d(E$_{DG}$ + $%
\Delta _{0}$)/dA = 0.74 meV/amu. The experimental results can be compared to
the Zollner et al. [298] calculations which are found to be of the correct
order of magnitude. The theoretical estimates for the contributions of the
linear isotope shifts of the minimum, indirect gaps which are caused by
electron-phonon interaction, are too large by a factor of $\sim 1.7$ and for
the smallest direct gap they are too large by a factor $\sim $ 3.2. The
analogous results were obtained by Zhernov [305], who have criticized the
approach of Cardona et al. in estimation of zero - point renormalization of
the band gap.

\bigskip

\bigskip

\bigskip \textbf{5. Some modern application of the stable and radioactive
isotopes.}

\bigskip

5.1. Stable isotopes.

5.1.1. Traditional applications.

\bigskip

Present part is devoted to description of different applications of the
stable isotope effect in solids. Capture of thermal neutrons by isotope
nuclei followed by nuclear decay produces new elements, resulting in a very
number of possibilities for isotope selective doping of solids [31]. There
are presented different facilities which use in this reactor technology. The
feasibility of constructing reactors dedicated to the production of \
neutron transmuting doped (NTD) silicon, germanium (and other compounds) was
analyzed in terms of technical and economic viability and the practicality
of such a proposal is examined [30]. The importance of this technology for
studies of the semiconductor doping as well as metal-insulator transitions
and neutral impurity scattering process is underlined. The introduction of
particle irradiation into processing of semiconductor materials and devices
creates a new need for additional understanding of
atomic-displacement-produced defects in semiconductors. The use of the
isotopes in a theory and technology of the optical fibers we briefly
considered too. For the first time it was shown also the influence of the
isotopes on properties of the optical fibers. There is short description of
theory and practice of semiconductor lasers. The discovery of the linear
luminescence of free excitons observed over a wide temperature range has
placed lithium hydride, as well as crystals of diamond in line as
prospective sources of coherent radiation in the UV spectral range. For LiH
isotope tuning of the exciton emission has also been shown [3].

In the next part we detail analyze the process of self-diffusion in isotope
pure materials and heterostructure. Interest in diffusion in solids is as
old as metallurgy or ceramics, but the scientific study of the phenomenon
may probably be dated some sixth-seven decades ago. As is well-known, the
measured diffusion coefficient depends on the chemistry and structure of the
sample on which it is measured. We are briefly discussed the SIMS (secondary
ion mass spectrometry) technique. In SIMS technique, the sample is bombarded
by reactive ions, and the sputtered-off molecules are ionized in a plasma
and fed into a mass-spectrometer. Self-diffusion is the migration of
constituent atoms in materials. This process is mediated by native defects
in solids and thus can be used to study the dynamics and kinetics of these
defects. The knowledge obtained in these studies is pivotal for the
understanding of many important mass transport processes such as impurity
diffusion in solids (see, e.g. reviews [306 - 308] and monographs [309 -
310]). Another area of interest in a detailing understanding of the self -
and foreign - atom diffusion behavior is self - and dopant diffusion in Si -
Ge alloys. Dopant diffusion in Ge is not well understood [311] and accurate
diffusion data are very limited [312]. Dopant diffusion in Ge and SiGe
isotope multilayer structures would significantly improve our understanding
of the mechanism of diffusion in Ge \ and SiGe alloys and how they change
with alloy composition and strain. This information is important in
particular for the integration of these materials in the next generation of
electronic devices. The advances in the understanding of atomic - transport
processes and defect reactions in semiconductors has contributed to the
remarkable increase of computer speed and capacity over the last two
decades. On the other hand, the higher computer power gives rise to more
reliable and predictive theoretical calculations of the properties of point
defects in semiconductors and hence also expedites their own technological
evolution [313, 314]. Very perspective is isotope-based quantum computer. We
should add here that the strength of the hyperfine interaction is
proportional to the probability density of the electron wavefunction at the
nucleus. In semiconductors, the electron wavefunction extends over large
distances through the crystal lattice. Two nuclear spins can consequently
interact with the same electron, leading to electron-mediated or indirect
nuclear spin coupling. Because the electron is sensitive to externally
applied electric fields, the hyperfine interaction and electron-mediated
nuclear spin interaction can be controlled by voltages applied to metallic
gates in a semiconductor device, enabling the external manipulation of
nuclear spin dynamics that is necessary for quantum computation in quantum
computers (for details see [3, 30]).

\bigskip

5.1.1.1. Diffusion.

\bigskip

\ \ One of the fundamental processes occuring in all matter is the random
motion of its atomic constituents. In semiconductors much has been learned
in recent years about the motion of host and impurity atoms as well as
native defects such as vacancies and interstitials. A number of excellent
review papers [306 - 308] and monographs [309 - 310] have been written .
When a concentration gradient dN/dx is introduced, random motion leads to a
net flux of matter J which is proportional to the gradient (Fick's first
law, see, e.g. [315])

J = -DdN/dx \ \ \ \ \ \ \ \ \ \ \ \ \ \ \ \ \ \ \ \ \ \ \ \ \ \ \ \ \ \ \ \
\ \ \ \ \ \ \ \ \ \ \ \ \ \ \ \ \ \ \ \ \ \ \ (136).

The diffusion coefficient D can in many (though not in all) cases be
described by a thermally activated constant (see, also [316]):

D = D$_{0}$exp(-E/k$_{B}$T). \ \ \ \ \ \ \ \ \ \ \ \ \ \ \ \ \ \ \ \ \ \ \ \
\ \ \ \ \ \ \ \ \ \ \ \ \ \ \ \ \ \ \ (137)

Impurity diffusion in semiconductors plays a key role in the fabrication of
electronic devices. For example, diffusion can be utilized as a desirable
process enabling the introduction of dopant impurities into areas defined by
a mask on a semiconductor wafer. Diffusion can act also as a determinal
process broadening narrow impurity implantation profiles or rapidly
admitting diffusion of undesirable impurities. There exists a very extensive
literature on diffusion studies for most semiconductors. The field is
extremely active especially for the semiconductors which have become
important recently for high-temperature electronics (e.g. C; SiC), light
emitting devices working in the green and blue range of the visible spectrum
(e.g. ZnSe, GaN) and IR records (Ge).

As simple as diffusion may appear to be, at least conceptually, there still
exist many basic unanswered questions. Results from supposedly identical
experiments conducted by different groups often scatter by significant
factors. This clearly indicates that there are still hidden factors which
need to be determined. Even for the most thoroughly studied crystalline
solid, Si (see, e.g. [311, 312] references therein), we still do not know
with certainty the relative contributions of vacancies and interstitials to
self- and impurity diffusion as a function of temperature, the position of
the Fermi level, and external effects such as surface oxidation and
nitridation (see also [313]).

Fuchs et al. [317] presented results of a very accurate method to measure
the self-diffusion coefficient of Ge which circumvents many of the
experimental problems encountered in the conventional methods. Fuchs et al
used Ge isotopic heterostructures (stable isotope), grown by molecular-beam
epitaxy (MBE). In general, isotope heterostructures consist of layers of
pure (e.g. $^{70}$Ge, $^{74}$Ge) or deliberately mixed isotopes of a
chemical element. In this paper Fuchs et al. [317] used the isotope
heterostructure growthing on the Ge substrate. At the interface only the
atomic mass is changing, while (to first order) all the other physical
properties stay the same. In the as-grown samples, this interface is
atomically flat with layer thickness fluctuations of about two atomic ML
[318]. Upon annealing, the isotopes diffuse into each other (self-diffusion)
with a rate which depends strongly on temperature. The concentration
profiles were measured with SIMS (secondary-ion-mass spectroscopy), after
pieces of the same samples have been separately annealed at different
temperature. This allows an accurate determination of the self-diffusion
enthalpy as well as the corresponding entropy. The isotopic heterostructures
are unique for the self-diffusion studies in several respects.

1) The interdiffusion of Ge isotopes takes place at the isotopic interface 
\textbf{inside} the crystal, unaffected by possible surface effects (e.g.,
oxidation, strains and impurities) encountered in the conventional technique
(see Fig. 55).

2) One sample annealed at one temperature provides five more or less
independent measurements: Germanium consists of five stable isotopes. Their
initial respective concentrations vary for the different layers of the
as-grown isotope heterostructure. After annealing, the concentration profile
of each of the five isotopes can be analyzed separately to obtain five data
points for each annealing temperature. Tan et al. [319] were the first to
make use of GaAs isotope superlattice ($^{69}$GaAs; $^{71}$GaAs) on a
Si-doped substrate to study Ga self-diffusiuon. Unfortunately however, their
analysis was only partially succesful because native defects and silicon out
diffusion from the doped substrate into the superlattice obscured their
results (see, also [317]).

Contrary to the short period superlattices required for Raman experiments,
Fuchs et al. [317] used sufficiently thick layers to access $\sqrt{Dt}$
products of one to several micrometers. This first studies as was mentioned
above will focus on Ge self-diffusion in undoped material. They are working
with bilayers of $^{70}$Ge and $^{74}$Ge (each 1000 or 2000 \AA\ thick)
which were grown by MBE on a natural substrate. Disregarding for the moment
the small differences in diffusity caused by different isotopes masses (see,
also [320]), they expected the Ge isotopes to diffuse symmetrically into
each other following complementary error functions. There exists no net flow
of Ge atoms and the atomic concentrations add up to unity at every point: $%
\left[ ^{70}\text{Ge}\right] $ + $\left[ ^{74}\text{Ge}\right] $ = 1. The
individual profiles are described by

$\left[ ^{70}Ge\right] _{x}$ = 0.5$\left[ ^{70}\text{Ge}\right] _{0}\left\{ 
\text{1 - erf}\left( x/2\sqrt{\text{Dt}}\right) \right\} $ .\ \ \ \ \ \ \ \
\ \ \ \ \ \ \ \ \ \ \ \ \ \ \ \ \ \ (138)

$\left[ ^{74}\text{Ge}\right] _{x}$ = 0.5$\left[ ^{74}\text{Ge}\right]
_{0}\left\{ \text{1 - erf}\left( \text{- x/2}\sqrt{\text{Dt}}\right)
\right\} $. \ \ \ \ \ \ \ \ \ \ \ \ \ \ \ \ \ \ \ \ \ \ \ (139)

The interface is located at x = 0 and $\left[ ^{70}\text{Ge}\right] _0$ = $%
\left[ ^{74}\text{Ge}\right] _0$ 4.4$\cdot $10$^{22}$ cm$^{-3}$.

For the experiments Fuchs et al. [317] have chosen five diffusion
temperatures and have adjusted the times so as to obtain similar $\sqrt{%
\text{Dt}}$ products. Fig. 56 shows SIMS results for the as-grown sample and
for the sample diffused at 586$^{0}$ C for the 55.55h. Because the isotopes
are only enriched into the high 90\% range, they obtained SIMS data from
some of the residual minor Ge isotopes. This redundancy in data is very
useful in the deconvolution of the SIMS instrument function and in improving
the accuracy of the data. The obtained results by Fuchs et al. are in
excellent agreement with previously published values.

Ga self-diffusion in GaAs [321] and GaP [322] was measured directly in
isotopically controlled GaAs and GaP heterostructures. In the case of GaP,
for the experiment, $^{71}$GaP and $^{69}$GaP epitaxial layers 200 nm thick
were grown by solid source MBE at 700$^{0}$ C on undoped GaP substrates. The
natural Ga isotope composition in the GaP substrates is 60.2\% $^{69}Ga$ and
39.8\% $^{71}$Ga. The compositions in the isotopically controlled epilayers,
on the other hand, were 99.6\% $^{69}$Ga ($^{71}$Ga) and 0.4\% $^{71}$Ga($%
^{69}$Ga). In the SIMS measurement, the primary ion beam was formed with 5.5
keV Cs$^{+}$ ions. GaCs$^{+}$ molecules were detected as secondary species
as the sputtering proceeded.

As before, assuming Fick's equations describe the self-diffusion process and
the diffusion coefficient D is constant [316], the concentrations of the Ga
isotopes can be expressed as

C(x) = $\frac{C_{1}\text{ + C}_{2}}{2}$ - $\frac{C_{1}\text{ - C}_{2}}{2}$%
erf(x/R), \ \ \ \ \ \ \ \ \ \ \ \ \ \ \ \ \ \ \ \ \ \ \ \ \ \ \ \ \ \ \ \ \
\ \ \ \ (140)

where x = 0 at the epitaxial interface, C$_{1}$ and C$_{2}$ are the initial
Ga isotope concentrations at the left- and right-hand side of the interface,
respectively, and erf(y) is the error function. The characteristic diffusion
length R was defined as

R = 2$\sqrt{\text{Dt }}$, \ \ \ \ \ \ \ \ \ \ \ \ \ \ \ \ \ \ \ \ \ \ \ \ \
\ \ \ \ \ \ \ \ \ \ \ \ \ \ \ \ \ \ \ \ \ \ \ \ \ \ \ \ \ \ \ \ \ \ \ \ \ \
\ \ \ \ \ \ \ (141)

where D is the Ga self-diffusion coefficient and t is the annealing time.

The SIMS data can then be compared with calculated values of C(x) .
Adjusting the diffusion length R, a fit of C(x) to the SIMS profile can be
made. Fig. 57 shows the SIMS profiles (sold lines) and the calculated C(x)
of $^{69}$Ga (circles) and $^{71}$Ga (continuous line) in a sample annealed
at T = 1111$^{0}$ C for3 h and 51 min. Excellent agreement was obtained
between the measured and the calculated profiles over two and a half orders
of magnitude in concentrations. From these results for GaP Wang et al. [322]
obtained the values of the activation enthalpy H$^{SD}$ and self-diffusion
entropy S$^{SD}$ equal to 4.5 eV and 4k$_{B}$, respectively. For comparison,
Wang et al. [322] obtained the activation enthalpy and entropy for GaAs are
4.24 eV and 7.5k$_{B}$, respectively. The significant difference in values
of S$^{SD}$, according Wang et al., indicates profound variations in the way
that the mediating native defects are formed or migrate in GaP as compared
to GaAs. The small value S$^{SD}$ in GaP may be connected to the stronger Ga
- P bond compared to the Ga - As bond (see, also [323]).

Bract et al. [324] studied (see, also Fig. 58) the Ga self-diffusion and Al
- Ga interdiffusion with isotope heterostructures of AlGaAs/GaAs. Ga
diffusion in Al$_{x}$Ga$_{1-x}$As with x = 0.41; 0.62; 0.68 and 1.0 was
found to decrease with increasing Al concentration. The intermixing observed
at AlGaAs/GaAs interfaces was accurately described if a
concentration-dependent interdiffusion coefficient was assumed. The higher
Al diffusity in GaAs as compared to Ga self-diffusion was attributed to the
higher jump frequency of $^{27}$Al as compared to $^{71}$Ga caused by the
difference in their masses. The lower Ga diffusity in AlAs compared to GaAs
was proposed to be due to lower thermal equilibrium concentrations of
vacancies (C$_{v}^{eq}$) in ALAs as compared to GaAs. The different values C$%
_{v}^{eq}$ in these materials were explained by the differences in the
electronic properties between AlAs and GaAs. We should add here that the
value of activation enthalpy Q of studied heterostructures lies in the range
3.6 $\pm $ 0.1 that is consistent with Wee et al. [325] results.

Very recently Bracht et al. [314] have reported the diffusion of boron,
arsenic and phosphorus in silicon isotope multilayer structures at
temperatures between 850$^{0}$ C and 1100$^{0}$C. The diffusion of all
dopants and self - atoms at a given temperature is modeled with the same
setting of all native - point - defect - related parameters. As an example
on the Fig. 58 the concentration profiles of $^{31}$P along with the
corresponding $^{30}$Si profiles measured with SIMS after diffusion are
shown.The diffusion of P in the isotope structure leads to an I
(interstitial) supersaturation and V (vacancy) undersaturation and therewith
suppresses the contribution of vacancies in P and Si diffusion compared to
the diffusion of As. This increases the sensitivity of the P and Si profiles
to negatively charged self - interstitials. The demand to describe the
diffusion of all dopants and the corresponding Si profiles at a given
temperature with the same energy levels of the native - point defects led to
the conclusion \ that negatively charged vacancies rather than negatively
charged self - interstitials dominate under n - type doping. Successful \
modeling of the simultaneous P and Si diffusion requires a contribution of a
singly positively charged mobile P defects to P diffusion (for details see,
also [313]).

\bigskip

5.1.1.2. Neutron transmutative doping of semiconductors.

\bigskip

The main advantage of the neutron transmutative doping (NTD) method, as we
know at present, is the precision doping which is connected with the linear
dependence of concentration of doping impurities on the doze of neutron
irradiation. Such dependence is numerous observed in the different
experiments (see, e.g. [326 - 328]). As an example, in Fig. 60 there is
shown the dependence of the concentrations doped phosphorus on the doze of
irradiation the Si crystal in nuclear reactor. This dependence was measured
with the help of Hall effect [329]. However, at the large doze of neutron
irradiation there is observed the non - linear dependence. On Fig. 45 in
Ref. 30 was shown \ the results of paper [330] \ where was observed the
deviation from linear law at the large doze of neutron irradiation of the
sample of $^{74}$Ge which was annealed after irradiation at 460$^{0}$C
during different time. \ More amazing effect was observed at the second
irradiation of the samples of $^{74}$Ge previously strong doped with As by
NTD method. Instead expectable increase of the concentration free charges
(electrons) n there is observed the decrease n. This decrease was direct
proportional to the neutron irradiation doze of $^{74}$Ge crystals. Both
effects are rather details analyzed in papers [327, 330]. The transmutation
of the stable germanium isotopes via capture of thermal neutrons is well
understood. Considering Ge and Si it should be note, that donors As and P
will be created , following neutron capture and $\beta $ - decay of isotopes
of these semiconductor elements. The new elements are, of course, the
prototypical donors. Neutron capture leads to NTD. Beside that, there are
elements which have light isotopes which upon neutron capture transmute to a
lower Z element either by electron capture or by positron decay. In this
case acceptors are created. A classical case is the transmutation of $%
_{32}^{70}$Ge into $_{31}^{71}$Ga (for details see [30]).

It is well-known that doping of silicon single crystals by incorporation of
impurities from the melt during solidification in most cases leads to an
inhomogeneous distribution of impurities in the solids [331 - 333]. This is
due to the fact that nearly all impurities in silicon have thermal
equilibrium distribution coefficients much less than unity and that the
solidification or crystal growths at each position of the interface is
characterized by a different state of thermal inequilibrium leading to
distribution coefficients that in space and time continuously change and
result in a nonuniform impurity distribution [334, 335]. In actual crystal
production the nonuniformity is further enhanced by lack of control of
exactly constant melt volume and feed of the doping impurity. The most
widely used doping elements in silicon are boron and phosphorus. Boron has a
distribution coefficient between 0.9 and 1 which makes a doping uniformity
of $\pm $10\% easily obtainable (see, e.g. [333]). The thermal equilibrium
distribution coefficient for phosphorus of approximately 0.3 leads in
general to the above mentioned large doping variations both on a microscale
(center to periphery) and on a microscale (striations). No other n-type
doping element has a larger distribution coefficient. Because fast diffusing
p-type dopants (Ga, Al) are available, because electron mobility is greater
than hole mobility, and because contact alloying technology is reasonable,
n-type silicon is generally used for solid state power devices [334, 336].
With avalanche breakdown voltages being determined from areas with lower
resistivities, use of a conventionally doped material results in hot-spot
formation prior to breakdown and too high forward voltage drop leading to
excessive heat dissipation because of a safe punch through design [331, 333,
337].

Phosphorus doping by means of NTD was suggested by Lark-Horovitz [338] and
Tanenbaum and Mills [339] for homogeneity purposes and has been applied for
high-power thyristor manufacturing in [331, 333, 336]. Hill et al. [336]
were demonstrated how such a homogeneous phosphorus doping may result in a
\textquotedblright theoretical design\textquotedblright\ possibility for
high-power components (see also below).

The process used for fractional transmutation of silicon into phosphorus and
thereby performing n-type doping

$_{14}^{30}$Si(n,$\gamma $) $=$ $\ _{14}^{31}$Si $_{2.62h}^{\beta
^{-}}\rightarrow $ $\ _{15}^{31}$P \ \ \ \ \ \ \ \ \ \ \ \ \ \ \ \ \ \ \ \ \
\ \ \ \ \ \ \ \ \ \ (142)

was first pointed out by Lark-Horovitz in 1951 [338]. Apart from special
applications and research, the above process was, however, not utilized to
any extent until the early seventies, at which time manufacturers of
high-power thyristors and rectifiers for high-voltage direct current
transmission lines, in particular, initiated usage of the transmutation
doping process [340, 336, 341]. The reasons for not using the neutron doping
method throughout the sixties may be found in the lack of a processing
technology which could benefit from a more uniform doping, insufficient
availability of high resistivity starting material, and the lack of nuclear
reactors with irradiation capacities in excess of that needed for testing
fuel and materials for nuclear power stations.

Let us, for the following discussion, assume that completely uniform neutron
doping may be accomplished. The homogeneity of the doped silicon is in this
case determined by the background doping, i.e., the distribution of
impurities in the starting material, where the net impurity concentration
may be of either donor or acceptor type. Let us further, for simplicity,
consider starting material of one conductivity type and assume complete n -
type conduction after irradiation and annealing. With C$_{S}$ being the net
impurity concentration of the starting material and C$_{D}$ the resulting
donor concentration after irradiation we have, for both n - and p - type
material,

C$_{D}^{\max }$ - C$_{D}^{\min }$ = C$_{S}^{\max }$ - C$_{S}^{\min }$. \ \ \
\ \ \ \ \ \ \ \ \ \ \ \ \ \ \ \ \ \ \ \ \ \ \ \ \ \ \ \ \ \ \ \ \ \ \ \ \ \
\ \ \ (143)

In such case we may define

1) the homogeneity factors for the starting material ($\alpha _S$) and for
the neutron doped material ($\alpha _D$), respectively

$\alpha _{S}$ = $\frac{\text{C}_{S}^{\min }}{C_{S}^{\max }}$ \ \ \ \ \ \ \ \
\ \ \ \ \ \ \ \ \ \ \ \ \ \ \ \ \ \ \ \ \ \ \ \ \ \ \ \ \ \ \ \ \ \ \ \ \ \
\ \ \ \ \ \ \ \ \ \ \ \ \ \ \ \ \ \ \ \ \ \ \ \ \ \ \ \ \ (144)

and

$\alpha _{D}$ = $\frac{\text{C}_{D}^{\min }}{_{{}}\text{C}_{D}^{\max }}$ \ \
\ \ \ \ \ \ \ \ \ \ \ \ \ \ \ \ \ \ \ \ \ \ \ \ \ \ \ \ \ \ \ \ \ \ \ \ \ \
\ \ \ \ \ \ \ \ \ \ \ \ \ \ \ \ \ \ \ \ \ \ \ \ \ \ \ \ \ \ \ \ \ (145) and

2) the doping factor

f$_{D}$ = $\frac{\text{C}_{D}^{\max }}{C_{S}^{\max }}$. \ \ \ \ \ \ \ \ \ \
\ \ \ \ \ \ \ \ \ \ \ \ \ \ \ \ \ \ \ \ \ \ \ \ \ \ \ \ \ \ \ \ \ \ \ \ \ \
\ \ \ \ \ \ \ \ \ \ \ \ \ \ \ \ \ \ \ \ \ \ \ \ \ \ \ \ (146)

From this is easily derived

1 - $\alpha _{D}$ = $\frac{\text{1 - }\alpha _{S}}{\text{f}_{D}}.$ \ \ \ \ \
\ \ \ \ \ \ \ \ \ \ \ \ \ \ \ \ \ \ \ \ \ \ \ \ \ \ \ \ \ \ \ \ \ \ \ \ \ \
\ \ \ \ \ \ \ \ \ \ \ \ \ \ \ \ \ \ \ \ \ \ \ \ (147).

Table 15 of Ref. 30 summarizes values of $\alpha _{D}$ as a function of $%
\alpha _{S}$ and f$_{D}.$ It is seen that in order to obtain neutron-doped
silicon with, for instance, a homogeneity factor greater than 0.9, it is
necessary to use a doping factor of at least 7 when starting from
\textquotedblright undoped\textquotedblright\ n-type material in which the
homogeneity factor is typically not greater than 0.3 when taking the
microcavitations (striations) into account. An examples of such
neutron-doped silicon are shown in Figs. 61 and 62. It should be noted that
in terms of resistivity, which is often used for impurity characterization,
a doping factor f$_{D}$ means use of starting material with minimum
resistivity a factor f$_{D}$ or 2.8f$_{D}$ greater than the resistivity
after neutron doping for n - and p-type starting material, respectively. The
difference is due to the electron mobility being 2.8 times greater than the
hole mobility. In conclusion of this section it should be generally noted
that in order to make neutron-doped silicon with significantly more uniform
resistivity than conventionally doped material, a doping factor f$_{D}$ = 5
or more should be applied.

Following Janus and Malmros [340] let us consider further the theoretical
case where a cylindrical silicon crystal is surrounded by a material with
the same neutron absorption and scattering efficiency as the silicon itself
(see Fig. 57 in Ref. [30].). Let us furthermore assume a thermal neutron
flux gradient along an x axis perpendicular to the crystal axis with the
neutrons coming from an external source. In this case the neutron flux will
have the form

$\Phi $ = $\Phi _{0}\cdot $ exp$\left( \text{-}\frac{\text{x}}{\text{b}}%
\right) $, \ \ \ \ \ \ \ \ \ \ \ \ \ \ \ \ \ \ \ \ \ \ \ \ \ \ \ \ \ \ \ \ \
\ \ \ \ \ \ \ \ \ \ \ \ \ \ \ \ \ \ \ \ \ \ \ \ \ \ (148)

where b, the decay length, may be obtained from the formula

b = (3 $\cdot $ $\sigma _{\limfunc{Si}}$ $\cdot $ $\sigma _{\text{Si,t}}$ $%
\cdot $ C$_{\text{Si}}^{2}$)$^{-0.5}$.\ \ \ \ \ \ \ \ \ \ \ \ \ \ \ \ \ \ \
\ \ \ \ \ \ \ \ \ \ \ \ \ \ \ \ \ \ \ \ \ \ \ \ \ \ \ \ (149)

$\sigma _{\limfunc{Si}}$ = 0.16 $\cdot $ 10$^{-24}$ cm$^{2}$ is the mean of
the absorption cross-sections for the 3 silicon isotopes, $^{28}$Si, $^{29}$%
Si and $^{30}$Si weighted with their abundance. $\sigma _{\text{Si,t}}$ =
2.3 $\cdot $ 10$^{-24}$ cm$^{2}$ is the total cross-section (absorption +
scattering) and C$_{\limfunc{Si}}$ = 4.96 $\cdot $ 10$^{22}$ cm$^{-3}$ is
the total number of silicon atoms in 1 cm$^{3}$. Hence b may be calculated:

b$_{silicon}$ = 19 cm. \ \ \ \ \ \ \ \ \ \ \ \ \ \ \ \ \ \ \ \ \ \ \ \ \ \ \
\ \ \ \ \ \ \ \ \ \ \ \ \ \ \ \ \ \ \ \ \ \ \ \ \ \ \ \ \ \ \ \ \ \ \ \ \ \
(150).

In order to improve the doping homogeneity in the cylindrical crystal this
will be slowly rotated around its axis. The time average of this flux at the
distance r from this axis is

$\bar{\Phi}$ = $\frac{\text{1}}{\pi }\int_{0}^{\pi }\Phi _{0}$exp$\left[ 
\text{-}\frac{\text{r}}{b}\text{cost}\right] $dt = $\Phi _{0}\left[ \text{1
+ }\frac{\text{1}}{\text{4}}\left( \frac{\text{r}}{\text{b}}\right) ^{2}%
\text{ + .....}\right] $. \ \ (151)

The ratio between the neutron dose at the periphery and at the axis of the
crystal cylinder will then be

$\frac{\bar{\Phi}\left( \text{a}\right) }{\bar{\Phi}\left( \text{0}\right) }%
\simeq $ 1 + $\frac{\text{1}}{\text{4}}\left( \frac{\text{a}}{\text{b}}%
\right) ^{2}$, \ \ \ \ \ \ \ \ \ \ \ \ \ \ \ \ \ \ \ \ \ \ \ \ \ \ \ \ \ \ \
\ \ \ \ \ \ \ \ \ \ \ \ \ \ \ \ \ \ \ \ \ \ \ \ \ \ \ \ \ (152)

where a is the crystal radius (Fig. 57 in Ref. [30]).

For intrinsic starting material the irradiation doped silicon will thus have
a homogeneity factor of

$\alpha _{D}$ $\simeq $ 1 - $\frac{\text{1}}{\text{4}}\left( \frac{\text{a}}{%
\text{b}}\right) ^{2}$ $\simeq $ 0.956 \ \ \ \ \ \ \ \ \ \ \ \ \ \ \ \ \ \ \
\ \ \ \ \ \ \ \ \ \ \ \ \ \ \ \ \ \ \ \ \ \ \ \ \ \ \ \ \ \ \ (153)

for an 80-mm-diameter crystal, i.e., the absorption limiting factor for the
obtainable radial variations.

In the above analysis we have neglected the effects of fast neutron
moderation in the silicon. By comparison, however, of irradiations performed
in reactors with fast neutron fluxes from 10$^{-4}$ to 1 times the thermal
flux and with different flux gradients, the authors [340] have observed no
influence on the resistivity homogeneity due to fast neutron moderation in
the silicon.

In irradiated silicon crystals for semiconductor device applications only
two isotopes $^{31}$Si and $^{32}$P are of importance in connection with
radioactivity of neutron doped material. For thermal neutron doses less than
10$^{19}$ neutron/cm$^{2}$, no other elements have been detected emitting
radiation. Futher more, $^{31}$Si, having a half - life of 2.62 h, decays to
an undetectable level in 3-5 days. For this reason, it will be discussed the
radioactivity only of the $^{32}$P isotope. Fig. 58 in Ref. [30] pictures
the $^{32}$P activity as a function of final resistivity for a variety of
thermal neutron flux levels typical for the nuclear test reactors in use. As
was shown in [340] absolute flux determination to 1\% accuracy has proven
obtainable for instance by means of calorimetric boron carbide monitors.

$_{15}^{31}$P(n,$\gamma $) $\ _{15}^{32}$P$_{14.3d}^{\beta ^{-}}$ $\
\rightarrow $ $\ _{16}^{32}$S \ \ \ \ \ \ \ \ \ \ \ \ \ \ \ \ \ \ \ \ \ \ \
\ \ \ \ \ \ \ \ \ \ \ \ \ \ \ \ \ \ \ \ \ \ \ \ \ \ \ \ (154)

as a secondary one with $^{31}$P concentration at each instant in time being
dependent on the neutron dose received and the time allowed for the $%
_{14}^{31}$Si $_{2.62h}^{\beta ^{-}}$ $\rightarrow $ $_{15}^{32}$P decay.

The use of NTD is of particular interest to thyristor manufacturers where
n-type starting material is required for the basic p - n - p structure [342,
335]. Some advantages for high power device design and performance include:

1) more precise control of avalanche breakdown voltage,

2) more uniform avalanche breakdown, i.e., greater capacity to withstand
overvoltages,

3) more uniform current flow in forward direction, i.e. greater surge
current capacity, and

4) narrower neutral zone and therefore narrower base and lower forward
voltage drop V$_f$.

The summary of some points concerning the preparation of NTD silicon for
special applications on an R and D scale describe in papers [335, 336]. The
production of large quantities of NTD silicon for power devices is described
in [343]. More recently (see, e.g. [344]) the NTD technique has been also
proposed for the effectual doping of P in a-Si:H films (see also [209]). The
results of [344] are shown that NTD technique is an excellent method for
doping of P in a-Si:H.

The NTD method have used with success in study of compound semiconductors:
GaAs [345 - 348] and GaP [349, 350]. NTD of GaAs is based on the following
thermal neutron capture nuclear reactions (see also [351]):

$^{69}$Ga (n,$\gamma $) $^{70}$Ga $_{21.1\text{ min}}^{\beta ^{-}}$ $%
\rightarrow $ $^{70}$Ge, \ \ \ \ \ \ \ \ \ \ \ \ \ \ \ \ \ \ \ \ \ \ \ \ \ \
\ \ \ \ \ \ \ \ (155)

$^{71}$Ga (n,$\gamma $) $^{72}$Ga $_{14.1h}^{\beta ^{-}}$ $\rightarrow $ $%
^{72}$Ge, \ \ \ \ \ \ \ \ \ \ \ \ \ \ \ \ \ \ \ \ \ \ \ \ \ \ \ \ \ \ \ \ \
\ \ \ \ \ (156)

$^{75}$As (n,$\gamma $) $^{76}$As $_{26.3\text{ h}}^{\beta ^{-}}$ $%
\rightarrow $ $^{76}$Se. \ \ \ \ \ \ \ \ \ \ \ \ \ \ \ \ \ \ \ \ \ \ \ \ \ \
\ \ \ \ \ \ \ \ \ \ \ \ \ (157)

The relative abundances of the isotopes involved in the reactions and the
cross-sections for these reactions are such that the ratio of Se and Ge
concentrations produced is

N$_{Se}$/N$_{Ge}$ = 1.46. \ \ \ \ \ \ \ \ \ \ \ \ \ \ \ \ \ \ \ \ \ \ \ \ \
\ \ \ \ \ \ \ \ \ \ \ \ \ \ \ \ \ \ \ \ \ \ \ \ \ \ \ \ \ \ \ \ \ \ \ \ \ \
(158)

Selenium is a typically shallow substitutional donor in GaAs with an
electronic energy level a few meV from the conduction bans edge [352].
Germanium in GaAs is an amphoteric impurity which acts as a shallow donor
(also a few meV from the conduction band) is situated on a Ga site and as an
acceptor level at E$_{V}$ + 0.04 eV if situated on an As site [351]. $%
\limfunc{Si}$nce$,$ if electronically active, all of the Se atoms and some
portion of the Ge atoms are expected to act as donors, NTD of GaAs is
expected to dope GaAs more n-type. The addition of donors moves the Fermi
level (E$_{F}$) away from the valence band (E$_{V}$) to the conduction band
(E$_{C}$). If a sufficiently high concentration of donors is added, E$_{%
\text{F}}$ will move to the upper half of the bandgap and the GaAs will be
converted to n-type. Analysis of Hall effect data as a function of
temperature provides a means of measuring the donor content in irradiated
GaAs samples. Young et al were thus able to compare electrically active
added donor content to the NTD-produced impurity concentrations determined
from nuclear measurements. The Hall effect analysis also allows them to
determine concentrations and energy levels (E) of impurities or defects in
the p-type GaAs samples if the Fermi level in the material moves near E at
some temperature over the range of measurements. This technique thus
provides a means of identifying and measuring undercompensated acceptor
content in the samples. The low temperature photoluminescence technique used
in paper [351] measured donor-to-acceptor or conduction-band-acceptor
luminescence. It provides an accurate determination of the position of
acceptor electronic levels in the GaAs, permitting positive identification
of impurities or defects with known luminescence lines. Identifications of
lines due to specific impurities or defects can be made using luminescence
techniques regardless of the position of the Fermi level in material. Little
detailed information concerning an acceptor level can be obtained from Hall
effect if that acceptor is overcompensated. However, the presence of
specific acceptors can be detected by luminescence techniques even in n-type
samples. On other hand, luminescence data do not provide the quantitative
information obtainable from Hall effect measurements.

Figures 63 and 64 show relative luminescence spectra for the four n-type
samples respectively. The spectral positions indicated by arrows for carbon
acceptor, the Ge acceptor, and 0.07 eV acceptor correspond to donor (or
band) to acceptor luminescence lines. The most important conclusion to be
drawn from a comparison of the spectra for the control and eight NTD samples
is that Ge acceptors not present in the \textquotedblright starting
material\textquotedblright\ control sample are introduce by the NTD process.
The increase in intensity of the Ge acceptor line with increasing dose
relative to both the carbon and 0.07 eV acceptor lines indicates that Ge
acceptor content increases with increasing transmutation doping. Therefore,
some of the Ge atoms produced by NTD in these samples are acting as
acceptors rather than donors. Photoluminescence measurement studies of the
control and eight annealed NTD samples at longer wavelengths indicate
another new line present only in NTD samples at about 9450 \AA . The
intensity of this line increases with increasing NTD dose.

The characteristic lifetimes of radioactive isotopes can be used to label
and identify defect levels in semiconductors which can be described in part
5.2.3.

\bigskip

5.1.1.3. Optical fiber.

\bigskip

The reflection and transmission of a plane wave, or ray, which is incident
on a planar interface between two semi - infinite, uniform media is
determined by Snell's laws (see, e.g. [353, 3]). In Fig. 65, the refractive
indices of the medium of incidence and the second medium are n$_{co}$ (core)
and n$_{cl}$ (cladding) \TEXTsymbol{<} n$_{co}$, respectively, and the
critical angle $\alpha _{c}$ = sin$^{-1}$(n$_{cl}$n$_{co}$). Further we
denote the angles of incidence, reflection and transmission, or refraction,
relative to the normal QN (see Fig. 65) by $\alpha _{i}$, $\alpha _{r}$, and 
$\alpha _{t}$, respectively. The incident, reflection and transmitted, or
refracted, rays and the normal QN are coplanar. If $\alpha _{i\text{ }}$ 
\TEXTsymbol{>} $\alpha _{c},$ the incident ray in Fig.65$^{a}$ undergoes
total internal reflection and $\alpha _{r}$ = $\alpha _{i}$, but if $\alpha
_{i\text{ }}$ \TEXTsymbol{<} $\alpha _{c}$ there is partial transmission, or
refraction, as shown in Fig. 65$^{b}$ and the angles satisfy

$\alpha _{i}$ = $\alpha _{r}$ \ \ \ \ \ \ \ \ \ \ \ \ \ \ \ \ \ \ \ \ \ \ \
\ \ \ \ \ \ \ \ \ \ \ \ \ \ \ \ \ \ \ \ \ \ \ \ \ \ \ \ (159) and

n$_{co}$sin$\alpha _{i}$ = n$_{cl}$sin$\alpha _{t}$. \ \ \ \ \ \ \ \ \ \ \ \
\ \ \ \ \ \ \ \ \ \ \ \ \ \ \ \ \ \ \ (160)

Usually for the planar waveguides it is convenient to express these laws in
terms of the complementary angles of incidence, reflection and transmission,
i.e. $\theta _{z}$ =$\pi /2$ - $\alpha _{i}$ = $\pi /2$ - $\alpha _{r}$ and $%
\theta _{t}$ = $\pi /2$ - $\alpha _{t}$ and the complementary critical angle 
$\theta _{c}$ = $\pi /2$ - $\alpha _{c}$.

One of the possible major applications of isotopic engineering it will be
considered isotopic fiber-optics and isotopic optoelectronics at large (see
also [3. 354]). It is known that for typical solids the lattice constant
variations of isotopically different samples are usually within the limits
[3]

$\frac{\Delta d}{d}$ $\sim $ 10$^{-3}$ $\div $ 10$^{-4}.$ (161)

Let us define an isotopic fiber as a structure in which core and cladding
have the same chemical content but different isotopic composition (see Fig.
66) The boundary between different isotopic regions form an isotopic
interface. The difference in the refractive index on both sides of the
isotopic interface could lead to the possibility of total internal
reflection of light and, consequently, could provide an alternative route to
the confinement of light. For a quantitative estimate let us consider a
boundary between SiO$_{2}$ (the main component of silica) where body sides
are identical chemically and structurally but have a different isotopic
content - e.g. $^{28}$Si$^{16}$O$_{2}$ and $^{30}$Si$^{18}$O$_{2}$
respectively (Fig. 66). In the first approximation the refractive index n is
proportional to the number of light scatterers in the unit volume. From the
Clausius-Mosotti relation (see, e.g. [355]) \ for the refractive index one
can deduce the following proportion ( at $\Delta n<<$ n)

$\frac{\Delta n}{n\text{ }}$ $\simeq $ 3c$\frac{\Delta d}{d}$, \ \ \ \ \ \ \
\ \ \ \ \ \ \ \ \ \ \ \ \ \ \ \ \ \ \ \ \ \ \ \ \ \ \ \ \ \ \ \ \ \ \ \ \ \
\ \ \ \ \ \ \ \ \ \ \ \ \ \ \ \ \ (162)

where c is a dimensionless adjustment factor of the order of unity.
Substituting Eq. (161) to Eq. (162) we can obtain

$\frac{\Delta n}{n}$ $\sim $ 3 $\cdot $ 10$^{-3}$ $\div $ 10$^{-4}$. \ \ \ \
\ \ \ \ \ \ \ \ \ \ \ \ \ \ \ \ \ \ \ \ \ \ \ \ \ \ \ \ \ \ \ \ \ \ \ \ \ \
\ \ \ \ \ (163)

Using the Snell law of light refraction we obtain the following expression
for the ray bending angle $\theta $ when the light travels through the
refractive boundary (see, also Fig. 67)

$\theta $ $\simeq \alpha _{0}$ - arc sin $\left( \frac{n_{1}}{n_{2}}\sin
\alpha _{0}\right) ,$ \ \ \ \ \ \ \ \ \ \ \ \ \ \ \ \ \ \ \ \ \ \ \ \ \ \ \
\ \ \ \ \ \ \ \ \ \ \ (164)

where $\alpha _{\text{0}}$ is the angle between the falling ray and the
direction normal to the interface. For a sliding ray $\left( \alpha _{\text{0%
}}\text{ }\simeq \text{ 90}^{\circ }\right) $, which is the control case for
light confinement in fibers, the combining of Eqs (163) and (164) leads to
an estimate

$\theta \sim $ 1.5 $\div $ 4.5$^{\circ }$. \ \ \ \ \ \ \ \ \ \ \ \ \ \ \ \ \
\ \ \ \ \ \ \ \ \ \ \ \ \ \ \ \ \ \ \ \ \ \ \ \ \ \ \ \ \ \ \ \ \ \ \ \ \ \
\ \ \ \ \ (165)

Thus, the isotopic fibers in which core and cladding are made of different
isotopes the half-angle of the acceptance-cone could be up to several
degrees [356]. The resulting lattice mismatch and strains at the isotopic
boundaries are correspondingly one part per few thousand [3] and, therefore,
could be tolerated. Further advancement of this \textquotedblright isotopic
option\textquotedblright\ could open the way for an essentially monolitic
optical chip with built-in isotopic channels inside the fully integrated and
chemically uniform structure.

Besides that we should pay attention to the fact that composition (different
isotopes) fluctuation are subject to the restoring force of the total free
energy of the glass system which will also seek to minimize itself. Using
isotope pure materials for core and cladding we should receive significant
less Rayleigh scattering (see e.g. [3]).

\bigskip

5.1.1.4. Laser materials.

\bigskip

\textbf{General considerations. }As is well-known, the word laser is an
acronym for \textquotedblright light amplification by the stimulated
emission of radiation', a phrase which covers most, though not all, of the
key physical processes inside a laser. Unfortunately, that concise
definition may not be very enlightening to the nonspecialist who wants to
use a laser but has less concern about the internal physics than the
external characteristics. A general knowledge of laser physics is as helpful
to the laser user as a general understanding of semiconductor physics is to
the circuit designer. From a practical standpoint, a laser can be considered
as a source of a narrow beam of monochromatic, coherent light in the
visible, infrared or UV parts of spectrum. The power in a continuous beam
can range from a fraction of a milliwatt to around 20 kilowatts (kW) in
commercial lasers, and up to more than a megawatt in special military
lasers. Pulsed lasers can deliver much higher peak powers during a pulse,
although the average power levels (including intervals while the laser is
off and on) are comparable to those of continuous lasers

The range of laser devices is broad. The laser medium, or material emitting
the laser beam, can be a gas, liquid, crystalline solid, or semiconductor
crystal and can range in size from a grain of salt to filling the inside of
a moderate-sized building. Not every laser produces a narrow beam of
monochromatic, coherent light. A typical laser beam has a divergence angle
of around a milliradian, meaning that it will spread to one meter in
diameter after traveling a kilometer. This figure can vary widely depending
on the type of laser and the optics used with it, but in any case it serves
to concentrate the output power onto a small area. Semiconductor diode
lasers, for example, produce beams that spread out over an angle of 20 to 40$%
^{0}$ hardly a pencil-thin beam. Liquid dye lasers emit at a range of
wavelengths broad or narrow depending on the optics used with them. Other
types emit at a number of spectral lines, producing light is neither truly
monochromatic nor coherent. Practically speaking, lasers contain three key
elements. One is the laser medium itself, which generates the laser light. A
second is the power supply, which delivers energy to the laser medium in the
form needed to excite it to emit light. The third is the optical cavity or
resonator, which concentrates the light to stimulate the emission of laser
radiation. All three elements can take various forms, and although they are
not always immediately evident in all types of lasers, their functions are
essential. Like most other light sources, lasers are inefficient in
converting input energy into light; efficiencies can range from under 0.01
to around 20\% [357].

In order to observe high excitation semiconductors (insulators) (HES)
effects, it is necessary that the electronic excitations interact with each
other at a sufficiently high rate. The short lifetime and diffusion path in
direct gap materials necessitate a rather high density n of bound (excitons)
or unbound electronic excitations. A value of n = 10$^{17}$ cm$^{-3}$ is a
reasonable average [357], though the onset of HES effects depends strongly
on the compounds under investigation, on the quality of the individual
sample and on the special conditions of the experiment (see, e.g. [358]).
The lifetime of excitons $\tau $ in direct gap materials is about 1 ns at
low densities. It is reduced at high densities by quadratic recombination
processes and stimulated emission to values around 0.1 ns (see [358, 359]
and references therein).

The generation rate G, which is needed to obtain a stationary concentration
of n = 10$^{17}$cm$^{-3}$ e -h pairs with a lifetime of 10$^{-10}$ s is at
least G $\simeq $ 10$^{27}$ cm$^{-3}$s$^{-1}$. G has now to be connected \
with the excitation intensity I$_{exc}$, where I$_{exc}$ is the energy per
units of area and time impinging on the sample. Some authors prefer to give
the photon flux density I$_{Ph}$ instead of I$_{exc}$ in the case of optical
excitation. They are connected by I$_{exc}$ = I$_{Ph}\cdot $ $\hslash \omega 
$. Thus, I$_{exc}$ $\simeq $ 10$^{6}$ Wcm$^{-2}$ with N$_{2}$ laser ($%
\hslash \omega _{exc}$ = 3.78 eV) corresponds to I$_{Ph}$ = 1.7 $\cdot $ 10$%
^{24}$cm$^{-2}$s$^{-1}$.

In the case of nonresonant two-photon band-to-band excitation of
semiconductors for which the absorption edge $\hslash \omega _{ed}$ fulfills
the inequality 2$\hslash \omega _{exc}$ \TEXTsymbol{>} $\hslash \omega _{ed}$
\TEXTsymbol{>} $\hslash \omega _{exc}$, the relation between G and I$_{exc}$
reads

I$_{exc}$ = $\left( \frac{\text{G2}\hslash \omega _{exc}}{K_{2}}\right)
^{1/2}$, \ \ \ \ \ \ \ \ \ \ \ \ \ \ \ \ \ \ \ \ \ \ \ \ \ \ \ \ \ \ \ \ \ \
\ \ \ \ \ \ \ \ \ \ \ (166)

where K$_{2}$ is the two-photon absorption coefficient. Since K$_{2}$ is of
the order of 10$^{-7}$W$^{-1}$cm for many direct materials (see, e.g. [360) I%
$_{exc}$ \ has to be about 2 $\cdot $ 10$^{7}$Wcm$^{-2}$ \ in order to
observe HES effects. In real experiments on semiconductors I$_{exc}$ is
varied from 10$^{5}$ Wcm$^{-2}$ to 10$^{8}$ Wcm$^{-2}$, where the upper
limit is given by the threshold for destroying the sample surface. The
excitation by two-photon absorption is especially useful if a homogeneous
bulk excitation of rather thick samples is desired. With a standard Q -
switched ruby laser volumes of several mm$^{3}$ can be pumped.

In the case of one-photon excitation in the exciton or band-to-band region,
an equation like

I$_{exc}$ = G$\hslash \omega _{exc}$/K$_{1}$ \ \ \ \ \ \ \ \ \ \ \ \ \ \ \ \
\ \ \ \ \ \ \ \ \ \ \ \ \ \ \ \ \ \ \ \ \ \ \ \ \ \ \ \ \ \ \ \ \ \ \ \ (167)

does not hold. The values of the one-photon absorption coefficient K$_{1}$
are of the order of 10$^{5}$ cm$^{-1}$ or even higher. However, the created
e-h pairs do not remain confined in a surface layer of a thickness of K$%
_{1}^{-1}\leqslant $ 0.1 $\mu $m but rapidly spread out into the volume of
the crystal, driven by the gradient of the chemical potential. The effective
penetration depth \ d$_{eff}$ is generally assumed to be one or a few
micrometers [358]. A value of this order of magnitude has been found in CdS,
where a thin platelet (above 4 $\mu $m thick) was excited by a N$_{2}$ -
laser from one side and the change of the excitonic reflection spectra with I%
$_{exc}$ was investigated on both sides of the sample [361]. Therefore,
equation (167) should rather be replaced by

I$_{exc}$ \ = G $\cdot $ $\hslash \omega _{exc}$ $\cdot $ d$_{eff}$. \ \ \ \
\ \ \ \ \ \ \ \ \ \ \ \ \ \ \ \ \ \ \ \ \ \ \ \ \ \ \ \ \ \ \ \ \ \ \ \ \ \
\ \ \ \ (168)

In the experiments I$_{exc}$ \ is typically varied from 10$^{3}$ Wcm$^{-2}$
up to 5 $\cdot $ 10$^{6}$ Wcm$^{-2}$, the upper limit is again given by the
damage threshold of the sample surface. With one-photon excitation it is
generally possible to reach higher G values. Because of the high diffraction
losses of the thin excited layers [362], it is partly possible to suppress
the optical stimulation of recombination processes, especially if small
diameters D of the excitation spot are used (D \TEXTsymbol{<} 100 $\mu $m).

It should be pointed out, that the excitation conditions for indirect gap
semiconductors are quite different from those described above. In these
materials the lifetime is several orders of magnitude larger than in direct
gap materials. Therefore, excitation sources with much lower values of I$%
_{exc}$ and even conventional incandescent lamps may be used, either pulsed
or in a continuously working mode.

\textbf{Diamond. }Diamond has an indirect band structure similar to silicon
[363], with six equivalent conduction band minima located on the \TEXTsymbol{%
<}100\TEXTsymbol{>} axes at 0.76k$_{\max }$. Quantitatively, however,
diamond is different from all standard cubic semiconductors [203] including
silicon in that the valence band spin-orbit interaction $\Delta _{0}$ at k =
0 is very much smaller than the excitonic interactions important in optical
experiments. The latter are the exciton binding energy E$_{B}$ = 80 meV and
the localization energy E$_{loc}$ = 51 meV of excitons to the acceptor in
p-type semiconducting diamonds [364]. Substitutional boron is the only
shallow impurity in diamond with an ionization energy E$_{i}$ = 370 meV
[28]. The spin-orbit splitting between the fourfold degenerate $\Gamma
_{8}^{+}$ band and the twofold $\Gamma _{7}^{+}$ \ band amounts to $\Delta
_{0}$ = 6 meV experimentally or to $\Delta _{0}$ = \ 13 meV theoretically in
linear muffin-tin orbital and k $\cdot $ p calculations [27]. In all other
standard cubic semiconductors this ordering \ of the interaction energies is
inverse, yielding $\Delta _{0}$ \TEXTsymbol{>}\TEXTsymbol{>} E$_{B}$ and E$%
_{loc}$. In silicon, e.g. $\Delta _{0}$ = \ 44 meV, E$_{B}$ \ = 14.7 meV and
E$_{loc}$ $\approx $ 0.1 E$_{i}$ according to Haynes' rule [365], with E$%
_{i} $ ranging from $\approx $ 45 meV for the shallow donor phosphorus and
the shallow acceptor boron up to $\approx $ 155 meV for the relatively deep
acceptor indium [365] (for details see [28]).

The diamond having a wide band gap can emit the intense ultraviolet (uv)
radiation due to recombination of the indirect free exciton [366].
Additionally, since the exciton has an extremely large binding energy (E$%
_{B} $ = 80 meV) [367], diamond is one of the most promising candidates to
realize opto-electronics devices, such as uv laser diode, available at a
higher temperature. Nevertheless, only a little work has been done on the
intrinsic recombination luminescence of indirect free excitons created only
by d.c. - operated electron beams with much higher energy relative to the
indirect band gap energy. \ Figure 68 shows typical photo-excited intrinsic
edge emission spectra high-pressure synthetic (HPS) type -IIa diamond at
temperatures of (a) 85K, (b) 125K and (c) 300K, which are almost the same as
cathodoluminescence spectra in natural type IIa [367, 368] and the chemical
vapor deposition (CVD) \ [369] (see, also Fig. 15 in Ref. [28]). \ \
Takiyama et al. [366] \ have been established that B$_{1}$ band consists of
two Maxwell - Boltzman components labeled B'$_{1}$ and B$_{1}$ with the
energy splitting ($\sim $ 7 meV) due to the spin-orbit interaction of the
hole state (see, also [367]). The obtained in paper [366] \ intensity ratio
of B'$_{1}$ to B$_{1}$ \ shows that termalization occurred between the
components in the free exciton state at each temperature (thermal
equilibrium). Luminescence intensities of all these bands increase with
increasing temperature and reach a maximum causes the remarkable reduction
of the intensity. The intensities at 125K and 300K are approximately twice
and 1/10 of that at 85K, respectively. \ No extrinsic emission due to bound
excitons was found in the indirect edge region (5.5 - 4.8 eV).

In Ref. [366] decays of B$_{1}$ emission in the HPS type-IIa diamond were
measured at the temperature of 75 - 300K and were found to be exponential
over more than one order of magnitude, as typically illustrated in Fig. 69.
The lifetime of free excitons at 85K is estimated to be 40 ns from the slope
of curve (a). With increasing temperature (125K), the lifetime (70 ns)
increases as well as the intensity. Both the lifetime and the intensity
reach to a maximum at $\sim $ 150K. At higher temperature (300K - curve c),
they become shorter (20 ns) and much weaker than those at 125K. These
decrements are due to the thermally-activated exciton ionization. The
measured free exciton lifetime in the temperature range 75 - 300K are shown
in Fig. 69. As it can see from Fig. 70, authors [366] successfully
reproduced the temperature dependence of the lifetime of the exciton, where
the best fit values obtained assuming E$_{B}$ = 80 meV.

As is well-known natural diamond is 98.9\% $^{12}$C and 1.1\% $^{13}$C;
synthetic diamonds, however, may be grown with $^{13}$C concentrations in
the range \TEXTsymbol{<}0.1\% to 99\%\TEXTsymbol{>}. \ \ The change of the
indirect gap of diamond between pure $^{12}$C and $\ ^{13}$C has been
determined by Collins et al. [367], using for this purpose the luminescence
spectra of diamond. The luminescence spectra of the natural ($^{12}$C ) and
synthetic ($^{13}$C) diamond were investigated by Collins et al. [367], Ruf
et al. [370]. Fig. 15 in Ref. 28 compares the edge luminescence for a
natural diamond with that for a synthetic diamond. The peaks labelled A, B
and C are due, respectively, to the recombination of a free exciton with the
emission of transvers - acoustic, transverse-optic and longitudinal-optic
phonons having wavevector $\pm $ k$_{\min }$ and quanta (in $^{12}$C
diamond) [371].

$\hbar \omega _{TA}$ = 87 $\pm $ 2, $\hbar \omega _{TO}$ = 141 $\pm $ 2, $%
\hbar \omega _{LO}$ = 163 $\pm $ 1 meV. \ \ \ \ \ \ \ \ \ \ \ \ \ \ \ \ \ \
\ \ \ \ (169)

Features B$_2$ and B$_3$ are further free-exciton processes involving the
above TO phonon with one and two zone-centre optic phonons, respectively.

Boron forms an effective-mass-like acceptor in diamond, and both specimens
used in Fig. 15 are slightly semiconducting with uncompensated boron
concentrations around 5 $\cdot $ 10$^{16}$ cm$^{-3}$ in the natural diamond
and 3 $\cdot $ 10$^{16}$ cm$^{-3\text{ }}$in the synthetic diamond. Peaks
labelled D are associated with the decay of excitons bound to the boron
acceptors (for details see Collins et al. [367]). Comparison of the data
from the two diamonds shows that the zero-phonon lines D$_{0}$ and D$%
_{0}^{\imath }$ are 14 $\pm $ 0.7 meV higher for $^{13}$C than for C$^{12}$
diamond, and that the LO and TO phonon energies are lower by a factor of
0.96 , equal within experimental error to the factor (12/13)$^{1/2}$
expected to first order when the lattice is changed from $^{12}$C to $^{13}$%
C. The low-energy thresholds of the free-exciton peaks A, B and C are given
by [367]

E$_{th}$(A) = E$_{gx}$ - $\hbar \omega _{TA}$; E$_{th}(B)=$ E$_{gx}$ - $%
\hbar \omega _{TO}$ and E$_{th}$ (C) = E$_{gx}$ - $\hbar \omega _{LO}$. \ \
\ \ \ \ \ \ \ \ \ \ \ (170)

As was shown by Collins et al. the predicted threshold are entirely
consistent with the experimental data. From the results of Collins et al. it
was concluded that the dominant contribution arises from electron-phonon
coupling, and that there is a smaller contribution due to a change in volume
of the unit cell produced by changing the isotope. These two terms were
calculated as 13.5 $\pm $ 2.0 and 3.0 $\pm $ 1.3 meV respectively. The more
detailed and quantitative investigations of E$_{g}$ $\sim $ f$\left(
x\right) $, where x is the isotope concentration, were done by Ruf et al.
[370], where was studied five samples of diamond with different
concentrations x (see Fig. 16 in Ref. [28]). From these data Ruf et al.
[370] determined the linear variation of E$_{g}$ $\sim $ f$\left( x\right) $
for diamond. Linear fits the experimental data of Ruf et al. , (solid line
in Fig. 17 in Ref. 28) yield a slope of 14.6 $\pm $ 0.5 meV/amu, close to
the theoretical predictions. To concluding this part we should note that
Okushi et al. [370$^{a}$] systematic studied of cathodoluminescence spectra
of free exciton emission from diamond under high excitation conditions.

\textbf{LiH. }The possibility of using indirect transitions for obtaining
light generation in semiconductors was probably to Basov et al. [372]. The
stimulating emission on the LO replicas (see, Fig. 71) of a zero - phonon
line free - exciton emission in CdS under two - photon excitation was first
obtained by Kulevsky and Prokhorov [374] (see also [360, 375, 376]). As was
shown by Liu and Liboff [377] taking into account the mixed crystals lattice
potential relief it could not be excluded absolutely the possibility on the
zero - phonon line emission (see, also below).

The detection of LO phonon replicas of free - exciton luminescence in wide -
gap insulators attracted (see Fig. 72) considerable attention to these
crystals (see e.g. [378]). At the same time it is allowed one to pose a
question about the possibility of obtaining stimulated emission in UV (VUV)
region (4 - 6 eV) of the spectrum, where no solid state sources for coherent
radiation exist yet. In the first place this related to the emitters working
on the transitions of the intrinsic electronic excitation. The last one
provides the high energetical yield of the coherent emission per unit volume
of the substance. The results obtained on solidified xenon [372] and argon
[379] under electron beam excitation with following excimer molecules
emission form an exception.

In this part we will discuss the investigation results of the influence of
the excitation light density on the free excitons emission spectra in the
wide - gap insulator LiH (LiH - LiF) crystals. \ As was shown above the
cubic LiH crystals are typical wide - gap ionic insulator with relatively
weak exciton - phonon interaction however: E$_{B}$/$\hbar \omega _{LO}$ =
0.29 where E$_{B\text{ }}$and $\hbar \omega _{LO}$ are exciton binding
energy and longitudinal optical phonon's energy, respectively. Besides it
might be pointed out that the analogous relation for CdS, diamond and NaI is
0.73; 0.45 and 12.7, respectively (see e.g. [221] and references therein).

The reflectance spectra of the investigated crystals with clean surface
(cleaved in LHeT) had a distinctly expressed excitonic structure (see also
chapter 3 of book [29]). However, despite of the identical structure of all
free - excitons luminescence spectra (Fig. 72), it is necessary to note a
rather big variation of the luminescence intensity of the crystals from the
different batches observed in the experiment. Therefore the crystals
possessing the maximum value of the free exciton luminescence quantum yield
were chosen for measurements of the density effects.

As it was shown early that the exciton luminescence is observed in LiH
crystals \ Fig. 72 excited by the energy photons from the depth of the
fundamental absorption. As an example, the 1LO and 2LO assisted luminescence
lines at the low excitation density are shown in Fig. 64 (curve 1) in Ref.
28. With the increase of the excitation density an additional emission
appears on the longwavelength side of both lines (curve 2, Fig. 64 in Ref.
28) These results show that the increase of the excitation density the
luminescence intensity of this new feature rises more quickly in the
vicinity of the 1LO replicas than in the vicinity of the 2LO replicas. This
is common regularity. At the same time it is necessary to note that for
obtaining of the spectrum demonstrated in Fig. 64 (curve 2) in Ref. 28 the
excitation light intensity was varied from three to twenty times depending
on the crystal origin (see also [372, 378]).

In more detail and with better resolution was investigated the influence of
the excitation light intensity on the shape and line intensity of 2LO
replicas [380]. The maximum sensitivity of the experimental equipment at the
maximum spectral resolution in these experiments was achieved by the sharp
focusing not only excitation but also the registrated channel. Results
presented in Fig. 73 witness that with the growth of the excitation light
intensity a little narrowing can be observed at the beginning which is
followed by an ignition of the luminescence intensity on the longwavelength
side of 2LO replicas line, as it is repeatedly observed for other ionic
semiconducting compounds (in details see, also [360]). Simultaneously with
this the appearance of a characteristic probably mode structure (see curves
2 and 3, Fig. 73) is observed the width of which is determined by power
surpassing the threshold. The divergence angle of the generation emission is
simply connected with it and for different semiconductors lies in the
interval 2 $\div $ 25$^{0}$ (see, e. g. [360]).

As is known at moderate excitations levels a linear dependence of
luminescence intensity on the excitation density is observed (see, e.g.
[381]). Such dependence is also observed in LiH crystals and at more
substantial excitation light intensities (see Fig. 66, curve 1 in Ref. 28).
Such linear coupling is usually considered to be unequivocal and testifying
about exciton formation only as a results of an indirect light absorption
process ([382]). At the same time the measurements of the respective
dependence on the long - wavelength side of 2LO replica line (see Fig 66,
curve 2 in Ref. 28) shown that the coupling between the luminescence and
excitation intensities is not only linear but, in fact, of a threshold
character as in case of other crystals (see, also [360, 383]).

\bigskip

5.1.2. New applications - quantum information.

\bigskip

The current rapid progress in the technology of high-density optical storage
makes the mere announcing of any other thinkable alternatives a rather
unthankful task. An obvious query "who needs it and what for?" has,
nevertheless, served very little purpose in the past and should not be used
to veto the discussion of non-orthodox technological possibilities. One such
possibility, namely the technology of isotopic information storage (IIS) is
discussed in this paragraph.

Isotopic information storage may consist in assigning the information 'zero'
or 'one' to mono-isotopic microislands (or even to a single atoms) within a
bulk crystalline (or thin film) structure. This technique could lead to a
very high density of ROM-type (read-only memory or permanent storage)
information storage, probably up to 10$^{20}$ bits per cm$^{3}$. The details
are discussed in papers [384, 261, 191, 28]: here it notes only that the use
of tri-isotopic systems (e.g. $^{28}$Si; $^{29}$Si; $^{30}$Si) rather than
di-isotopic (e.g.$^{12}$C; $^{13}$C) could naturally lead to direct three
dimensional color imaging without the need for complicated redigitizing (it
is known that any visible color can be simulated by a properly weighted
combination of three prime colors, but not of two).

Indeed, let us assume that the characteristic size of one
information-bearing isotopic unit (several atoms) is 100 \AA . Then 1 cm$^{3%
\text{ }}$ of crystalline structure, e.g. diamond, is able to store roughly
(10$^{8}$)$^{3}$/100$^{3}$ = 10$^{18}$ bits of information [384,28]. This
capacity greatly exceeds that need to store the information content of all
literature ever published ($\cong 10^{17}$ bits), including all newspapers.

The main potential advantage of isotope-mixed crystals lies in the fact that
the information is incorporated in the chemically homogeneous matrix. There
are no chemically different impurities (like in optical storage with color
centres) or grain boundaries between islands of different magnetization
(like in common magnetic storage). The information in isotope-mixed crystals
exists as a part of the regular crystals lattice. Therefore, the stored
information in isotope-mixed crystals is protected by the rigidity of the
crystal itself. There are no \textquotedblright weak
points\textquotedblright\ in the structure (impurities, domain wells,
lattice strain etc.) which can lead to the information loss due to bond
strains, enhanced diffusion, remagnetization, etc. Differences in the bond
lengths between different isotopes (e.g. $^{28}$Si - $^{29}$Si or $^{29}$Si
- $^{30}$Si; H - D and so on) are due to the anharmonicity of zero-point
vibrations (see, e.g. [26, 151] ). This is not enough for the development of
any noticeable lattice strains although these differences are sufficiently
large to be distinguishably detected in IIS - reading) (for details see
[191]).

The development of efficient quantum algorithms for classically hard
problems has generated interest in the construction of a quantum computer. A
quantum computer uses superpositions of all possible input states. By
exploiting this quantum parallelism, certain algorithms allow one to
factorize [396] large integers with astounding speed, and rapidly search
through large databases, and efficiently simulate quantum systems [398]. In
the nearer term such devices could facilitate secure communication and
distributed computing. In any physical system, bit errors will occur during
the computation. In quantum computing this is particularly catastrophic,
because the errors cause decoherence [399] and can destroy the delicate
superposition that needs to be preserved throughout the computation. With
the discovery of quantum error correction [400] and fualt-tolerant
computing, in which these errors are continuously corrected without
destroying the quantum information, the construction of a real computer has
became a distinct possibility (see, also [401]). The task that lie ahead to
create an actual quantum computer are formidable: Preskill [402] has
estimated that a quantum computer operating on 10$^{6}$ qubits with a 10$%
^{-6}$ probability of error in each operation would exceed the capabilities
\ of contemporary conventional computers on the prime factorization problem.
To make use of error-correcting codes, logical operations and measurement
must be able to proceed in parallel on qubits throughout the computer.
Several systems have been recently been proposed to obtain a physical
implementation of a quantum computer (for detail see, e,g. [28, 191] and
references therein).

As was shown for the first time by Schr\H{o}dinger [385] fundamental
properties of quantum systems, which \ might be include to information
processes are:

1. Superposition: a quantum computer can exist in an arbitrary complex
linear combination of classical Boolean states, which evolve in parallel
according to a unitary transformation [386].

2. Interference: parallel computation paths in the superposition, like paths
of a particle through an interferometer, can reinforce or cancel one
another, depending on their relative phase [387].

3. Entanglement: some definite states of \ complete quantum system do not
correspond to definite states of its parts (see, also [388]).

4. Nonlocality and uncertainty: an unknown quantum state cannot be
accurately copied (cloned) nor can it be observed without being disturbed
[389, 390].

These four elements are very important in quantum mechanics, and as we'll
see below in information processing. All (classical) information can be
reduced to elementary units, what we call bits. Each bit is a yes or a no,
which we may represent it as the number 0 or the number 1. Quantum
computation and quantum information are built upon analogous concept, the
quantum bit [391], or qubit for short. It is a two-dimensional quantum
system (for example, a spin 1/2, a photon polarization, an atomic system two
relevant states, etc.) with Hilbert space. In mathematical terms, the state
of quantum state (which is usually denoted by $\mid \Psi >$ [385]) is a
vector in an abstract Hilbert space of possible states for the system. The
space for a single qubit is spanned by a basis consisting of the two
possible classical states, denoted, as above, by $\mid 0>$ and $\mid $1%
\TEXTsymbol{>}. This mean that any state of qubit can be decomposed into the
superposition

$\mid \Psi >$ = $\alpha \mid 0>$ + $\beta \mid 1>$ \ \ \ \ \ \ \ \ \ \ \ \ \
\ \ \ \ \ \ \ \ \ \ \ \ \ \ \ \ \ \ \ \ \ \ \ \ \ \ \ \ \ \ \ \ \ \ (171)

with suitable choices of the complex coefficients \textit{a }and \textit{b}.
The value of a qubit in state $\mid \Psi >$ is uncertain; if we measure such
a qubit, we cannot be sure in advance what result we will get. Quantum
mechanics just gives the probabilities, from the overlaps between $\mid \Psi
>$ and the possible outcomes, rules due originally by Max Born (see, e.g.
[392]). Thus the probability of getting 0 is $\mid <0\mid \Psi >\mid ^{2}$ = 
$\mid $a$\mid ^{2}$and that for 1 is $\mid <1\mid \Psi >\mid ^{2}$ = $\mid $b%
$\mid ^{2}$. Quantum states are therefore normalized; \TEXTsymbol{<}$\Psi
\mid \Psi $\TEXTsymbol{>} = (b*a*)$\cdot \left( 
\begin{array}{c}
b \\ 
a%
\end{array}%
\right) $ = 1 (where $\mid \Psi >$ is represented by the vector $\left( 
\begin{array}{c}
b \\ 
a%
\end{array}%
\right) $) and the probabilities sum to unity. Quantum mechanics also tells
us that (assuming the system is not absorbed or totally destroyed by the
action of measurement) the qubit state of Eq. (171) suffers a projection to $%
\mid 0>$ ($\mid 1>$) when we get the result 0(1). Because $\mid \alpha \mid
^{2}$ + $\mid \beta \mid ^{2}$ = 1 we may rewrite Eq. (171) as (see, e.g.
[393])

$\mid \Psi >$ = cos$\theta \mid 0>$ + e$^{i\varphi }$sin$\theta \mid 1>$ \ \
\ \ \ \ \ \ \ \ \ \ \ \ \ \ \ \ \ \ \ \ \ \ \ \ \ \ \ \ \ \ \ \ \ \ \ \ \ \
\ \ \ \ (172)

where $\theta $, $\varphi $ are real numbers. Thus we can apparently encode
an arbitrary large amount of classical information into the state of just
one qubit (by coding \ the information into the sequence of digits of $%
\theta $ and $\varphi $). However in contrast to classical physics, quantum
measurement theory places severe limitations on the amount of information we
can obtain about the identity of a given quantum state by performing any
conceivable measurement on it. Thus most of the quantum information is
"inaccessible" but it is still useful - for example \ it is necessary \ in
its totality to correctly predict any future evolution of the state and to
carry out the process of quantum computation (see, e.g. [395]).

The numbers $\theta $ and $\varphi $ define a point on the unit three -
dimensional sphere, as shown in Fig. 74. This sphere is often called the
Bloch (Poinkare) sphere [394]; it provides a useful means of visualizing the
state of a single qubit. A classical bit can only site at the north or the
south pole, whereas a qubit is allowed to reside at any point on the surface
of the sphere [28].

Recently Kane [403] proposed a very interesting and elegant design for a
spin resonance transistor (SRT). He proposed to use the nuclear spins of $%
^{31}$P dopant atoms, embedded in a silicon host, as the qubits. At low
temperatures (T \TEXTsymbol{<} 0.1mK) \ the dopant atoms do not ionize, and
the donor electron remains bound to the $^{31}$P nucleus. The control over
the qubits is established by placing a gate-electrode, the so-called A-gate
(see Fig. 75), over each qubit. By biasing the A-gate, one control the
overlap of the bound electron with the \ nucleus and thus the hyperfine
interaction between nuclear spin and electron spin, which allows two nuclear
spins to interact by electron spin-exchange, which provides the required
controlled qubit-qubit interaction (Fig. 76). The rate of loss of phase
coherence between qubits in a quantum system is typically characterized by
the dephasing time T$_{2}$. At sufficiently low $^{31}$P concentrations at
temperature T = 1.5K, the electron-spin relaxation time is thousands of
seconds and the $^{31}$P nuclear spin relaxation time exceeds 10 hours. It
is likely that at millikelvin temperatures the phonon limited $^{31}$P
relaxation time is of the order of 10$^{18}$ [404], making this system ideal
for quantum computation. \ In the work done by Kane at al. [405] a method
for probing the spin quantum numbers of a two-electron system using a
single-electron transistor is presented. In spite of significant
technological difficulties for the realization of Kane's ingenious idea,
e.g. it requires phosphorus atoms ordered as a regular array with spacing of
20 nm as well as a gate spacing of 10 nm , his model has triggered an
enormous interest in semiconductor realizations of the nuclear spin quantum
computers. We should add, that very perspective system of GaAs/Al$_{1-x}$Ga$%
_{x}$As has a huge number of active nuclear spin ($\sim $ 10$^{4}$) , this
reason does not permit this system used in quantum computation. Vrijen et
al. [406] suggest using the full power of modern electronic band structure
engineering and epitaxial growth techniques, to introduce a new, more
practical, field effect SRT transistor design that might lend itself to a
near term demonstration of qubits on a silicon-germanium hetero-structures
(for details see [407, 191]).

The last ides was developed by Shlimak et al. [408, 409]. The first step
consists of the growth of isotopically engineered Si and Si - Ge epitaxial
layers with lateral modulation of nuclear spin isotope content in the layer.
The simplest case is a sequence of stripes of the spinless isotope (for
example, $^{28}$Si) and nonzero spin \ isotope ($^{29}$Si or natural Si,
which contains 4.7\% of $^{29}$Si, or a controlled mixture of them). The
following method, based on molecular - beam - epitaxy (MBE) growth on
vicinal planes, can be suggested for preparation of such a striped layer.
Usually, the substrate surface of Si cut at a small angle $\theta $ to some
crystallographic direction consists of atomic size steps [410], followed by
a relatively long plateau (see Fig. 77$^{a}$). The typical size of the steps
a is of order 0.1 - 0.5 nm (depending on the crystallographic orientation),
which gives the following value for the length of plateau: d = a/$\theta $ =
100 - 500 nm for $\theta $ = 10$^{-3}$. The atoms deposited on the hot
substrate are mobile and move toward steps in the corner where more dangling
bonds are utilized (see Fig. 77$^{b}$ and Fig. 77$^{c}$). As a result,
during the process of deposition and formation of a new layers, the steps
move from left to right continuously. For a given deposition of one
monolayer, i.e. \ to cover the entire plateau. If one isotope is deposition
during the rest of this time $\tau $ (Fig. 77$^{b}$), followed by the other
isotope being deposition during the rest of this time interval (t - $\tau $%
), then a periodical striped layer will be obtained with a controlled ratio
of the strip widths: $\ell $/(d - $\ell $) = $\tau $/(t - $\tau $), Fig 77$%
^{c}$. Taking into account that the typical growth rates in MBE reach values
of about o.5 \AA s$^{-1}$ [411], the time t will permit the proposed
procedure. The required number of monolayers with some amount of isotope $%
^{29}$Si is determined by the minimal number of nuclear spins in a quantum
dot which can be detected electronically. According [408] the next step of
fabricating a nanosize structure on top of this layer. For this purpose, the
scanning probe microscopy (SPM) technique is usually used, which can go
lower than the limits obtainable by conventional methods, such as photon -
and electron - beam litography [412, 413]. In the case for fabrication of a
nuclear - spin qubit device on top of an isotopically engineered Si layer,
the authors [408] suggest using the atomic - force - microscopy - (AFM) -
assisted local oxidation technique, which consists of the following steps.
During the nanostructure patterning, the silicon surface will be firstly H
passivated by treatment in buffered HF acid, which strips the native oxide
and terminates the surface bonds with a hydride layer. This passivating
hydride layer is robust and protects the surface for days from oxidation.
Under suitable bias between conducting the AFM tip and the surface, the high
electric field oxidizes the Si surface in the immediate vicinity of the tip.
Feature down to 10 nm were obtained with this technique [408]. The more
details of describing technology presents in Fig. 78.

Authors [409] believe that the proposed source - drain (SD) channel can also
be used for the read - out operation, i.e. for the detection of a single
nuclear spin state. The direct control of a nuclear spin state via nuclear
magnetic resonance (NMR) measurements is a difficult problem. In Ref 414,
coherent control of the local nuclear spin was demonstrated, base on pulsed
NMR in a quantum Hall device. In Ref. 415, a self - contained semiconductor
device is described \ \ that can nuclear spins in a nano - scale region.
Measurements of the electron spin state are much easier taking into account
the possibility \ of a spin - to - charge conversion for electrons. In
accordance with the Kane model [403], the state of the nuclear spin $^{31}$P
is mediated by the spin of donor electron via the hyperfine interaction.
Therefore the task is to determine the spin orientation of the corresponding
donor electron. The suggested method [406] is based on the fact that at low
temperatures, a donor atom can capture the second electron with small
ionization energy, about 1 meV, which results in the appearance of
negatively charged donor (D$^{-}$ - center [371]). However, this process is
possible only when the spin orientation of the second electron is opposite
to that of the first electron. The appearance of the charged donor in the
vicinity of the narrow SD channel will affect the current [416, 417] and can
therefore be detected. As a result, one can determine the spin orientation
of two neighboring donor electrons if one applies a potential difference
between the corresponding A - gates which will cause the electron to jump
from one donor to another. If we choose the spin orientation of the given
donor as a reference, one can determine the spin state of the neighboring
qubits on the right and left sides.

Following [409] we consider further a new mechanism of entanglement for
distant qubits and discuss, first the principles of two - qubit operation.
It has been shown [418] that two - bit gates applied to a pair of electron
or nuclear spins are universal for the verification of all principles of
quantum computation.

Because direct overlap of wavefunctions for electrons localized on P donors
is negligible for distant pairs, authors [409] proposed another principle of
coupling \ based \ on the placement of qubits at fixed positions in a quasi
- one - dimensional Si nanowire and using the indirect interaction of $^{31}$%
P nuclear spins with spins of electrons localized in the nanowire \ which
they called hereafter a "1D electrons". This interaction dependes on the
amplitude \ of the wavefunction \ of the "1D electrons" estimated at the
position of the given donor nucleus $\Psi _{n}$(r$_{i}$) and can be
controlled by the change in the number of "1D electrons" N in the wire. At N
= 0, the interqubit coupling is totally suppressed, each $^{31}$P nuclear
spin interact only with its own donor electron. Vey recently it was shown
[409$^{a}$], that the optical detection of the nuclear spin state, and
selective pumping and ionization of donors in specific electronic and
nuclear spin states, suggests a number of new possibilities which could be
useful for the realization of silicon - based quantum computers.

\bigskip

5.2. Radioactive isotopes.

\bigskip

Radioactive isotopes (RI) \ are radioactive atoms of common elements like
carbon, cobalt, or sodium etc. Usually RI are located in atomic "ash" that
is left behind after uranium atoms are split in a "nuclear pile". Some RI
are produced from exposure of common elements to powerful radiation inside a
nuclear reactor during fission. Fission occurs when an atoms nucleus splits
into two or more smaller nuclei, producing a large amount of energy. RI
release radiation in the form of alpha, beta and gamma rays. The strength of
the radiation is relative to the rate where radioactive material decays.
Because of this, different radioisotopes can be used for different purposes,
depending on their strength.

Some radioactive elements, such as radium - 224, radium - 226, radon 222,\
polonium - 210, tritium ($^{3}$H), carbon - 14, etc. are found in nature,
but most radioactive materials are produced commercially in nuclear reactors
or cyclotrons (see, e.g. [8,9, 13]) - also called particle accelerators.
With nuclear reactors and cyclotrons, it is possible to make useful amounts
of radioactive material safely and at low cost. Usually only one type of
radionuclide can be produced at a time in a cyclotron, while a reactor can
produce many different radionuclides simultaneously. As for unstable
isotopes, there are over 1,000 some of which exist in nature, but most of
which have been created synthetically in laboratories \ in nuclear reactors
or cyclotrons. Although accelerators do create small quantities of lingering
radioactivity, they do not pose the staggering high - level waste and
proliferation problems associated with nuclear reactors, nor do they have
any potential for catastrophic accidents of any kind, nor are they capable
of producing weapons materials in militarily significant amounts.

It should be recognized that RI have been used in nuclear medicine, industry
and scientific research (solids).

\bigskip

5.2.1. Human health.

\bigskip

Nuclear Medicine is a \ branch of medicine that uses radiation to \ provide
information about the functioning of a person's specific organs or to treat
disease. In most cases, the information is used by physicians to make a
quick, accurate diagnosis of the patient's illness. The thyroid, bones,
heart, liver and many other organs can be easily imaged, and disorders in
their function revealed. In some cases radiation can be used to treat
diseased organs, or tumors (see, e.g. 34, 35]). Five Nobel Laureates have
been intimately involved with the use of radioactive tracers in medicine.

In the developed countries (26\% of world population) the frequency of
diagnostic nuclear medicine is 1.9\% per year, and the frequency of therapy
with radioisotopes is about one tenth of this. In Europe there are some 10
million nuclear medicine procedure per year. The use of radiopharmaceuticals
in diagnosis of growing at over 10\% per year. Nuclear medicine was
developed in the 1950s by physicians with an endocrine emphasis, initially
using iodine - 131 to diagnose and then treat thyroid disease. In recent
years specialists have also come from radiology, as dual CT/PET (see below)
procedures have become established.

As is well - known, diagnostic techniques in nuclear medicine use
radioactive tracers which emit gamma rays from within the body. These
tracers are generally short - lived isotopes linked to chemical compounds
which permit specific physiological processes to be scrytinized. They can be
given by injection, inhalation or orally. The first type where single
photons are detected by a gamma camera which can view organs from many
different angles. The camera builds up an image from the points from which
radiation is emitted; this image is enhanced by a computer and viewed by a
physician on a monitor for indications of abnormal conditions [419, 420].

A more recent development is Positron Emission Tomography (PET [ 36, 421,
422]) which is a more precise and sophisticated technique using isotopes
produced in a cyclotron. A positron - emitting radionuclide is introduced,
usually by injection, and accumulates in the target tissue. As it decay \ it
emits a positron, which promptly combines with a nearby electron resulting
in the simultaneous emission of two identifiable gamma rays in opposite
directions. These are detected by a PET camera and give very precise
indication of their origin. PET's most important clinical role in oncology,
with fluorine - 18 as the tracer, since it has proven to be the most
accurate non - invasive method of detecting and evaluating most cancers. It
is also used in cardiac and brain imaging [36]. This particular image shows
brain activity of a patient with Alzheimer's disease in Fig. 80.

New procedure combine PET with computed X - ray tomography (CT) scans to
give coregistration of the two images (PETCT), enabling 30\% better
diagnosis than with traditional gamma camera alone. It is powerful and
significant tool which provides unique information on a wide variety of
disease from dementia to cardiovascular disease and cancer (oncology).
Positioning of the radiation source within the body makes the fundamental
difference between nuclear medicine and other imaging techniques such as x -
rays. Gamma imaging by either method described provides a view of the
position and concentration of the radioisotope within body. Organ
malfunction can be indicated if the isotope is either partially taken up in
the organ (cold spot), or taken up in excess (hot spot). If a series of
images is taken over a period of time, an unusual pattern or rate of isotope
movement could indicate malfunction in the organ. A distinct advantage of
nuclear imaging over x - ray techniques is that both bone and soft tissue
can be imaged very successfully. This has led to its common use in developed
countries where the probability of anyone having such a test is about one in
two and rising.

Besides diagnosis the RI is very effective used in radiotherapy. Rapidly
dividing cells are particularly sensitive to damage by radiation. For this
reason, some cancerous growths can be controlled or eliminated by
irradiating the area containing the growth. External irradiation can be
carried out using a gamma beam from a radioactive cobalt - 60 source, though
in developed countries the much more versatile linear accelerators are now
being utilized as a high - energy x - ray source (gamma and x- rays are much
the same).

Internal radiotherapy is by administering or planting a small radiation
source, usually a gamma or beta emitter, in the target area. Iodine - 131 is
commonly used to treat thyroid cancer, probably the most successful kind of
cancer treatment. It is also used to treat nonmalignant thyroid disorders.
Iridium - 192 implants are used especially in the head and breast. They are
produced in wire form and are introduced through a catheter to the target
area. After administering the correct dose, the implant wire is removed to
shielded storage. This brachitherapy \ (short - range) procedure gives less
overall radiation to the body, is more localized to the target tumor and is
cost effective.

Treating leukemia may involve a bone marrow transplant, in which case the
defective bone marrow will first be killed off with a massive (and otherwise
lethal) dose of radiation before being replaced with healthy bone marrow
from a donor. Many therapeutic procedures are palliative, usually to relieve
pain. For instance, a strontium - 89 and increasingly samarium - 153 are
used for the relief of cancer - induced bone pain. Rhenium - 186 is a never
product for this (see, also [420,423]).

For some medical conditions, it is useful to destroy \ or weaken
malfunctioning cells using radiation. The radioisotope that generates the
radiation can be localized in the required organ in the same way it is used
for diagnosis - through a radioactive element following its usual biological
path, or through the element being attached to a suitable biological
compound. in most cases, it is beta radiation which causes the destruction
of the damaged cells. This is radiotherapy. Short - range radiotherapy is
known as brachytherapy, and this is becoming the main means of treatment.

Although radiotherapy is less common than diagnostic use of radioactive
material in medicine, it is nevertheless widespread, important and growing.
An ideal theraeutic radioisotope is a strong beta emitter with just enough
gamma to enable imaging, e.g. lutetium - 177. This is prepared from
ytterbium - 176 which is irradiated to become Yb - 177 which decays rapidly
to Lu - 177. Yttrium - 90 is use for treatment of cancer, particularly non -
Hodgkin's lymphoma, and its more widespread use is envisaged, including for
arthritis treatment.

Iodine - 131 and phosphorus - 32 are also used for therapy. Iodine - 131 is
used to treat the thyroid for cancers and other abnormal conditions such as
hyperthyroidism (over - active thyroid). In a disease called Polycythemia
vera, an excess of red blood cells is produced in the bone marrow.
Phosphorus - 32 is used to control this excess. A new and still experimental
procedure uses boron - 10 which concentrates in the tumor. The patient is
then irradiated with neutrons which are strongly absorbed by the boron, to
produce high - energy alpha particles which kill the cancer. Some example of
RI very effective used in everyday life presented in Table 19.

\bigskip

5.2.2. Geochronology.

\bigskip

The main data of the geochronology from which all conclusions are based are
elemental isotopic abundance of the radionuclides which are largely deduced
from meteorites. Meteorites are the prime source of information concerning
the earliest stage of the solar system since they have undergone virtually
no physical or chemical change since their time of formation approximately
4.6 $\cdot $ 10$^{9}$ years ago. The most primitive meteorites are believed
to be carbonaceous chondrites which are largely heterogeneous agglomerations
of particles which have undergone little heating since their formation.
Other sources of elemental abundance determinations include the Earth, Moon,
cosmic rays, Sun, and other stellar surfaces. A key result is the high
degree of isotopic homogeneity among the aforementioned sources which
supports the nebular hypothesis for the formation of the solar system. The
essence of the nebular hypothesis is that the Sun, planets, comets,
asteroids and meteorites formed from a common gaseous nebula which was well
mixed. The formation of our solar system from a gaseous nebula is not
precisely understood and the most extensive work in this area see in [424].
However, the recent determination of small variations in isotopic abundance
indicates that some elements of different nucleosynthetic histories were not
completely mixed. The understanding of these variations is already
constraining the detailed steps of solar system formation by imposing mixing
time scales (see, e.g. [2, 22]).

Although it is no surprise that determination of the age of the Earth
(geochronology) is based on physical phenomena it is less expected that this
is also the case for the chronology of a substantial part of the
archaeological record. Of course on Roman sites the layer - by - layer
finds, particularly of coins with inscriptions, often permit dating by
reference to the enduring writings of contemporary authors, for example
Julius Caesar, and in any case such writings establish the basic chronology
of the period. To some extent the same is true further back in time, notably
by relating to the king - lists giving the reign durations of the Egyptian
pharaohs; these lists extend back to the First Dynasty and the earliest
pyramid at about 5000 years ago - though even this age is not science -
independent since, because of missing sections of the lists, it is reliant
on an astronomical calculation of the date of a \ recorded stellar event
(see, also, chapter 2). Beyond 5000 years ago all was conjecture until \ the
so - called "radiocarbon revolution" in the early 1950s [425] ; from then on
the 'deeper the older' was replaced by ages based on the laboratory -
measured half - life (5568 - 5730 years) of $^{14}$C. Today the main dating
tool for the last 50,000 years or so is based on the radiocarbon method [426
- 428] (see, also [429]). The main radioactive methods for the periods
before the time span of radiocarbon are potassium - argon, uranium - series
dating, and fission - track dating. Thermoluminescence (TL) [430 - 433]
overlaps with radiocarbon in the time period for which is useful, but also
has potential for dating earlier epochs - as do optical dating [431] and
electron spin resonance (EPR) - all trapped electron dating methods that
rely indirectly on radioactive decay.

The journal \textit{Radiocarbon }publishes the most up - date curves which
in principle permit the conversion of radiocarbon dates to calibrated dates.
The calibration curve \ (see Fig. 81) produced by Stuiver et all. [426, 427]
combines the available \ data from tree - rings, uranium - thorium dated
corals, and varve - counted marine sediment, to give a curve from 24,000 to
0 Cal BP. Calibration programs and curves can be obtained directly from the
Radiocarbon website at www.radiocarbon.org. Several programs are now
available which use a statistical methodology, termed Bayesian, to generate
probability distributions of age estimations for single $^{14}$C. The
crucial point is that in any publication it should be indicated whether or
not the radiocarbon determination has been calibrated, and if it has been,
by which particular system or curve. Radiocarbon dating by the accelerator
mass spectrometry (AMS) technique is opening up new possibilities. Precious
objects and works of art can now be dated because minute samples are all
that is required. In 1988 AMS dating resolved the long - standing
controversy over the age of the Turin Shroud (Fig. 82), a piece of cloth
with the image of a man's body on it that many genuinely believed to be the
actual imprint of the body of Christ. Laboratories at Tuscon, Oxford and
Zurich all placed it in the 14$^{th}$ century AD (present time), not from
the time of Christ at all, although this remains a matter of controversy.
Radiocarbon looks set to maintain its position as the main dating tool back
to 50,000 years ago for organic materials. For inorganic materials, however,
TL and other, new, techniques are very useful (see below).

The preceding techniques are nuclear in the strict sense of the world: the
essence of the dating clock is the build - up of a dauther product, as with
potassium - argon and uranium - series, or the gradual disappearance of a
radioactive isotope, as with radiocarbon. Unlike the preceding techniques
luminescence dating is remarkable in utilizing \ a phenomena of which
variants can be seen with the naked eye. There are two variants of
luminescence dating: TL and optically stimulated luminescence (OSL), the
latter also being referred to as optical dating [431]. For both variants,
the latent dating information is carried in the form of trapped electrons;
these are electrons which have been ionized by nuclear radiation and which
have diffused into the vicinity of a defect in the lattice that is
attractive to electrons, for example, such as a negative - ion vacancy, and
have become trapped there (see Fig. 83). The nuclear radiation is from
radioelements in the sample and in its surroundings; there is also a small
contribution from cosmic rays. The more prolonged the exposure to ionizing
radiation the greater number of trapped electrons, which hence increases
with years that have elapsed since the last event at which the traps were
emptied. This setting of the clock to zero is the event dated and it can be
due to the agency of heat, as with pottery, or of light, as with geological
sediment. A measure of the number of trapped electrons is obtained by
stimulation - by heat in the case of TL and by light on the case of OSL. In
either case stimulation causes the eviction of electrons from their traps
where upon they diffuse around the crystal until some form of recombination
centre is found, such as a defect activated by being charged with a hole.
The time spent in diffusion is very short and recombination can be regarded
as instaneous. In the case of a luminescence centre there is emission of
light, the color being characteristic of the type of centre. Fig. 83 gives
an indication of the overall mechanism; it is an over - simplified
representation of reality but forms a useful basis for discussion (for
details see also [434]). It is presumed that there is no shortage of
activated luminescence centres and also that the radiation flux is not
sufficient to cause any significant increase in the number of centres over
the age span of the sample. An alternative to the picture given is to
consider the process to be dominated by trapped holes; however, although
this may represent reality in some cases it is irrelevant to the discussion
of most phenomenon and it is convenient to use a description based on
trapped electrons (see, also [430]). A similar description is relevant to
dating by EPR except that there is then no eviction.

The basis of the evaluation of age is summarized in Fig. 84. The 'natural'
signal, that resulting from the natural irradiation during burial, is
compared with signals, from the sample, resulting from known doses of
nuclear radiation; these are controlled by a calibrated radioisotope source.
This procedure allows evaluation of the paledose, the laboratory dose of
nuclear radiation needed to induce 'artificial' luminescence equal to the
natural signal. \ According to [430] the evaluation of age is given by

T = $\frac{\text{Paleodose}}{\text{Dose - rate}}$. \ \ \ \ \ \ \ \ \ \ \ \ \
\ \ \ \ \ \ \ \ \ \ \ \ \ \ \ \ \ \ \ \ \ \ \ \ \ \ \ \ \ \ \ \ \ \ \ \ \ \
\ \ \ \ \ \ \ \ \ \ \ \ \ \ \ \ \ (173)

The dose - rate represents the rate at which energy is absorbed from the
flux of nuclear radiation; it is evaluated by assessment of the
radioactivity of the sample and its surrounding burial material, this is
carried out both in laboratory and in the field.

At present, together with flint and calcite, the minerals of dominant
interest archaeologically are quartz and feldspar, whether from pottery (to
which mineral grains are added as temper), from sediment, or from volcanic
products (see also [429]). The age range covered by the various types of
sample and technique is remarkable - from few tens of years \ to around half
a million. The limitation with quartz and flint is usually due to the onset
of saturation - when all traps \ have become occupied; with feldspar it is
more likely to be due to inadequate electron retention in the traps.

Fig. 85 shows an example of a TL glow - curve. A crucial feature of TL
measurement is suppression of so - called 'spurious' TL. This is not induced
by radiation and is a surface phenomenon which is not well understood -
prior inter - grain friction plays a part but there are other influences as
well. Fortunately it can be avoided if the TL oven is flushed with high
purity nitrogen or argon, after removal of air; elimination is also enhanced
by red - rejection color filters. The glow - curve from a sample in which
there is only one trap type consists of a broad peak but in practice several
trap types are usually present in a sample and the glow - curve consists of
a number of overlapping peaks [434]. For archaeological or geological dating
the glow - curve region of interest is upwards of 300$^{0}$C; below this
temperature the TL is from traps that so shallow that they will have
suffered serious loss of electrons during the centuries of burial.\bigskip

5.2.3. Solid state physics.

\bigskip

The first application of radioactive isotopoes in solid state physics
research dates back almost century , when radioactive lead atoms were used
to study self - diffusion in lead [435]. The 'radiotracer diffusion'
technique was born . Nowadays it is common method for investigating atomic
diffusion processes in solids (see, also above). Up to now approximately 100
different radioactive isotopes have been used (see, Fig. 86) in nuclear
solid state physics [31], ranging \ from $^{8}$Li up to $^{213}$Fr. They are
produced by nuclear reactions in reactors or at accelerators and the doping
of the host lattice is performed either by nuclear reactions inside the
material, by recoil implantation or by diffusion or implantation after
nuclear production and chemical separation. The radioactive nuclei are used
as probes of their structural or electronic environment either in metals
[436], insulators [437], semiconductors or superconductors [438] and also on
surfaces and interfaces (see above) [439 - 441]. However, a major part of
the activity is focused on the investigation of defects and impurities in
semiconductors such as Si, Ge, III - V or II - VI compounds.

The characteristic lifetimes of radioactive isotopes can be used to label
and identify defect levels in semiconductors which can be detected by
photoluminescence [347] and Raman-scattering spectroscopy [348]. Magerle et
al. [347] show photoluminescence spectra of GaAs doped with $^{111}$In that
decays to $^{111}$ Cd. $^{111}$In is isoelectronic to Ga and hence occupies
Ga lattice sites in GaAs. It decays to $^{111}$Cd with a lifetime $\tau
_{111_{In}}$ = 98 h by electron capture [37] . Since the recoil energy of
the Cd nucleus due to the emission of the neutrino is much smaller than the
typical displacement energy in GaAs [30], $^{111}$Cd atoms on Ga sites (Cd$%
_{Ga}$) are created by the decay of $^{111}$In on Ga sites ($^{111}$In$_{Ga}$%
) and act there as shallow acceptors. This chemical transmutation was
monitored by photoluminescence spectroscopy. Figure 87 shows successively
taken photolumunescence spectra from the $^{111}$In doped sample. A spectrum
from the undoped part is also shown. The photoluminescence spectrum of the
undoped part of the sample shows the features well known for undoped
MBE-grown GaAs [30]. The peaks FX and AX around 819 nm are due to the
recombination of free and bound excitons. The peak (e,C) at 830 nm and its
LO phonon replica (e,C) - LO at 850 nm are due to recombination of electrons
from the conduction band into C acceptor states. The recombination of
electrons from donor states into C acceptor states appears as a small
shoulder at the right-hand sides of either of these two peaks. C is a
residual impurity in GaAs present in MBE-grown material with a typical
concentration between 10$^{14}$ and 10$^{15}$ cm$^{-3}$ [3]. Magerle et al.
determined the height I$_{Cd}$/I$_{C}$ of the (e, Cd) peak normalized to I$%
_{C}$ as the function of time after doping. These was done by substracting
the normalized spectrum of the undoped part from the normalized spectra of
the $^{111}$In doped part. The height I$_{Cd}$/I$_{C}$ of the (e,Cd) peak
remaining in these difference spectra is displayed in the insert of Fig. 87.
Indicated authors fitted these data by

$\frac{\text{I}_{Cd}}{\text{I}_{C}}\left( \text{t}\right) $ = $\frac{\text{I}%
_{Cd}}{\text{I}_{C}}\left( \text{t = }\infty \right) \left( \text{1 - e}^{-%
\frac{\text{t}}{\tau }}\right) $ \ \ \ \ \ \ \ \ \ \ \ \ \ \ \ \ \ \ \ \ \ \
\ \ \ \ \ \ \ \ \ \ \ \ \ \ \ (174)

and obtained a time constant $\tau $ = 52(17) h, which is not the nuclear
lifetime $\tau _{111_{In}}$ = $98$ h of $^{111}$In. Evidently I$_{Cd}$/I$%
_{C} $ is not proportional to N$_{Cd}$ . The photoluminescence intensity I$%
_{Cd}$ is proportianal to the recombination rate of excess carriers per unit
area through Cd acceptors states $\Delta $n$_{L}$B$_{Cd}$N$_{Cd}$, where B$%
_{Cd}$ is a recombination coefficient. The excess sheet carrier
concentration in the implanted layer $\Delta $n$_{L}$ can be expressed in
terms of the total carrier lifetime in the implanted layer $\tau _{L}$ and
the generation rate of excess carriers per unit area in the implanted layer f%
$_{L}$G by using the first of the two equilibrium conditions

f$_{L}$G = $\frac{\Delta n_{L}}{\tau _{L}}$ and f$_{B}$G = $\frac{\Delta
n_{B}}{\tau _{B}}$. \ \ \ \ \ \ \ \ \ \ \ \ \ \ \ \ \ \ \ \ \ \ \ \ \ \ \ \
\ \ \ \ \ \ \ \ \ \ \ \ \ \ \ \ (175)

The second one describes the balance between the generation rate f$_{B}$G
and the recombination rate of excess carriers $\frac{\Delta n_{B}}{\tau _{B}}
$ in the bulk. The total generation rate G is proportional to the incident
photon flux and f$_{L}$ + f$_{B}$ = 1. To get an expression for $\tau _{L}$,
cited authors assumed two additional recombination processes in the
implanted layer: the radiative recombination via Cd acceptors and
nonradioactive recombination due to residual implantation damage, and write
the recombination rate in the small single approximation (see, e.g., [3] and
references therein) \ as

$\frac{\Delta n_{L}}{\tau _{L}}$ = $\frac{\Delta n_{L}}{\tau _{B}}$ + $%
\Delta $n$_{L}$B$_{Cd}$N$_{Cd}$ + $\Delta $n$_{L}$B$_{nr}$f$_{nr}$N$_{Cd}$ \
\ \ \ \ \ \ \ \ \ \ \ \ \ \ \ \ \ \ (176)

Here $\Delta $n$_{L}$B$_{nr}$f$_{nr}$N$_{Cd}$ is the nonradioactive
recombination rate per unit area due to residual implantation damage, f$%
_{nr} $N$_{Cd}$ is the concentration of these nonradioactive recombination
centers, and B$_{nr}$ is the corresponding recombination coefficient. Hence $%
\Delta $n$_{L}$ and $\Delta $n$_{B}$ can be expressed as a function of N$%
_{Cd}$ and the recombination rates through all the different recombination
channels and thereby the relative photoluminescence peak intensities can be
deduced. I$_{C}$ is proportional to the sum of the (e,C) recombination rates
per unit area in the implanted layer and the bulk and within this model $it$
can obtain

I$_{C}$ $\propto $ $\frac{\Delta n_{L}\text{ + }\Delta n_{B}}{\tau _{C}}$ = G%
$\frac{\tau _{B}}{\tau _{C}}\left( \frac{\text{f}_{L}}{\text{1 + }\Phi _{Cd}%
\text{/f}_{B}\text{b}}\text{ + f}_{B}\right) $. \ \ \ \ \ \ \ \ \ \ \ \ \ \
\ \ \ \ \ \ \ \ \ \ \ \ (177)

Here $\Phi _{Cd}$ is the dose between 10$^{9}$ and 10$^{13}$ cm$^{2}$.
Thereby $\tau _{C}$ = 1/B$_{C}$N$_{C}$ is an effective lifetime describing
the recombination probability through C acceptor states and b is a constant
defined below. With help of Eqs. (175) and (177) it can be obtain (assuming
that the detection efficiencies of both peaks are equal) the following
relation between I$_{Cd}$/I$_{C}$ and $\Phi _{Cd}$ :

$\frac{\text{I}_{Cd}}{\text{I}_{C}}$ = $\frac{\Delta n_{L}B_{Cd}N_{Cd}}{%
\left( \Delta n_{L}\text{ + }\Delta n_{B}\right) \text{/}\tau _{C}}$ = $%
\frac{\text{a}}{\text{1 + b/}\Phi _{Cd}}$, \ \ \ \ \ \ \ \ \ \ \ \ \ \ \ \ \
\ \ \ \ \ \ \ \ \ \ \ \ \ \ \ \ \ \ \ \ \ \ \ \ (178)

with

a = $\frac{\text{f}_{L}}{\text{f}_{B}}\frac{\text{B}_{\text{Cd}}}{\left( 
\text{B}_{\text{nr}}\text{f}_{nr}\text{ + }B_{Cd}\right) }\frac{\tau _{\text{%
C}}}{_{\tau _{\text{B}}}}$ and b = $\frac{\text{d}}{\text{f}_{\text{B}%
}\left( \text{B}_{\text{nr}}\text{f}_{nr}\text{ + }B_{Cd}\right) \tau _{%
\text{B}}}$. \ \ \ \ \ \ \ \ \ \ \ \ \ \ (179)

This model describes quantitatively the dependence of (e, Cd) intensity of N$%
_{Cd}$ and cited authors use it to describe the increase of I$_{Cd}$/I$_{C}$
with time in the $^{111}$In-doped sample. Authors [347] model the change of
the carrier lifetime $\tau _{L}$ with time t in the $^{111}$In doped sample
as

$\frac{\text{1}}{\tau _{\text{L}}}$ = $\frac{\text{1}}{\tau _{\text{B}}}$ + B%
$_{Cd}$N$_{In}\left( \text{1 - e}^{-\frac{\text{t}}{\tau }}\right) $B$_{nr}$f%
$_{nr}$, \ \ \ \ \ \ \ \ \ \ \ \ \ \ \ \ \ \ \ \ \ \ \ \ \ \ \ \ \ \ \ \ \
(180)

where N$_{In}$ = $\Phi _{In}$/d is the initial $^{111}$In concentration, $%
\tau $ = $\tau _{111_{In}}$ = 98.0 h is the nuclear lifetime of $^{111}$In,
and B$_{Cd}$, B$_{nr}$ and f$_{nr}$ are the same constants as above. Thereby
we assume following Magerle et al. that the same kinds of nonradioactive
recombination centers are produced by In doping as by Cd doping and that the
Cd concentration are identical to the $^{111}$In concentration profile.
Taking into account all above saying we can write

$\frac{\text{I}_{Cd}}{\text{I}_{C}}$ = $\frac{\text{a}}{\text{1 + b/ }\Phi
_{In}\left( \text{1 - e}^{-\frac{\text{t}}{\tau }}\right) \text{ + c/}\left( 
\text{ e}^{-\frac{\text{t}}{\tau }}\text{ - 1}\right) },$ \ \ \ \ \ \ \ \ \
\ \ \ \ \ \ \ \ \ \ \ \ \ \ \ \ \ \ \ \ \ \ \ \ \ \ \ \ (181)

where a and b are the same constants as above and c = B$_{\text{nr}}$f$_{nr}$%
/$\left( \text{B}_{\text{nr}}\text{f}_{nr}\text{ + }B_{Cd}\right) .$ This c
term accounts for the fact that the $^{111}$In doped sample the
concentration of nonradioactive centers is not changing with Cd
concentration. Magerle et al. fitted Eq. (181) to the data shown in inset of
Fig. 87, keeping $\tau $ = 98.0 h, a = 1.25 and b= 3.0 x 10$^{11}cm^{-2}$ ,
and obtained $\Phi _{In}$ = 4.49 x 10$^{11}$ cm$^{-2}$ and c = 0.5 (2). This
fit is shown as a solid line and agrees perfectly with the experimental
data. In conclusion of this part it should note that this identification
technique is applicable to a large variety of defect levels since for most
elements suitable radioactive isotope exist (details see [3]).

Coupling between the LO phonon mode and the longitudinal plasma mode in NTD
semi-insulating GaAs was studied in paper [348] using Raman-scattering
spectroscopy and a Fourier-transform infrared spectrometer. Raman spectra
are shown in Fig. 88 for unirradiated, as-irradiated and annealed samples.
The remarkable feature is the low intensity and asymmetric linewidth of the
LO-phonon spectrum observed in annealed samples, which are annealed above 600%
$^{0}$C. The behavior is not understood by considering the only LO phonon.
We should pay attention to the electrical activation of NTD impurities,
which begin to activate electrically around 600$^{0}$C. In the
long-wavelength limit, the valence electrons, the polar lattice vibrations,
and the conduction electrons make additive contributions to the total
dielectric response function [3]:

$\varepsilon _{T}$(0, $\omega $) = $\varepsilon _{\infty }$ + $\left(
\varepsilon _{0}\text{ - }\varepsilon _{\infty }\right) $/$\left[ \left( 
\text{1 - }\omega ^{2}\text{/}\omega _{t}^{2}\right) \text{ - }\omega
_{p}^{2}\varepsilon _{\infty }\text{/}\omega ^{2}\right] $. \ \ \ \ \ \ \ \
\ \ \ \ \ \ \ \ \ \ \ \ \ \ (182)

The high - frequency value (L$_{+}$) of the mixed LO - phonon - plasmon
modes is calculated from the roots of the dielectric constant of Eq. (182).
The frequencies of the L$_{+}$ $\limfunc{mod}e$ and of the longitudinal
plasma mode $\omega _{p}$ = $\left( \text{4}\pi \text{ne}^{2}\text{/}%
\varepsilon _{\infty }\text{m*}\right) ^{1/2}$ for various annealing
temperatures are listed in Table 20. Here n is the electron concentration,
m* the effective mass in the conduction band (= 0.07m$_{0}$), and $%
\varepsilon _{\infty }$ (= 11.3) the optical dielectric constant. The mixed
LO-phonon-plasma mode appears around 300 cm$^{-1}$ for electron
concentration of (0.8-2)x10$^{17}$ cm$^{-3}$. The phonon strength [3] for
the high-frequency mode (L$_{+}$) of the interacting plasmon - LO - phonon
mode is about 0.95 for an electron concentration of 1x10$^{17}$ cm$^{-3}$,
while that for the low-frequency mode (L\_) is below 0.1. Therefore, the
asymmetric linewidth of the Raman spectrum observed in the annealed NTD GaAs
arises from both the LO-phonon and L$_{+}$ modes, but the L\_ mode is not
observed because of a very weak phonon strength. As a result, the LO-phonon
intensity decreases with increasing coupling, and L$_{+}$ mode appears
beside the LO-phonon peak.

The absorption spectra in the various annealing temperatures for NTD GaAs
are shown in Fig. 89. \ In unirradiated samples, an absorption around 2350 cm%
$^{-1}$ is assigned as the antisymmetric stretching vibration of CO$_{2}$
arising from CO$_{2}$ in an ambient atmosphere. The absorption peaks
observed around 500 cm$^{-1}$ are also assigned as a two-phonon overtone
scattering [348$^{a}$] of transverse optical phonons (TO); these were
observed at 493 cm$^{-1}$ $\left[ \text{2TO}\left( \text{X}\right) \right] $%
, 508 cm$^{-1}\left[ \text{2TO}\left( L\right) \right] $, and 524 cm$^{-1}%
\left[ \text{2TO}\left( \Gamma \right) \right] $, respectively. In as -
irradiated samples, a continuous absorption extending to the higher energy
was observed. Although this origin cannot be attributed to interstitial
anion clusters as discussed in neutron irradiated GaP [348$^{b}$]. In
samples annealed above 600$^{0}$C, the remarkable absorption was observed at
wave numbers below 1450 cm$^{-1}$. The absorption increases with increasing
annealing temperature (see Fig. 89). This behavior arises from the free -
electron absorption due to the activation of NTD impurities, which occur at
annealing temperatures above 600$^{0}$C. The free-electron absorption
observed is consistent with a collective motion as a plasmon mode described
in Raman-scattering studies.

Kuriyama et al. [350] were studied by a photoluminescence method the
transmuted impurities Ge and S in NTD semi-insulating GaP. In NTD GaP, Ge
and S impurities are transmuted from Ga and P atoms by (n,$\gamma $)
reactions, respectively. Ge in GaP is an amphoteric impurity for which both
the donor and acceptor states appear to be deep. The ratio between
transmuted impurities Ge and S is about 16 : 1. Unfortunately, after the
transmutation reactions, the transmuted atoms are usually not in their
original positions but displaced into interstitial positions due to the
recoil produced by the $\gamma $ and $\beta $ particles in the nuclear
reactions. In addition, the defects induced by the fast neutron irradiation
disturb the electrical activation of transmuted impurities. However, Frenkel
type defects [349] in NTD GaP were annealed out between 200 and 300$^{0}$C,
while P antisite (P$_{Ga}$) defects of $\sim $ 10$^{18}$ cm$^{-3}$
annihilated at annealing temperatures between 600 and 650$^{0}$C. therefore,
transmuted impurities, Ge and S, would be substituted on Ga and/or P lattice
sites by annealing at around 650$^{0}$C.

Fig. 90 shows the photoluminescence (PL) spectra of unirradiated and NTD
GaP. The PL spectrum (peak 1) of unirradiated samples shows signature of the
DA pair recombination involving S donor and carbon acceptor [348$^{c}$]. Two
(peaks 2 and 3) of the replicas occur at energies consistent with electronic
transitions accompanied by zone-center optical phonons with energies 50.1
meV (LO$_{\Gamma }$) and 100.2 (2LO$_{\Gamma }$). Sulfur, silicon and carbon
in GaP are the most common as the residual impurities [348$^{c}$]. In NTD -
GaP the main transition energy was observed 1.65 eV. Since Ge in GaP is the
amphoteric impurity with deep acceptor and donor levels, strong phonon
co-operation will also occur. but optical transition rates will be
significant only for associates. Similar situation has been proposed for Si
in GaP [371], forming a nearest-neighbor Si$_{Ga}$ - Si$_{P}$ $\limfunc{%
complex}.$ Therefore, the broad emission would be expected to arise from a
nearest-neighbor Ge$_{Ga}$ - Ge$_{P}$ coupled strongly to the lattice. To
confirm the presence of the Ge$_{Ga}$ - Ge$_{P}$ compose, the temperature
dependence of the half-width, W, of the broad emission was measured. If the
localized electron transitions from the excited state to the ground state of
this complex center produce the characteristics luminescence, the dependence
would be appear to follow the configuration-coordinate (CC) [371, 3] model
equation:

W = A$\left[ \coth \left( \text{h}\nu \text{/2kT}\right) \right] ^{\text{1/2}%
}$, \ \ \ \ \ \ \ \ \ \ \ \ \ \ \ \ \ \ \ \ \ \ \ \ \ \ \ \ \ \ \ \ \ \ \ \
\ \ \ \ \ \ \ \ \ \ \ \ \ \ \ \ \ \ (183)

where A is a constant whose value is equal to W as the temperature
approaches 0K and h$\nu $ is the energy of the vibrational mode of the
excited state. In Fig. 91, Eq. (183) has been fitted to the experimental
value for NTD-GaP. For the estimation of W, the spectrum of the 1.65 eV band
was substracted from that of the 1.87 eV band. The value of h$\nu $ used was
0.025 eV. The good fit to this equation that was found for the Ge$_{Ga}$ - Ge%
$_{P}$ center in NTD-GaP shows the validity of applying the CC model.
Results of paper [350] indicate that NTD method is a useful one for
introducing Ge donor, resulting from a fact the Ge atoms are transmuted from
Ga lattice sites in GaP. The obtained results are consistent with the
presence of the Ge$_{Ga}$ - Ge$_{P}$ complex as described earlier (see, e.g.
[371]).

To concluding this part we briefly describe the present definition of the
kilogram with mass of a certain number of silicon atoms [442 - 451]. \ To
determine a new value of the Avogadro constant with a relative combined
standard uncertainty of 2 $\cdot $ 10$^{-8}$, the mass determination of a 1
kg $^{28}$Si sphere is crucial and should be determination to an
unprecedented level of accuracy. By this year, the laboratories involved in
the International Avogadro Project\ (the list of the participants in this
Project see e. g. in [446]) should be able to determine \ the mass of a 1 kg
silicon sphere under vacuum with combined standard uncertainty of 4 $\mu $g.

The value derived from the slope of the function M$_{\func{Si}}$ = f($\rho $%
) as shown in Fig. 92 leads to what can be considered our best knowledge for
the molar volume in single - crystal silicon [445], M$_{\func{Si}}$/$\rho $
= 12.0588207(54) cm$^{3}$mol$^{-1}$. The combination of this value with the
1998 the Committee on Data for Science and Technology (CODATA) recommended
value of the Si lattice parameter [452] leads to an Avogadro constant N$_{A}$
6.0221330(27) $\cdot $ 10$^{23}$ mol$^{-1}$, a candidate for consideration
in future adjustments of the values of the constants by CODATA put forward
by the CCM Working Group on the Avogadro constant. This value disagrees by
more than 1 part in 10$^{6}$ with the CODATA 1998 recommended value for N$%
_{A}$ based on Planck's constant as determined mainly by watt balance
experiments (for details see, also [445, 450]).

\bigskip

\textbf{Conclusion.}

\bigskip

We have discussed the manifestation and origin of the isotope effect. For
the fist time we review from one point of view the current status
manifestation of the isotope effect in a nuclear, atomic \ and molecular as
well as solid state physics. It was shown that although these manifestations
of the isotope effect vary, they all have one common feature - they all
depend on mass.

In nuclear physics the mass of free particles (proton, neutron) doesn't
conserve in the weak interaction process. This contradiction is removed
although partly if we take into account the modern presentation that the
mass of proton (neutron) is created from quark condensate (not from
constituent quarks) which is the coherent superposition of the states with
different chirality. Thus the elucidation of the reason of origin of the
nucleon mass is taken down to elucidation of the reason to break down the
chiral symmetry in Quantum Chromodynamics. In this context we should note
that we do not know why the observed mass pattern (M$_{n}$, M$_{p}$, m$_{u}$%
, m$_{d}$, etc.) looks like indicated in text, but nuclear physics analyze
the consequence of this empirical fact, but not the reasons of such pattern.

The concordance in the value \ of binding electron energy between theory and
experiments have needed to take into account the nucleus motion. The binding
energy corrections due to motion of the nucleus decrease rapidly with
increasing nuclear mass. Due to the motion of nucleus,different isotopes of
the same element have slightly different the frequency of spectral line.
This so - called isotope displacement to the discovery of heavy hydrogen
with the mass number A = 2 (deuterium). It was shown that each line in the
spectrum of hydrogen was actually double. Really, the isotope shift of an
optical transition is the sum of two terms: the mass effect and the field
effect.

Perhaps the most surprising thing about molecular vibrations (diatomic
molecule) is that the frequencies of vibration may be correctly calculated
by means of classical mechanics. When an atom of a molecule is replaced by
an isotopic atom of the same element, it is assumed that the potential
energy function and configuration of the molecule are changed by negligible
amount. The frequencies of vibration may, however, be appreciably altered
because of the change in mass involved. This especially true if hydrogen is
the atom in question because of the large percentage change in mass. This
vibration frequency shift or isotopic effect is very simple observed. The
direct correlation between vibration frequency and isotope's mass was
studied on the example of "fullerene" molecule C$_{60}$. The observation of
essentially equal effects of isotopic substitution of $^{18}$O and $^{17}$O
for $^{16}$O on the rates of formation of ozone was called a "non - mass
dependent" or "mass - independent" isotopic effect. We should underline that
the origin of "mass - independent" isotopic effect doesn't understand well.

In the fourth part of our review we have discussed the effects of isotopic
substtution on the physical properties of solids and possible applications
we have considered in the last part. Three types of effects have been
identified: effects corresponding to the average isotopic mass, effects
corresponding to the random distribution of the isotopic masses, and in few
cases, effects due to the nuclear spin. In the case thermal expansion we
have emphasized the zero - point renormalization of band gap energy, which
is strongly affected by the isotopic mass. The energy of zero - point
vibrations obtained from experiments in isotope - mixed crystals turned out,
as a rule, to be compared (excluding $^{12}$C$_{x}^{13}$C$_{1-x}$, LiH$_{x}$D%
$_{1-x}$ mixed crystals) to the energy of the longitudinal optical phonons.
For thermal conductivity, we discussed the effect of mass disorder, which is
particularly important at low temperature. We next discussed the direct
effect of average isotopic masses and their random distribution on phonon
and exciton (electron) states.The harmonic approximation must be modified to
include anharmonicity which depends on the average isotopicmasses of each of
the constituent elements. This approach is necessary for the first step for
the system with strong scattering potential at isotope substitution (for
example, LiH$_{x}$D$_{1-x}$mixed crystals), which induces LO($\Gamma $)
phonons localization in such systems. From this, there follows a necessity
of developing such an approach that would lead to a self - consistent model
of lattice dynamics, within which a unified description of not only local
(small conentration range), but also crystal vibrations of mixed crystals in
the whole range of the component's concentrations will be possible. A more
consistent way of accounting anharmonicity, beginning, probably, already
from the isotope - defect model, is also required. Without such an approach
it is impossible to describe neither elastic nor vibrational properties of
isotopically mixed crystals. Our view is that it is precisely \ the
consistent way of treating anharmonicity that will allow us to develop such
a model of lattice dynamics and will make it possible to describe not only
weak, but strong scattering of phonons due to isotopic disorder.

Isotope control and engineering of different solids offers almost limitness
possibilities for solid - state investigations and novel divices. First of
all we should indicate the "new" reactor tehnology - neutron transmutative
doping (NTD) of solids. Capture of thermal neutrons by isotope nuclei
followed by nuclear decay produces new elements, resulting in a large number
of possibilities for isotope selective doping of solids. The importance of
this technology for syudies of the semiconductors for doping materials (for
different devices) is underlined.

One of the fundamental processes occuring in all matter is the random motion
of its atomic constituents. As simple as diffusion may appear to be, at
least conceptually, there still exist many basic unanswered questions.
Results from supposedly identical experiments by different groups often
scatter by significant factors. This clearly indicates that there are still
hidden factors which need to be determined. Even the most throughly studied
crystals of Si, we will do not know with certainty the relative
contributions of vacancies and interstitials to self - and impurity
diffusion as a function of temperature, the positionof the Fermi level, and
external effects such as surface oxidation or nitridation. The isotope
control can contribute with new experimental approaches to the study of
diffusion and can help in impoving our understanding of the many interacting
factors. A vey accurate method to measure the self - diffusion coefficient
uses the isotope heterostructure growing on the same substrate. At the
interface only the atomic mass is changing, while all other physical
properties sray the same. In as - grown samples, this interface is
atomically flat with layer thickness fluctuations about two atomic layers.
Upon annealing, the isotopes diffuse into each other (self - diffusion) with
a rate that depends strongly on temperature.

The discvery of the linear luminescence of free excitons observed over wide
temperature range has placed lithium hydride, as well as diamond crystals,
in the row of possible sources of coherent radiation in the UV spectral
range. The elements of fiber - optics using the different values of
refractive index of the different isotopes, from which it is easy to produce
the core and cladding of the fiber should be mentioned.

The development of efficient quantum algorithms for classicaly hard problems
has generated interest in the construction of a quantum computer. A quantum
computer uses superposition of all possible input states. By exploiting this
quantum parallelism, certain algorithms allow one to factirize large
integers with astounding speed, and rapidly search through large databases,
and efficiently simulate quantum systems. In the nearer term such devices
could facilitate communication and distributed computing. In any physical
systems, bit errors will occur during computation. In quantum computing this
is particularly catastrophic, because the errors cause decoherence, and can
destroy the delicate superposition that needs to be preserved throughout the
computation. With the discovery of quantum error correction and fault -
tolerant computing, in which these errors are continuously corrected without
destroying the quantum information, the construction of a real computer has
became a distinct possibility. In last two decades it was proposed some
scheme of quantum processor based on nuclear spin impurity or some isotope
of Si.

Finally we mention an application of highly pure $^{28}$Si which is now in
progress as a part of an international cooperation. This work is motivated
by the fact that nearly all fundamental units (meter, second, amper, volt,
etc.) cam nowadays be based on atomic properties, the only exception being
the kilogram. For the purpose of redifining the unit of mass in terms of the
atomic mass, an extremely perfect sphere of approximately 1 kg weight is
being made of highly pure $^{28}$Si.

\bigskip

\subsubsection{\textbf{Acknowledgments.}}

I would like to express my deep thanks to many authors and publishers whose
Figures and Tables I used in my review. Many thanks are due to Prof. W.
Reder for carefully reading of my manuscript as well as Dr. P. Knight for
improving my English and O. Tkachev for technical assistance. I wish to
express my deep gratitude my family for a patience during long preparation
of this review.

\bigskip

\textbf{\bigskip }

\textbf{References.}

\bigskip

1. Burbidge E.M., Burbidge G.R., Fowler W.A. and Hoyle F., 1957, Rev. Mod.
Phys. \textbf{29}, 547.

2. Wallerstein G., Jhen I., Jr, Parker P. et al., 1997, ibid, \textbf{69},
995; Esposito S., Primordial Nucleosynthesis: Accurate Prediction for Light
Element Abundances, ArXiv:astro-ph/ 9904411.

3. Plekhanov V.G., 2004, Applications of the Isotopic Effect in Solids
(Springer, Heidelberg).

4. Gell - Mann M., 1997,The Quark and the Jaguar (Adventures in the Simple
and the Complex) (W.H. Freeman and Co., NY).

5. Green A.E.S.,1955, Nuclear physics (McCraw-Hill, NY).

6. Kaplan I., 2002, Nuclear Physics, 2$^{nd}$ ed. (Addison-Wesley, NY).

7. Burcham W.E., 1973, Nuclear Physics. An Introduction. 2$^{nd}$ ed.
(Longman, NY).

8. Krane K.S., 1988, Introductory Nuclear Physics (Wiley and Sons, NY -
Chichester)

9. Lilley J., 2001, Nuclear Physics (Wiley and Sons, Chichester - NY).

10. Wong S.M., 1998, Introductory Nuclear Physics (Wiley and Sons,
Chichester - NY).

11. Hodgston P.E., Gadioli E., and Gadioli-Erba E., 2000, Introductory
Nuclear Physics (Oxford University Press, Oxford - NY).

12. Heyde K., 2004, Basic Ideas and Concepts in Nuclear Physics (IOP,
Bristol - Philadelphia).

13.Schirokov Ju.M. and Judin N.P., 1980, Nuclear Physics (Science, Moscow)
(in Russian).

14. Taylor R.E., 1991, Rev. Mod. Phys. \textbf{63}, 573.

15. Kendall H.W., 1991, ibid \textbf{63}, 597.

16. Friedman J.I., 1991, ibid \textbf{63}, 615.

17. Wilczek F.A., 2005, Uspekhi Fiz. Nauk \textbf{175}, 1337 (in Russian).

18. Halzen F. and Martin D., 1984, Quarks and Leptons (Wiley, NY).

19. Striganov A.P. and Donzov Ju.P., 1955\bigskip , Usp. Fiz. Nauk \textbf{%
55,} 314 (in Russian).

20. Frish S.E., 1963, Optical Spectra of Atoms ( Fizmatgiz, M.-L). (in
Russian).

21. Sobel'man I.I., Itroduction in Theory of Atomic Spectra, 1977, 2$^{nd}$
Ed.\ (Science, Moscow) (in Russian).

22. King W.H., 1984, Isotope Shift in Atomic Spectra, Plenum, NY.

23. Eliashevich M.A., 1962, Atomic and Molecular Spectroscopy ( Fizmatgiz,
Moscow). (in Russian).

24. Herzberg G., 1951, Molecular Spectra and Molecular Structure (D. van
Nostrand, NY).

25. Wilson E.B., Jr., Decius J.C. and Gross P.C., 1955, Molecular
Vibrations. The Theory of Infrared and Raman Vibrational Spectra.
(McGraw-Hill, NY).

26. Plekhanov V.G., 2005, Phys. Reports \textbf{410}, 1.

27. Cardona M. and Thewalt M.L.W., 2005, Rev. Mod. Phys. \textbf{77}, 1173.

28. Plekhanov V.G., 2006, Progr. Mat. Science, \textbf{51}, 287.

29. Plekhanov V.G., 2004, Giant Isotope Effect in Solids (Stefan -
University Press, La Jola) (USA).

30. Plekhanov V.G., 2003, J. Mater. Science \textbf{38}, 3341.

31. Schatz G., Weidinger A. and Gardener A., 1996, Nuclear Condensed Matter
Physics, 2$^{nd}$ Ed. (Wiley, NY).

32. Forkel - Wirth D., 1999, Rep. Progr. Phys. \textbf{62,} 527.

33. Forkel - Wirth D. and Deicher M., 2001, Nuclear Physics \textbf{A693},
327.

34. Adelstein S.I. and Manning F.Y. (Eds.) 1995, Isotopes for Medicine and
Life Science, (National Academy Press, Washington).

35. Gol'din L.L., 1973, Uspekhi-Phys (Moscow) \textbf{110}, 77 (in Russian);
Amaldi U. and Kraft G., 2005, Rep. Progr. Phys. \textbf{68}, 1861.

36. Ter - Pogossian M.M., 1985, in Positron Emission Tomography, ed. by I.
Reivich, A. Alovi. (Alan R. Press, NY).

37. Baranov V.Ju., (Ed.) 2005, Isotopes, Vol. 1 and 2 (Fizmatlit, Moscow
)(in Russian).

38. Manuel. O., 2001, Origins of Elements in the Solar Systems (Kluwer
Academic Press, NY).

39. Simmons E.H., 2000, Top Physics, ArXiv, hep - ph/0011244.

40. Froggatt C.D., 2003, Surveys High Energ. Phys. \textbf{18}, 77.

41. Ioffe B.L., 2006, Usp. Fiz. Nauk (Moscow) \textbf{176}, 1103 (in
Russian).

42. Dolgov A.D. and Zel'dovich Ja.B., 1981, Rev. Mod. Phys. \textbf{53}, 1.

43. Linde A.D., 1984, Usp. Fiz. Nauk \textbf{144}, 177 (in Russian).

44. Ioffe B.L., 2001, ibid \textbf{171}, 1273 (in Russian).

45. Langacker P., 2003, Structure of the Standard Model, ArXiv: hep - ph.
0304186; Novaes S.F., 2000, Standart Model: An Introduction, ArXiv:
hep-ph/00012839; Froggatt C.D., Nielsen H.B., Trying to Understand the
Standard Model Parameters, ArXiv: hep-ph/0308144; Donoghue J.F., Golowich E.
and Holstein B.R., 1992, Dynamics of Standard Model (Cambridge University
Press, Cambridge); Burgess C. and Moore G., 2006, Standard Model A Primer,
(Cambridge University Press, Cambridge).

46. Royzen I.I., Feinberg E.L., Chernavskaya, 2004, Usp. Fiz. Nauk \textbf{%
174}, 473 (in Russian).

47. Marciano W., and Pagels H., 1978, Phys. Reports \textbf{36}, 137; Lee
D.W., 1972, Chiral Dynamics (Gordon and Breach, NY; Coleman S., 1985,
Aspects of Symmetry (Cambridge University Press, Cambridge).

48. Plekhanov V.G., 1997, Uspekhi - Phys. (Moscow) \textbf{167}, 577 (in
Russian).

49. Dirac P.A.M., 1958, The Principles of Quantum Mechanivs ( Oxford
University Press, U.K); \ Feyman R.P., Leighton R.P.and Sands M., 1965, The
Feyman Lecture in Physics, vol.3 (Addison - Wesley, Reading, MA); Landau
L.D. and Lifshitz E.M., 1977, Quantum Mechanics (Nonrelativistic Theory)
(Pergamon, NY).

50. Grabert H. and Horner H., Eds.1991, Special issue on single charge
tunneling, Z. Phys. \textbf{B85}, No 3, pp. 317 - 467; Basche T., Moerner
W.E., Wild U.P. (Eds.) 1996 Single-Molecule Optical Detection. Imaging and
Spectroscopy (VCH, Weinheim).

51. Bell J.S., 1964, Physics \textbf{1}, 195; 1966, Rev. Mod. Phys. \textbf{%
38}, 447.

52. Einstein A., Podolsky B. and Rosen N., 1935, Phys. Rev. \textbf{47}, 777.

53. Gisin N., Ribvordy G., Tittel W. and Zbinden H., 2002, Rev. Mod. Phys. 
\textbf{74}, 145.

54. Gibbert G., Amrick M., 2000, Practical Quantum Cryptography; A
Comprehensive Analysis, ArXiv: quant - ph/ 0009027.

55. Soddy F., 1913, Nature (London) \textbf{92}, 399.

56. Frauenfelder H. and Henley E.M., 1991, Subatomic Physics (Prentice Hall,
NY).

57. Kelly J.J., 2002, Phys. Rev. \textbf{C66}, 065203.

58. Lederer C.M. and Shirley V.S., 1978, Table of Isotopes (Wiley, NY).

59. Aston F.W., 1948, Mass-spectra and Isotopes (Science, Moscow) (in
Russian); Blaum K., 2006, Phys. Reports \textbf{425,} 1.

60. Cahn R.N. and Goldhater G.G., 1989, The Experimental Foundations of
Particle Physics (Camdridge University Press, Cambridge).

61. Shurtleft R. and Derringh E., 1989, Am. J. Phys. \textbf{47}, 552.

62. Barrett R.C. and Jackson D.F., 1977, Nuclear Sizes and Structure
(Clarendon, Oxford); Waraquier M., Morean J., Heyde K., 1987, Phys. Reports 
\textbf{148}, 249.

63. Hornyack W.F., 1975, Nuclear Structurs (Academic, NY); Brink B.M. 1965,
Nuclear Forces (Pergamon, NY).

64. Carlson J. and Schiavilla R., 1998, Rev. Mod. Phys. \textbf{70}, 743.

65. Davies C. and Collins S., 2000, Physics World, August, 35 - 40.

66. Schiff L.I., 1937, Phys. Rev.\textbf{\ 52}, 149.

67. Share S.S. and Stehn J.R., 1937, Phys. Rev. \textbf{52}, 48.

68. Schwinger J. and Teller E., 1937, ibid \textbf{52}, 286.

69. Greene G.L, Kessler E.G., 1986, Phys. Rev. Lett. \textbf{56}, 819.

70. Gartenhaus S., 1955, Phys. Rev. \textbf{100}, 900.

71. Adair R.K., 1950, Rev. Mod. Phys. \textbf{22}, 249.

72. Wilson R., 1963, The Nucleon - Nucleon Interaction (Wiley, NY); Henley
E.M. and \ \ \ Schiffer J.P., 1998, Nuclear Physics, ArXiv: nucl-th /
9807041.

73. Segre E., 1982, Nuclei and Particles (Reading MA, Benjamin, NY).

74. Arndt R.A., Roper L.D., 1987, Phys. Rev.\textbf{\ D35}, 128.

75.Vinh Mau R., 1979, in, Mesons in Nuclei, Rho M., and Wilkinson D.H. ,
eds, (North-Holland, Amsterdam).

76. Machleidt R., Holinde K. and Elster Ch., 1987, Phys. Rev. \textbf{149},
1.

77. Heisenberg W., 1932, Zs. Physik \textbf{77}, 1.

78. Renton P., 1990, Electroweak Interactions: An Introduction to the
Physics of Quarks and Leptons (Cambridge University Press, Cambridge);
Holstein B.R., 1989, Weak Inreactions in Nuclei (Princeton University Press,
Princeton); Paschoes E.A., 2007, Electroweak Theory (Cambridge University
Press, Cambridge).

79. Edelman S., 2004, Phys. Lett. \textbf{B 592,} 1.

80. Roy D.P., 1999, Basic Constituents of Matter and Their Interactions - A
Progress Report: ArXiv: hep - ph/9912523.

81. Wilczek F., 2004, The Universe is a Strange Place, ArXiv: astro -
ph/0401347.

82. Aitchison I.J.R. and Hey A.J.G., 1990, Gauge Theories in Particle
Physics: A Practical Introduction (Adam Hilger, Bristol).

83. Griffiths D.J., 1987, Introduction to Elementary Partcles (J.Wiley \&
Sons, NY).

84. Volcarce A. , Fernandez and Gonzalez P., 2005, Rep. Progr. Phys. \textbf{%
68}, 965.

85. Wagner W., 2005, ibid \textbf{68}, 2409.

86. Gross D.J., 2005, Rev. Mod. Phys. \textbf{77}, 837.

87. Politzer D., 2005, ibid, \textbf{77}, 851.

88. Wilczek F., 2005, ibid, \textbf{77}, 857.

89. Abbas A., 2001, Mod. Phys. Lett. \textbf{A16}, 755; 2004, \textbf{A19}%
,2365.

90. Yndurain F.I., 1999, The theory of Quark and Gluon Interactions
(Springer, Berlin).

91. Satz H., 1985, Ann. Rev. Nucl. Sci. \textbf{35}, 245; \ Celik T., Engles
J. and Satz H., 1985, Nuclear Physics \textbf{B256}, 670.

92. Walecka J.D., 1995, Theoretical Nuclear and Subnuclear Physics (Oxford
University Press, Inc., NY).

93. Glashow S.L., 1995, Nucl. Phys. \textbf{22}, 579; Weinberg S., 1967,
Phys. Rev. Lett. \textbf{19}, 1264; Salam A., 1968, in, Elementary Particle
Theory, Proc. Eigth Nobel Symposium, ed. Svartholm N., p. 367; Weinberg S.,
2004, The Making of the Standard Model, ArXiv: hep - ph /0401010.

94. Weinberg S., 1994, Strong Interactions at Low -Energies, ArXiv: hep -
ph/ 9412326.

95. Epelbaum E., Meissner U.-G., Gl\"{o}ckle W., 2003, Nucl. Phys. \textbf{%
A714}, 535; Meissner U.-G., 2005, Quark Mass Dependence of Baryon
Properties, ArXiv: hep-ph/0509029; M\"{u}ller B., 2007, From Quark - Gluon
Plasma to the Perfect Liquid, ArXiv: nucl - th/ 0710.3366; Weise W., 2008,
Overview and Perspectives in Nuclear Physics, ArXiv: nucl - th/ 0801.1619.

96. Ecker G., 1995, Chiral Perturbation Theory, ArXiv:hep - ph/9501357; Pich
A., 1995, Chiral Perturbation Theory, ArXiv:hep - ph/9502366; \ Bean S.R.,
Bedaque P.F., Haxton W.C., 2001, \ in, "At the Frontier of Particle \_
Handbook of QCD", ed. by Shifman M. (World Scientific, Singapore); Weise W.,
2007, Yukawa's Pion, Low - Energy QCD and Nuclear \ Chiral Dynamics, ArXiv:
nucl - th/ 0704.1992; Bijens J., Chiral Perturbation Theory Beyon One Loop,
ArXiv: hep-ph/0604043; 2007, Progr. Part. Nucl. Phys. \textbf{58}, 521.

97. Yukawa Y., 1935, Proc. of the Physico - Mathematical Society, \textbf{17}%
, 48.

98. Frosch R.F., Hofstadter R., McCarthy J.R., 1968, Phys. Rev.\textbf{\ 174}%
, 1380.

99. Heilig K. and Steudel A., 1974, At. Data Nucl. Data Tables \textbf{14},
613; \ Brandt H.W., Heilig K. and Steudel A., 1977; Phys. Lett. \textbf{A64}%
, 29; Aufmuth F., Heilig K. and Steudel A., 1987, At. Data Nucl. Data Tables 
\textbf{37}, 445.

100. Kelic A., Schmidt K.-H., Enqvist T., 2004, Phys. Rev. \textbf{C70},
064608.

101. Shpol'sky E.V., 1974, Atomic Physics, Part One (Fiz-Mat. Lit., Moscow)
(in Russian).

102. Shera E.B., Ritter E.T., Perkins R.B., 1976, Phys. Rev. \textbf{C14},
731.

103. Lee P.L., Boehm F. and Hahn A.A., 1978, Phys. Rev. \textbf{C17}, 1859.

104. Bertolozii W., Friar J., Heisenberg J., 1972, Phys. Lett. \textbf{B41},
408.

105. The prehistory of quark masses is reviewed \ in Grasser J. and
Leutwyller H., 1982, Phys. Rep. \textbf{87}, 77.

106. Leutwyller H., Masses of the Light Quarks, ArXiv: hep-ph/9405330;
ArXiv: hep-ph/ 9602255; Insights and Puzzles in Light Quark Physics,
ArXiv:hep-ph/ 07063138.

107. Froggatt C.D., The Problem of Mass, ArXiv: hep-ph/0312220.

108. Narison S., 2002, in, QCD as a Theory of Hadrons: from Partons to
Confinment, (Cambridge University Press, Cambridge).

109. Procura M., Nusch B.U., Weise W., 2006, Phys. Rev. \textbf{D73},
114510; Musch B., Hadron Masses, ArXiv: hep-ph/0602029.

110. Beane S.R., Orginos K., and Savage M.J., 2007, Nucl. Phys. \textbf{B768}%
, 38.

111. Ioffe B.L., 1995, Nucleon Spin Structure: Sum Rules,
Arxiv:hep-ph/9511401.

112. Scherer S. and Schindler M.R., 2005, Chiral Perturbation Theory Primer,
ArXiv:hep-ph/0505265.

113. Ioffe B.L., 2006, Progr. Part. Nucl. Phys. \textbf{56}, 232.

114. Gell-Mann M., Oakes R.J. and Renner B., 1968, Phys. Rev. \textbf{175},
2195.

115. Pich A and Prades J., 2000, Nucl. Phys. Proc. Suppl. \textbf{86}, 236.

116. Ioffe B.L. 1981, Nucl. Phys. \textbf{B188}, 317.

117. Altarelli G., 2004, Nucl. Instr. and Methods, \textbf{A518}, 1.

118. Pospelov M. and Ritz A., 2005, Electric Dipole Moment as Probes of New
Physics, ArXiv: hep-ph/0504231; 2005, Annal. Phys. (NY) \textbf{318}, 119.

119. Khriplovich I.B. and Lamoreaux S.K., 1997, CP Violation Without
Strangeness: Electric Dipole Moments of Particles, Atoms and Molecules
(Springer, Berlin).

120. Djouadi A., 2008, Phys. Rep. \textbf{457}, 1.

121. Ramsey-Musolf M.J. and Su S., 2008, ibid, \textbf{456}, 1.

122. Weinberg C.S., 1989, Rev. Mod. Phys. \textbf{61}, 1; The Cosmological
Constant Problems, ArXiv:astro-ph/0005265; Tegmark M., Rees M.J.,Wilczek F.,
Dimensionless Constants Cosmology and Other Dark Matters,
ArXiv:astro-ph/0511774; Dolgov A.D., Cosmology and New Physics,
ArXiv:hep-ph/0606230; Padmanabhan T., 2003, Phys. Rep. \textbf{380}, 235;
Dark Energy: Mystery of the Millenium, ArXiv:astro-ph/0603114; Fuller G.,
Hime A., Ramsey-Musolf M., Dark Matter in Nuclear Physics,
ArXiv:nucl-ex/0702031.

123. Milton K.A., 2004, J. Phys. A: Math.Gen. \textbf{37}, R209; Lamoreaux
S.K., 2005, Rep. Prog. Phys. \textbf{68}, 201; Mahajan G., Sarkar S.
Padmanbhan T., Casimir Effect Confronts Cosmological Constants, ArXiv:
astro-ph/0604265.

124. Stacey D.N., 1966, Rep. Progr. Phys. \textbf{29,} 171.

125. Bausche J. and Champeau R.-J., 1976, Adv. At. and Mol. Physics, \textbf{%
12, }39.

126. Arnikar H.J., 1989, Isotopes in Atomic Age (Wiley, NY).

127. See, e.g. Ramsden E.N. 1985, A - Level Chemistry (Stanley Thornes
Publishers, Hull); Malone L.J., 2003, Basic Concepts of Chemistry (Wiley,
NY).

128. Kogan V.I., 2000, Uspekhi Fiz. Nauk \textbf{170}, 1351 (in Russian).

129. Condon E.U., Shortly G.H., 1953, The theory of Atomic Spectra
(Cambridge University Press, Cambridge).

130. Rudzikas Z., 2006 Theoretical Atomic Spectroscopy (Cambridge University
Press, Cambridge). \ \ 

131. Striganov A.P., 1956, Uspekhi Fiz. Nauk \textbf{58, }365 (in Russian).

132. Huber A., Udem Th., Gross B., 1998, Phys. Rev. Lett. \textbf{80,} 468.

133. Udem Th. Gross B., Kourogi M., 1997, ibid, \textbf{79,} 2646.

134. Vinti J.P., 1939, Phys. Rev. \textbf{56, }1120.

135. Rosenrhal J. and Breit G., 1932, Phys. Rev. \textbf{41, }459.

136. Racah G., 1932, Nature (London) \textbf{129,} 723.

137. Brix P. and Kopferman H., 1958, Rev. Mod. Phys. \textbf{30,} 517.

138. Goorvitch D. Davis S.P. and Kleinman H., 1969, Phys. Rev. \textbf{188,}
1897.

139. Martensson - Pendrill A. - M., Gough D.S. and Hannaford P., 1994, Phys.
Rev. \textbf{A49, }3351; Berzinsh U., Gustafsson M., Hanstrop D., 1998,
Isotope Shift in the Electron Affinity of Chlorine, ArXiv: physics 9804028.

140. Hughes D.S. and Eckart C., 1930, Phys. Rev. \textbf{36,} 694.

141. Seltzer E.C., 1969, Phys. Rev. \textbf{188,} 1916.

142. Broch E.K., 1945, Arch. Math Natuv. \textbf{48, }25.

143. Haken H., Wolf H.Ch., 2005, The Physics of Atoms and Quanta (Springer,
Berlin-Heidelberg).

144. Lee P.L. and Boehm F., 1973, Phys. Rev. \textbf{C8,} 819.

145. Eliashevich M.A., 1946, Uspekhi Fiz. Nauk \textbf{48, }482 (in Russian).

146. Anderson A. (ed), 1973, The Raman Effect (Marcell Dekker, Inc. NY).

147. Long D.A., 1977, Raman Spectroscopy (McGraw - Hill, Inc. UK).

148. Grasselli J.G., Snavely M., Bulkin B.J., 1981, Chemical Application of
Raman Spectroscopy (Wiley, NY - Toronto).

149. Shymanski H.A. (ed.) 1967, Raman Spectroscopy (Plenum Press, NY).

150. Measures R.M., 1984, Laser Remote Sensing - Fundamentals and
Applications (Wiley - Interscience, NY - Singapore).

151. Plekhanov V.G, 2007, J. Phys. Condens. Matter \textbf{19}, 086221 (9pp).

152. Banwell C.N.,1983, Fundamentals of Molecular Spectroscopy (Mc - Graw -
Hill, London - NY).

153. Bhagavantam S and Venkatarayudu T., 1951, Theory of Groups and Its
Applications to Physical Problems (Adha University Press, Waltair).

154. Danielewicz - Ferchmin I and Ferchmin A.R., 2004, Phys. Chem. Liquids 
\textbf{42}, 1 -36.

155. Walrafen G.E., 1964, J. Chem. Phys. \textbf{40}, 3249.

156. Chaplin M.F., 2000, Biophys. Chem.\textbf{\ 83}, 211.

157. Kroto H.W., 1985, Nature (London) \textbf{318, }162.

158. Kratschmer W., Fositropolous and Hoffman D.R., 1990, Chem. Phys. Lett. 
\textbf{170}, 167.

159. Martin M.C. and Fabian J., 1995, Phys. Rev. \textbf{B51}, 2844.

160. Rosenberg A. and Kendziora C., 1995, ibid, \textbf{B51}, 9321.

161. \ Menendez J., Page J.B., and Guha S., 1994, Phil. Magazine \textbf{70}%
, 651.Guha S., Menendez J., Page J.B., 1997, Phys. Rev. \textbf{B56}, 15431.

162. Mauersberger K., 1981, Geophys. Res. Lett. \textbf{8}, 935.

163. Thiemens M.H. Heidenreich J.E., 1983, Science \textbf{219}, 1073.

164. Thornton E.K. and Thornton E.R. 1970, Origin and Interpretation of
Isotope Effect, in Collins C.J. and Bowman N.S. (eds.) Isotope Effects in
Chemical Reaction (American Chemical Society Monograph) (Van Nostrand
Reinhold Co., NY - London) \ p.p. 213 - 285.

165. Biegelsen J., Lee M.W. and Mandel F., 1973, Ann. Rev. Phys. Chem. 
\textbf{24}, 407 -440.

166. Weston R.E., 1999, Chem. Rev. \textbf{99}, 2115.

167. Thiemens M.H., 1999, Science \textbf{283}, 341.

168. Mauersberger K., Krankowsky D., Janssen C. and Schinke R., 2005, Adv.
At. Mol. and Optical Physics \textbf{50,} 1.

169. Johnston H.S., 1966, Gas Phase Reaction Rate Theory, (The Ronald Press
Company, NY).

170. Criss R.E., 1995, Stable Isotope Distribution in Global Earth Physics,
A Handbook of Physical Constants (American Geophysical Union, NY).

171. Weston R.E., 2006, J. Nucl. Sci. and Technol. (Japan) \textbf{43}, 295.

172. Anderson S.M., Hulsebusch D. and Mauersberger K., 1997, J. Chem. Phys. 
\textbf{107}, 5385.

173. Janssen Ch., Guenther J. and Mauersberger K., 1999, ibid, \textbf{111},
7179.

174. Mauersberger K., Erbacher K. and Krankowsky D., 1999, Science, \textbf{%
283}, 270.

175. Hathorn B.C. and Marcus R.A., 1999, J. Chem. Phys. \textbf{111}, 4087.

176. Hathorn B.C. and Marcus R.A.,2000, ibid, \textbf{113}, 9497.

177. Hathorn B.C. and Marcus R.A., 2001, J. Phys. Chem. \textbf{A105}, 5586.

178. Gao Y.Q. and Marcus R.A. 2002, J. Chem. Phys. \textbf{116}, 137.

179. Gao Y.Q., Chen W -Ch. and Marcus R.A. , J. Chem. Phys. \textbf{117},
1536.

180. Babikov D., Kendrick B.K., Walker R.B., 2002, Chem. Phys. Lett. \textbf{%
272}, 686.

181. Babikov D., Kendrick B.K., Walker R.B., 2003, J. Chem. Phys. \textbf{118%
}, 6298.

182. Babikov D., Kendrick B.K., Walker R.B., 2003, J. Chem. Phys. \textbf{119%
}, 2577.

183. Hulston J.R. and Thode H.G., 1965, J. Geophys. Res. \textbf{70}, 3475.

184. Clayton R.N., Grossman L., Mayeda T.K., 1973, Science \textbf{182}, 485.

185. Gellene G.I., 1996, Science \textbf{274}, 1344.

186. Valentini J.J., 1987, J. Chem. Phys. \textbf{86}, 6757; Bates D.R.,
1988, Geophys. Res. Lett. \textbf{15}, 13; Griffith K.S. and Gellene G.I.,
1992, J. Chem. Phys. \textbf{96}, 4403.

187. Sehestedy J., Nielsen O.J., Egsgaard H., 1995, J. Geophys. Res. \textbf{%
100,} 20979; 1998, ibid, \textbf{103}, 3545.

188. Pines D., 1963, Elementary Excitations in Solids (W.A. Benjamin, Inc,
NY - Amsterdam).

189. Pekar S.I., 1982, Crystaloptics and Addition Waves (Naukova Dumka,
Kiev) (in Ruaaian).

190. Srivastava G.P., 1990, The Physics of Phonons (Hilger, Bristol).

191. Plekhanov V.G., 2007, Isotopes and Their Applications in Quantum
Information, Preprint N002 Computer Science College, Tallinn (in Russian).

192. CallawayJ., 1964, Energy Band Structure (Academic, NY).

193. Martin R.M., 2004, Electronic Structure - Basic Theory and Practical
Methods (Cambridge University Press, Cambridge).

194. Jones H., 1962, The Theory of Brillouin Zones and Electronic States in
Crystals (North - Holland Co., Amsterdam).

195. Ziman J.M. , 1963, Electrons and Phonons (Oxford University, London).

196. Ashcroft N.W., Mermin N.D., 1975, Solid State Physics (Holt, Rinehhart
and Winston, NY - Sydney).

197. Kittel Ch., 1972, Introduction to Solid State Physics (J. Wiley and
Sons, Inc., NY - London).

198. Lax M., Burstein E., 1955, Phys. Rev. \textbf{97}, 39.

199. Loudon R., 1964, Adv. Phys. \textbf{13}, 423; Cowley R.A., 1971, in
Raman Effect, ed. Anderson A (Marcel Dekker, NY).

200. Birman J.L., 1974, in Handbuch f\={u}r Physik, Vol \textbf{25/2b}
(Springer, Berlin - NY).

201. Agekyan V.F., Asnin A.M., Kryukov \ V.M., And Markov I.I., 1989, Fiz.
Tverd. Tels \textbf{31}, 101 (in Russian).

202. Fuchs H.D., Etchegoin P. and Cardona M., 1993, Phys. Rev. Lett. \textbf{%
70}, 1715.

203. Plekhanov V.G., 2001, Isotope Effect in Solid State Physics, in,
Semiconductors and Semimetals, Vol. \textbf{68 }(Eds. Willardson R.K. and
Weber E.) (Academic, San Diego).

204. Hanzawa H., Umemura N., Nisida Y. and Kanda H., 1996, Phys. Rev. 
\textbf{B54,} 3793.

205. Plekhanov V.G., 2003, Physics - Uspekhi, \textbf{46,} 689.

206. Chrenko R.M., 1988, J. Appl. Phys. \textbf{63}, 5873.

207. Hass K.C., Tamor M.A., Anthony T.R., and Banholzer W.F., 1991, Phys.
Rev. \textbf{44}, 12046.

208. Solin S.H., Ramdas A.K., 1970, Phys. Rev. \textbf{B1}, 1687.

209. Plekhanov V.G., 2001, Materials Sci. \& Eng. \textbf{R35}, 141.

210. Elliott R.J., Krumhansl J.A., Leath P.L., 1974, Rev. Mod. Phys. \textbf{%
46}, 465.

211. Barker A.S. (Jr), Sievers A.J., 1975, Rev. Mod. Phys. \textbf{47}
(Suppl. 2) s1.

212. Chang I.F., Mitra S.S., 1971, Adv. Phys. \textbf{20}, 360.

213. Ipatova I.P., 1988, in Optical Properties of Mixed Crystals (Modern
Problems in Condensed Matter Sciences, Vol. 23, eds. Elliott R.J., Ipatova
I.P) Ch.1 (North - Holland,Amsterdam) p. 1.

214. Plekhanov V.G., 1997, Opt. Spectrosc. \textbf{82}, 95.

215. Plekhanov V.G., 1995, Phys. Rev. \textbf{B51}, 8874.

216. Plekhanov V.G., 1993, Opt. Spectrosc. \textbf{75}, 31.

217. Plekhanov V.G., 2001, J. Raman Spectrosc. \textbf{32}, 631.

218. Plekhanov V.G., 2006, J. Nucl. Sci. \& Technol. (Japan)\textbf{\ 43},
375.

219. Kapustinsky A.F., Shamovsky L.M., Bayushkina K.S., 1937, Acta
Physicochem. (USSR) \textbf{7}, 799.

220. Plekhanov V.G., Betenekova T.A., Pustovarov V.A., 1976, Sov. Phys.
Solid State \textbf{18}, 1422.

221. Plekhanov V.G., 1998, Rep. Progr. Phys. \textbf{61}, 1045.

222. Kreingol'd F.I., Lider K.F., Shabaeva M.B., 1984, Fiz. Tverd. Tela 
\textbf{26}, 3940 (in Russian).

223. Onodera Y., Toyozawa Y., 1968, J. Phys. Soc. Japan\textbf{\ 24}, 341.

224. Kreingol'd F.I., Lider and Solov'ev K.I., 1976, JETP Lett, (Moscow) 
\textbf{23}, 679 (in Russian).

225.Kreingol'd F.I., Lider K.F, Sapega V.F., 1977, Fiz. Tverd. Tela \textbf{%
19}, 3158 (in Russian).

226. Kreingol'd F.I. and Kulinkin B.S., 1986, ibid \textbf{28}, 3164 (in
Russian).

227. Kreingol'd F.I., 1978, ibid, \textbf{20}, 3138 (in Russian).

228. Bobrysheva A.I., Jeru I.I., Moskalenko S.A., 1982, Phys. Stat. Solidi
(b) \textbf{113}, 439.

229. Zhang J.M., Giehler M., Ruf. T., 1998, Phys. Rev. \textbf{B57}, 1348;
Zhang J.M., Ruf T., Lauck R, 1998, ibid \textbf{B57}, 9716.

230. Meyer T.A., Thewalt M.L.W. and Lauck R., 2004, Phys. Rev. \textbf{B69},
115214 - 5.

231. Bir G.L. and Picus G.E., 1972, Symmetry and Deformation in
Semiconductors (Science, Moscow) (in Russian).

232. Thomas D.G., (ed.) 1967, II - VI Semiconducting Compounds (Benjamin,
NY).

233. Lastras - Martionez L.F., Ruf. T., Konuma M. and Aspnes D., 2000, Phys.
Rev. \textbf{B61}, 12946.

234. Karaskaja \ D., Thewalt M.L.w., Ruf. T., 2003, Solid State Commun. 
\textbf{123}, 87; Karaskaja D., Thewalt M.L.W., Ruf. T., 2003, Phys. Stat.
Sol. (b)\ \textbf{235}, 64; Thewalt M.L.W., 2005, Solid State Commun. 
\textbf{133}, 715.

235. Tsoi S., Alawadhi H., Lu X., Haller E.E., 2004, Phys. Rev. \textbf{B70}%
, 193201 - 4; Ramdas A.K., Rodriguez S., Tsoi S. Haller E.E., 2005, Solid
State Commun. \textbf{133}, \ 709; Tsoi S., Rodriguez S., Ramdas A.K., 2005,
Phys. Rev. \textbf{B72}, 153203 - 4.

236. Kim H., Rodriguez S. and Anthony \ T.R., 1997, Solid State Commun. 
\textbf{102}, 861.

237. Cardona M., 2002, Solid State Commun. \textbf{121}, 7.

238. Klochikhin A.A. and Plekhanov V.G., 1980, Sov. Phys. - Solid State 
\textbf{22}, 342.

239. Klemens P.G., 1958, in Solid State Physics: Advances in Research and
Applications, Vol.7 (Eds. F. Seitz, D. Turnbull) (Academic, NY).

240. Holland M.G., 1966, in Physics of III - V Copounds (Semiconductors and
Semimetals, Vol. 2, eds. Willardson R.K., Beer A.C.) (Academic, NY) p. 3.

241. Touloukian Y.S. Powell R.W., Ho C.Y. Klemens P.G., 1970, Thermophysical
Properties of Materials, Vol. 1 (IFI.Plenum Press, NY - Washington).

242. Berman R., 1976, Thermal Conduction in Solids (Clarendon Press, Oxford).

243. Berman R.Z.,1958, Phys. Chemie. Neue Fol. \textbf{16}, 10.

244. Callaway J., 1959, Phys. Rev. \textbf{113}, 1046.

245. Peierls R., 1955, Quantum Theory of Solids (Clarendon Press, Oxford).

246. Ziman J.M., 1979, Models of Disorder (Cambridge University Press,
Cambridge).

247. Pomeranchuk I.Ya., 1942, J. Phys. (USSR) \textbf{6}, 237.

\bigskip 248.GeballeT.H., Hull G.W., 1958, Phys. Rev. \textbf{110}, 773.

249. Onn D.G., Witek A., Qiu Y.Z, Anthony T.R. and Banholzer W.F.,1992,
Phys. Rev. Lett. \textbf{68}, 2806.

250. Olson J.R., Pohl R.O., Vandersande J.W., 1993, Phys. Rev. \textbf{B47},
14850.

251. Wei L., Kuo P.K.,Thomas R.L., 1993, Phys. Rev. Lett. \textbf{70}, 3764.

252. Debye P., 1912, Ann. Phys. $\mathbf{39}$, 789.

253. Cardona M., Kremer R.K., Sanat M., Esreicher S.K., Anthony T.R., 2005,
Solid State Commun. \textbf{133}, 465.

254. Slack G.A., 1973, J. Phys. Chem. Solids \textbf{34}, 321.

255. Vandersande J.W., Zolton A., Olson J.R., 1992, Proc. Seventh Int. Conf.
on Scattering of Phonons in Condensed Matter (Cornell University, Ithaca).

256. Asen - Palmer M., Bartkowsky K., Gmelin E., Phys. Rev. \textbf{B56},
9431.

257. Capinski W.S., 1997, Appl. Phys. Lett. \textbf{71}, 2109.

258. Capinski W.S., Maris H.J., Tamura S., 1999, Phys. Rev. \textbf{B59},
10105.

259. Ruf T., Henn R.W., Asen - Palmer M., 2000, Solid State Commun \textbf{%
115}, 243; Erratum, 2003, ibid, \textbf{127}, 257.

260. Inyushkin A.V., 2002, Inorganic Materials \textbf{38}, 427; Broido
D.A., Ward A., Mingo N., 2005, Phys. Rev. \textbf{B72}, 014308/1 - 8.

261. Plekhanov V.G., 2000, Physics - Uspekhi \textbf{43}, 1147.

262. Omini M., Sparavigna A., 1997, Nuovo Cimento \textbf{D19}, 1537.

263. Sparavigna A., 2002, Phys. Rev. \textbf{B65}, 064305; 2003, ibid 
\textbf{B67}, 144305.

264. Hass K.C., Tamor M.A., Anthony T.R. and Banholzer W.F., 1992, Phys.
Rev. \textbf{B45,} 7171.

265. Fuchs H.D., Grein C.H., Thomsen C., 1991, ibid, \textbf{B43}, 4835.

266. Wang D.T., G\H{o}bel A., Zegenhagen J., 1997, ibid, \textbf{B56}, 13167.

267. Tamura S., 1983, ibid, \textbf{B27}, 858.

268. Tamura S., 1984, ibid, \textbf{B30}, 849.

269. Widulle F., Serrano J. and Cardona M., 2002, Phys. Rev. \textbf{B65},
075206/1 - 10.

270. Spitzer J., Etchegoin P., Bangolzer W.F., 1993, Solid State Commun. 
\textbf{88}, 509.

271. Vogelgesand R., Ramdas A.K., Anthony T.R., 1996, Phys. Rev. \textbf{B54}%
, 3989.

272. Widulle F., Ruf. T., Ozhogin V.I., 2001, Solid State Commun. \textbf{118%
}, 1.

273. Vast N., Baroni S., 2000, Phys. Rev. \textbf{B61}, 9387.

274. Vast N., Baroni S., 2000, Comput. Mater. Sci. \textbf{17}, 395.

275. Rohmfeld S., Hundhausen M., Ley L., Pensl G., 2001, Phys. Rev. Lett. 
\textbf{86}, 826.

276. Plekhanov V.G., Altukhov V.I., 1985, J. Raman Spectrosc. \textbf{16},
358.

277. Plekhanov V.G., 2001, Progr. Solid State Chem. \textbf{27}, 71.

278. Plekhanov V.G., 1996, Phys. Rev. \textbf{B54}, 3869.

279. Elliott R.J., Ipatova I.P., (Eds), 1988, Opical Properties of Mixed
Crystals (North - Holland, Amsterdam).

280. Parks C., Ramdas A.K., Rodriguez S., Itoh K.M., 1994, Phys. Rev. 
\textbf{B49}, 14244.

281. Davies G., 1993, Semicond. Sci. Technol. \textbf{8,} 127.

282. Permogorov S., Reznitsky A., 1992, J. Lumin. \textbf{52}, 201.

283. Efros A.L., Raikh M.E., 1988, in [279], Chapter 5.

284. Lifshitz I.M., 1987, Selected Works (Science, Moscow) (in Russian).

285. Kanehisa V.A., Elliott R.J., 1987, Phys. Rev. \textbf{B35,} 2228.

286. Schwabe N.F., Elliott R.J., 1996, ibid, \textbf{B54}, 5318.

287. Nelson R.J., Holonjak N., Groves W., 1976, ibid, \textbf{B13}, 5415.

288. Mahanti S.D., Varma C.M., 1972, ibid, \textbf{B6}, 2209.

289. Mahanti S.D., 1974, ibid, \textbf{B10}, 1384.

290. Bethe H.A., Salpiter E., 1957, Quantum Theory of One and Two Elctron
Atoms (Academic, NY).

291. Hama J., Kawakami N., 1988, Phys. Lett. \textbf{A126}, 348.

292. Betenekova T.A., Shabanova I.N., Gavrilov F.F., 1978, Fiz. Terd. Tela 
\textbf{20}, 3470 (in Russian).

293. Ichikawa K., Susuki N., Tsutsumi K., 1981, J. Phys. Soc. Japan \textbf{%
50}, 3650.

294. Rechenberg H., 1999, in, The Casimir Effect 50 Years Later, Bordag M.,
(Ed.) (World Dcientific, Singapore) p. 10.

295. Planck M., 1911, Verh. d. Deutsch Phys. Ges. (2) \textbf{13}, 138.

296. Nernst W., and Lindemann F.A., 1911, Z. Electrochem. \textbf{17}, 817;
Einstein A., and Stern O., 1913, Ann. Phys. \textbf{40}, 551.

297. Fan H.Y., 1951, Phys. Rev. \textbf{82}, 900.

298. Zollner S., Cardona M and Gopalan S., 1992, ibid, \textbf{B45}, 3376.

299. Baym G., 1968, Lectures on Quantum Mechanics (Benjamin, NY) p. 98.

300. Lautenschlager P., Garriga M., Vina L., 1987, Phys. Rev. \textbf{B36},
4821.

301. Plekhanov V.G., Phys. Solid Stat. (St. Petersburg) \textbf{35}, 1493.

302. Cardona M., 2005, Solid State Commun. \textbf{133}, 3.

303. Plekhanov V.G., 2007, Unpublished results.

304. Alawadhi H., Tsoi S., Lu X., 2007, Phys. Rev. \textbf{B75}, 205207.

305. Zhernov A.P., 2002, Fiz. Tverd. Tela (St. Petersburg) \textbf{44}, 992.

306. Diffusion in Semiconductors and Non - Metallic Solids, 1998, ed. by
Beke D.L., Landolt - B\"{o}rnstein, New Series, Group III, Vol. 33A
(Springer, Berlin).

307. Impurities and Defects in Group IV Elements, IV - IV and III - V
Compounds, 2002, ed. by Schulz M., Landolt - B\"{o}rnstein, New Series,
Group III, Vol. 41A2, Part $\alpha $ (Springer, Berlin).

308. Impurities and Defects in Group IV Elements, IV - IV and III - V
Compounds, 2003, ed. by Schulz M., Landolt - B\"{o}rnstein, New Series,
Group III, Vol. 41A2, Part $\beta $ (Springer, Berlin).

309. Mehrer H., 2007, Diffusion in Solids (Fundamentals, Methods, Materials,
Diffusion - Controlled Processes) (Springer Series in Solid - State
Sciences, Vol. 155) (Springer, Berlin) 654 p.

310. Heitjans P., K\"{a}rger J., 2008 (Eds), Diffusion in Condensed Matter
(Methods, Materials, Models) (Springer, Berlin) 970 p.

311. Bracht H. and Stolwijk N.A., 2002, Solubility in Silicon and Germanium,
Landolt - B\"{o}rnstein, New Series, Group III, Vol. 41A2, Part $\alpha $
(Springer, Berlin).

312. Stolwijk N.A. and Bracht H., 1998, Diffusion in Silicon, Germanium
andTheir Alloys, Landolt - B\"{o}rnstein, New Series, Group III, Vol. 33A
(Springer, Berlin).

313. Bracht H., 2007, Phys. Rev. \textbf{B75}, 035210/1 - 16.

314. Bracht H., Silvester H.H., Sharp I.D., 2007, ibid, \textbf{B75},
035211/1 - 21.

315. Tan T.Y., G\"{o}sele U.M. and Yu S., 1991, Crit. Rev. Solid State Phys. 
\textbf{17}, 47.

316. Crank J., 1975, The Mathematics of Diffusion (Oxford University Press,
London - New York).

317. Fuchs H.D., Walukewicz W., Dondl W., 1995, Phys. Rev. \textbf{B51},
16817.

318. Itoh K., Hansen W.L., Ozhogin V.I., 1993, J. Mater. Res. \textbf{8},
1341.

319. Tan T.Y., You H.M., Yu S., G\"{o}sele U.M., 1992, J. Appl. Phys. 
\textbf{72,} 5206.

320. Campbell D.R., 1975, Phys. Rev. \textbf{B12}, 2318.

321. Wang L., Hsu L., Erickson J.W., Cardona M., Phys. Rev. Lett. \textbf{76}%
, 2342.

322. Wang L., Wolk J.A., Hsu L., 1997, Appl. Phys. Lett. \textbf{70}, 1831.

323. Beernik K.J., Sun D., Treat D.W., 1995, Appl. Phys. Lett. \textbf{66},
3597.

324. Bracht H., Haller E.E., Eberl K and Cardona M., 1999, Appl. Phys. Lett. 
\textbf{74}, 49.

325. Wee S.F., Chai M.K., Gillin W.P., 1997, J. Appl. Phys. \textbf{82},
4842.

326. Larrabee R.D., 1984, (ed) Neutron Transmutation Doping of Semiconductor
Materials (Plenum Press, New York - London).

327. Schlimak I., 1999, Fiz. Tverd. Tela \textbf{41}, 837 (in Russian).

328. Ionov A.N., Matveev M.N. and Shmik D.V., 1989, J. Techn. Phys. (St.
Petersburg) \textbf{59}, 169 (in Russian).

329. Brice D.K., 1971, Radiat. Effects \textbf{11}, 227.

330. Schlimak I., Ionov A.N., Rentzsch R., 1996, Semicond. Sci. Technol. 
\textbf{11}, 1826.

331. Sn\"{o}ller M.S., 1974, IEEE Trans. Electron. Devices \textbf{ED - 21},
313; Haas W. and Sn\"{o}ller M.S., 1976, J. Electron. Mater. \textbf{5,} 57.

332. High - Power Semiconductor Devices, 1976, IEEE Trans. Electron. Devices 
\textbf{ED - 23}.

333. Hass W. and Sn\"{o}ller M.S., 1976, in [332] p. 803.

334. Baliga B.I., 1984, in [326] p. 167.

335. Farmer J.W and Nugent J.C., 1984, in [326] p. 225.

336. Hill M.J. Van Iseghem P.M., Zimmerman W., 1976, in [332] p. 809.

337. Muhlebauer A., Seldak F. and Voss P., 1975, J. Electrochem Soc. \textbf{%
122}, 1113.

338. Lark - Horowitz K., 1951, in, Proc. Conf. on Semicond. Materials, ed.
by Henish H.K. (Butterworth, London).

339. Tanenbaum M. and Mills A.D., 1961, J. Electrochem. Soc. \textbf{108},
171.

340. James H.M., Malmr\"{o}s O., 1976, in [332] p. 797.

341. Van Iseghem P.V., 1976, in [332] p. 823.

342. Dzhakeli V.G. and Kachlishvili G., 1984, Sov. Phys. Semicond. \textbf{18%
}, 926.

343. Gulberg J., 1981, (ed), Neutron - Transmutation - Doped Silicon (Plenum
Press, NY).

344. Hamanaka H., Kuriyama K., Yahagi M., 1984, Appli. Phys. Lett. \textbf{45%
}, 786.

345. Klahr C.N., Cohen M., 1964, Nucleonics \textbf{22}, 62.

346. Rentzsch R., Friedland K.J. and Ionov A.N., 1988, Phys. Stat. Solidi
(b) \textbf{146}, 199.

347. Magerle R., Buchard A. Deicher M., 1995, Phys. Rev. Lett. \textbf{75},
1594.

348. Kuriyama K. and Sakai K., 1996, Phys. Rev. \textbf{B53}, 987.

348$^{a}$. Sekine T., Uchinokura K. and Marzuura E., 1977, J. Phys. Chem.
Solids \textbf{48}, 109.

348$^{b}$. Kawakubo T. and Okada M., 1990, J. Appl. Phys. \textbf{67}, 3111.

348$^{c}$. Alawadhi H., Vogelgesang R., Chin T.P.,1997, J. Appl. Phys. 
\textbf{82}, 4331.

349. Kuriyama K., Miyamoto Y., Koyama T and Ogawa O., 1999, J. Appl. Phys. 
\textbf{86}, 2352.

350. Kuriyama K., Ohbora K. and Okada M., 2000, Solid State Commun. \textbf{%
113}, 415.

351. Young M.H., Hunter A.T., Baron R., 1984, in [326] p.1.

352. Sze S.M., 1969, Physics of Semiconductor Devices (Wiley, NY).

353. Allan W.B., 1973, Fibre optics - Theory and Practice, (Plenum Press,
London - New York).

354. Berezin A.A., 1987, J. Phys. Chem. Solids \textbf{48}, 863; ibid, 1989, 
\textbf{50}, 5.

355. Pohl R.W., 1947, Itroduction into Optics (Moscow) (in Russian).

356. Zhuravleva L.M., Plekhanov V.G., 2007, Method of Fiber's Manufacture,
Patent of Russian Federation N 2302381.

357. Thyagarajan K., Ghatok A.K., (ed) 1982, Lasers Theory and Applications
(Plenum Press, NY).

358. Klingshirn C., 1987, in Spectrosc. Solid - State Laser Type Matter,
Proc. Course Enrico Fermi, Erice (London - New York).

359. Rossi F., Kuhn T., 2002, Rev. Mod. Phys. \textbf{74}, 895.

360. Klingshitn C., Haug H., 1981, Phys. Rep. \textbf{70}, 315.

361. Lyssenko V.G., Revenko V.I., 1978, Sov. Phys. Solid State \textbf{20,}
1238.

362. Johnson Jr. W.D., 1971, J. Appl. Phys. \textbf{42,} 2732.

363. Pankov Z., 1971, Optical Processes in Semiconductors (Prentice Hill,
Englewood Cliffs, NJ).

364. Dean P.J., Lightowlers E.C., Wright D.R., 1965, Phys. Rev. \textbf{140}%
, A352.

365. Haynes J.R., 1960, Phys. Rev. Lett. \textbf{4}, 361.

366. Takiyama K., Abd - Elrahman M.I., Fujita T., Oda T., 1996, Solid State
Commun. \textbf{99}, 793.

367. Collins A.T., Lawson S.C., Davies G., Kanda H., 1990, Phys. Rev. Lett. 
\textbf{65}, 891.

368. Sauer R., Sternschulte H., Wahl S., Thonke K., Anthony T.R., 2000,
Phys. Rev. Lett. \textbf{84}, 4172.

369. Kawarada H., Yokota T., Hiraki A., 1994, Appl. Phys. Lett. \textbf{64},
451.

370. Ruf T., Cardona M., Anthony T.R., 1998, Solid State Commun. \textbf{105}%
, 311.

370$^{a}$. Okushi H., Watanabe H., Kanno S., 202, Phjys. Stat. Solidi (a) 
\textbf{202}, 2051; \ 2003, ibid, \textbf{203}, 3226.

371. Dean P.J., 1973, Progr. Solid State Chem. \textbf{8}, 1.

372. Basov N.G., Danilychev V.A., Popov Yu.M., 1970, JETP Lett. \textbf{132}%
, 473.

373. Packard J.R., Campbell D.A., Tait W.C. , 1967, J. Appl. Phys. \textbf{38%
}, 5255.

374. Kulevsky L.A., Prokhorov A.M., 1966, IEEE - QE \textbf{2}, 584.

375. Haug H., 1982, Adv. Solid State Phys. \textbf{22}, 149.

376. Packard J.R., Tait W.C., Campbell D.A., 1969, IEEE - QE \textbf{5}, 44.

377. Liu K.C., Liboff R.L., 1983, J. Appl. Phys. \textbf{54}, 5633.

378. Plekhanov V.G., 1981, Proc. Int. Conf. LASERS - 80 \ (STS Press,
McClean, VA, USA).

379. Schwenter N., Dossel O., Nahme N., 1982, Laser Technique for Extreme UV
Spectroscopy (ATPT, NY).

380. Plekhanov V.G., 1989, Quantum Electron. (Moscow) \textbf{16}, 2156 (in
Russian).

381. Gross E.F., 1976, Selected Papers (Science, Leningrad).

382. Knox R.S. 1963, Theory of Excitons (Academic Press, London - New York).

383. Brodin M.S., Reznitchenko V.Ya., 1986, in Georgabiani G., Sheinkman
MK., (eds), Physics of A$_{2}$B$_{6}$ Compounds (Science, Moscow) (in
Russian).

384. Berezin A.A., 1986, Kybernetes \textbf{15}, 15; 1988, Chemtronics 
\textbf{3}, 116.

385. Schr\"{o}dinger E., 1935, Naturwissenschaften \textbf{23}, 807.

386. Deutsch D., 1985, Proc. Roy. Soc. (London) \textbf{400}, 97.

387. Deutsch D., 1998, The fabric of Reality (Penguin Press, Allen Line).

388. Kilin S.Ya., 1999, Uspekhi Fiz. Nauk \textbf{169}, 507 (in Russian).

389. Wooters W.K., Zurek W.H., 1982, Nature, \textbf{299}, 802.

390. Dieks D., 1982, Phys. Lett. \textbf{A92}, 271.

391. Schumacher B., 1995, Phys. Rev. \textbf{A51,} 2738.

392. Kadomzev B.B., 1999, Dynamics and Information (UFN, Moscow) (in
Russian).

393. Josza R., 1998, in, Lo H K., Spiller T., Popescu S., (eds) Introduction
to Quantum Information aqnd Quanrum Computation (World Scientific, London).

393$^{a}$. Galindo A., Martin - Delgado M.A., 2002, Rev. Mod. Phys. \textbf{%
74}, 347.

394. Spiller T., 1998, in [393] p. 1 - 28.

395. Barenco A.,1998, in [393] p. 143 - 184.

396. Shor P.W., 1994, in Proc. 35 th Annual Symposium on Foundations of
Computer Science (Santa Fe, NM, USA).

397. Grove L.K., 1997, Phys. Rev. Lett. \textbf{79}, 325.

398. Feinman R., 1982, Int. J. theor. Phys. \textbf{21}, 467.

399. Unruch P.W., 1995, Phys. Rev. \textbf{A51}, 992.

400. Shor P.W., 1996, Phys. Rev. \textbf{A52}, 2493.

401. Valiev K.A., Kokin A.A., 2001, Quantum Computers (RHD, Moscow) (in
Russian).

402. Preskill J., 1998, Proc. Roy. Soc. (London) \textbf{A454}, 385.

403. Kane B.E., 1998, Nature \textbf{393}, 133.

404. Waugh J.S., Slichter C.P., 1988, Phys. Rev. \textbf{B37}, 4337.

405. Kane B.E., 2000, Fortschr. Phys. \textbf{48}, 1023.

406. Vrijen R., Yablonovich E., Wang K., 2000, Phys. Rev \textbf{A52},
012306/1 - 10.

407. DiVinchenzo D.P., 2000, Fortschr. Phys. \textbf{48}, 771.

408. Shlimak I., Safarov V.I. and Vagner I.D., 2001, J. Phys.: Condens
Matter \textbf{13}, 6959.

409. Shlimak I., and Vagner I.D., 2006, ArXiv: cond - matter/0612065.

409$^{a}$. Yang A., Steger M., Karaiskaj D., 2006, Phys. Rev. Lett.\textbf{\
97}, 227401.

410. Bowden C.M. and Pethel S.D., 2000, Laser Phys. \textbf{10}, 35.

411. Sch\"{a}fler F., 1992, Semicond. Sci. Technol. \textbf{7}, 260.

412. Dagata J.A., 1995, Science 270, 1625.

413. Marchi F., 1998, J. Vac. Sci. Technol. \textbf{B16}, 2952.

414. Machida T., Yamazaki T., Ikushina K., 2003, Appl. Phys. Lett. \textbf{82%
}, 409.

415. Yusa G., Muraki K., Takashina K., 2005, Nature \textbf{434}, 1001.

416. Kurten M.J. and Uren M.J., 1989, Adv. Phys. \textbf{38}, 367.

417. Xiao M., Martin I., Jiang H.W., 2003, Phys. Rev. Lett. \textbf{91},
078301.

418. DiVinchenzo D.P., 1995, Phys. Rev. \textbf{A51}, 1015.

419. http://www. nupece.org/iai 2001/report/B43.pdf.

420. http://www. cbvp.com/nmrc/mia.html.

421. Huang S. - C., 1986, Principles of tracer kinetic modeling in positron
emission tomography and autoradiography in, Positron Emission Tomography and
Autoradiography: Principles and Applications for the brain and Heart, ed. by
M.E. Phelps, M.C. Mazotta, M.R. Schelbert (Raven, NY).

422. Berman S.R., 1997, Positron Emission Tomography of Heart in, Cardiac
Nuclear Medicine, 3rd ed, ed. by M.C. Gerson, Health Professons Division
(McGraw-Hill, NY); Valk P.E., Bailey D.L., Townsend D.W., 2004, Positron
Emission Tomography: Basic Science and Clinical Practice (Springer, New
York); Bender H., Palmelo H., Valk P.E., 2000, Atlas of Clinical PET in
Oncology: PET versus CT and MRI (Springer, New York).

423. http//www.vbvp.com/nmrc/nucmed.html.

424. Gehrels T. (ed), 1978 Protostars and Planets (University of Arizona
Press, Tuscon).

425. Libby W.F., 1952, Radiocarbon Dating (University of Chicago Press,
Chicago).

426. Stuiver M., Pearson C.W., 1993, Radiocarbon \textbf{35}, 1.

427. Stuiver M., Reimer P.J., 1993, ibid, \textbf{35}, 215.

428. Taylor R.E., 1987, Radiocarbon Dating: An Archaeological Perspective
(Academic Press, New York and London; Taylor R.E. and Aitken M.J. (eds.),
1997, Chronometric Dating in Archaeology (Plenum Press, New York).

429. Symabalisty E.M.D. and Schramm D.N., 1981, Rep. Progr. Phys. \textbf{44}%
, 293.

430. Aitken M.J., 1985, Thermoluminescence (Academic, London).

431. Aitken M.J., 1985, Introduction in Optical Dating (Oxford University
Press, Oxford).

432. Aitken M.J., Stinger C.B. and Mellars P.A. (eds), 1993, The Origin of
Modern Humans and the Impact of Chronometric Dating (Princeton University
Press, Princeton, NJ).

433. Aitken M.J., 1999, Rep. Progr. Phys.\textbf{\ 62}, 1333.

434. McKeever S.W.S., 1985, Thermoluminescence of Solids (Cambridge
University Press, Cambridge).

435. Groh J. and Hevesey G.V., 1920, Ann. Phys. \textbf{65}, 318.

436. Lindros M., Hass H., Pattyn H., 1992, Nucl. Instrum. Methods \textbf{B64%
}, 256.

437. Restle M., Quintel H., Ronning C., 1995, Appl. Phys. Lett. \textbf{66},
2733.

438. Amarel V.S., 1998, J. Magn. Mater. \textbf{177 - 81}, 511.

439. Lohmuller J., Hass H., Schatz G., 1996, Hyperfine Interact. \textbf{%
97/98}, 203.

440. Granzer H., Hass H., Schatz G., 1996, Phys. Rev. Lett. \textbf{77},
4261.

441. Bertschat H.H., Hass H., Kowallik R., 1997, Phys. Rev. Lett. \textbf{78}%
, 342.

442. Tarbeyev Yu.V., Kaliteyevsky A.K., Sergeyev V.I., 1994, Metrologia 
\textbf{31}, 269.

443. Kuegens U. and Becker P., 1998, Meas. Sci. Technol. \textbf{9}, 1072.

444. Bulanov A.D., Devyatych G.G., Gusev A.V., 2000, Cryst. Res. Technol. 
\textbf{35}, 9.

445. Becker P., 2001, Rep. Prog. Phys. \textbf{64}, 1945.

446. Becker P., 2001, Metrologia \textbf{38}, 85.

447. Becker P. and Gl\"{a}ser M., 2003, Meas. Sci. technol.\textbf{\ 14},
1249.

448. Becker P., 2003, Metrologia \textbf{40}, 366.

449. Mills I.M., Mohr P.J., Taylor B.N., 2005, Metrologia \textbf{42}, 71.

450. Picard A., 2006, Metrologia\textbf{\ 43}, 46.

451. Becker P., Godisov O.N., Taylor P., 2006, Meas. Sci. Technol. \textbf{17%
}, 1854.

452. Mohr P.J., Taylor B.N., 2000, Rev. Mod. Phys.\textbf{\ 72}, 351.

\bigskip

\textbf{Figure captions.}

\bigskip

Fig.1. Structure within the atom. If the protons and neutrons in this
picture were 10 cm across, then the quarks and electrons would be less than
0.1 mm in size and the entire atom would be about 10 km across (after
http://www.lbl.gov/abc/wallchart/).

Fig. 2. Atomic nomenclature.

Fig. 3. Comparison between charge ($\rho _{ch}$) and magnetization ($\rho
_{m}$) for the proton (a) and neutron (b). Both densities are normalized to $%
\int $drr$^{2}\rho $(r) = 1 (after [57]).

Fig. 4. A mass-spectrum analysis of krypton. The ordinates for the peaks at
mass positions 78 and 80 should be divided by 10 to show these peaks in
their true relation to the others (after [8]).

Fig. 5. The binding energy per nucleon B/A as a function of the nuclear mass
number A (after [12]).

Fig. 6. The radial charge distribution of several nuclei determined from
electron scattering. The skin thickness value t is roughly constant at 2.3
fm. The central density changes vey little from the lightest nuclei to
theheaviest (afer[62]).

Fig. 7. Coulomb potential used for defining the nuclear radius R.

Fig. 8. The spherical square-well potential, adjusted to describe correctly
the binding enrgy E$_{B}$ of the deuteron. The full depth is also given and
amounts to V$_{0}$ = U = 38.5 MeV (after [12]).

Fig. 9. The deuteron wave function for R = 2.1 fm (after [8]).

Fig. 10 The potential fot deutron triplet states with even L. the distance
is in units of deuteron radius R = 4.31 fm (after [70]).

Fig. 11. Schematic diagram of a scattering arrangement. The scattering angle 
$\theta $ is between wave vector $\overrightarrow{k}$, along the direction
of the projectile, and $\overrightarrow{k}$', that of the scattered
particle. The result is independent of the azimuthal angle $\Phi $ unless
the orientation of the spin of one of the particles involve d is known
(details see in text).

Fig. 12. The effect of a scattering potential is to shift the phase of the
scattered wave at points beyond the scattering regions, where the wave
function is that of a free particle (after [9]).

Fig. 13. Real part of NN - scattering phase shifts in degrees for low-order
partial waves [74]: a - \textit{pp} - scattering with contribution from the
Coulomb potential removed; b - isovector \textit{np} - scattering, and c -
isoscalar \textit{np} scattering. Filled circles in the $^{1}$S$_{0}$ and $%
^{3}$S$_{1}$ phase shifts of np - scattering are the calculated results
using a Paris potential (after [75]).

Fig. 14. Very small changes in the NN wave function near r = R can lead to
substabtial differences in the scattering length when the extrapolation is
made (after [13]).

Fig. 15. Building blocks of matter, \ fermions have three quarks qqq and $\ $%
antiquarks $\overline{q}q\overline{q}$ as well as bosons are quarks and
antiquarks.

Fig. 16. Comparison of a collection in hadronic or nuclear matter phase and
within quark - gluon plasma description (after [91]).

Fig. 17. Neutron decay via weak interaction into proton (a) and Feyman
diagram of this process (b).

Fig. 18. Typical spectra showing the muonic 2p - 1s x-ray doublet for three
isotopes of Fe. The two peaks show the 2p$_{3/2}$ $\rightarrow $ 1s$_{1/2}$
and 2p$_{1/2}$ $\rightarrow $ 1s$_{1/2}$ transitions in the ratio 2 : 1
determined by the the statistical weight (2j + 1) of the initial initial
state. The isotope shift can clearly be seen as the change in energy of the
transitions (after [102]).

Fig. 19. Balmer series in the hydrogen emission spectrum. The convergence of
the lines to the series limit H$_{\infty }$ is clearly seen.

Fig. 20. $\beta $ lines of the Balmer series in a mixture of equal parts
hydrogen and deterium. One sees the isotope effect, which is explained by
motion of the nucleus (see text).

Fig. 21. Experimental 1s - 2s spectra in hydrogen and deuterium as measured
relatively to selected modes of the optical comb generator (after [132]).

Fig. 22. Alternating measurements of the 1s - 2s transition frequency in
hydrogen and deuterium (after [132]).

Fig. 23. The potential energy curve of typical diatomic molecule (solid
line) as a function of internuclear distance. The broken line and dotted
curves are those of the quadratic and cubic functions that give the best
approximation to the curve at the minimum.

Fig. 24. Idealized model of Rayleigh scattering and Stokes and anti - Stokes
Raman scattering. Here $\nu _{v}$ is vibration frequency of molecule.

Fig. 25. Spring and ball model - three modes of vibration for H$_{2}$O.

Fig. 26. IR (1) and Raman spectra (2) of bensoic acid.

Fig. 27. IR spectra of ordinary and heavy water. Weak bands in the spectrum
of 99.5\% D$_{2}$O, which coincide with characteristic HDO frequencies shown
by arrows, are indicative of the sensitivity of the IR spectrum to
impurities (after [155]).

Fig. 28. a - unpolarized Raman spectrum in the frequency region of the
pentagonal - pinch mode, for a frozen sample of nonisotopically enriched C$%
_{60}$ in CS$_{2}$ at 30 K. The points are the experimental data, and the
solid curve is a three - Lorentzian fit. The highest - frequency peak is
assigned to the totally symmetric pentagonal - pinch A$_{g}$ mode in
isotopically pure $^{12}$C$_{60}$. The other two peaks are assigned to the
perturbed pentagonal - pinch mode in molecules having one and two $^{13}$C -
enriched C$_{60}$, respectively. The insert shows the evolution of these
peaks as the solution is heated. b - the points give the measured
unpolarized raman spectrum in the pentagonal - pinch region for a frozen
solution of $^{13}$C - enriched C$_{60}$ in CS$_{2}$ at 30 K. The solid line
is a theoretical spectrum computed using the sample's mass spectrum, as
described in the text (after [161]).

Fig. 29. Pressure dependence of the rate constant for the reaction O + O$%
_{2} $ $\longrightarrow $ O$_{3}$ (after [171]). \ 

Fig. 30. a - First - order Raman scattering spectra Ge with different
isotope contents [27] and b - First - order Raman scattering in isotopically
mixed diamond crystals $^{12}$C$_{x}^{13}$C$_{1-x}$. The peaks A, B, C, D, E
and F correspond to x = 0.989; 0.90; 0.60; 0.50; 0.30 and 0.001 [204].

Fig. 31. Second - order Raman scattering spectra in synthetic diamond with
different isotope concentration at room temperature [207].

Fig. 32. Second - order Raman scattering spectra in the isotopically mixed
crystals LiH$_{x}$D$_{1-x}$ at room temperature. 1 - x = 0; 2 - 0.42; 3 -
0.76; 4 - 1. The arrows point out a shift of LO($\Gamma $) phonons in the
mixed crystals [218].

Fig. 33. The dependence of ln($\delta $\%) $\sim $ f[ln($\frac{\partial
\omega }{\partial \text{M}}$)]: points are experimental values and
continuous line - calculation on the formuls (113) [218].

Fig. 34. Mirror reflection spectra of crystals: 1 - LiH; 2 - LiH$_{x}$D$%
_{1-x}$; 3 - LiD; at 4.2 K. 4 - source of light without crystal. Spectral
resolution of the instrument is indicated on the diagram (after [221]).

Fig. 35. a - Reflection spectra in the A and B excitonic polaritons region
of Cd$^{nat}$S and Cd$^{34}$S at 1.3K with incident light in the $%
\overrightarrow{\text{E}}$ $\perp $ $\overrightarrow{C}$. The broken
vertical lines connecting peaks indicate measured enrgy shifts reported in
Table 18. In this polarization, the n = 2 and 3 excited \ states of the A
exciton, and the n = 2 excited state of the B exciton, can be observed. b -
Polarized photoluminescence spectra in the region of the A$_{\text{n = 2}}$
and A$_{\text{n = 3}}$ free exciton recombination lines of Cd$^{nat}$S and Cd%
$^{34}$S taken at 1.3K with the $\overrightarrow{\text{E}}$ $\perp $ $%
\overrightarrow{C}$. The broken vertical lines connecting peaks indicate
measured enrgy shifts reported in Table 18 (after [230]).

Fig. 36. a -Signatures of the E$_{\text{0}}$' and E$_{1}$ excitonic band
gaps of $^{28}$Si observed (dots) in photomodulated reflectivity. The solid
line is a theoretical fit using the excitonic line shape. b - Photomodulated
reflectivity spectra of isotopically enriched Si exhibiting isotopic shifts
of the E$_{\text{0}}$' and E$_{1}$ gaps (after [235]).

Fig. 37. A - Photoluminescence (PL) and wavelength - modulated transmission
(WMT) spectra of isotopically enriched $^{30}$Si recorded at 20K ; B - The
excitonic indirect band gap and the associated phonon energies as a function
of M (after [235]).

Fig. 38. Thermal conductivity of natural abundance (1.1\% of $^{13}$C)
diamons (lower squares), isotopically enriched (0.1\% $^{13}$C) diamond
(upper squares), together with the low - temperaturw data of Slack (see
[254]) (circles) and high - temperature data of Vandersande et al. (see
Ref.[253]). \ The solid curves are the result of fitting the Callaway theory
[244] to the data, using tha same set fitting parameters. The inset shows
the calculated thermal conductivity corresponding to 1\%, 0.1\% and 0.001\% $%
^{13}$C concentration according to Callaway theory (after [251]).

Fig. 39. Thermal conductivity of the highly isotopically enriched $^{28}$Si
sample SI284 (filled circles) and the natural Si reference SINI (open
circles). The filled and open triangles are other measurements for highly
isotopically \ enriched $^{28}$Si and natural Si, respectively (from Ref.
[241]); "plus" symbols denote the "standard" curve for natural Si (from
[241]). The thin solid and dashed lines are the theoretical results of [257]
for $^{28}$Si and natural Si, respectively. The thick solid line has been
calculated with the same theory using the actual mass variance g$_{2}$ of
sample SI284 (see, also Table 19) (after [259]).

Fig. 40. Raman shift of isotopically disordered diamond. The open symbols
represent experimental values [204, 264, 270, 271]. The dashed line
indicates the harmonic scaling of the phonon frequency within the VCA ($%
\omega \sim $ $\overline{\text{m}}^{\text{-1/2}}$). The solid line
corresponds to a fit with Eq. (2.40) for n = 2, 3 of Ref. [26] to all
experimental data, added to the VCA scaling (after [269]).

Fig. 41. Disorder-induced shift of the Raman phonon of diamond as a function
of the $^{\text{13}}$C concentration. The open symbols are Raman
experimental data , whereas the asterisks correspond to ab initio
calculations [27, 273]. The solid line is a fit with Eq. (2.40) for n = 2, 3
of Ref. [26] to all experimental data. The dotted and dot-dashed lines
represent the fits to theoretical values obtained from ab initio and CPA
calculations, respectively (after [269]).

Fig. 42. Disorder-induced broadening of the Raman phonon of diamond as a
function of the $^{\text{13}}$C concentration. The filled circles have been
obtained from the Raman data by taking into account the corresponding
instrumental resolutions and substracting the anharmonic broadening $\Gamma
_{\text{anh}}$ $\approx $ 2 cm$^{\text{-1}}$ (FWHM). The solid line is a fit
with Eq. (2.40) for n = 2, 3 \ of Ref. [26] to these points. The dotted and
dot-dashed lines are the corresponding fits to the values obtained from ab
initio [272, 273] and CPA [207, 264] calculations, respectively (after
[269]).

Fig. 43. Disorder-induced shift of the Raman phonon of Si as a function of
the $^{\text{30}}$Si concentration [272]. The solid line is a fit with Eq.
(2.40) for n = 2, 3 of Ref. [26] to the experimental data. The dot-dashed
line represents the corresponding fit to the values obtained from CPA
calculations (after [269]).

Fig. 44. Disorder-induced broadening of the Raman phonon of Si as a function
of the $^{\text{30}}$Si concentration [272]. The solid line is a fit with
Eq. (2.40) for n = 2, 3 of Ref. [26] to the experimental data. The
dot-dashed line represents the corresponding fit to the values obtained from
CPA calculations (after [269]).

Fig. 45. Disorder - induced broadening of the Raman modes of the 6H - SiC
polytype versus the $^{\text{13}}$C concentration of the carbon sublattice.
The data are taken from [275]. the solid lines represent fits with Eq.
(2.40) for n = 2, 3 of Ref. [26] to the data points that correspond to the
TO(0), TO(2/6) and TO(6/6) phonon modes (after [269]).

Fig. 46. Dependence of the half - width of the line of 2LO($\Gamma $)
phonons in Raman scattering spectra of (2) the pure crystal LiH and \ (3) LiH%
$_{x}$D$_{1-x}$. Curve (1) shows the profile of the line of exciting light
(after [205]).

Fig. 47. Energy of band - to - band transitions E$_{g}$ as function of
isotope concentration in mixed crystals LiH$_{x}$D$_{1-x}$ at 2 K: 1 -
linear dependence \ of E$_{g}$ on x in virtual crystal model; 2 -
calculation according to Eq. (121), points derived from reflection spectra
indicated by croses, and those from luminescence spectra by triangles (after
[221]).

Fig. 48. Concentration dependence of half - width of the lineof ground state
of exciton in mirror reflection spectra at 2K: 1 - approximation of virtual
crystal model; 2 - calculation according to Eq. (122); experimental points
indicated by croses (after [221]).

Fig. 49. Concentration dependence of binding energy of Wanier - Mott
exciton; 1 - approximation \ of vitual crystal model; 2 - calculation
according to Eq. (127): experimental points indicated by triangles (after
[277]).

Fig. 50. Temperature dependence of the indirect gap of silicon. The points
are experimental, the solid curve represents a single Einstein oscillator
fit to the experimental points. The dashed line represents the asymptotic
behavior at high temperature: its intersept with the vertical axis allows us
to estimate the bare gap and thus the zero - point renormalization due toe
electron - phonon interaction (after [302]).

Fig. 51. Temperature dependence of the indirect gap measured for Ge (dots).
The solid line (through the points and at higher temperature) represents a
fit a single oscillator. The thin line below 200 K represents the linear
extrapolation of the single - oscillator fit to T = 0K, used to determine
the zero - point renormalizaton \ of - 53 meV (after [27]).

Fig. 52. Energy of the indirect exciton of dimond versus temperature. The
points are experimental. The extrapolation of the linear behavior of the
fitted curve at high T as a straight line to T = 0K leads to a zero - point
gap renormalization of - 370 meV, much larger than the accuracy of 100 meV
often claimed for full - blown ab initio calculatuions (after [27]).

Fig. 53. Zero - point energy renormalization of exciton state in LiH
crystals (after [303]).

Fig. 54. Temperature shift of location of maximum of 1s exciton state in the
reflection spectrum of crystals, 1 : LiH; 2 : LiH$_{0.25}$D$_{0.75}$; 3 :
LiD. Experimental values are shown by points and the calculated values by
full curves (after [221]).

Fig. 55. Schematic of the isotope heterostructure used by Fuchs et al.
(after [317]).

Fig. 56. a - Experimental depth of the atomic fraction of $^{70}$Ge, $^{72}$%
Ge, $^{73}$Ge, $^{74}$Ge and $^{76}$Ge of a diffusion - annealed sample at
586$^{0}$C for 55.55h. The solid line is a theoretical fit of $^{70}$Ge; b -
experimental depth profiles of the same samples as in a) but before
annealing (after [317]).

Fig. 57. SIMS depth profiles of $^{69}$Ga and $^{71}$Ga in GaP isotope
epilayers annealed at 1111$^{0}$C for 231 min. The filled circles represent
the calculated $^{69}$Ga concentration profile (after [322]).

Fig. 58. SIMS depth profiles of AL(o), Ga($\square )$ and $^{71}$Ga(+) in
the as - grown Al$^{71}$Ga/Al$^{69}$Ga/$^{71}$GaAs heterostructure (b) [see
(a)] and after annealing at 1050$^{0}$C for 1800 s [see(b)]. Solid lines in
(a) connect the data to guide the eye. Solid lines in (b) show best fits to
the experimental profiles (after [324]).

Fig. 59. SIMS concentration profiles of $^{31}$P and $^{30}$Si after
annealing of the P - implanted Si isotope structure at temperatures and
times indicated in the figures.The solid line in (a) - (f) represent
theoretical best fit. The lower dashed lines show the corresponding super -
and undersaturation of self - interstitials and vacancies, respectively. The
upper dashed line in (d) is the Si profile that is expected in the case Si
diffusion proceeds under intrinsic and thermal - equilibrium conditions
(after [314]).

Fig. 60. The dependence of the phosphorus atoms concentration on the neutron
irradiation doze of Si crystals and followed annealing at 800$^{0}$C during
1 h. The dependence was measured by Hall effect (after \ [328]).

Fig. 61. Spreading resistance measurements of a thermal neutron irradiation
doped silicon slice. Step - length on scan 1 and 2 is 250 $\mu $m and on
scan 3 step - length is 50 $\mu $m. Starting material has been selected
greater than 1500 $\Omega $cm n - type (after [340]).

Fig. 62. Typical lateral microscopic resistivity distributions in
conventionally doped silicon and in silcon doped by neutron irradiation
(after [333]).

Fig. 63. Photoluminescence spectra for p - type NTD samples; displayed with
intensity of luminescence due to .07 eV acceptor held constant forthe four
spectra (after [351]).

Fig. 64. Relative photoluminescence spectra for four n - type NTD samples.
The four spectra are not normalized with respect to each other (after [351]).

Fig. 65. Reflection at a planar interface between unbounded regions of
refractive indices n$_{co}$ and n$_{cl}$ \TEXTsymbol{<} n$_{co}$ showing (a)
total internal reflection and (b) partial reflection and refraction (after
[30]).

Fig. 66. Isotopic fiber in which the core and cladding are both pure SiO$%
_{2} $, but with a different isotopic composition (after [354]).

Fig. 67. Light guidance in an optical waveguide by total \ internal
reflection (after [30]).

Fig. 68. Photoluminescence spectra of the indirect - free exciton in the
type - IIa diamond under the 215 nmlaser excitation with a power density $%
\approx $ 100kWcm$^{-2}$ at a temperature of (a) 85K, (b) 125K and (c) 300K.
The insert is the enlarged spectrum near the B$_{3}$ band of (b0 (after
[366]).

Fig. 69. Logarithmic plots of the time - decay spectra of the B$_{1}$
emission in the type - IIa diamond under the 215 nm laser excitation with
power a density of $\approx $ 100kWcm$^{-2}$ at temperatures of 85K ($%
\blacktriangle $); 125K ($\bullet $) and 300K ($\circ $) (after [366]).

Fig. 70. Experimental (closed circles) and theoretical (solid line)
variation of the indirect - free exciton lifetimes with temperature in the
type IIa diamond (after [366]).

Fig. 71. Schematic representation of pump relaxation and laser transition
for (1) a three level system and (2) a four level system (after [28]).

Fig. 72. Photoluminescence spectra of free excitons at 4.2K: (1) LiH; (2) LiH%
$_{x}$D$_{1-x}$ and (3) LiD crystals (after [221]).

Fig. 73. Dependence of the line shape of the 2LO replica at 4.2K on the
excitation light intensity I$_{0}$: (1) 0.03; (2) 0.09; (3) 0.40 and (4) 1
(after [370]).

Fig. 74. Bloch sphere representation of a qubit (after [393$^{a}$]).

Fig. 75. Illustration of two cells in one - dimensional array containing $%
^{31}$P donors and electrons in a Si host, separated by a berrier from metal
gates on the surface. "A gates" control the resonance frequency of the
nuclear spin qubits, "J gates" control the electron - mediated coupling
between adjacent nuclear spins. The ledge over which the gates cross
localizes the gate electric field in the vicinity of the donors (after
[403]).

Fig. 76. An electric field applied to an A gatepulls the electron wave
function away from the donor and towards the barrier, reducing the hyperfine
interaction and the resonance frequency of the nucleus. The donor nucleus -
electron system is a voltage - controlled oscillator with a tuning parameter 
$\alpha $ of the order 30 MHz (after [403]).

Fig. 77. Scheme of fabrication of a striped Si layer with a modulated
concentration of nuclear spin isotopes (after \ [408]).

Fig. 78. Schematics of the proposed device. After NTD, $^{31}$P donors
appear only inside the $^{30}$Si - spots and underlying $^{74}$ - strips
will be heavily with $^{75}$As donors. All sizes are shown in nm (after
[409]).

Fig. 79. Schematics of a $^{28}$Si nanowire L with an array of $^{30}$Si -
spots (qubits and non - qubits after NTD). Each spot is supplied by
overlying A - gate, underlying SD - channel and lateral N - gate. This
device architecture allows to realize an indirect coupling between any
distant qubits (after [409]).

Fig. 80. Present picture (PET) shows brain activity of a patient with
Alzheimer's disease (after [422]).

Fig. 81. Calibrated radiocarbon timescale based on Irish oak. The straight
line indicates the ideal radiocarbon/calender age timescale (after [426]).

Fig. 82. Part of the Turin Shroud, bearing the image of a man's head.
Radiocarbon AMS dating has given a calibrated age range for the cloth of a
1260 - 1390 years of present time (see, however, text).

Fig. 83. Energy - level representation of the thermoluminescence. i -
ionization due to exposure of the crystal to the flux of nuclear radiation,
with trapping of electrons and holes at defects, T and L, respectively. ii -
Storage during antiquity; the lifetime \ of the electrons in the traps needs
to be much longer than the age span of the sample in order that leakage be
negligible. The lifetime is determined by the depth E of the trap below \
the conduction band, and for dating purposes we are interested in those deep
enough ($\sim $ 1.5 eV) for the lifetime to be the order of a million years
or more. iii - To observe thermoluminescence the sample is heated and there
is a certain temperature at which the thermal vibration of the crystal
lattice causes eviction Some of these evicted electrons reach luminescence
centres, and if so, light is emitted in the process of recombining at those
centres. Alternatively, the electron may recombine at a non - luminescence
centre (a 'killer' centre) or be captured by a deeper trap ( after [430]).

Fig. 84. The event dating, whether in thermoluminescence dating or in
optical dating, is the setting to zero, or near zero, of the latent
luminescence acquired at some time in the past. \ With sediment this zeroing
occurs through exposure to daylight ('bleaching') during erosion, transport
and deposition, whereas with fired materials, it is through heating.
Subsequently the latent signal builds up again through exposure to the weak
natural flux of nuclear radiation. For OSL the dating signal is obtained by
exposure of the grains from the sample to a beam of light; for TL it is
obtained by heating (after [433]).

Fig. 85. Natural glow - curve for burned flint from the lower Paleolitrhic
site at Belvedere, Holland (after [430]).

Fig. 86. Radioactive isotopes used for solid state physics experiments
(after [32]).

Fig. 87. Photoluminescence spectra of undoped and $^{111}$In doped GaAs
successively taken 4h; 7h; 12h; 22h; 2d; 4d and 9d after doping. all spectra
are normalized to the intensity of the (e, C) peak. In the inset, the height
I$_{Cd}$/I$_{C}$ of the (e,C) peak in these spectra is shown a function of
time after doping with $^{111}$In. The solid line is a theoretical fit (see
text) (after [347]).

Fig. 88. Raman spectra at room temperature taken for the various annealing
temperatures of (100) oriented NTD GaAs irradiated withneutron dozes. The
coupling L$_{+}$ mode is observed at annealing temperature above 600$^{0}$C
(after [348]).

Fig. 89. Infrared - absorption spectra at room temperature taken for the
various annealing temperatures of the NTD GaAs used for the Raman scattering
experiments (after [348]).

Fig. 90.Photoluminescence (PL) spectra taken at 15K for unirradiated and NTD
- GaP. PL peaks 1, 2 and 3 in unirradiated GaP represent S$_{p}$ - C$_{p} $
DA pair recombination, its LO - phonon replica, and 2LO - phonon replica,
respectively. 1.65 and 1.87 eV emissions in NTD - GaP are attributed to Ge$%
_{Ga}$ - Ge$_{P}$ complex and S$_{p}$ - C$_{p}$ DA pair recombination,
respectively (after [350]).

Fig. 91. Variation of the half - width W with the square root of the
temperature T for the 1.65 eV in NTD - GaP. solid line is a theoretical fitr
with h$\nu $ 0.025 eV (after [350]).

Fig. 92. Molar mass versus density of Si single crystal samples. The result
for the molar volume derived from the linear data is M$_{\func{Si}}$/$\rho $
= 12.0588207 (54) cm$^{3}$mol$^{-1}$ (after [446]).

\bigskip

\bigskip Table 1. Fundamental interactions

\begin{tabular}{llllllll}
Interaction & FQ & Mass & Range (m) & RS & Spin & T C-S (m$^{2}$) & TTS (s)
\\ 
Strong & Gluon & 0 & 10$^{-15}$ & 1 & 1 & 10$^{-30}$ & 10$^{-23}$ \\ 
Weak & W$^{\mp }$; Z & 81; 93 GeV/c$^{2}$ & 10$^{-18}$ & 10$^{-5}$ & 1;1 & 10%
$^{-44}$ & 10$^{-8}$ \\ 
Electromagnetic & Photon & 0 & $\infty $ & $\alpha $ = 1/137 & 1 & 10$^{-33}$
& 10$^{-20}$ \\ 
Gravity & Graviton & 0 & $\infty $ & 10$^{-30}$ & 2 & - & -%
\end{tabular}

\bigskip Here - FQ field quant; RS relative strength; TC-S - typical cross -
section; TTS - typical time scale.

\ \ \ \ \ \ \ \ \ \ \ Table. 2. The basic properties of the atomic
constituents.

\begin{tabular}{lllll}
Particle & Charge & Mass (u) & Spin ($\hslash $) & Magnetic Moment (JT$^{-1}$%
) \\ 
Proton & e & 1.007276 & 1/2 & 1.411$\cdot $10$^{-26}$ \\ 
Neutron & 0 & 1.008665 & 1/2 & - 9.66$\cdot $10$^{-27}$ \\ 
Electron & - e & 0.000549 & 1/2 & 9.28$\cdot $10$^{-24}$%
\end{tabular}

Table 3. Sample values of nuclear magnetic dipole moments (after [58]).

\begin{tabular}{ll}
Nuclide & $\mu $($\mu _{N}$) \\ 
n & - 1.9130418 \\ 
p & + 2.7928456 \\ 
$^{2}$H(D) & + 0.8574376 \\ 
$^{17}$O & - 1.89379 \\ 
$^{57}$Fe & + 0.09062293 \\ 
$^{57}$Co & + 4.733 \\ 
$^{93}$Nb & + 6.1705%
\end{tabular}

Table 4. Some values of nuclear electric quadrupoe moments (after [58]).

\begin{tabular}{ll}
$^{2}$H(D) & + 0.00288 \\ 
$^{17}$O & - 0.02578 \\ 
$^{59}$Co & + 0.40 \\ 
$^{63}$Cu & - 0.209 \\ 
$^{133}$Cs & - 0.003 \\ 
$^{161}$Dy & + 2.4 \\ 
$^{176}$Lu & + 8.0 \\ 
$^{209}$Bi & - 0.37%
\end{tabular}

Table 5. Masses of electron, nucleons and some nuclei (after [56]).

\begin{tabular}{llll}
Particle & Number of Protons & Number of Neutrons & Mass (MeV) \\ 
e & 0 & 0 & 0.511 \\ 
p & 1 & 0 & 938.2796 \\ 
n & 0 & 1 & 939.5731 \\ 
$_{1}^{2}$H & 1 & 1 & 1876.14 \\ 
$_{1}^{3}$H & 1 & 2 & 2808.920 \\ 
$_{2}^{3}$He & 2 & 1 & 2808.391 \\ 
$_{2}^{4}$He & 2 & 2 & 3728.44 \\ 
$_{3}^{7}$Li & 3 & 4 & 6533.832 \\ 
$_{4}^{9}$Be & 4 & 5 & 8392.748 \\ 
$_{6}^{12}$C & 6 & 6 & 11174.860 \\ 
$_{8}^{16}$O & 8 & 8 & 14895.077 \\ 
$_{92}^{238}$U & 92 & 146 & 221695.831%
\end{tabular}

Table 6. Table of main families of particles.

\begin{tabular}{lll}
Family & Particle & Fundamental \\ 
Lepton & electron & yes \\ 
Lepton & neutrino & yes \\ 
Hadron & proton & no \\ 
Hadron & neutron & no \\ 
Hadron & delta & no \\ 
Hadron & sigma & no \\ 
Hadron & many & more%
\end{tabular}

Table 7. Properties of nucleons and few - nucleonssystems. 
\begin{tabular}{llllllll}
Particle & Symbol & Spin & Parity & BE(MeV) & MM ($\mu _{0}$) & QM(fm$^{2}$)
& RMS CR(fm) \\ 
Proton & p & 1/2 & + &  & 2.79284739$\pm $6$\cdot $10$^{-8}$ &  & 0.88 \\ 
Neutron & n & 1/2 & + &  & -1.9130428$\pm $5$\cdot $10$^{-7}$ &  &  \\ 
Deuteron & $^{2}$H & 1 & + & 2.2246 & 0.8574376$\pm $4$\cdot $10$^{-7}$ & 
0.288$\pm $10$^{-3}$ & 1.963 \\ 
Triton & $^{3}$H & 1/2 & + & 8.482 & 2.978960$\pm $10$^{-6}$ &  & 1.63$\pm $%
0.03 \\ 
Helion & $^{3}$He & 1/2 & + & 7.718 & -2.127624$\pm $1.12$\cdot $10$^{-6}$ & 
& 1.97$\pm $0.0015 \\ 
Alpha & $^{4}$He & 0 & + & 28.28 &  &  & 1.671$\pm $0.014%
\end{tabular}

Here - BE - binding energy; MM - magnetic moment; QM - quadrupole moment;
RMS CR - RMS charge radius.

Table 8. Nucleon - nucleon scattering length (a) \ and effective range (r$%
_{e}$)

\begin{tabular}{lll}
& s = 0; T = 1 (fm) & s = 1; T = 0 (fm) \\ 
pp: \ a & -17.1$\pm $0.2 &  \\ 
pp: \ r$_{e}$ & 2.794$\pm $0.015 &  \\ 
nn: \ a & -16.6$\pm $0.6 &  \\ 
nn: \ r$_{e}$ & 2.84$\pm $0.03 &  \\ 
np: \ a & -23.715$\pm $0.15 & 5.423$\pm $0.005 \\ 
np: \ r$_{e}$ & 2.73$\pm 0.03$ & 1.73$\pm $0.02%
\end{tabular}

Table 9. Characteristics of the quarks.

\begin{tabular}{lll}
Flawor & Electric Charge (e) & Mass (GeV/c$^{2}$) \\ 
\textbf{u} \ - \ up & + 2/3 & 0.004 \\ 
\textbf{d} \ - down & - 1/3 & 0.008 \\ 
\textbf{c} \ - \ charm & + 2/3 & 1.5 \\ 
\textbf{s} \ - strange & - 1/3 & 0.15 \\ 
\textbf{t} \ - \ top & + 2/3 & 176 \\ 
\textbf{b} \ - beaty (bottom) & - 1/3 & 4.7%
\end{tabular}

Table 10. Characteristics of the leptons.

\begin{tabular}{llll}
Flawor & Electric charge (e) & Spin & Mass (Gev/c$^{2}$) \\ 
$\nu _{e}$ - electron neutrino & 0 & 1/2 & \TEXTsymbol{<}7$\cdot 10^{-9}$ \\ 
e$^{-}$ - electron & -1 & 1/2 & 0.000511 \\ 
$\nu _{\mu }$ - muon neutrino & 0 & 1/2 & \TEXTsymbol{<} 0.00027 \\ 
$\mu ^{-}$ - muno (mu - minus) & -1 & 1/2 & 0.106 \\ 
$\nu _{\tau }$ - tau neutrino & 0 & 1/2 & \TEXTsymbol{<}0.03 \\ 
$\tau ^{-}$ - tau (tau - minus) & -1 & 1/2 & 1.771%
\end{tabular}

Table 11. The root - mean - square radius (\TEXTsymbol{<}r$^{2}$\TEXTsymbol{>%
}$^{1/2}$) for different calcium isotopes.

\begin{tabular}{lllll}
Nucleus & $^{40}$Ca & $^{42}$Ca & $^{44}$Ca & $^{48}$Ca \\ 
\TEXTsymbol{<}r$^{2}$\TEXTsymbol{>}$^{1/2}$ (fm) & 3.4869 & 3.5166 & 3.5149
& 3.4762%
\end{tabular}

Table 12. The wavelength of some lines of the various spectral series in
hydrogen. The series with m = 3 was observed in 1924 by Pfund; it begins
with a line of $\lambda $ = 74000\AA , but is not shown in the table.

\begin{tabular}{lllll}
n$\smallsetminus $m & Lyman & Balmer & Paschen & Bracket \\ 
2 & 1216 \AA  &  &  &  \\ 
3 & 1026\AA  & 6563\AA  &  &  \\ 
4 & 973\AA  & 4861\AA  & 18751\AA  &  \\ 
5 & 950\AA  & 4340\AA  & 12818\AA  & 40500\AA  \\ 
Year of discovery & 1906 & 1885 & 1908 & 1922%
\end{tabular}%
\newline
Table 13.Energy corrections for motion of the nucleus for the Rydberg
numbers of several one - electron atoms.

\begin{tabular}{llllll}
Atom & H($^{1}$H) & D($^{2}$H) & T($^{3}$H) & He$^{+}$ & Li$^{2+}$ \\ 
A & 1 & 2 & 3 & 4 & 7 \\ 
-$\frac{\Delta \text{E}}{\text{E}}\cdot 10^{4}$ & 5.45 & 2.75 & 1.82 & 1.36
& 0.78%
\end{tabular}%
\newline

Table 14. The symmetry elements of the C$_{2V}$ point group (see, e.g.
[200]).

\begin{tabular}{lllllll}
C$_{2V}$ & E & C$_{2}$ & $\sigma _{V}$(xz) & $\sigma _{V}$(yz) &  &  \\ 
A$_{1}$ & 1 & 1 & 1 & 1 & z & x$^{2}$; y$^{2}$z$^{2}$ \\ 
A$_{2}$ & 1 & 1 & -1 & -1 & R$_{z}$ & xy \\ 
B$_{1}$ & 1 & -1 & 1 & -1 & x, R$_{y}$ & xz \\ 
B$_{2}$ & 1 & -1 & -1 & 1 & y,R$_{x}$ & yz%
\end{tabular}

Table 15. Main vibrations of water isotopologues.

\begin{tabular}{llll}
Gas & $\nu _{1\text{, }}$cm$^{-1}$ & $\nu _{2}$, cm$^{-1}$ & $\nu _{3}$, cm$%
^{-1}$ \\ 
H$_{2}^{16}$O & 3657.05 & 1594.75 & 3755.93 \\ 
H$_{2}^{17}$O & 3653.15 & 1591.32 & 3748.32 \\ 
H$_{2}^{18}$O & 3649.69 & 1588.26 & 3741.57 \\ 
HD$_{{}}^{16}$O & 2723.68 & 1403.48 & 3707.47 \\ 
D$_{2}^{16}$O & 2669.40 & 1173.38 & 2787.92 \\ 
T$_{2}^{16}$O & 2233.9 & 995.37 & 2366.61%
\end{tabular}

Table 16. Main vibrations of liquid ordinary and heavy water (cm$^{-1}$).

\begin{tabular}{lll}
Vibrations & H$_{2}$O & D$_{2}$O \\ 
combination of $\nu _{2}$ + libration & 2127.5 & 1555 \\ 
$\nu _{2}$ & 1643 & 1209.4 \\ 
$\nu _{1}$, $\nu _{2}$ and overtone of $\nu _{2}$ & 3404 & 2504%
\end{tabular}

Table 17. Values of the energy of maxima (in meV) in exciton reflection
spectra of pure and mixed crystals at 2K, and energies of exciton binding E$%
_{\text{b}}$, band-to-band transitions E$_{\text{g}}$ (after [221]).

\begin{tabular}{|cccccc|}
\hline
\multicolumn{1}{|c|}{Energy, meV} & \multicolumn{1}{c|}{LiH} & 
\multicolumn{1}{c|}{LiH$_{\text{0.82}}$D$_{\text{0.18}}$} & 
\multicolumn{1}{c|}{LiH$_{\text{0.40}}$D$_{\text{0.60}}$} & 
\multicolumn{1}{c|}{LiD} & $^{\text{6}}$LiH (78K) \\ \hline
\multicolumn{1}{|c|}{E$_{\text{1s}}$} & \multicolumn{1}{c|}{4950} & 
\multicolumn{1}{c|}{4967} & \multicolumn{1}{c|}{5003} & \multicolumn{1}{c|}{
5043} & 4939 \\ \hline
\multicolumn{1}{|c|}{E$_{\text{2s}}$} & \multicolumn{1}{c|}{4982} & 
\multicolumn{1}{c|}{5001} & \multicolumn{1}{c|}{5039} & \multicolumn{1}{c|}{
5082} & 4970 \\ \hline
\multicolumn{1}{|c|}{E$_{\text{b}}$} & \multicolumn{1}{c|}{42} & 
\multicolumn{1}{c|}{45} & \multicolumn{1}{c|}{48} & \multicolumn{1}{c|}{52}
& 41 \\ \hline
\multicolumn{1}{|c|}{E$_{\text{g}}$} & \multicolumn{1}{c|}{4992} & 
\multicolumn{1}{c|}{5012} & \multicolumn{1}{c|}{5051} & \multicolumn{1}{c|}{
5095} & 4980 \\ \hline
\end{tabular}

Table 18. The energy shifts of all of the transitions studied in [230] are
given in terms of the Cd$^{34}$S minus the Cd$^{nat}$S energy, $\Delta $E.

\begin{tabular}{lll}
Transition & Method & $\Delta $E (cm$^{-1}$) \\ 
I$_{2}$ & PL & 10.6$\pm $0.1 \\ 
I$_{2}^{z}$ & PL & 11.1$\pm $0.1 \\ 
I$_{2}^{a}$ & PL & 10.6$\pm $0.1 \\ 
A$_{n\text{ =1}}$ ($\Gamma _{6}$) & A$\parallel $ & 10.8$\pm $0.2 \\ 
A$_{n\text{ =1}}$ ($\Gamma _{5}^{L}$) & PL & 11.0$\pm 0.2$ \\ 
A$_{n\text{ =1}}$ ($\Gamma _{5}^{L}$) & R$\perp $ & 10.9$\pm 0.2$ \\ 
A$_{n=2}$ & PL$\parallel $ & 11.3$\pm $0.4 \\ 
A$_{n=2}$ & PL$\perp $ & 11.1$\pm 0.4$ \\ 
A$_{n=2}$ & R$\perp $ & 10.2$\pm $0.5 \\ 
A$_{n=3}$ & PL$\parallel $ & 11.8$\pm $1.1 \\ 
A$_{n=3}$ & PL$\perp $ & 10.9$\pm $0.6 \\ 
A$_{n=3}$ & R$\perp $ & 10.7$\pm $0.6 \\ 
B$_{n=1}$($\Gamma _{1}$) & R$\parallel $ & 10.9$\pm $0.3 \\ 
B$_{n=1}$($\Gamma _{5}^{L}$+$\Gamma _{5}^{T}$) & R$\perp $ & 10.6$\pm $0.4
\\ 
B$_{n=2}$ & R$\parallel $ & 9.4$\pm $1.2 \\ 
B$_{n=2}$ & R$\perp $ & 9.8$\pm 1.2$ \\ 
C$_{n=1}$($\Gamma _{1}$) & R$\parallel $ & 15$\pm $6 \\ 
C$_{n=1}$($\Gamma _{5}$) & R$\perp $ & 14$\pm $5%
\end{tabular}

The methods used were photoluminescence spectroscopy (*PL) and reflection
spectroscopy (R). For measurements made using polarized light, the $%
\parallel $ or $\perp $ specifies the orientation of the \textbf{E} vector
vs the c axis.

Table 19. Some very often used in everyday life RI (see text)

$%
\begin{array}{ccc}
\text{Isotope} & \text{Half - life} & \text{Applications} \\ 
\text{Actinium - 225} & \text{10.0 d} & 
\begin{array}{c}
\text{An alpha emitter that shows promise in the } \\ 
\text{treatment \ of certain types of cancer}%
\end{array}
\\ 
\text{Californium - 252} & \text{2.64 y} & 
\begin{array}{c}
\text{Used to treat cervical cancer melanoma, brain} \\ 
\text{cancer treatment}%
\end{array}
\\ 
\text{Cobalt - 60} & \text{5.27 y} & 
\begin{array}{c}
\text{A gamma emitter used in irradiations food and } \\ 
\text{medical equipment sterilization}%
\end{array}
\\ 
\text{Tungsten - 188} & \text{69.8 d} & 
\begin{array}{c}
\text{Used to prevent the re - closure (restenosis) of coronary} \\ 
\text{arteries folowing heart surgery}%
\end{array}
\\ 
\text{Copper - 67} & \text{61.9 h} & 
\begin{array}{c}
\text{Used to label monolocal antibodies and destroy } \\ 
\text{target tumors; PET scanning}%
\end{array}
\\ 
\text{Strontium - 82} & \text{25.6 d} & \text{PET scanning} \\ 
\text{Calcium - 42} & \text{Stable} & \text{Along with calcium - 44 used in
human calcium retention studies} \\ 
\text{Lithium - 6} & \text{Stable} & \text{Neutron capture terapy research}
\\ 
\text{Carbon - 11} & \text{20.3 m} & \text{Radiotracer in PET scans to study
normal/abnormal brain function} \\ 
\text{Germanium - 68} & \text{271 d} & \text{PET imaging} \\ 
\text{Bismuth -213} & \text{46 m} & \text{Used for targeted alpha therapy
(TAT) especiallycancers} \\ 
\text{Holmium - 213 } & \text{ 26 h} & \text{Being developed for diagnosis
and treatment of liver tumours} \\ 
\text{Iodine - 125 } & \text{60 d} & \text{ Used in cancer brachytherapy
(prostate and brain)} \\ 
\text{Iodine -131 } & \text{8d} & 
\begin{array}{c}
\text{Widely used in treating thyroid cancer. A strong gamma emitter, but }
\\ 
\text{used for beta therapy}%
\end{array}
\\ 
\text{Iridium - 192} & \text{74 d} & 
\begin{array}{c}
\text{Supplied in wire form for use as an internal radiotherapy source for}
\\ 
\text{cancer treatment (used then removed)}%
\end{array}
\\ 
\text{H - 3} & \text{12.3 y} & \text{Labeling PET imaging}%
\end{array}%
$

Table 20. Electron concentrations and the coupling modes of NTD GaAs (after
[348]).

\begin{tabular}{lllll}
Sample & EC (cm$^{-3}$) & LO - phonon frequency (cm$^{-1}$) & L$_{+}$ mode
(cm$^{-1}$) & PF (cm$^{-1}$) \\ 
unirradiated & 1 $\sim $ 2 $\times $ 10$^{7}$ & 296.6 &  &  \\ 
as - irradiated & a & 295.6 &  &  \\ 
500$^{0}$C annealed & a & 297.8 &  &  \\ 
600$^{0}$C annealed & 8.2 $\times $ 10$^{16}$ & 296.0 & 299 & 96.4 \\ 
650$^{0}$C annealed & 2.2 $\times $ 10$^{17}$ & 296.6 & 304 & 158 \\ 
700$^{0}$c annealed & 2.5 $\times $ 10$^{17}$ & 296.2 & 305 & 168%
\end{tabular}

a) Since the conduction is dominated by Mott-type hopping conduction (M.
Satoh and K. Kuriyama, Phys. Rev. B40, 3473 (1989), the electron
concentration can not be measured by the van der Pauw method.

EC = Electron Concentration;

PF = Plasma Frequency.

\end{document}